\documentclass[10pt]{book}
\usepackage{sectsty}
\usepackage[english]{babel}

\usepackage{graphicx}

\usepackage{hyperref}
\usepackage{amsmath,amssymb}
\usepackage{bm}
\usepackage{enumerate}
\usepackage{amsthm}
\usepackage[all,2cell]{xy}
\usepackage{mathrsfs}
\usepackage{mathtools}
\usepackage{tikz}
\usepackage{tikz-cd}

\usetikzlibrary{cd}

\usepackage[paperwidth=175mm, paperheight=250mm, left=23mm, right=23mm, top=25mm, bottom=25mm]{geometry}

\usepackage{marginnote}

\newskip\zzz \def\allowhyphens{\nobreak\hskip\zzz}     
{\catcode`p=12 \catcode`t=12 \gdef\removePT#1pt{#1}}   
\def\slantfactor#1{\expandafter\removePT\the\fontdimen1#1 }
\def\slantedraise#1#2{\dimen0=#1\dimen0=\slantfactor\font \dimen0
  \kern\dimen0\raise#1#2\kern-\dimen0}
\newdimen\middledotspace \newdimen\middledotraise
\def\middledot{\setbox0=\hbox{.}%
  \ifdim\fontdimen3\font=0pt \middledotspace=0pt
    \else  \middledotspace=0.333\wd0\fi
  \middledotraise=1ex  \advance\middledotraise by -\ht0
  \slantedraise{\middledotraise}{\hbox to\middledotspace{\hss.\hss}}}
\def\ll{\ifmmode{\mathchar"321C}\else
  l\allowhyphens\discretionary{-}{}{\middledot}\allowhyphens l\fi}
\def\LL{\ifmmode\else
  L\allowhyphens\discretionary{-}{}{\middledot}\allowhyphens L\fi}

\usepackage{imakeidx}
\makeindex[intoc]

\usepackage[nottoc]{tocbibind}

\newtheorem{thm}{Theorem}[section]
\newtheorem{dfn}[thm]{Definition}
\newtheorem{prop}[thm]{Proposition}

\newtheorem{lem}[thm]{Lemma}
\newtheorem{exmpl}[thm]{Example}
\newtheorem{rmrk}[thm]{Remark}

\newcommand\restr[2]{{
  \left.\kern-\nulldelimiterspace 
  #1 
  \right|_{#2} 
}}

\newcommand*{\transp}[2][-3mu]{\ensuremath{\mskip1mu\prescript{\smash{\mathrm t\mkern#1}}{}{\mathstrut#2}}}%

\newcommand{\R}{\mathbb{R}}
\renewcommand{\d}{\mathrm{d}}

\newcommand{\Cinfty}{\mathscr{C}^\infty}
\newcommand{\T}{\mathrm{T}}
\newcommand{\cT}{\mathrm{T}^\ast}
\newcommand{\Id}{\mathrm{Id}}

\newcommand{\Lie}{\mathscr{L}}

\newcommand{\X}{\mathfrak{X}}

\renewcommand{\L}{\mathcal{L}}
\newcommand{\F}{\mathcal{F}}
\renewcommand{\H}{\mathcal{H}}
\renewcommand{\P}{\mathcal{P}}

\newcommand{\W}{\mathcal{W}}
\newcommand{\Reeb}{\mathcal{R}}

\newcommand{\C}{\mathcal{C}}
\newcommand{\D}{\mathcal{D}}
\newcommand{\V}{\mathcal{V}}
\renewcommand{\O}{\mathcal{O}}
\renewcommand{\W}{\mathcal{W}}

\newcommand{\rmC}{\mathrm{C}}
\newcommand{\rmR}{\mathrm{R}}

\newcommand{\bfX}{\mathbf{X}}
\newcommand{\bfY}{\mathbf{Y}}
\newcommand{\bfZ}{\mathbf{Z}}

\newcommand{\parder}[2]{\frac{\partial #1}{\partial #2}}
\newcommand{\dparder}[2]{\dfrac{\partial #1}{\partial #2}}
\newcommand{\tparder}[2]{\partial #1/\partial #2}
\newcommand{\parderr}[3]{\frac{\partial^2 #1}{\partial #2\partial #3}}
\newcommand{\dparderr}[3]{\dfrac{\partial^2 #1}{\partial #2\partial #3}}

\let\graph\relax
\DeclareMathOperator{\graph}{graph}

\DeclareMathOperator{\Ima}{Im}
\DeclareMathOperator{\tr}{tr}

\DeclareMathOperator{\Hom}{Hom}
\DeclareMathOperator{\Diff}{Diff}
\DeclareMathOperator{\rk}{rank}
\DeclareMathOperator{\Div}{div}
\DeclareMathAlphabet{\mathpzc}{OT1}{pzc}{m}{it}
\def\d{\mathrm{d}}

\let\ds\displaystyle

\usepackage{fancyhdr}
\title{Geometrical aspects of contact mechanical systems and field theories}
\author{Xavier Rivas}
\date{\today}

\renewcommand\maketitle{
\thispagestyle{empty}
\noindent \rule{\textwidth}{1pt}
}

\usepackage{titlesec}

\titleformat{\chapter}[display]
  {\normalfont\huge\bfseries}{\chaptertitlename\ \thechapter}{20pt}{\Huge}
\titlespacing*{\chapter}{0pt}{30pt}{20pt}

\usepackage{setspace}

\begin{document}

\setstretch{1.1}

\frontmatter


\maketitle

\begin{center}
\sffamily
\large 

Ph.\,D.\,Thesis

\vspace{10mm}
{
\LARGE G\Large\uppercase{eometrical aspects of 
\\[1mm]
contact mechanical systems\\[2.5mm] and field theories}}

\vspace{10mm}
\large 
Author \\ 
\textbf{Xavier Rivas Guijarro}

\vspace{10mm}
Thesis Advisors 
\\ 
\textbf{Xavier Gràcia Sabaté\\Narciso Román Roy}

\vspace{10mm}
October 2021

\vspace{15mm} 
\normalsize
Doctoral Programme in Applied Mathematics

\vspace{3mm} 
Departament de Matemàtica Aplicada

\vspace{1.5mm} 
Facultat de Matemàtiques i Estadística

\vspace{1.5mm} 
Universitat Politècnica de Catalunya

\end{center}

\noindent \rule{\textwidth}{1pt}

\vfill

\chapter*{Abstract}\markboth{Abstract}{Abstract}


Many important theories in modern physics can be stated using the tools of differential geometry. It is well known that symplectic geometry is the natural framework to deal with autonomous Hamiltonian mechanics. This admits several generalizations for nonautonomous systems and classical field theories, both regular and singular. Some of these generalizations are the subject of the present dissertation.

In recent years there has been a growing interest in studying dissipative mechanical systems from a geometric perspective by using contact structures. In the present thesis we review what has been done in this topic and go deeper, studying symmetries and dissipated quantities of contact systems, and developing the Lagrangian--Hamiltonian mixed formalism (Skinner--Rusk formalism) for these systems.

With regard to classical field theory, we introduce the notion of $k$-precosymplectic manifold and use it to give a geometric description of singular nonautonomous field theories. We also devise a constraint algorithm for $k$-precosymplectic systems.

Furthermore, field theories with damping are described through a modification of the De Donder--Weyl Hamiltonian field theory. This is achieved by combining both contact geometry and $k$-symplectic structures, resulting in what we call the $k$-contact formalism. We also introduce two notions of dissipation laws, generalizing the concept of dissipated quantity. The preceding developments are also applied to Lagrangian field theory. The Skinner--Rusk formulation for $k$-contact systems is described in full detail and we show how to recover both the Lagrangian and Hamiltonian formalisms from it.

Throughout the thesis we have worked out several examples both in mechanics and field theory. The most remarkable mechanical examples are the damped harmonic oscillator, the motion in a constant gravitational field with friction, the parachute equation and the damped simple pendulum. On the other hand, in field theory, we have studied the damped vibrating string, the Burgers' equation, the Klein--Gordon equation and its relation with the telegrapher's equation, and the Maxwell's equations of electromagnetism with dissipation.

\paragraph{Keywords:} 
contact manifold, 
$k$-contact structure, 
De Donder--Weyl theory, 
dissipation law, 
field theories, 
Hamiltonian formalism, 
Lagrangian formalism, 
$k$-symplectic structure,
Skinner--Rusk formulation, 
symmetries, 
$k$-precosymplectic structure

\paragraph{MSC2020:}
70S05; 
35R01, 
53D05, 
53D10, 
53Z05, 
70H03, 
70H05, 
70H33, 
70H45, 
70S10 

%
%
%
%
%
%

\chapter*{Agraïments}\markboth{Agraïments}{Agraïments}



En primer lloc, vull agrair la tasca dels meus directors de tesi: el Narciso Román, que m'ha guiat des del treball de final de grau fins al dia d'avui; i el Xavier Gràcia, que es va incorporar codirigint el meu treball de final de màster. Sempre m'han donat suport i han estat disposats a ajudar-me desde el primer dia. Tots dos han anat infinitament més enllà del que es pot demanar a un director de tesi, i per això sempre els estaré agraït.

També vull expressar el meu agraïment al professor Miguel C. Muñoz per la seva disposició a ajudar-me sempre que ho he necessitat des del primer dia. De les converses mantingudes amb ell sempre he après alguna cosa. No puc deixar de donar les gràcies a tots els companys del Departament de Matemàtiques de la UPC amb qui he compartit docència a la EPSEVG, EETAC, EEBE, FME i EPSEB per l'acollida que m'han brindat i l'ajuda en la docència que sempre m'han ofert.

Voldria mostrar el meu agraïment a la Red de Geometría, Mecánica y Control per tota l'ajuda i suport financer per poder assistir a trobades, congressos i, especialment, a les escoles d'estiu, de les quals conservo grans records. D'entre els companys de la xarxa, vull fer especial esment als meus coautors Jordi Gaset, Manuel de León i Manuel Lainz, amb qui hem co\ll aborat els últims anys en els temes de contacte i de qui he après moltíssim.

Al llarg de tots aquests anys he tingut un vincle especial amb la Facultat de Matemàtiques i Estadística i amb la seva gent. Un agraïment especial per als companys Marta Arnau i Daniel Torres, sempre presents, des del principi fins al dia d'avui.

A l'Anna li he de donar les gràcies per acompanyar-me aquests anys i no només pel suport incondicional, sinó també per donar-me l'impuls necessari per eixamplar les meves aspiracions, encara que no sempre sigui fàcil.

No podria acabar aquests agraïments sense esmentar els meus pares, sense els quals no estaria escrivint aquesta tesi. Sempre m'han empès a fer allò que em fes feliç i m'han convençut que, amb esforç, podria assolir tot el que em proposés.


\tableofcontents

\newpage

\chapter{Introduction}\markboth{Introduction}{Introduction}


\subsection*{Geometric mechanics and field theories}

The study of dynamical systems has always had a great impact on some branches of mathematics, physics and engineering. Until the second half of the twentieth century, the main advances in this field were based on analytical and numerical methods. However, in the 60's, J. Klein \cite{Kle1962}, A. Lichnerowicz \cite{Lic1951,Lic1975}, J. M. Souriau \cite{Sou1970} and W. M. Tulczyjew \cite{Tul1976,Tul1976b,Tul1977} among others began to use modern methods in differential geometry in order to study physical systems.

This \emph{geometrization} of physical systems provides a correspondence between physical concepts and well-known intrinsically defined geometric objects. For instance, the differential equations defining a physical system can be thought as vector fields in the phase manifold of the system, symmetries can be identified with actions of Lie groups on the phase manifold and the constraints arising in a physical system provide submanifolds of the phase space. Ultimately, physical concepts must not depend on coordinates, which justifies the use of geometric, coordinate-free formulations of physical theories.

Mechanical systems can be formulated in terms of differential geometry. For instance, the natural framework for autonomous mechanical systems is symplectic geometry \cite{Abr1988,Arn1989,Car2015,Cho1978,DeLeo1989,God1969,Gui2011,Lib1987,Man1998,Mar1985}. One of the main results in symplectic geometry is the so-called Darboux theorem \cite{Abr1978,Bry1991,Dar1882}. The proof can be extended to presymplectic manifolds \cite{Cra1986}. Time-dependent mechanical systems can be described by using cosymplectic and precosymplectic or contact geometry \cite{Abr1978,Ban2016,Car1993,DeLeo2017,Ech1991,Gei2008,Kho2013,Man1998}.

Geometric covariant descriptions of first-order classical field theories can be performed by appropriate generalizations of these structures. The simplest one is {\it $k$-symplectic geometry} introduced by A. Awane \cite{Awa1992,Awa2000}, and used later by M. de León et al. \cite{DeLeo1997,DeLeo1988,DeLeo1988b}, and L. K. Norris \cite{Nor2000,Nor1993} to describe first-order field theories. They coincide with the {\it polysymplectic manifolds} described by G. C. Günther \cite{Gun1987} (although these last ones are different from those introduced by G. Sardanashvily et al. \cite{Gia1997,Sar1995} and I. V. Kanatchikov \cite{Kan1998}, that are also called {\it polysymplectic}). This structure is applied to first-order regular autonomous field theories \cite{Bua2015,Ech1996,Rom2007}. The degenerate case can be dealt with through the notion of {\it $k$-presymplectic structures}, which allows to describe the corresponding field theories where the Lagrangian is singular \cite{Gra2009}.

A natural extension of the above are {\it $k$-cosymplectic manifolds}, which allow to generalize the cosymplectic description of non-autonomous mechanical systems to regular field theories whose Lagrangian and/or Hamiltonian functions, in local description, depend on the space-time coordinates \cite{DeLeo1998,DeLeo2001}. The singular case of these theories is described in \cite{Gra2020}. See \cite{DeLeo2015,Gra2004,Mun2010} for more details on the $k$-symplectic and $k$-cosymplectic formalisms.

Finally, one can consider the multisymplectic formalism, which is a more general formalism for classical field theories that can be constructed using multisymplectic geometry, which was first introduced by J. Kijowski, W. M. Tulczyjew and other authors \cite{Gol1973,Kij1973,Kij1979}. See also \cite{Ald1980,Car1991,DeLeo1996,Ech1996,Ech2000,Gar1974,Got1999,Hel2004,Kru2002,Pau2002,Rom2009,Rom2011,Sau1989}. Although there are some partial results \cite{Can1999,DeLeo2003b,Mar1988}, a Darboux-type theorem for multisymplectic manifolds in general is not available.

\subsection*{Contact mechanics and field theories}

The interest in dissipative systems has grown significantly in the recent years. In part, this is because of the adjunction of contact geometry \cite{Ban2016,Gei2008,Kho2013} to the study of non-conservative Lagrangian and Hamiltonian dynamical systems \cite{Bra2017a,Bra2017b,Car2019,DeLeo2019b,DeLeo2020b,Gas2019,Liu2018}. It has been seen that this geometric approach using contact geometry is very useful in many different areas such as thermodynamics \cite{Bra2018,Sim2020}, quantum mechanics \cite{Cia2018}, circuit theory \cite{Got2016} and control theory \cite{Ram2017} among others \cite{Bra2020,DeLeo2021c,DeLeo2021b,DeLeo2021d,DeLeo2017,Ese2021,Kho2013,Sus1999}. In \cite{DeLeo2019}, a constraint algorithm to deal with singular contact Hamiltonian systems is developed. This constraint algorithm is used in \cite{DeLeo2020} to describe a generalization of the Skinner--Rusk formalism for contact systems. A generalization of this formalism to higher-order contact systems is developed in \cite{DeLeo2021}. The Herglotz principle \cite{DeLeo2021,Gue1996,Her1930} gives a variational formulation for contact Hamiltonian systems. There have been several attempts to generalize this variational principle to field theories \cite{Geo2003,Laz2018}.

\subsection*{Constraint algorithms}

Singular systems play a very important role in modern physics, both in mechanics and, particularly, in classical field theory. Actually, many of the most important physical theories are singular. For instance, Maxwell's theory of electromagnetism, Einstein's general relativity and, in general, every gauge theory. Singular theories have a main problem: the failure of usual existence or uniqueness theorems for the solutions of the differential equations describing them. However, sometimes we can solve this problem by finding a submanifold of the phase manifold where we can ensure the existence of solutions by means of a constraint algorithm.

The first constraint algorithm for the Hamiltonian formalism of singular autonomous mechanics was developed by P. A. M. Dirac and P. G. Bergmann \cite{And1951,Dir1950}. These articles were written using local coordinates and were later generalized \cite{Bat1986,Dir1964,Han1976,Kam1982,Sud1974,Sun1982}. Many contributions in the geometric version of this algorithm have been done for autonomous mechanical systems \cite{Got1979,Got1980,Got1978,Gra1991,Gra1992b,Mar1997,Mun1992,Vig2000}. The constraint algorithm was also generalized to deal with nonautonomous mechanical systems \cite{Car1993,Chi1994,DeLeo2002,Gra2005} (see also \cite{DeLeo1996b} for the formulation using jet bundles). This constraint algorithms were also adapted to work with field theories described by singular Lagrangians in both the $k$-symplectic \cite{Gra2009} and multisymplectic \cite{DeLeo1996,DeLeo2005} formalisms.

\subsection*{Skinner--Rusk formalism}

R. Skinner and R. Rusk developed a unified formalism to deal with singular systems more efficiently by combining in a single description both the Lagrangian and Hamiltonian formalisms of mechanical systems \cite{Ski1983} (although a previous description using local coordinates had been made in \cite{Kam1982}). The main goal of this formalism is to obtain a common framework for both regular and singular systems. This formalism is sometimes called {\it unified} formalism because it describes simultaneously the Lagrangian and Hamiltonian formulations of the dynamics. The Skinner--Rusk formalism has been generalized to time-dependent systems \cite{Bar2008,Can2002,Gra2005}. In \cite{Cor2002} it was used to study vakonomic systems and compare the solutions of vakonomic and nonholonomic mechanics. This formalism has also been extended to higher-order autonomous and nonautonomous mechanical systems \cite{Gra1991b,Gra1992,Pri2011,Pri2012}, to describe control systems \cite{Bar2007,Col2010} and to field theory \cite{Cam2009,DeLeo2003,Ech2004,Pri2015,Rey2005,Rey2012,Vit2010}. In particular, in \cite{Cap2018,Cap2020,Gas2018} it was used to describe different models of gravitational theories.

The Skinner--Rusk formalism, in its original version for first-order autonomous mechanical systems, is based on the Whitney sum of the tangent and the cotangent bundles (the velocity and momentum phase spaces of the system respectively) $\W = \T Q\times_Q\cT Q$, called the Pontryagin bundle of $Q$. The bundle $\W$ is endowed with a canonical presymplectic structure $\omega$, which is the pull-back of the canonical symplectic form of the contangent bundle $\cT Q$. Given a Lagrangian function $L$ in the tangent bundle $\T Q$, we can construct a Hamiltonian function $\H = \C - L$ in the Pontryagin bundle $\W$, where $\C$ is the so-called coupling funcion. Thus, we have a presymplectic Hamiltonian system $(\W,\omega,\H)$. Now we can write the Hamiltonian equation $i(X)\omega = \d\H$, where $X$ is the vector field containing the dynamics of the system.

This formalism has several advantages with respect to the Lagrangian and Hamiltonian formalism. In the first place, we recover the Legendre map $\F L$ as constraints from the compatibility condition. We also obtain the holonomy condition as a direct consequence of applying the constraint algorithm (even if the Lagrangian is singular). Finally, it is important to point out that both the Lagrangian and Hamiltonian formalisms can be easily recovered from the Skinner--Rusk formalism. The main drawback of this formalism is that the presymplectic system obtained is always singular, and hence we need a suitable constraint algorithm \cite{Got1979,Got1980,Got1978}.

\subsection*{Goals of the thesis}

In the preceeding paragraphs we have mentioned several problems in geometric mechanics and field theories. The present thesis aims to offer advances on those topics. In particular, the main goals of this thesis are the following:

\begin{itemize}
	\item To develop a constraint algorithm to deal with singular nonautonomous field theories in the $k$-cosymplectic framework, generalizing the one given in \cite{Gra2009} for singular autonomous field theories using the $k$-symplectic framework.
	\item To complete the contact Lagrangian formalism for dissipative mechanical systems. In particular, to give a complete study of contact Lagrangian functions with holonomic dissipation term and study the different notions of symmetry and infinitesimal symmetry for contact systems and see how we can find dissipated and conserved quantities from these symmetries.
	\item Generalize the Skinner--Rusk formalism to contact systems and apply it to different examples of both regular and singular dissipative mechanical systems.
	\item Extend the notion of contact manifold and $k$-symplectic structure to deal with dissipative field theories. This new geometry has been called $k$-contact geometry.
	\item Use the framework of $k$-contact geometry to develop a Hamiltonian formalism and a Lagrangian formalism for dissipative field theories and study their symmetries and dissipation laws.
	\item Generalize the Skinner--Rusk mixed formalism to the case of $k$-contact systems and apply it to many different examples, both regular and singular.
\end{itemize}

\subsection*{Structure of the dissertation}

This thesis is divided in two different parts. The first one, Chapters \ref{ch:survey-on-contact-mechanics}--\ref{ch:contact-examples}, is devoted to the study of contact mechanical systems.

In Chapter \ref{ch:survey-on-contact-mechanics} we study contact systems. We begin by reviewing the most important notions of contact geometry. We define the notions of contact manifold, Reeb vector field and state the existence and uniqueness of the Reeb vector field and give the Darboux theorem for contact manifolds. With this geometric background, we can define the notion of contact Hamiltonian system and write the contact Hamilton equations for both vector fields and integral curves in many different ways. One of them, which is partially equivalent to the others, is formulated without using the Reeb vector field. The last section of this chapter is devoted to the study of contact Lagrangian systems, paying special attention to the Lagrangians with a holonomic dissipation term.

Chapter \ref{ch:symmetries-contact-systems} is devoted to present several kinds of symmetries for both contact Hamiltonian and Lagrangian systems. It is well-known that symmetries of symplectic systems yield conserved quantities. In this case, some symmetries of contact systems give dissipated quantities. We also analyze some properties of symmetries and dissipated quantities. In particular, we see that the quotient of two dissipated quantities is a conserved quantity. We finish this chapter by studying the relation between the symmetries of a symplectic systems and its conserved quantities and the corresponding contactified system.

Chapter \ref{ch:Skinner-Rusk-contact} is devoted to generalize the formalism developed by R. Skinner and R. Rusk in \cite{Ski1983} to the case of contact mechanical systems. We begin by defining the phase bundle of the Skinner--Rusk formalism: the extended Pontryagin bundle $\W$. We describe the precontact structure of this Pontryagin bundle and define the Hamiltonian function in $\W$ associated to a Lagrangian function. Thus, we can state the Lagrangian-Hamiltonian problem. As this system is singular, we need to apply a suitable constraint algorithm in order to deal with it. We study the constraints that arise and, in particular, we recover the holonomy condition and the Legendre map. Finally we show that both the Lagrangian and Hamiltonian formalisms described in the previous chapters can be recovered from the Skinner--Rusk formalism.

In Chapter \ref{ch:contact-examples} we analyze several examples of dissipative mechanical systems. In every example we study different aspects of the theory developed in the previous chapters. The list of examples treated in this chapter is:
\begin{itemize}
	\item the damped harmonic oscillator,
	\item the motion of a particle in a constant gravitational field with friction,
	\item the parachute equation,
	\item Lagrangians with holonomic dissipation term,
	\item central force with dissipation,
	\item the damped simple pendulum using the Lagrange multipliers method,
	\item Cawley's Lagrangian with dissipation.
\end{itemize}

The second part, consisting of Chapters \ref{ch:k-symplectic-k-cosymplectic-formalisms}--\ref{ch:examples-k-contact}, is devoted to the study of field theories. In particular, we deal with field theories described by singular Lagrangians and we develop a geometric formalism to work with dissipative field theories.

Chapter \ref{ch:k-symplectic-k-cosymplectic-formalisms} is a review of the $k$-symplectic and $k$-cosymplectic formulations of first-order classical field theories. We describe both the Lagrangian and Hamiltonian descriptions. We define the notions of $k$-symplectic and $k$-cosymplectic manifold and state their corresponding Darboux theorems.

In Chapter \ref{ch:constraint-algorithms} we summarize the constraint algorithm for autonomous field theories and develop a generalization of this algorithm to deal with nonautonomous field theories. We begin by defining the notion of $k$-presymplectic manifold and prove its corresponding Darboux theorem. Then, we describe the constraint algorithm for singular autonomous field theories. In order to develop a constraint algorithm for singular nonautonomous field theories, we first define the notion of $k$-precosymplectic manifold and prove the existence of global Reeb vector fields in them. Next, we generalize the constraint algorithm previously described to nonautonomous field theories. Finally, we present some examples in order to illustrate how does the constraint algorithm work. In particular, we deal with Lagrangians affine in the velocities and a singular quadratic Lagrangian.

Chapter \ref{ch:k-contact-Hamiltonian} is devoted to present the $k$-contact Hamiltonian formalism for dissipative field theories. We begin by introducing the framework of $k$-contact geometry, proving the existence and uniqueness of Reeb vector fields and the existence of two types of coordinates: adapted coordinates and Darboux coordinates. Then, we use the geometric framework of $k$-contact geometry to develop the Hamiltonian formalism for dissipative field theories. Finally, we generalize the different notions of symmetry introduced in Chapter \ref{ch:symmetries-contact-systems} for contact systems to $k$-contact Hamiltonian systems.

In Chapter \ref{ch:k-contact-Lagrangian} we develop the Lagrangian formalism for dissipative autonomous field theories. In particular, we write the $k$-contact Euler--Lagrange equations. We also give a brief summary on how to deal with dissipative field theories described by singular Lagrangians and the correspondending constraint algorithm. Finally, we define several notions of symmetry for $k$-contact Lagrangian systems and relate them to dissipation laws. In particular, we pay attention to the symmetries of the Lagrangian function.

Chapter \ref{ch:Skinner-Rusk-k-contact} generalizes the Skinner--Rusk formalism for contact systems introduced in Chapter \ref{ch:Skinner-Rusk-contact} to the framework of $k$-contact systems. First of all, we define the extended Pontryagin bundle $\W$ and give a complete description of its canonical $k$-precontact structure. This allows us to state the Lagrangian-Hamiltonian problem and apply to it the constraint algorithm described in the previous chapter. In particular, we give a complete descriptions of the constraints arising, including the {\sc sopde} condition and the Legendre map. We also see how to recover both the Lagrangian and Hamiltonian formalisms from the Skinner--Rusk formalism.

Finally, Chapter \ref{ch:examples-k-contact} is devoted to analyze several examples of dissipative field theories. In each example we study different topics of the theory developed in the previous chapters. The list of examples studied in this chapter is:
\begin{itemize}
	\item the damped vibrating string,
	\item two coupled vibrating strings with damping,
	\item Burgers' equation as a contactification of the heat equation,
	\item the inverse problem for a type of elliptic and hyperbolic partial differential equations,
	\item a comparison between Lorentz-like forces and dissipative forces on a vibrating string,
	\item Klein--Gordon and the telegrapher's equation,
	\item Maxwell's equations with dissipation and damped electromagnetic waves.
\end{itemize}

Throughout this thesis, all the manifolds are real, second countable and of class $\Cinfty$. Manifolds and mappings are assumed to be smooth and the sum over crossed repeated indices is understood.

\mainmatter

\part{Mechanics}
\label{pt:mechanics}

	\chapter{Survey on contact mechanics}
	\label{ch:survey-on-contact-mechanics}
		

This first chapter is devoted to review the main notions on contact geometry and contact mechanics and to detail some of our contributions on these topics. In Section \ref{sec:contact-geometry} we recall the notion of contact manifold, which is the main geometrical object when dealing with contact dynamical systems. We state the existence and uniqueness of the Reeb vector field and we also give the Darboux theorem for contact manifolds. This theorem states that every contact manifold is locally diffeomorphic to the product manifold $\cT Q\times\R$. In Section \ref{sec:contact-Hamiltonian-systems} we define de concept of contact Hamiltonian system and we write the contact Hamiltonian equations for vector fields and integral curves in many different ways. One of them, which is partially equivalent to the others, is formulated without using the Reeb vector field. Section \ref{sec:contact-Lagrangian-systems} is the one devoted to the study of contact Lagrangian systems. We begin by extending the canonical structures of the tangent bundle $\T Q$ of a manifold to $\T Q\times\R$.

We see how these structures allow us to construct a contact structure (if the Lagrangian is singular, the structure is precontact \cite{DeLeo2019}) in $\T Q\times\R$. We can also define the notion of second-order differential equation. With all these geometric tools, we can define the concept of contact Lagrangian system and write the contact Euler--Lagrange equations. We will pay special attention to a particular case of contact Lagrangian functions: the Lagrangians with holonomic dissipation term. These Lagrangians are of great interest as they appear in many applications, as we will see in Chapter \ref{ch:contact-examples}. Some references on these topics are \cite{Bra2017a,Bra2018,Bra2017b,Cia2018,DeLeo2019b,DeLeo2017,Gas2019,Gei2008,Got2016,Liu2018}.

\section{Contact geometry}
\label{sec:contact-geometry}

In this section we define some geometric structures that will be necessary to describe the contact formalism of dissipative mechanical systems. 

\begin{dfn}\label{dfn:contact-manifold}
	Consider a smooth manifold $M$ of odd dimension $2n+1$. A differential form $\eta\in\Omega^1(M)$ such that $\eta\wedge(\d\eta)^{\wedge n}$ is a volume form in $M$ is a \textbf{contact form}\index{contact!form}. In this case, $(M,\eta)$ is said to be a \textbf{contact manifold}\index{contact!manifold}.
\end{dfn}

\begin{rmrk}\rm
    If $\eta$ is a contact form, $\eta' = f\eta$ is also a contact form for every nonvanishing function $f\in\Cinfty(M)$:
    $$ \eta'\wedge(\d\eta')^{\wedge n} = f\eta\wedge(\d f\wedge\eta + f\d\eta)^{\wedge n} = f^{n+1}\eta\wedge(\d\eta)^{\wedge n} \neq 0. $$
\end{rmrk}

Notice that the condition $\eta\wedge(\d\eta)^{\wedge n}\neq 0$ implies that the contact form $\eta$ induces a decomposition of the tangent bundle $\T M$ in the form $\T M = \ker\eta\oplus\ker\d\eta \equiv \D^{\rm C}\oplus\D^{\rm R}$.

\begin{prop}
	Given a contact manifold $(M,\eta)$, there exists a unique vector field $\Reeb\in\X(M)$, called \textbf{Reeb vector field}\index{Reeb!vector field}, such that
	\begin{equation}\label{eq:Reeb-condition}
		\begin{cases}
			i(\Reeb)\d\eta = 0\,,\\
			i(\Reeb)\eta = 1\,.
		\end{cases}
	\end{equation}
\end{prop}
The Reeb vector field $\Reeb$ generates the distribution $\D^{\rm R}$, called the \textbf{Reeb distribution}\index{Reeb!distribution}.

\begin{rmrk}\rm
    It is easy to check that $\Lie_\Reeb\eta = 0$ and hence, $\Lie_\Reeb\d\eta = 0$.
\end{rmrk}

\begin{prop}
	Let $(M,\eta)$ be a contact manifold. Around every point $p\in M$ there exist local coordinates $(x^I, s)$ such that the contact form $\eta$ and the Reeb vector field $\Reeb$ are written
	$$ \Reeb = \parder{}{s}\,,\quad\eta = \d s - f_I(x)\d x^I\,, $$
	where the functions $f_I$ depend only on the $x^I$. These coordinates are called \textbf{adapted coordinates}\index{coordinates!adapted, for a contact manifold} of the contact structure.
\end{prop}
\begin{proof}
	Consider the coordinates $(x^I, s)$, $I = 1,\dotsc,2n$ rectifying the Reeb vector field $\Reeb$ in an open set $U\subset M$. In these coordinates, $\restr{\Reeb}{U} = \tparder{}{s}$.

	Then, $\eta = a\d s - f_I(x, s)\d x^I$. Imposing conditions \eqref{eq:Reeb-condition} we see that $a = 1$ and $\tparder{f_I}{s} = 0$, and hence the result is proved.
\end{proof}

Nevertheless, one can go even further and show that for every contact manifold there exist Darboux-type coordinates:

\begin{thm}[Darboux theorem for contact manifolds]
	Consider a contact manifold $(M,\eta)$ of dimension $2n+1$. Then, around every point $p\in M$ there exists a local chart $(U,q^i,p_i,s)$, $i = 1,\dotsc,n$, such that
	$$ \restr{\eta}{U} = \d s - p_i\d q^i\,. $$
	These coordinates are called \textbf{Darboux}\index{coordinates!Darboux, for a contact manifold}, \textbf{natural}\index{coordinates!natural, for a contact manifold} or \textbf{canonical coordinates}\index{coordinates!canonical, for a contact manifold} of the contact manifold $(M,\eta)$.
\end{thm}
Notice that Darboux coordinates are a particular case of adapted coordinates and hence, in Darboux coordinates, the Reeb vector field is
\begin{equation}\label{eq:coordinates-Reeb}
	\restr{\Reeb}{U} = \parder{}{s}\,.
\end{equation}

Now we are going to introduce a couple of specially relevant examples of contact manifolds.

\begin{exmpl}[Canonical contact structure]\label{ex:canonical-contact-structure}\index{canonical contact structure}\rm
	Let $Q$ be a smooth manifold of dimension $n$. Then, the product manifold $\cT Q\times\R$ has a canonical contact structure given by the 1-form $\eta = \d s - \theta$, where $s$ is the canonical coordinate of $\R$ and $\theta$ is the pull-back of the Liouville 1-form $\theta_\circ\in\Omega^1(\cT Q)$ by the projection $\cT Q\times\R\to\cT Q$. Taking coordinates $(q^i)$ on $Q$ and natural coordinates $(q^i,p_i)$ on $T^\ast Q$, the local expression of the contact 1-form is
	$$ \eta = \d s - p_i\d q^i\,. $$
	We also have that $\d\eta = \d q^i\wedge\d p_i$ and hence, the Reeb vector field is $\Reeb = \tparder{}{s}$.
\end{exmpl}

\begin{exmpl}[Contactification of a symplectic manifold]\label{ex:contactification-symplectic-manifold}\index{contactification of a symplectic manifold}\rm
	Consider a symplectic manifold $(N,\omega)$ such that $\omega = -\d\theta$. Let us define the product manifold $M = N\times\R$. Denoting also by $\theta$ the pull-back of $\theta$ to the product manifold, the 1-form $\eta\in\Omega^1(M)$ given by
	$$ \eta = \d s - \theta\,, $$
	where $s$ is the cartesian coordinate in $\R$, is a contact form on $M$. Thus, $(M,\eta)$ is a contact manifold called the \textbf{contactification} of $N$. The previous example, \ref{ex:canonical-contact-structure}, is just the contactification of the cotangent bundle $\cT Q$ with its canonical symplectic structure.
\end{exmpl}

Given a contact manifold $(M,\eta)$, we have the vector bundle isomorphism
$$
	\begin{array}{rccl}
		\flat\colon & \T M & \longrightarrow & \cT M \\
		& v & \longmapsto & i(v)\d\eta + \left(i(v)\eta\right)\eta
	\end{array}
$$
which can be extended to a $\Cinfty(M)$-module isomorphism
$$
	\begin{array}{rccl}
		\flat\colon & \X(M) & \longrightarrow & \Omega^1(M) \\
		& X & \longmapsto & i(X)\d\eta + \left(i(X)\eta\right)\eta
	\end{array}
$$
\begin{rmrk}\rm
	Notice that with this isomorphism in mind, we can define the Reeb vector field in an alternative way as $\Reeb = \flat^{-1}(\eta)$.\index{Reeb!vector field}
\end{rmrk}

\section{Contact Hamiltonian systems}
\label{sec:contact-Hamiltonian-systems}

This section introduces the concept of contact Hamiltonian system and gives three different characterizations of the contact Hamiltonian vector field. We also offer a new way of writing the contact Hamilton equations without using the Reeb vector field $\Reeb$. This can be useful when dealing with singular systems, where we do not have a uniquely determined Reeb vector field.

\begin{thm}
	Given a contact manifold $(M,\eta)$, for every $H\in\Cinfty(M)$, there exists a unique vector field $X_H\in\X(M)$ such that
	\begin{equation}\label{eq:contact-hamilton-equations-fields}
		\begin{dcases}
			i(X_H)\d\eta = \d H - (\Lie_\Reeb H)\eta\,,\\
			i(X_H)\eta = -H\,.
		\end{dcases}
	\end{equation}
	The integral curves $\bm{\gamma}\colon I\subset\R\to M$ of $X_H$ satisfy equations
	\begin{equation}\label{eq:contact-hamilton-equations-curves}
		\begin{dcases}
			i(\bm{\gamma}')\d\eta = (\d H - (\Lie_\Reeb H)\eta)\circ\bm{\gamma}\,,\\
			i(\bm{\gamma}')\eta = -H\circ\bm{\gamma}\,,
		\end{dcases}
	\end{equation}
	where $\bm{\gamma}'\colon I\subset\R\to\T M$ is the canonical lift of the curve $\bm{\gamma}$ to the tangent bundle $\T M$.
\end{thm}

\begin{dfn}
	The vector field $X_H$ defined by equations \eqref{eq:contact-hamilton-equations-fields} is the \textbf{contact Hamiltonian vector field}\index{contact!Hamiltonian vector field} associated to the Hamiltonian function\index{Hamiltonian function} $H$. Equations \eqref{eq:contact-hamilton-equations-fields} and \eqref{eq:contact-hamilton-equations-curves} are the \textbf{contact Hamiltonian equations}\index{contact!Hamiltonian equations} for vector fields and integral curves, respectively.

	The triple $(M,\eta,H)$ is a \textbf{contact Hamiltonian system}\index{contact!Hamiltonian system}.
\end{dfn}

\begin{prop}\label{prop:dissipation-Hamiltonian-mechanics}
	Given a contact Hamiltonian system $(M,\eta,H)$, the contact Hamiltonian vector field satisfies
	\begin{equation}\label{eq:dissipation-Hamiltonian-mechanics}
		\Lie_{X_H}H = -(\Lie_\Reeb H)H\,,
	\end{equation}
	which expresses the \textsl{dissipation of the Hamiltonian function}.
\end{prop}
\begin{proof}
	\begin{align*}
		\Lie_{X_H}H &= -\Lie_{X_H}i(X_H)\eta\\
		&= -i(X_H)\Lie_{X_H}\eta\\
		& = -i(X_H)\left( \d(i(X_H)\eta + i(X_H)\d\eta) \right)\\
		&= -i(X_H)\left( -\d H + \d H - (\Lie_\Reeb H)\eta \right)\\
		&= i(X_ H)((\Lie_\Reeb H)\eta)\\
		&= -(\Lie_\Reeb H)H\,.
	\end{align*}
\end{proof}
The following proposition gives us two equivalent ways of writing equations \eqref{eq:contact-hamilton-equations-fields}:
\begin{prop}\label{prop:equivalent-contact-Hamiltonian-equations}
	Consider the contact Hamiltonian system $(M,\eta,H)$ and a vector field $X\in\X(M)$. The following are equivalent:
	\begin{enumerate}[{\rm(1)}]
		\item $X$ is the contact Hamiltonian vector field of the contact Hamiltonian system $(M,\eta,H)$ (i.e., it satisfies equations \eqref{eq:contact-hamilton-equations-fields}).
		\item $X$ satisfies that
		$$ \flat(X) = \d H - (\Lie_\Reeb H + H)\eta\,. $$
		\item $X$ is a solution to the equations
		$$
			\begin{cases}
				\Lie_X\eta = -(\Lie_\Reeb H)\eta\,,\\
				i(X)\eta = -H\,.
			\end{cases}
		$$
	\end{enumerate}
\end{prop}

Taking Darboux coordinates $(q^i,p_i,s)$ in the contact manifold $(M,\eta)$, the contact Hamiltonian vector field has local expression
\begin{equation*}
	X_H = \parder{H}{p_i}\parder{}{q^i} - \left(\parder{H}{q^i} + p_i\parder{H}{s}\right)\parder{}{p_i} + \left(p_i\parder{H}{p_i} - H\right)\parder{}{s}\,.
\end{equation*}
Let $\gamma(t) = (q^i(t),p_i(t),s(t))$ be an integral curve of $X_H$. Then, it is a solution to equations \eqref{eq:contact-hamilton-equations-curves}, which in Darboux coordinates read
\begin{equation}\label{eq:contact-hamiltonian-equations-darboux-coordinates}
	\begin{dcases}
		\dot q^i = \parder{H}{p_i}\,,\\
		\dot p_i = -\left(\parder{H}{q^i} + p_i\parder{H}{s}\right)\,,\\
		\dot s = p_i\parder{H}{p_i} - H\,.
	\end{dcases}
\end{equation}

\begin{exmpl}\rm
	Consider the Hamiltonian system $(\cT Q\times\R,\eta, H)$ where $\eta = \d s - p_i\d q^i$ and the Hamiltonian function $H$ is given by
	$$ H = \frac{p^2}{2m} + V(q) + \gamma s\,, $$
	where $m$ represents the mass of a particle and $\gamma$ is a constant. This Hamiltonian functions corresponds to a mechanical system with a friction force linear with respect to the momenta. Writing equations \eqref{eq:contact-hamiltonian-equations-darboux-coordinates} we obtain the dynamical equations
	\begin{equation*}
		\begin{dcases}
			\dot q^i = \frac{p_i}{m}\,,\\
			\dot p_i = -\parder{V}{q^i} - \gamma p_i\,,\\
			\dot s = \frac{p^2}{2m} - V(q) - \gamma s\,,
		\end{dcases}
	\end{equation*}
	which are the damped Newtonian equations.
\end{exmpl}
Now we are going to see a different way of writing equations \eqref{eq:contact-hamilton-equations-fields} without using the Reeb vector field. This might be useful when dealing with systems defined by singular Lagrangians, because in this case we do not have a uniquely determined Reeb vector field.
\begin{prop}\label{prop:contact-hamilton-equations-fields-alt}
	Consider the contact Hamiltonian system $(M,\eta,H)$ and let $U$ be the open set defined as $U = \{ p\in M\,\vert\, H(p)\neq 0\}\subset M$. Consider the 2-form $\Omega\in\Omega^2(U)$ given by $\Omega = -H\d\eta + \d H\wedge\eta$. A vector field $X\in\X(U)$ is the contact Hamiltonian vector field if, and only if,
	\begin{equation}\label{eq:contact-hamilton-equations-fields-alt}
		\begin{dcases}
			i(X)\Omega = 0\,,\\
			i(X)\eta = -H\,.
		\end{dcases}
	\end{equation}
\end{prop}
\begin{proof}
	Let $X$ be a vector field in $M$ satisfying equations \eqref{eq:contact-hamilton-equations-fields-alt}. Then,
	$$ 0 = i(X)\Omega = -H i(X)\d\eta + (i(X)\d H)\eta + H\d H\,. $$
	Hence,
	\begin{equation}\label{eq:proof-alternate-1}
		Hi(X)\d\eta = (i(X)\d H)\eta + H\d H\,.
	\end{equation}
	Contracting with the Reeb vector field $\Reeb$,
	$$ 0 = Hi(\Reeb)i(X)\d\eta = (i(X)\d H)i(\Reeb)\eta + Hi(\Reeb)\d H\,, $$
	and $i(X)\d H = -Hi(\Reeb)\d H$. Combining this with equation \eqref{eq:proof-alternate-1}, we obtain
	$$ Hi(X)\d\eta = H(\d H - (i(\Reeb)\d H)\eta) = H(\d H - (\Lie_\Reeb H)\eta)\,, $$
	and then $i(X)\d\eta = \d H - (\Lie_\Reeb H)\eta$.
	Conversely, suppose that $X$ satisfies equations \eqref{eq:contact-hamilton-equations-fields}. Then,
	\begin{align*}
		i(X)\Omega &= i(X)(-H\d\eta + \d H\wedge\eta)\\
		&= -Hi(X)\d\eta + (i(X)\d H)\eta + H\d H\\
		&= H(\Lie_\Reeb H)\eta + (\Lie_X H)\eta\\
		&= (H\Lie_\Reeb H + \Lie_X H)\eta\,,
	\end{align*}
	and thus $i(X)\Omega = 0$ bearing in mind the dissipation of the Hamiltonian \eqref{eq:dissipation-Hamiltonian-mechanics}.
\end{proof}
	Consider $p\in M$ such that $H(p) = 0$. The second equation in both \eqref{eq:contact-hamilton-equations-fields} and \eqref{eq:contact-hamilton-equations-fields-alt} implies that $X_p\in\ker\eta_p$. The remaining equation in \eqref{eq:contact-hamilton-equations-fields} is $i(X_p)\d_p\eta = \d_pH - (\Lie_{\Reeb_p}H)\eta_p$, while the corresponding one in \eqref{eq:contact-hamilton-equations-fields-alt} is $i(X_p)\Omega_p = (\Lie_{X_p}H)\eta_p = 0$. It is clear that these equations are not equivalent. However, the first one implies the second using the dissipation of the Hamiltonian \eqref{eq:dissipation-Hamiltonian-mechanics}, but not conversely.

\begin{prop}\label{prop:contact-hamilton-equations-curves-alt}
	Let $(M,\eta, H)$ be a contact Hamiltonian system and consider the open subset of $M$ $U = \{p\in M\,\vert\, H(p)\neq 0\}\subset M$. A curve $\bm{\gamma}\colon I\subset\R\to M$ is an integral curve of the contact Hamiltonian vector field $X_H$ if, and only if, it is a solution to equations
	\begin{equation*}
		\begin{cases}
			i(\bm{\gamma}')\Omega = 0\,\\
			i(\bm{\gamma}')\eta = -H\circ \bm{\gamma}\,.
		\end{cases}
	\end{equation*}
\end{prop}

\section{Contact Lagrangian systems}\label{sec:contact-Lagrangian-systems}

\subsection*{Lagrangian phase space and geometric structures}

Let $Q$ be a manifold with dimension $n$ and coordinates $(q^i)$. Consider the product manifold $\T Q\times\R$ with the canonical projections
$$ s\colon \T Q\times\R\to\R\ ,\quad\tau_1\colon\T Q\times\R\to\T Q\ ,\quad\tau_0\colon\T Q\times\R\to Q\times\R\,. $$
Notice that $\tau_1$ and $\tau_0$ are the projection maps of two different vector bundle structures. In what follows, we will usually have the second one in mind. In fact, with this structure, $\T Q\times\R$ is the pull-back of the tangent bundle $\T Q$ with respect to the projection $Q\times\R\to Q$. We will denote by $(q^i,v^i,s)$ the natural coordinates in $\T Q\times\R$.

We want to develop a contact Lagrangian formalism. First of all, we need to extend the usual geometric structures of Lagrangian mechanics to the contact Lagrangian phase space $\T Q\times\R$. We can write $\T(\T Q\times\R) = (\T(\T Q)\times\R)\oplus(\T Q\times\T\R)$ and hence every operation acting on tangent vectors of $\T Q$ can act on tangent vectors of $\T Q\times\R$.

For instance, the vertical endomorphism of $\T(\T Q)$ yields a \textbf{vertical endomorphism}\index{vertical endomorphism}
$$ \mathcal{J}\colon \T(\T Q\times\R)\to\T(\T Q\times\R)\,.$$
In a similar way, the Liouville vector field $\Delta_\circ$ on $\T Q$ yields a \textbf{Liouville vector field}\index{Liouville vector field} $\Delta\in\X(\T Q\times\R)$, which coincides with the Liouville vector field of the vector bundle structure defined by $\tau_0$. The local expressions of these objects in Darboux coordinates are
$$ \mathcal{J} = \parder{}{v^i}\otimes\d q^i\ ,\quad \Delta = v^i\parder{}{v^i}\,. $$

\begin{dfn}
	Consider a path $\bm{\gamma}\colon I\subset\R\to Q\times\R$, where $\bm{\gamma} = (\gamma_1,\gamma_0)$. The \textbf{prolongation}\index{prolongation!of a path to $\T Q\times\R$} of $\bm{\gamma}$ to $\T Q\times\R$ is the path
	$$ \bm{\gamma}' = (\gamma_1',\gamma_0)\colon I\subset\R\to\T Q\times\R\,, $$
	where $\gamma_1'$ is the prolongation of $\gamma_1$ to $\T Q$. The path $\bm{\gamma}'$ is said to be \textbf{holonomic}.
\end{dfn}

\begin{dfn}
	A vector field field $\Gamma\in\X(\T Q\times\R)$ is said to satisfy the \textbf{second-order condition}\index{second order condition} or to be a \rm{\textsc{sode}} if all its integral curves are holonomic.
\end{dfn}
The following proposition gives an alternative characterization of \textsc{sode}s using the canonical structures defined above:
\begin{prop}
	A vector field $\Gamma\in\X(\T Q\times\R)$ is a \rm{\textsc{sode}} if and only if $\mathcal{J}\circ\Gamma = \Delta$.
\end{prop}
If a path has local expression $\bm{\gamma}(t) = (\gamma^i(t), s(t))$, then its prolongation to $\T Q\times\R$ has local expression
$$ \bm{\gamma}'(t) = \left(\gamma^i(t),\frac{\d\gamma^i}{\d t}(t), s(t)\right)\,. $$
The local expression of a \textsc{sode} is
$$ \Gamma = v^i\parder{}{q^i} + f^i\parder{}{v^i} + g\parder{}{s}\,. $$
Hence, in coordinates, a \textsc{sode} defines a system of differential equations of the form
$$
	\begin{dcases}
		\frac{\d^2q^i}{\d t^2} = f^i(q, \dot q, s)\,,\\
		\frac{\d s}{\d t} = g(q, \dot q, s)\,.
	\end{dcases}
$$

\subsection*{Contact formalism for Lagrangian systems}

\begin{dfn}
	Given a Lagrangian function $\L\colon\T Q\times\R\to\R$, we define its associated \textbf{Lagrangian energy}\index{Lagrangian!energy} as $E_\L = \Delta(\L) - \L\in\Cinfty(\T Q\times\R)$. The \textbf{Cartan forms}\index{Cartan forms} associated to $\L$ are
	\begin{equation*}
		\theta_\L = \transp{\mathcal{J}}\circ\d\L\in\Omega^1(\T Q\times \R)\ ,\quad \omega_\L = -d\theta_\L\in\Omega^2(\T Q\times\R)\,.
	\end{equation*}
	The \textbf{contact Lagrangian form}\index{contact!Lagrangian form} is
	$$ \eta_\L = \d s - \theta_\L\in\Omega^1(\T Q\times\R)\,, $$
	and satisfies that $\d\eta_\L = \omega_\L$. The couple $(\T Q\times\R,\L)$ is called a \textbf{contact Lagrangian system}\index{contact!Lagrangian system}.
\end{dfn}

In natural coordinates $(q^i, p_i, s)$ on $\T Q\times\R$, the contact Lagrangian form $\eta_\L$ is
$$ \eta_\L = \d s - \parder{\L}{v^i}\d q^i\,, $$
and hence,
$$ \d\eta_\L = -\parderr{\L}{s}{v^i}\d s\wedge\d q^i - \parderr{\L}{q^j}{v^i}\d q^j\wedge\d q^i - \parderr{\L}{v^j}{v^i}\d v^j\wedge\d q^i\,. $$

\begin{dfn}\label{dfn:fibre-derivative}
	Let $E,F$ be two vector bundles over a manifold $B$. Given a bundle map $f\colon E\to F$, the \textbf{fibre derivative}\index{fiber derivative} of $f$ is the map
	$$ \F f\colon E\to\Hom(E,F)\cong F\otimes E^\ast\,, $$
	obtained by restricting the map $f$ to the fibers $f_b\colon E_b\to F_b$, and computing the usual derivative:
	$$ \F f(e_b) = D f_b(e_b)\,. $$
\end{dfn}

In particular, when the second vector bundle is trivial of rank 1, that is for a function $f\colon E\to\R$, then $\F f\colon E\to E^\ast$. This fiber derivative also has a fiber derivative $\F(\F f) = \F^2 f\colon E\to E^\ast\otimes E^\ast$, which is called the \textbf{fiber Hessian}\index{fiber Hessian} of $f$. For every $e_b\in E$, $\F^2 f(e_b)$ is a symmetric bilinear form on $E_b$. It can be checked that $\F f$ is a local diffeomorphism at a point $e\in E$ if, and only if, the Hessian $\F^2 f(e)$ is non-degenerate (see \cite{Gra2000} for details).

\begin{dfn}\label{dfn:contact-legendre-map}
	Let $\L\colon \T Q\times\R\to\R$ be a Lagrangian function. The \textbf{Legendre map}\index{Legendre map} of $\L$ is the fiber derivative of $\L$, considered as a function on the vector bundle $\tau_0\colon \T Q\times\R\to Q\times\R$.
\end{dfn}

The Legendre map of a Lagrangian function $\L\colon \T Q\times\R\to\R$ is the map $\F\L\colon\T Q\times\R\to\cT Q\times\R$ given by
$$ \F\L(v,s) = (\F\L(\cdot,s)(v),s)\,, $$
where $\L(\cdot, s)$ is the Lagrangian function with $s$ freezed.

Notice that taking into account the Legendre map of $\L$, we can alternatively define the Cartan forms as
$$ \theta_\L = \F\L^\ast(\pi_1^\ast\theta)\ ,\quad \omega_\L = \F\L^\ast(\pi_1^\ast\omega)\,. $$
\begin{prop}\label{prop:contact-regular-lagrangian}
	Let $\L$ be a Lagrangian function on $\T Q\times\R$. Then, the following are equivalent:
	\begin{enumerate}[{\rm(1)}]
		\item The Legendre map $\F\L$ is a local diffeomorphism.
		\item The fiber Hessian $\F^2\L\colon \T Q\times\R \to (\cT Q\times\R)\otimes_{Q\times\R} (\cT Q\times\R)$ of $\L$ is everywhere non-degenerate.
		\item The couple $(\T Q\times\R,\eta_\L)$ is a contact manifold.
	\end{enumerate}
\end{prop}

The previous proposition can be easily proved using natural coordinates $(q^i, v^i, s)$ in $\T Q\times\R$ and taking into account that
$$ \F\L(q^i,v^i,s) = \left(q^i,\parder{\L}{v^i}, s\right)\,, $$
and hence,
$$ \F^2\L(q^i,v^i, s) = (q^i, W_{ij}, s)\,, $$
where $W_{ij} = \dparderr{\L}{v^i}{v^j}$.

\begin{dfn}
	A Lagrangian function is \textbf{regular}\index{Lagrangian!regular} if the equivalent statements in Proposition \ref{prop:contact-regular-lagrangian} hold. Otherwise, the Lagrangian $\L$ is \textbf{singular}. In particular a regular Lagrangian is \textbf{hyperregular} if its Legendre map $\F\L$ is a global diffeomorphism.
\end{dfn}

From the previous definitions and results, we get that every regular contact Lagrangian system $(\T Q\times\R, \L)$ has associated a contact Hamiltonian system $(\T Q\times\R, \eta_\L, E_\L)$. From \eqref{eq:Reeb-condition}, we have that the Reeb vector field $\Reeb_\L\in\X(\T Q\times\R)$ for this contact Hamiltonian system is given by the conditions
\begin{equation*}
	\begin{cases}
		i(\Reeb_\L)\d\eta_\L = 0\,,\\
		i(\Reeb_\L)\eta_\L = 1\,.
	\end{cases}
\end{equation*}
Its local expression in natural coordinates $(q^i,v^i,s)$ is
$$ \Reeb_\L = \parder{}{s} - W^{ji}\parderr{\L}{s}{v^j}\parder{}{v^i}\,, $$
where $W^{ij}$ is the inverse of the matrix of fiber Hessian of the Lagrangian $W_{ij}$, that is, $W^{ij}W_{jk} = \delta^i_k$.

Notice that the Reeb vector field is not as simple as in the Hamiltonian case \eqref{eq:coordinates-Reeb}. This is because the natural coordinates in $\T Q\times\R$ are not adapted coordinates for the contact Lagrangian form $\eta_\L$.

\begin{rmrk}\rm
	The Lagrangian energy satisfies the relation
	$$ \Lie_{\Reeb_\L}E_\L = -\parder{\L}{s}\,. $$
\end{rmrk}

\subsection*{The contact Euler--Lagrange equations}

\begin{dfn}
	Consider a regular contact Lagrangian system $(\T Q\times\R,\L)$. The \textbf{contact Euler--Lagrange equations} for a holonomic curve $\bm{\tilde \gamma}\colon I\subset\R\to\T Q\times\R$ are
	\begin{equation}\label{eq:contact-euler-lagrange-holonomic-curve}
		\begin{cases}
			i(\bm{\tilde\gamma}')\d\eta_\L = (\d E_\L - (\Lie_{\Reeb_\L}E_\L)\eta_\L)\circ\bm{\tilde\gamma}\,,\\
			i(\bm{\tilde\gamma}')\eta_\L = -E_\L\circ\bm{\tilde\gamma}\,,
		\end{cases}
	\end{equation}
	where $\bm{\tilde\gamma}'\colon I\subset\R\to\T(\T Q\times\R)$ is the canonical lift of $\bm{\tilde\gamma}$ to $\T(\T Q\times\R)$.

	The \textbf{contact Lagrangian equations} for a vector field $X\in\X(\T Q\times\R)$ are
	\begin{equation}\label{eq:contact-Lagrangian-equations}
		\begin{cases}
			i(X)\d\eta_\L = \d E_\L - (\Lie_{\Reeb_\L}E_\L)\eta_\L\,,\\
			i(X)\eta_\L = -E_\L\,.
		\end{cases}
	\end{equation}
	A vector field $X_\L\in\X(\T Q\times\R)$ solution to these equations is a \textbf{contact Lagrangian vector field} (it is a contact Hamiltonian vector field for the function $E_\L$).
\end{dfn}

Taking into account Propositions \ref{prop:contact-hamilton-equations-fields-alt} and \ref{prop:contact-hamilton-equations-curves-alt}, in the open subset $U = \{p\in M\,\vert\, E_\L(p) \neq 0\}\subset M$, the above equations are equivalent to
\begin{equation*}
	\begin{cases}
		i(\bm{\tilde\gamma}')\Omega_\L = 0\,,\\
		i(\bm{\tilde\gamma}')\eta_\L = -E_\L\circ\bm{\tilde\gamma}\,,
	\end{cases}
\end{equation*}
and
\begin{equation}\label{eq:contact-Lagrangian-equations-Reeb-independent}
	\begin{cases}
		i(X)\Omega_\L = 0\,,\\
		i(X)\eta_\L = -E_\L\,,
	\end{cases}
\end{equation}
where $\Omega_\L = -E_\L\d\eta_\L + \d E_\L\wedge\eta_\L$.

Let $\bm{\tilde\gamma}(t) = (q^i(t), \dot q^i(t), s(t))$ be a holonomic curve on $\T Q\times\R$ in natural coordinates. Then, equation \eqref{eq:contact-euler-lagrange-holonomic-curve} reads
\begin{align}
	\parderr{\L}{v^j}{v^i}\ddot q^j + \parderr{\L}{q^j}{v^i}\dot q^j + \parderr{\L}{s}{v^i}\dot s - \parder{\L}{q^i} = \frac{\d}{\d t}\left(\parder{\L}{v^i}\right) - \parder{\L}{q^i} &= \parder{\L}{s}\parder{\L}{v^i}\,,\label{eq:contact-Euler-Lagrange-1}\\
	\dot s &= \L\,,\label{eq:contact-Euler-Lagrange-2}
\end{align}
which coincide with the generalized Euler--Lagrange equations stated in \cite{Her1930}.

Consider a vector field $X\in\X(\T Q\times\R)$ with local expression
$$ X = f^i\parder{}{q^i} + F^i\parder{}{v^i} + g\parder{}{s}\,. $$
Then, equations \eqref{eq:contact-Lagrangian-equations} for the vector field $X$ read
\begin{align}
	(f^j - v^j)\parderr{\L}{v^j}{s} &= 0\,,\label{eq:contact-Lagrangian-equations-vector-field-1}\\
	(f^j - v^j)\parderr{\L}{v^i}{v^j} &= 0\,,\label{eq:contact-Lagrangian-equations-vector-field-2}\\
	(f^j - v^j)\parderr{\L}{q^i}{v^j} + \parder{\L}{q^i} - \parderr{\L}{s}{v^i}g - \parderr{\L}{q^j}{v^i}f^j - \parderr{\L}{v^j}{v^i}F^j + \parder{\L}{s}\parder{\L}{v^i} &= 0\,,\label{eq:contact-Lagrangian-equations-vector-field-3}\\
	\L + \parder{\L}{v^i}(f^i - v^i) - g &= 0\,,\label{eq:contact-Lagrangian-equations-vector-field-4}
\end{align}
where we have used the relation
\begin{equation}\label{eq:Lagrangian-energy-dissipation-contact}
	\Lie_{\Reeb_\L}E_{\L} = -\parder{\L}{s}\,.
\end{equation}

\begin{prop}\label{prop:contact-Euler-Lagrange-vector-field}
	Consider a regular Lagrangian function $\L$ and let $X$ be its contact Lagrangian vector field. Then $X$ is a {\sc sode} and equations \eqref{eq:contact-Lagrangian-equations-vector-field-3} and \eqref{eq:contact-Lagrangian-equations-vector-field-4} become
	\begin{align}
		\parderr{\L}{v^j}{v^i} F^j + \parderr{\L}{q^j}{v^i}v^j + \parderr{\L}{s}{v^i}\L - \parder{\L}{q^i} &= \parder{\L}{s}\parder{\L}{v^i}\,,\label{eq:regular-contact-Lagrangian-equations-vector-field-1}\\
		g &= \L\,,\label{eq:regular-contact-Lagrangian-equations-vector-field-2}
	\end{align}
	which, for the integral curves of $X$, are the Euler--Lagrange equations \eqref{eq:contact-Euler-Lagrange-1} and \eqref{eq:contact-Euler-Lagrange-2}. This {\sc sode} $\Gamma_\L\equiv X$ is called the \textbf{Euler--Lagrange vector field} associated to the Lagrangian function $\L$.
\end{prop}
\begin{proof}
	It follows from the coordinate expressions. If $\L$ is a regular Lagrangian, equations \eqref{eq:contact-Lagrangian-equations-vector-field-2} lead to $v^i = f^i$, which are the {\sc sode} conditions for the vector field $X$. Then, \eqref{eq:contact-Lagrangian-equations-vector-field-1} holds identically and \eqref{eq:contact-Lagrangian-equations-vector-field-3} and \eqref{eq:contact-Lagrangian-equations-vector-field-4} give equations \eqref{eq:regular-contact-Lagrangian-equations-vector-field-1} and \eqref{eq:regular-contact-Lagrangian-equations-vector-field-2} or, equivalently, for the integral curves of $X$, the Euler--Lagrange equations \eqref{eq:contact-Euler-Lagrange-1} and \eqref{eq:contact-Euler-Lagrange-2}.
\end{proof}

In this way, the local expression of the Euler--Lagrange vector field $X$ for a regular Lagrangian $\L$ is
$$ X = \L\parder{}{s} + v^i\parder{}{q^i} + W^{ik}\left( \parder{\L}{q^k} - \parderr{\L}{q^j}{v^k}v^j - \L\parderr{\L}{s}{v^k} + \parder{\L}{s}\parder{\L}{v^k} \right)\parder{}{v^i}\,. $$

\begin{rmrk}\rm
	It is interesting to point out how, in the Lagrangian formalism of dissipative systems, the expression in coordinates \eqref{eq:contact-Euler-Lagrange-2} relates the variation of the ``dissipative coordinate'' $s$ to the Lagrangian function and, from here, we can identify this coordinate with the Lagrangian action, $s = \int\L\d t$.
\end{rmrk}

\begin{rmrk}\rm
	If the Lagrangian function $\L$ is singular $(\T Q\times\R,\eta_{\L})$ is not a contact manifold, but a precontact one. Hence, the Reeb vector field is not uniquely defined. It can be proved that the contact Lagrangian equations \eqref{eq:contact-Lagrangian-equations} are independent on the Reeb vector field chosen \cite{DeLeo2019}. Alternatively, Proposition \ref{prop:contact-hamilton-equations-fields-alt} holds also in this case and hence, the Reeb-independent equations \eqref{eq:contact-Lagrangian-equations-Reeb-independent} can be used instead. In any case, solutions to the contact Lagrangian equations are not necessarily {\sc sode} and, in order to obtain the Euler--Lagrange equations \eqref{eq:regular-contact-Lagrangian-equations-vector-field-1} (or \eqref{eq:contact-Euler-Lagrange-1}), the condition $\mathcal{J}(X) = \Delta$ must be added to the above contact Lagrangian equations. Furthermore, these equations are not necessarily consistent everywhere on $\T Q\times\R$ and a suitable {\it constraint algorithm} must be implemented in order to find a {\it final constraint submanifold} $S_f\hookrightarrow \T Q\times\R$ (if it exists) where there are {\sc sode} vector fields $X\in\X(\T Q\times\R)$, tangent to $S_f$, which are (not necessarily unique) solutions to the above equations on $S_f$. All these problems are studied in detail in \cite{DeLeo2019}.
\end{rmrk}

\subsection*{The canonical Hamiltonian formalism for contact Lagrangian systems}

In the (hyper)regular case, we have a diffeomorphism between $(\T Q\times\R, \eta_\L)$ and $(\cT Q\times\R, \eta)$ such that $\F\L^\ast\eta = \eta_\L$. Furthermore, there exists (at least locally) a function $H\in\Cinfty(\cT Q\times\R)$ such that $\F\L^\ast H = E_\L$. Then, we have the conctact Hamiltonian system $(\cT Q\times\R,\eta,H)$, for which $\F\L_\ast\Reeb_\L = \Reeb$. Then, if $X_H\in\X(\cT Q\times\R)$ is the contact Hamiltonian vector field associated to $H$, we have that $\F\L_\ast\Gamma_\L = X_H$, where $\Gamma_\L$ is the Euler--Lagrange vector field defined in Proposition \ref{prop:contact-Euler-Lagrange-vector-field}.

For singular Lagrangians, following \cite{Got1979} we define

\begin{dfn}
	A singular Lagrangian $\L$ is \textbf{almost-regular} if $\P := \Ima(\F\L)=\F\L(\T Q\times\R)$ is a closed submanifold of $\cT Q\times\R$, the Legendre map $\F\L$ is a submersion onto its image, and the fibers $\F\L^{-1}(\F\L(v_q,s))\subset \T Q\times\R$ are connected submanifolds for every $(v_q,s)\in \T Q\times\R$.
\end{dfn}

In the almost-regular case, we have $(\P, \eta_\P)$, where $\eta_\P = j_\P^\ast\eta\in\Omega^1(\P)$ and $j_\P\colon\P\hookrightarrow\cT Q\times\R$ is the natural embedding. Furthermore, the Lagrangian energy $E_\L$ is $\F\L$-projectable; i.e. there is a unique $H_\P\in\Cinfty(\P)$ such that $E_\L = \F\L_\circ^\ast H_\P$, where $\F\L_\circ\colon\T Q\times\R\to\P$ is the restriction of $\F\L$ to the closed submanifold $\P$, defined by $\F\L = j_\P\circ\F\L_\circ$. Then, there exists a Hamiltonian formalism associated to the original Lagrangian system, which is developed on the submanifold $\P$, and the contact Hamiltonian equations for $X_\P\in\X(\P)$ are \eqref{eq:contact-hamilton-equations-fields} adapted to this situation or, equivalently,
\begin{equation}\label{eq:contact-canonical-without-Reeb}
	\begin{cases}
		i(X_\P)\Omega_\P = 0\,,\\
		i(X_\P)\eta_\P = -H_\P\,,
	\end{cases}
\end{equation}
where $\Omega_\P = -H_\P\d\eta_\P + \d H_\P\wedge\eta_\P$. As in the Lagrangian formalism, these equations are not necessarily consistent everywhere on $\P$ and we must implement a suitable \textit{constraint algorithm} in order to find a \textit{final constraint submanifold} $\P_f\hookrightarrow\P$ (if it exists) where there exist vector fields $X\in\X(\P)$, tangent to $\P_f$, which are (not necessarily unique) solutions to \eqref{eq:contact-canonical-without-Reeb} on $\P_f$. See \cite{DeLeo2019} for a deeper analysis.

\subsection*{Lagrangians with holonomic dissipation term}

In a recent paper \cite{Cia2018} by Ciaglia \textit{et al.} a Lagrangian description for some systems with dissipation was given using a modification of the Lagrangian formalism inspired by the contact Hamiltonian formalism. In this section we are going to prove that this description coincides with the general formalism developed in section \ref{sec:contact-Lagrangian-systems} when applied to a particular class of contact Lagrangians.

\begin{dfn}
	A \textbf{Lagrangian with holonomic dissipation term} in $\T Q\times\R$ is a function $\L = L + \phi\in\Cinfty(\T Q\times\R)$, where $L = \tau_1^\ast L_\circ$, for a Lagrangian function $L_\circ\in\Cinfty(\T Q)$ and $\phi = \tau_0^\ast\phi_\circ$, for $\phi_\circ\in\Cinfty(Q\times\R)$.
\end{dfn}

Taking natural coordinates $(q^i, v^i, s)$, a Lagrangian with holonomic dissipation term has the form
$$ \L(q^i, v^i, s) = L(q^i, v^i) + \phi(q^i, s)\,. $$
Notice that this implies that the momenta defined by the Legendre map are independent of the coordinate $s$. In addition, for these Lagrangians the conditions $\parderr{\L}{v^i}{s} = 0$ hold. This motivates the name given in the definition.

\begin{rmrk}\rm
	The Lagrangian formalism developed in \cite{Cia2018} is a little less general than the one treated here. In \cite{Cia2018} only the case $\phi = \phi(s)$ is taken into consideration.
\end{rmrk}

\begin{prop}
	Consider the Lagrangian with holonomic disspation term $\L = L + \phi$. Then, its Cartan 1-form, contact form, Lagrangian energy and Reeb vector field as a contact Lagrangian are
	$$ \theta_\L = \theta_L\ ,\quad \eta_\L = \d s - \theta_L\ ,\quad E_\L = E_L - \phi\ ,\quad \Reeb_\L = \parder{}{s}\,, $$
	whre $\theta_L$ is the Cartan 1-form of $L$ considered (via pull-back) as a 1-form on $\T Q\times\R$, and $E_L$ is the energy of $L$ as a function on $\T Q\times\R$.

	The Legendre map of $\L$, $\F\L\colon\T Q\times\R\to\cT Q\times\R$, can be expressed as $\F\L = \F L\times\Id_\R$, where $\F L$ is the Legendre map of $L$. The Hessians are related by $\F^2\L(v_q,s) = \F^2L(v_q)$. Moreover, $\L$ is regular if, and only if, $L$ is regular.
\end{prop}
\begin{proof}
	The proof of this proposition is immediate taking coordinates. In particular, the assertion about the Legendre map is a direct consequence of the fact that
	$$ \parder{\L}{v^i} = \parder{L}{v^i}\,.$$
	In a similar way, the relation between the Hessians is expressed in coordinates as
	$$ \parderr{\L}{v^i}{v^j} = \parderr{L}{v^i}{v^j}\,.$$
	This shows that the Lagrangian $\L$ is regular if, and only if, $L$ is regular.
\end{proof}

It is also clear that $\L$ is hyperregular if, and only if, $L$ is also hyperregular. This means that the Legendre map $\F\L$ is a diffeomorphism, and the canonical Hamiltonian formalism for the Lagrangian with holonomic dissipation term can be formulated as stated above.

Consider the contact Lagrangian system $(\T Q\times\R, \eta_\L, E_\L)$ where $\L = L + \phi$ is a Lagrangian function with holonomic dissipation term. The dynamical equations for vector fields of this system are
$$
	\begin{cases}
		i(X)\d\eta_\L = \d E_\L - (\Lie_{\Reeb_\L}E_\L)\eta\L\,,\\
		i(X)\eta_\L = -E_\L\,.
	\end{cases}
$$
Taking coordinates $(q^i, v^i, s)$ in $\T Q\times\R$, if $X = f^i\parder{}{q^i} + F^i\parder{}{v^i} + g\parder{}{s}$, the second contact Lagrangian equation for $X$ reads
$$ \L + \parder{L}{v^i}(f^i - v^i) - g = 0\,, $$
and this is equation \eqref{eq:contact-Lagrangian-equations-vector-field-4} for the Lagrangian $\L = L + \phi$. The first contact Lagrangian equation is
\begin{equation}\label{eq:contact-lagrangian-holonomic-dissipation-term-first-eq}
	(f^i - v^i)\parderr{L}{v^j}{v^i} = 0\,,
\end{equation}
and
$$ \left( \parderr{L}{q^i}{v^j} - \parderr{L}{q^j}{v^i} \right)f^j + \parderr{L}{q^i}{v^j}v^j - \parderr{L}{v^j}{v^i}F^j = -\parder{L}{q^i} - \parder{\phi}{q^i} - \parder{\phi}{s}\parder{L}{v^i}\,, $$
which corresponds to equation \eqref{eq:contact-Lagrangian-equations-vector-field-3} for the Lagrangian $\L$. Notice that equations \eqref{eq:contact-Lagrangian-equations-vector-field-1} are identities as
$$ \parderr{\L}{v^j}{s} = 0\,. $$
Finally, as in Proposition \ref{prop:contact-regular-lagrangian}, if the Lagrangian $\L$ is regular (i.e., if $L$ is regular), equation \eqref{eq:contact-lagrangian-holonomic-dissipation-term-first-eq} implies that $f^i = v^i$, that is, the vector field $X$ is a {\sc sode} and the equations of motion become
\begin{align*}
	\dot s &= \L\,,\\
	\parderr{L}{v^j}{v^i}\ddot q^j + \parderr{L}{q^j}{v^i}\dot q^j - \parder{L}{q^i} = \frac{\d}{\d t}\left( \parder{L}{v^i} \right) - \parder{L}{q^i} &= \parder{\phi}{q^i} + \parder{\phi}{s}\parder{L}{v^i}\,.
\end{align*}
These are the expression in coordinates of the contact Euler--Lagrange equations \eqref{eq:contact-Euler-Lagrange-1} and \eqref{eq:contact-Euler-Lagrange-2} for the Lagrangian $\L = L + \phi$.

	\chapter[Symmetries, conserved and dissipated quantities in contact systems]{Symmetries, conserved and \phantom{m} dissipated quantities in contact systems}
	\label{ch:symmetries-contact-systems}


In this chapter we will introduce the notions of symmetry, conserved and dissipated quantity for both Hamiltonian and Lagrangian contact systems. We will see that these concepts are closely related. In Section \ref{sec:symmetries-contact-Hamiltonian-systems} we define two different types of symmetries: dynamical symmetries and contact symmetries and we prove that every contact symmetry is also a dynamical symmetry. In Section \ref{sec:dissipated-conserved-quantities-contact-Hamiltonian} we introduce the concepts of conserved and dissipated quantity. Then, we prove that every infinitesimal dynamical symmetry gives rise to a dissipated quantity. In particular, we see that the Hamiltonian function of a contact Hamiltonian system is a dissipated quantity and that the quotient of two dissipated quantities is a conserved quantity. Section \ref{sec:symmetries-canonical-contact-Hamiltonian-systems} we deal with the symmetries of canonical contact Hamiltonian systems. In particular, we prove the momentum dissipation theorem. In Section \ref{sec:symmetries-contact-Lagrangian-systems} we study the symmetries of contact Lagrangian systems. Finally, in Section \ref{sec:symmetries-contactified-system} we analyze the relations between the symmetries and the conserved and dissipated quantities of a symplectic Hamiltonian system and its corresponding contactified system. This chapter is based in \cite{Gas2019}. See \cite{DeLeo2020b} for another approach to these topics.

\section{Symmetries of contact Hamiltonian systems}
\label{sec:symmetries-contact-Hamiltonian-systems}

Given a dynamical system, there are many different concepts of symmetry depending on the underlying structure they preserve. Thus, one can consider the transformations that preserve the geometric structure of the dynamical system, or those preserving its solutions \cite{Gra2002,Rom2020}. In this chapter, we discuss these subjects for contact systems. See also \cite{DeLeo2020b} for a complementary approach on these topics.
\begin{dfn}\label{dfn:dynamical-symmetry-contact}
	Consider a contact Hamiltonian system $(M, \eta, H)$ and let $X_H$ be its contact Hamiltonian vector field. A \textbf{dynamical symmetry} of this system is a diffeomorphism $\Phi\colon M\to M$ such that $$ \Phi_\ast X_H = X_H\,. $$
\end{dfn}
According to the definition above, a dynamical symmetry maps solutions into solutions.
\begin{dfn}
	An \textbf{infinitesimal dynamical symmetry} of a contact Hamiltonian system $(M,\eta,H)$ is a vector field $Y\in\X(M)$ whose local flow is a dynamical symmetry; that is, $\Lie_YX_H = [Y,X_H] = 0$.
\end{dfn}
There are other kinds of symmetries that leave the geometric structures invariant. They are the following:
\begin{dfn}
	A \textbf{contact symmetry} of a contact Hamiltonian system $(M,\eta,H)$ is a diffeomorphism $\Phi\colon M\to M$ satisfying
	$$ \Phi^\ast\eta = \eta\ ,\quad \Phi_\ast H = H\,. $$
	An \textbf{infinitesimal contact symmetry} is a vector field $Y\in\X(M)$ whose local flow is a contact symmetry; i.e.,
	$$ \Lie_Y\eta = 0\ ,\quad \Lie_Y H = 0\,. $$
\end{dfn}
Furthermore, we have:
\begin{prop}\label{prop:contact-symmetry-preserves-Reeb}
	Every (infinitesimal) contact symmetry preserves the Reeb vector field; that is, $\Phi^\ast\Reeb = \Reeb$ (or $[Y,\Reeb] = 0$).
\end{prop}
\begin{proof}
	We have that
	\begin{align*}
		& i(\Phi_\ast^{-1})(\Phi^\ast\d\eta) = \Phi^\ast(i(\Reeb)\d\eta) = 0\,,\\
		& i(\Phi_\ast^{-1})(\Phi^\ast\eta) = \Phi^\ast(i(\Reeb)\eta) = 1\,,
	\end{align*}
	and, as $\Phi^\ast\eta = \eta$ and the Reeb vector field is unique, from these equalities we get that $\Phi_\ast^{-1}\Reeb = \Reeb$. The proof for the infinitesimal case is immediate from the definition.
\end{proof}
Now, taking into account everything stated above, we can see the relation between contact symmetries and dynamical symmetries:
\begin{prop}\label{prop:contact-symmetries-dynamical-symmetries}
	(Infinitesimal) contact symmetries are (infinitesimal) dynamical symmetries.
\end{prop}
\begin{proof}
	If $X_H$ is the contact Lagrangian vector field,
	\begin{align*}
		i(\Phi_\ast X_H)\d\eta &= i(\Phi_\ast X_H)(\Phi^\ast\d\eta) = \Phi^\ast(i(X_H)\d\eta)\\
		&= \Phi(\d H - (\Lie_\Reeb H)\eta) = \d H - (\Lie_\Reeb H)\eta\,,\\
		i(\Phi_\ast X_H)\eta &= i(\Phi_\ast X_H)(\Phi^\ast\eta) = \Phi^\ast(i(X_H)\eta) = \Phi^\ast(-H) = -H\,.
	\end{align*}
	The proof for the infinitesimal case is straightforward from the definition.
\end{proof}

\section{Dissipated and conserved quantities of contact Hamiltonian systems}
\label{sec:dissipated-conserved-quantities-contact-Hamiltonian}

Associated to symmetries of contact Hamiltonian systems are the concepts of dissipated and conserved quantities.

\begin{dfn}\label{dfn:dissipated-quantity-contact}
	A \textbf{dissipated quantity} of a contact Hamiltonian system $(M, \eta, H)$ with contact Hamiltonian vector field $X_H$ is a function $F\in\Cinfty(M)$ such that
	\begin{equation}\label{eq:contact-dissipated-quantity}
		\Lie_{X_H}F = -(\Lie_\Reeb H)F\,.
	\end{equation}
\end{dfn}

In a contact Hamiltonian system, symmetries and dissipated quantities are related as follows.

\begin{thm}[Dissipation theorem for contact Hamiltonian systems]\label{thm:dissipation-contact}
	Let $Y\in\X(M)$ be a vector field. If $Y$ is an infinitesimal dynamical symmetry, $[Y, X_H] = 0$, then the function $F = -i(Y)\eta$ is a dissipated quantity.
\end{thm}
\begin{proof}
	This is a consequence of
	\begin{align*}
		\Lie_{X_H}F &= -\Lie_{X_H}i(Y)\eta\\
		&= -i(Y)\Lie_{X_H}\eta - i(\Lie_{X_H}Y)\eta\\
		&= (\Lie_\Reeb H)i(Y)\eta + i([Y, X_H])\eta\\
		&= -(\Lie_\Reeb H)F + i([Y,X_H])\eta\\
		&= -(\Lie_\Reeb H)F\,,
	\end{align*}
	where we have used the third statement in Proposition \ref{prop:equivalent-contact-Hamiltonian-equations}.
\end{proof}

\begin{rmrk}\rm
	The last equality reveals that $[Y, X_H]\in\ker\eta$ is a necessary and sufficient condition for $F$ to be a dissipated quantity. This fact has been noted in \cite{DeLeo2020b}. Nevertheless, it is important to point out that such transformations are not dynamical symmetries in the sense of Definition \ref{dfn:dynamical-symmetry-contact}, since in general they do not map solutions into solutions.
\end{rmrk}

In particular, as we pointed out in Proposition \ref{prop:dissipation-Hamiltonian-mechanics}, the contact Hamiltonian vector field $X_H$ is trivially an infinitesimal dynamical symmetry and its associated dissipated quantity is the energy, $F = -i(X_H)\eta = H$:

\begin{thm}[Energy dissipation theorem for contact Hamiltonian systems]\label{thm:energy-dissipation-contact}
	$$ \Lie_{X_H}H = -(\Lie_\Reeb H)H\,. $$
\end{thm}

Notice that these are non-conservation theorems. We are dealing with dissipative systems and hence, dynamical symmetries are not associated to conserved quantities but to dissipated quantities. In particular, as stated in the theorem above, the energy is not a conserved quantity.

\begin{dfn}
	A \textbf{conserved quantity} of a contact Hamiltonian system $(M, \eta, H)$ is a function $G\in\Cinfty(M)$ such that
	$$ \Lie_{X_H}G = 0\,. $$
\end{dfn}

Notice that every dissipated quantity changes with the same rate $-\Lie_\Reeb H$. Hence, we have the following:

\begin{prop}\phantomsection\label{prop:conserved-dissipated-contact}
	\begin{enumerate}[{\rm(1)}]
		\item If $F_1$ and $F_2$ are two dissipated quantities and $F_2\neq 0$, $F_1/F_2$ is a conserved quantity.
		\item If $F$ is a dissipated quantity and $G$ is a conserved quantity, $FG$ is a dissipated quantity.
	\end{enumerate}
\end{prop}
\begin{proof}
	\begin{enumerate}[{\rm(1)}]
		\item
		\begin{align*}
			\Lie_{X_H}\left(\frac{F_1}{F_2}\right) &= \frac{F_2\Lie_{X_H}F_1 - F_1\Lie_{X_H}F_2}{F_2^2}\\
			&= -\frac{F_1\Lie_\Reeb H}{F_2} + \frac{F_1F_2\Lie_\Reeb H}{F_2^2} = 0\,,
		\end{align*}
		\item
		$$ \Lie_{X_H}(FG) = G\Lie_{X_H}F + F\Lie_{X_H}G = -(\Lie_\Reeb H)FG\,. $$
	\end{enumerate}
\end{proof}

\begin{rmrk}\rm
	Taking into account the previous theorem, if $H\neq 0$, it is possible to assign a conserved quantity to an infinitesimal dynamical symmetry $Y$. Indeed, using Theorem \ref{thm:energy-dissipation-contact} and Proposition \ref{prop:conserved-dissipated-contact}, it is clear that
	$$ G = -\frac{1}{H}i(Y)\eta $$
	is a conserved quantity.
\end{rmrk}

Finally, contact symmetries can be used to generate new dissipated quantities from a given dissipated quantity. The following result is a direct consequence of Definitions \ref{dfn:dynamical-symmetry-contact} and \ref{dfn:dissipated-quantity-contact}.

\begin{prop}
	If $\Phi\colon M\to M$ is a contact symmetry and $F\in\Cinfty(M)$ is a dissipated quantity, then so is $\Phi^\ast F$.
\end{prop}
\begin{proof}
	We have
	$$ \Lie_{X_H}(\Phi^\ast F) = \Phi^\ast\Lie_{\Phi_\ast X_H}F = \Phi^\ast\Lie_{X_H}F = \Phi^\ast(-\Lie_\Reeb H)F = -(\Lie_\Reeb H)(\Phi^\ast F)\,. $$
	The proof for the infinitesimal case is straightforward from the definition.
\end{proof}

\section{Symmetries of canonical contact Hamiltonian systems}
\label{sec:symmetries-canonical-contact-Hamiltonian-systems}

Consider the canonical contact manifold $(\cT Q\times\R, \eta)$ with contact form
$$ \eta = \d s - p_i\d q^i\,, $$
as in Example \ref{ex:canonical-contact-structure}. If $\varphi\colon Q\to Q$ is a diffeomorphism, we can construct the diffeomorphism
$$ \Phi = (\cT\varphi, \Id_\R) \colon \cT Q\times\R\to\cT Q\times\R\,, $$
where $\cT\varphi\colon \cT Q\to\cT Q$ is the canonical lift of $\varphi$ to $\cT Q$. Then $\Phi$ is said to be the \textbf{canonical lift} of $\varphi$ to $\cT Q\times\R$. Any transformation $\Phi$ of this kind is called a \textbf{natural transformation} of $\cT Q\times\R$.

In the same way, consider a vector field $Z\in\X(Q)$. Its \textbf{complete lift} to $\cT Q\times\R$ is the vector field $Y\in\X(\cT Q\times\R)$ whose local flow is the canonical lift of the local flow of $Z$ to $\cT Q\times\R$; that is, the vector field $Y = Z^{\rm C\ast}$ where $Z^{\rm C\ast}$ denotes the complete lift of $Z$ to $\cT Q$, identified in a natural way as a vector field in $\cT Q\times\R$. Any infinitesimal transformation $Y$ of this kind is called a \textbf{natural infinitesimal transformation} of $\cT Q\times\R$.

It is well known that the canonical forms $\theta_\circ\in\Omega^1(\cT Q)$ and $\omega_\circ = -\d\theta_\circ\in\Omega^2(\cT Q)$ are invariant under the action of canical lifts of diffeomorphisms and vector fields from $Q$ to $\cT Q$. Now, taking into consideration the definition of the contact form $\eta\in\Omega^1(\cT Q\times\R)$, we have the following results.

\begin{prop}
	If $\Phi\in\Diff(\cT Q\times\R)$ (resp. $Y\in\X(\cT Q\times\R)$) is a canonical lift to $\cT Q\times\R$ of a diffeomorphism $\varphi\in\Diff(Q)$ (resp. of $Z\in\X(Q)$), then
	\begin{enumerate}[{\rm(1)}]
		\item $\Phi^\ast\eta = \eta$ (resp. $\Lie_Y\eta = 0$).
		\item If, in addition, $\Phi^\ast H = H$ (resp. $\Lie_YH = 0$), then it is a (infinitesimal) contact symmetry.
	\end{enumerate}
\end{prop}

In particular, we have

\begin{thm}[Momentum dissipation theorem]\label{thm:momentum-dissipation-contact}
	If $\tparder{H}{q^i} = 0$, then $\parder{}{q^i}$ is an infinitesimal contact symmetry, and its associated dissipated quantity is the corresponding momentum $p_i$; that is,
	$$ \Lie_{X_H}p_i = -(\Lie_\Reeb H)p_i\,. $$
\end{thm}
\begin{proof}
	A simple computation in coordinates shows that $\Lie\left(\tparder{}{q^i}\right)\eta = 0$ and that $\Lie\left(\tparder{}{q^i}\right)H = 0$. Therefore, it is a contact symmetry and, in particular, a dynamical symmetry. The other results are a consequence of the dissipation theorem.
\end{proof}

\section{Symmetries of contact Lagrangian systems}
\label{sec:symmetries-contact-Lagrangian-systems}

Let $(\T Q\times\R, \L)$ be a regular contact Lagrangian system with Reeb vector $\Reeb_\L$ and contact Euler--Lagrange vector field $X_\L$ (i.e. solution to equations \eqref{eq:contact-Lagrangian-equations}).

Everything said above about symmetries and dissipated quantities for contact Hamiltonian systems holds when it is applied to the contact system $(\T Q\times\R, \eta_\L, E_\L)$. Thus, we have the same definitions for dynamical and contact symmetries and the dissipation theorem states that $-i(Y)\eta_\L$ is a dissipated quantity for every infinitesimal dynamical symmetry $Y$. In particular, the energy dissipation theorem \ref{thm:energy-dissipation-contact} applied to the contact system $(\T Q\times\R, \eta_\L, E_\L)$ states that
$$ \Lie_{X_\L}E_\L = -(\Lie_{\Reeb_\L}E_\L)E_\L\,. $$

If $\varphi\in\Diff(Q)$ is a diffeomorphism, we can construct the diffeomorphism
$$ \Phi = (\T\varphi,\Id_\R)\colon\T Q\times\R\to\T Q\times\R\,, $$
where $\T\varphi$ is the canonical lift of $\varphi$ to $\T Q$. Under these hypotheses, the map $\Phi$ is said to be a \textbf{natural transformation} of $\T Q\times\R$.

On the other hand, given a vector field $Z\in\X(Q)$, we can define its \textbf{complete lift} to $\T Q\times\R$ as the vector field $Y\in\X(\T Q\times\R)$ whose local flow is the canonical lift of the local flow of $Z$ to $\T Q\times\R$; that is, the vector field $Y = Z^{\rm C}$, where $Z^{\rm C}$ denotes the complete lift of $Z$ to $\T Q$, identified in the natural way as a vector field in $\T Q\times\R$. An infinitesimal transformation of this type is called a \textbf{natural infinitesimal transformation} of $\T Q\times\R$.

It is well known that the vertical endomorphism $J$ and the Liouville vector field $\Delta_\circ$ in $\T Q$ are invariant under the action of canonical lifts of diffeomorphisms $\varphi\in\Diff(Q)$ and vector fields $Z\in\X(Q)$. Taking into account the definitions of the canonical endomorphism $\mathcal{J}$ and the Liouville vector field $\Delta$ in $\T Q\times\R$, it can be seen that canonical lifts of diffeomorphisms and vector fields from $Q$ to $\T Q\times\R$ preserve these canonical structures. They also preserve the Reeb vector field $\Reeb_\L$.

As an immediate consequence, we get a relation between Lagrangian-preserving natural transformations and contact symmetries:

\begin{prop}\label{prop:canonical-lift-Lagrangian-invariant}
	If $\Phi\colon\Diff(\T Q\times\R)$ (resp. $Y\in\X(\T Q\times\R)$) is the canonical lift of $\varphi\in\Diff(Q)$ to $\T Q\times\R$ (resp. of $Z\in\X(Q)$) that leaves the Lagrangian invariant, then it is a (infinitesimal) contact symmetry, that is,
	$$ \Phi^\ast\eta_\L = \eta_\L\ ,\quad \Phi^\ast E_\L = E_\L\qquad \mbox{(resp. $\Lie_Y\eta_\L = 0\ ,\quad \Lie_YE_\L = 0 $).} $$
	As a consequence, it is also a (infinitesimal) dynamical symmetry.
\end{prop}

As a corollary of the previous result, we have a similar result to the momentum dissipation Theorem \ref{thm:momentum-dissipation-contact}:

\begin{thm}\label{thm:contact-Lagrangian-not-depending-position}
	If $\tparder{\L}{q^i} = 0$, then $\parder{}{q^i}$ is an infinitesimal contact symmetry and its associated dissipated quantity is the momentum $\tparder{\L}{v^i}$:
	$$ \Lie_{X_\L}\left(\parder{\L}{v^i}\right) = -(\Lie_{\Reeb_\L}E_\L)\parder{\L}{v^i} = \parder{\L}{s}\parder{\L}{v^i}\,. $$
\end{thm}

In \cite{Geo2002}, a similar problem is considered where the dissipation factor used is $\parder{\L}{s}$ which, as we have seen in \eqref{eq:Lagrangian-energy-dissipation-contact}, is the same that we have obtained.

\section{Symmetries of a contactified system}
\label{sec:symmetries-contactified-system}

The dissipation theorem \ref{thm:dissipation-contact} provides a dissipated quantity from an infinitesimal dynamical symmetry $Y$ with no additional hypotheses, in contrast to Noether symmetries, where the generator of the symmetry is required to fulfill some additional conditions in order to yield a conserved quantity. Our will is to understand this different behaviour.

Consider a Hamiltonian system $(P, \omega, H_\circ)$ on an exact symplectic manifold $P$, with symplectic form $\omega = -\d\theta\in\Omega^2(P)$ and Hamiltonian function $H_\circ\in\Cinfty(P)$. Its associated Hamiltonian vector field $X_\circ$ is defined by
$$ i(X_\circ)\omega = \d H_\circ\,. $$
The contactified of $(P,\omega)$ is the contact manifold $(M,\eta)$, where $M = P\times\R$ is endowed with the contact form $\eta = \d s - \theta$; here $s$ is the canonical coordinate of $\R$ and we use $\theta$ for the pull-back of the 1-form $\theta\in\Omega^1(P)$ to the manifold $M$ (see Example \ref{ex:contactification-symplectic-manifold}).

The pull-back of the Hamiltonian function $H_\circ$ to $M$ defines a contact Hamiltonian function $H = H_\circ$ on $M$. The corresponding contact Hamiltonian vector field can be written as
$$ X_H = X_\circ + \ell\parder{}{s}\,, $$
where $X_\circ$ is the Hamiltonian vector field of $H_\circ$ as a vector field on the product manifold $M$, and $\ell = \langle\theta,X_\circ\rangle - H_\circ$.

Consider now a vector field $Y_\circ\in\X(P)$ and build the vector field $Y = Y_\circ + b\parder{}{s}$ with $b\in\Cinfty(P)$.

\begin{lem}
	The vector field $Y$ is a dynamical symmetry of the contact Hamiltonian system $(M,\eta, H)$ ($\Lie_YX_H = 0$) if, and only if, the vector field $Y_\circ$ is a dynamical symmetry of the symplectic Hamiltonian system $(P, \omega, H_\circ)$ ($\Lie_{Y_\circ}X_\circ = 0$) and $\Lie_Y\ell = \Lie_{X_H}b$.
\end{lem}
\begin{proof}
	The proof of this lemma is a simple computation:
	$$ [Y,X_H] = [Y_\circ, X_\circ] + (\Lie_Y\ell - \Lie_{X_H}b)\parder{}{s}\,. $$
\end{proof}

Now let us consider the quantity
$$ G = -i(Y)\eta = -i(Y_\circ)\eta - b\,. $$
Computing,
\begin{align*}
	\Lie_{X_H}G &= -i(Y)\Lie_{X_H}\eta - i([X_H, Y])\eta\\
	&= (\Lie_\Reeb H)i(Y)\eta - i(Y)(\Lie_{X_H}\eta + (\Lie_\Reeb H)\eta) + i([Y, X_H])\eta\\
	&= (\Lie_\Reeb H)i(Y)\eta - i(Y)(\Lie_{X_H}\eta + (\Lie_\Reeb H)\eta) + i([Y_\circ,X_\circ])\eta + (\Lie_Y\ell - \Lie_{X_H}b)\,.
\end{align*}
Let us analyze the vanishing of these summands: the first one is zero because $H = H_\circ$ does not depend on $s$, the second one also is because $X_H$ is the contact Hamiltonian vector field (see Theorem \ref{prop:equivalent-contact-Hamiltonian-equations}), and, according to the previous lemma, the third and fourth ones vanish if $\Lie_YX_H = 0$. Hence, we can conclude that if $Y$ is a dynamical symmetry, $G$ is a conserved quantity of the contact Hamiltonian system $(M, \eta, H)$. It is conserved instead of dissipated because the Hamiltonian $H$ does not depend on the variable $s$. Now,
$$ \Lie_{X_H}G = \Lie_{X_\circ}i(Y_\circ)\theta - \Lie_{X_H}b\,. $$
We obtain the following: when $Y_\circ$ is a dynamical symmetry of the symplectic Hamiltonian system, the function $G_\circ = i(Y)\theta$ is not necessarily a conserved quantity because the hypotheses of Noether's theorem are not necessarily fulfilled. However, in the contactified Hamiltonian system $(M, \eta, H)$ with $Y$ a dynamical symmetry under the conditions of the previous lemma, we have
$$ \Lie_{X_H}G_\circ = \Lie_{X_H}(G + b) = \Lie_{X_H}b = \Lie_Y\ell\,, $$
and the time-derivative of $G_\circ$ is compensated with the time-derivative of $b$.

Now the question is when is $G_\circ$ conserved under the action of $X_\circ$? The answer is when $\Lie_{Y_\circ}\ell = 0$. This happens, for example, when $Y_\circ$ is an exact Noether symmetry, i.e. when $\Lie_{Y_\circ}\theta = 0$ and $\Lie_{Y_\circ}H_\circ = 0$, since
$$ \Lie_{Y_\circ}\ell = \Lie_{Y_\circ}(i(X_\circ)\theta - H_\circ) = 0\,. $$

	\chapter[Skinner--Rusk formalism for contact systems]{Skinner--Rusk formalism for \phantom{n} contact systems}
	\label{ch:Skinner-Rusk-contact}
		

This chapter is devoted to generalize the formalism first introduced in local coordinates in \cite{Kam1982} and later developed by R. Skinner and R. Rusk in \cite{Ski1983} to the case of contact mechanical systems. The Skinner--Rusk formalism was developed in order to find an alternative way to deal with both regular and singular Lagrangians. This can be achieved by mixing in one unified formalism both Lagrangian and Hamiltonian formalisms.

In Section \ref{sec:extended-Pontryagin-bundle-contact} we define the extended Pontryagin bundle $\W$ and describe its canonical precontact structure, introducing its canonical contact 1-form, the coupling function and the canonical 1- and 2-forms. This will allow us to introduce in Section \ref{sec:contact-dynamical-equations} the contact dynamical equations for the precontact Hamiltonian system $(\W,\eta,\H)$. Then, as the system is singular, we need to apply the constraint algorithm described in \cite{DeLeo2019}. We study the constraints that arise and, in particular, we recover the holonomy condition and the Legendre map. Finally, in Section \ref{sec:SR-contact-recovering} we show that both the Lagrangian and Hamiltonian formalisms can be recovered from the Skinner--Rusk formalism. See \cite{DeLeo2020}.

\section{The extended Pontryagin bundle: precontact canonical structure}
\label{sec:extended-Pontryagin-bundle-contact}

For a contact dynamical system, the configuration space is $Q\times\R$, where $Q$ is an $n$-dimensional manifold, with coordinates $(q^i, s)$. Consider now the bundles $\T Q\times\R$ and $\cT Q\times\R$ endowed with natural coordinates $(q^i, v^i, s)$ and $(q^i, p_i, s)$ adapted to the bundle structures. Consider also the canonical projections
\begin{gather*}
	\tau_1\colon\T Q\times\R\to\T Q\quad,\quad\tau_0\colon\T Q\times\R\to Q\times\R\,,\\
	\pi_1\colon\cT Q\times\R\to\cT Q\quad,\quad\pi_0\colon\cT Q\times\R\to Q\times\R\,.
\end{gather*}
We denote by $\d s$ the volume form in $\R$, and its pull-backs to all the product manifolds. Let $\theta_\circ\in\Omega^1(\cT Q)$ and $\omega_\circ = -\d\theta_\circ\in\Omega^2(\cT Q)$ be the canonical forms of the cotangent bundle $\cT Q$, whose expressions in coordinates are $\theta_\circ = p_i\d q^i$ and $\omega_\circ = \d q^i\wedge\d p_i$. Denote $\theta = \pi_1^\ast\theta_\circ\in\Omega^1(\cT Q\times\R)$ and $\omega = \pi_1^\ast\omega_\circ\in\Omega^2(\cT Q\times\R)$. Notice that $\omega = -\d\theta$.

\begin{dfn}\label{dfn:contact-pontryagin-bundle}
	We define the extended \textbf{Pontryagin bundle}
	$$ \W = \T Q\times_Q\cT Q\times\R\,, $$
	endowed with the natural submersions
	\begin{align*}
		& \rho_1\colon\W\to\T Q\times\R\,,\\
		& \rho_2\colon\W\to\cT Q\times\R\,,\\
		& \rho_0\colon\W\to Q\times\R\,,\\
		& s\colon\W\to\R\,.
	\end{align*}
\end{dfn}

The extended Pontryagin bundle has natural coordinates $(q^i, v^i, p_i, s)$.

\begin{dfn}
	Consider a path $\gamma\colon\R\to\W$. The path $\gamma$ is said to be \textbf{holonomic} in $\W$ if the path $\rho_1\circ\gamma\colon\R\to\T Q\times\R$ is holonomic.

	A vector field $X\in\X(\W)$ is said to satisfy the \textbf{second-order condition} or to be a {\sc sode} in $\W$ if its integral curves are holonomic in $\W$.
\end{dfn}

A holonomic path in $\W$ has local expression
$$ \gamma = \left(q^i(t), \dot q^i(t), p_i(t), s(t)\right)\,. $$
A \textsc{sode} in $\W$ reads as
$$ X = v^i\parder{}{q^i} + F^i\parder{}{v^i} + G_i\parder{}{p_i} + f\parder{}{s}\,. $$
The extended Pontryagin bundle $\W$ defined in \ref{dfn:contact-pontryagin-bundle} has the following natural structures:
\begin{dfn}
	\begin{enumerate}[{\rm(1)}]
		\item The \textbf{coupling function} in $\W$ is the map $\C\colon\W\to\R$ defined as
		$$ \C(w) = \langle {\rm p}_q,{\rm v}_q\rangle\,, $$
		where $w = ({\rm v}_q,{\rm p}_q, s)\in\W$, $q\in Q$, ${\rm p}_q\in\cT Q$, and ${\rm v}_q\in\T Q$.
		\item The \textbf{canonical 1-form} is the $\rho_0$-semibasic form $\Theta = \rho_2^\ast\theta\in\Omega^1(\W)$. The \textbf{canonical 2-form} is $\Omega = -\d\Theta = \rho_2^\ast\omega\in\Omega^2(\W)$.
		\item The \textbf{canonical contact 1-form} is the $\rho_1$-semibasic form $\eta = \d s - \Theta\in\Omega^1(\W)$.
	\end{enumerate}
\end{dfn}
Taking natural coordinates,
$$ \Theta = p_i\d q^i\ ,\qquad \eta = \d s - p_i\d q^i\ ,\qquad \d\eta = \d q^i\wedge\d p_i = \Omega \,.$$

\begin{dfn}\label{dfn:SR-contact-hamiltonian-function}
	Let $L\in\Cinfty(\T Q\times\R)$ be a Lagrangian function and consider the Lagrangian $\L = \rho_1^\ast L\in\Cinfty(\W)$. The \textbf{Hamiltonian function associated to} $\L$ is the function
	$$ \H = \C - \L = p_iv^i - \L(q^j, v^j, s)\in\Cinfty(\W)\,. $$
\end{dfn}

\begin{rmrk}\rm
	Notice that the canonical contact 1-form $\eta$ is a precontact form in $\W$. Thus, $(\W,\eta)$ is a precontact manifold and $(\W,\eta,\H)$ is a precontact Hamiltonian system. These concepts where introduced in \cite{DeLeo2019}. Then, equations \eqref{eq:Reeb-condition} do not have a unique solution and the Reeb vector field is not uniquely defined. Actually, in natural coordinates, the general solution to equations \eqref{eq:Reeb-condition} is
	$$ \Reeb = \parder{}{s} + F^i\parder{}{v^i}\,, $$
	for arbitrary coefficients $F^i$. Nevertheless, the formalism is independent on the choice of the Reeb vector field. In this particular case, as $\W$ is a trivial bundle over $\R$, the canonical vector field $\tparder{}{s}\in\X(\R)$ can be canonically lifted to $\W$ and used as Reeb vector field.
\end{rmrk}

\section{Contact dynamical equations}
\label{sec:contact-dynamical-equations}

\begin{dfn}
	The \textbf{Lagrangian--Hamiltonian problem} associated to the precontact system $(\W,\eta,\H)$ consists in finding the integral curves of a vector field $X_\H\in\X(\W)$ such that
	$$ \flat(X_\H) = \d\H - (\Lie_\Reeb\H + \H)\eta\,, $$
	that is, which is a solution of the contact Hamiltonian equations \eqref{eq:contact-hamilton-equations-fields}:
	\begin{equation}\label{eq:SR-contact-hamilton-equations-fields}
		\begin{cases}
			i(X_\H)\d\eta = \d\H - (\Lie_\Reeb\H)\eta\,,\\
			i(X_\H)\eta = -\H\,,
		\end{cases}
	\end{equation}
	or, what is equivalent,
	\begin{equation*}
		\begin{cases}
			\Lie_{X_\H}\eta = -(\Lie_\Reeb\H)\eta\,,\\
			i(X_\H)\eta = -\H\,.
		\end{cases}
	\end{equation*}
	Then, the integral curves $\gamma\colon I\subset\R\to\W$ of $X_\H$ are the solutions to the equations
	\begin{equation}\label{eq:SR-contact-hamilton-equations-curves}
		\begin{cases}
			i(\gamma')\d\eta = (\d\H - (\Lie_\Reeb\H)\eta)\circ\gamma\,,\\
			i(\gamma')\eta = -\H\circ\gamma\,,
		\end{cases}
	\end{equation}
\end{dfn}

As $(\W,\eta,\H)$ is a precontact Hamiltonian system, equations \eqref{eq:SR-contact-hamilton-equations-fields} are not necessarily consistent everywhere in $\W$. Hence, we need to implement the standard constraint algorithm in order to find the final constraint submanifold (if it exists) in which there exist consitent solutions to equations \eqref{eq:SR-contact-hamilton-equations-fields}. In what follows, we will detail this procedure.

Take a natural chart $(q^i, v^i, p_i, s)$ in $\W$. The vector field $X_\H\in\X(\W)$ has local expression
$$ X_\H = f^i\parder{}{q^i} + F^i\parder{}{v^i} + G_i\parder{}{p_i} + f\parder{}{s}\,. $$

Therefore we have
\begin{align*}
	i(X_\H)\eta &= f - f^ip_i\,,\\
	i(X_\H)\d\eta &= f^i\d p_i - G_i\d q^i\,,
\end{align*}
and
\begin{align*}
	\d\H &= v^i\d p_i + \left(p_i - \parder{\L}{v^i}\right)\d v^i - \parder{\L}{q^i}\d q^i - \parder{\L}{s}\d s\,,\\
	(\Lie_\Reeb\H)\eta &= -\parder{\L}{s}(\d s - p_i\d q^i)\,.
\end{align*}
Then, the second equation in \eqref{eq:SR-contact-hamilton-equations-fields} gives
\begin{equation}\label{eq:SR-contact-second}
	f = (f^i - v^i)p_i + \L\,,
\end{equation}
while the first equation in \eqref{eq:SR-contact-hamilton-equations-fields} leads to
\begin{align}
	f^i &= v^i & &(\mbox{coefficients in }\d p_i)\,,\label{eq:SR-contact-one}\\
	p_i &= \parder{\L}{v^i} & &(\mbox{coefficients in }\d v^i)\,,\label{eq:SR-contact-two}\\
	G_i &= \parder{\L}{q^i} + p_i\parder{\L}{s} & &(\mbox{coefficients in }\d q^i)\,,\label{eq:SR-contact-three}
\end{align}
and the conditions from the coefficients in $\d s$ hold identically. From these conditions, we have:
\begin{itemize}
	\item Equations \eqref{eq:SR-contact-one} are the holonomy conditions. This implies that $X_\H$ is a {\sc sode}. As usual, the {\sc sode} condition arises straightforwardly from the Skinner--Rusk formalism. This reflects the fact that this geometric condition in the Skinner--Rusk formalism is stronger than in the standard Lagrangian formalism.
	\item Conditions \eqref{eq:SR-contact-two} are algebraic equations defining a submanifold $\W_1\hookrightarrow\W$, the \textbf{first constraint submanifold} of the Hamiltonian precontact system $(\W,\eta,\H)$. $\W_1$ is the graph of the Legendre map $\F L$ introduced in Definition \ref{dfn:contact-legendre-map}:
	$$ \W_1 = \{({\rm v}_q, \F L({\rm v}_q, s))\in\W\ \vert\ ({\rm v}_q,s)\in\T Q\times\R\}\,. $$
	Notice that this implies that the Skinner--Rusk formalism includes the definition of the Legendre map as a consequence of the constraint algorithm.
\end{itemize}

Hence, the vector fields solution to equations \eqref{eq:SR-contact-hamilton-equations-fields} have the form
$$ X_\H = v^i\parder{}{q^i} + F^i\parder{}{v^i} + \left( \parder{\L}{q^i} + p_i\parder{\L}{s} \right)\parder{}{p_i} + \L\parder{}{s}\qquad\mbox{(on $\W_1$),} $$
where $F^i$ are arbitrary functions.

Now, the constraint algorithm continues by imposing the tangency of $X_\H$ to $\W_1$, to ensure that the dynamic trajectories remain in $\W_1$. The constraints defining $\W_1$ are
$$ \xi_j^1 = p_j - \parder{\L}{v^j}\in\Cinfty(\W)\,. $$
The tangency condition $X_\H\left(\xi_j^1\right) = 0$ on $\W_1$ reads
\begin{equation}\label{eq:SR-contact-tangency-W1}
	0 = -\parderr{\L}{q^i}{v^j}v^i - \parderr{\L}{v^i}{v^j}F^i - \L\parderr{\L}{s}{v^j} + \parder{\L}{q^j} + p_j\parder{\L}{s}\quad\mbox{(on $\W_1$).}
\end{equation}
Once we get to this point, we have to distinguish two different cases:
\begin{itemize}
	\item If the Lagrangian $L$ is regular, equations \eqref{eq:SR-contact-tangency-W1} allow us to determine all the functions $F^i = \frac{\d v^i}{\d t}$. In this case, the algorithm ends and the solution is unique.
	\item On the other hand, if the Lagrangian $L$ is singular, these equation establish relations among the coefficients $F^i$: some of them may remain undetermined and the solutions may not be unique. Moreover, new constraints $\xi^2_\mu\in\Cinfty(\W)$ may appear. These new constraints define a submanifold $\W_2\hookrightarrow\W_1\hookrightarrow\W$. The algorithm continues by demanding that $X_\H$ must be tangent to the new submanifold $\W_2$ and so on until we find a final constraint submanifold $\W_f$ (if it exists) where we have tangent solutions $X_\H$.  
\end{itemize}

Let $\gamma(t) = (q^i(t), v^i(t), p_i(t), s(t))$ be an integral curve of $X_\H$. We have that $f^i = \dot q^i$, $F^i = \dot v^i$, $G_i = \dot p_i$ and $f = \dot s$. Then, equations \eqref{eq:SR-contact-second}, \eqref{eq:SR-contact-one}, \eqref{eq:SR-contact-two} and \eqref{eq:SR-contact-three} lead to the coordinate expression of equations \eqref{eq:SR-contact-hamilton-equations-curves}. In particular,
\begin{itemize}
	\item Equation \eqref{eq:SR-contact-one} implies that $v^i = \dot q^i$, that is, the holonomy condition.
	\item Using \eqref{eq:SR-contact-one}, equation \eqref{eq:SR-contact-second} gives
	\begin{equation}\label{eq:SR-contact-Euler-Lagrange-2}
		\dot s = \L\,,
	\end{equation}
	which is equation \eqref{eq:contact-Euler-Lagrange-2}.
	\item Conditions \eqref{eq:SR-contact-three} are
	$$ \dot p_i = \parder{\L}{q^i} + p_i\parder{\L}{s} = -\left( \parder{\H}{q^i} + p_i\parder{\H}{s} \right)\,, $$
	which are the second group of Hamilton's equations \eqref{eq:contact-hamiltonian-equations-darboux-coordinates}. Now, using \eqref{eq:SR-contact-two}, that is, on the submanifold $\W_1$, these equations are
	$$ \frac{\d}{\d t}\left( \parder{\L}{v^i} \right) = \parder{\L}{q^i} + \parder{\L}{v^i}\parder{\L}{s}\,, $$
	which are the Euler--Lagrange equations \eqref{eq:contact-Euler-Lagrange-1}. The first group of Hamilton's equations \eqref{eq:contact-hamiltonian-equations-darboux-coordinates} arises from Definition \ref{dfn:SR-contact-hamiltonian-function}, taking into account the holonomy condition.
	\item Using conditions \eqref{eq:SR-contact-two} (i.e. on $\W_1$) and \eqref{eq:SR-contact-Euler-Lagrange-2}, the tangency condition \eqref{eq:SR-contact-tangency-W1} gives again the contact Euler--Lagrange equations \eqref{eq:contact-Euler-Lagrange-1}. Notice that if the Lagrangian $L$ is singular, these equation might be incompatible.
\end{itemize}

\section[Recovering the Lagrangian and Hamiltonian formalisms]{Recovering the Lagrangian and Hamiltonian \phantom{n} formalisms and equivalence}
\label{sec:SR-contact-recovering}

In this section we are going to the equivalence between the Skinner--Rusk formalism and the Lagrangian and Hamiltonian formalisms. First of all, notice that if we denote $\jmath_1\colon\W_1\hookrightarrow\W$ the natural embedding, we have
\begin{equation*}
	(\rho_1\circ\jmath_1)(\W_1) = \T Q\times\R\ ,\quad (\rho_2\circ\jmath_1)(\W_1) = P_1\subseteq\cT Q\times\R\,.
\end{equation*}

\begin{figure}[h]
	\centering
	\begin{tikzcd}
		& \W \arrow[ddr, bend left, "\rho_2"] \arrow[ddl, bend right,  swap, "\rho_1"] \\
		& \W_1 \arrow[dl] \arrow[dr] \arrow[u, hookrightarrow, "\jmath_1"] \\
		\T Q\times\R \arrow[rr, "\F L"] \arrow[rrd, pos=0.7, "\F L_1"] & & \cT Q\times\R \\
		& \W_f \arrow[dl] \arrow[dr] \arrow[uuu, bend right=40, hookrightarrow, crossing over, swap, pos=0.7, "\jmath_f"]  & P_1 \arrow[u, hookrightarrow, swap, "j_1"] \\
		S_f \arrow[rr] \arrow[uu, hookrightarrow] & & P_f \arrow[u, hookrightarrow]
	\end{tikzcd}
	\caption{Recovering the Lagrangian and Hamiltonian formalisms}
	\label{fig:diagram-contact-Skinner-Rusk}
\end{figure}
In particular, $P_1\subset\cT Q\times\R$ is a submanifold if the Lagrangian $L$ is almost-regular, an open subset of $\cT Q\times\R$ if $L$ is regular or $P_1 = \cT Q\times\R$ if $L$ is hyperregular. Furthermore, as the first constraint submanifold $\W_1$ is the graph of the Legendre map $\F L$, the restriction projection $\rho_1\circ\jmath_1\colon\W_1\to\T Q\times\R$ is a diffeomorphism. In the same way, if $L$ is almost-regular, for every submanifold $\jmath_\alpha\colon\W_\alpha\hookrightarrow\W$ obtained from the constraint algorithm, we have
\begin{equation*}
	(\rho_1\circ\jmath_\alpha)(\W_\alpha) = S_\alpha\hookrightarrow \T Q\times\R\ ,\quad (\rho_2\circ\jmath_\alpha)(\W_\alpha) = P_\alpha\hookrightarrow P_1\subseteq\cT Q\times\R\,.
\end{equation*}
Notice that $\W_\alpha\subset\W_1 = \graph\F L$ implies $\F L(S_\alpha) = P_\alpha$. Now, let $\jmath_f\colon\W_f\hookrightarrow\W$ be the final constraint submanifold and
\begin{equation*}
	(\rho_1\circ\jmath_f)(\W_f) = S_f\hookrightarrow \T Q\times\R\ ,\quad (\rho_2\circ\jmath_f)(\W_f) = P_f\hookrightarrow P_1\subseteq\cT Q\times\R\,.
\end{equation*}
This situation is represented in diagram in Figure \ref{fig:diagram-contact-Skinner-Rusk}.

Every function or differential form on $\W$ or vector field on $\W$ and tangent to $\W_1$ can be restricted to $\W_1$. Hence, they can be translated to the Lagrangian side using that $\W_1\cong\T Q\times\R$ or to the Hamiltonian side projecting to the second factors of the product bundle, $\cT Q\times\R$. With all this in mind, we have the following result:
\begin{thm}
	Let $\gamma\colon I\subseteq\R\to\W$ be a path taking values in $\W_1$. It can be split as $\gamma = (\gamma_L, \gamma_H)$, where $\gamma_L = \rho_1\circ\gamma\colon I\subseteq\R\to\T Q\times\R$ and $\gamma_H = \F L\circ\gamma_L\colon I\subseteq\R\to P_1\subseteq\cT Q\times\R$.

	Consider a path $\gamma\colon I\subseteq\R\to\W$, with $\Ima(\gamma)\subset\W_1$ satisfying equations \eqref{eq:SR-contact-hamilton-equations-curves} (at least on a submanifold $\W_f\subset\W_1$). Then, $\gamma_L$ is the prolongation to $\T Q\times\R$ of the projected curve $\sigma = \rho_0\circ\gamma\colon\R\to Q\times\R$ ($\gamma$ is holonomic), and it is a solution to \eqref{eq:contact-euler-lagrange-holonomic-curve}. In addition, the path $\gamma_H = \F L\circ\sigma'$ is solution to \eqref{eq:contact-hamilton-equations-curves} (on $P_f$).

	Conversely, if $\sigma\colon\R\to Q\times\R$ is a path such that $\sigma'$ is a solution to \eqref{eq:contact-euler-lagrange-holonomic-curve} (on $S_f$), then the path $\gamma = (\sigma', \F L\circ\sigma')$ is a solution to \eqref{eq:SR-contact-hamilton-equations-curves}. Moreover, $\F L\circ\sigma'$ is a solution to \eqref{eq:contact-hamilton-equations-curves} (on $P_f$).
\end{thm}

It is important to point out that in the case the Lagrangian $L$ is singular, these results hold on the submanifolds $\W_f$, $S_f$ and $P_f$.

Considering that the paths $\gamma\colon\R\to\W$ solution to \eqref{eq:SR-contact-hamilton-equations-curves} are the integral curves of {\sc sode}s $X_\H\in\X(\W)$ solution to \eqref{eq:SR-contact-hamilton-equations-fields} and that the paths $\gamma_L\colon\R\to\T Q\times\R$ are the integral curves of {\sc sode}s $X_L\in\X(\T Q\times\R)$ solution to \eqref{eq:contact-euler-lagrange-holonomic-curve}, then we have:

\begin{thm}
	Consider a vector field $X_\H\in\X(\W)$ solution to \eqref{eq:SR-contact-hamilton-equations-fields} (at least on $\W_f$) and tangent to $\W_1$ (resp. to $\W_f$). Then, $X_L\in\X(\T Q\times\R)$, defined as $X_L\circ\rho_1 = \T\rho_1\circ X_\H$, is a {\sc sode} tangent to $S_f$ solution to \eqref{eq:contact-Lagrangian-equations} (on $S_f$), with $\H = \rho_1^\ast E_L$.

	Furthermore, every {\sc sode} solution to \eqref{eq:contact-euler-lagrange-holonomic-curve} (on $S_f$) can be obtained in this way from a vector field $X_\H\in\X(\W)$ (tangent to $\W_f$) solution to \eqref{eq:SR-contact-hamilton-equations-fields} (on $\W_f$).
\end{thm}

We can also recover the Hamiltonian formalism in a similar way having in mind that the paths $\gamma_H\colon\R\to\cT Q\times\R$ are the integrals curves of vector fields $X_H\in\X(\cT Q\times\R)$ solution to \eqref{eq:contact-hamilton-equations-fields}.

\begin{thm}
	Consider a vector field $X_\H\in\X(\W)$ solution to equations \eqref{eq:SR-contact-hamilton-equations-fields} (at least on $\W_f$) and tangent to $\W_1$ (resp. to $\W_f$). The vector field $X_H\in\X(\cT Q\times\R)$ defined as $X_H\circ\rho_2 = \T\rho_2\circ X_\H$ is a solution to \eqref{eq:contact-hamilton-equations-fields} (on $P_f$ and tangent to $P_f$), whith $\H = \rho_2^\ast H$.
\end{thm}

These results correspond to those obtained from the Skinner--Rusk formalism for non-autonomous dynamical systems. See \cite{Bar2008,Can2002}.

It is important to remark that, in the case of singular Lagrangians, we only have equivalence between the constraint algorithms in the Skinner--Rusk and in the Lagrangian formalisms if we impose the second-order condition to the Lagrangian formalism as an additional condition. This is because, unlike in the Skinner--Rusk formalism, the holonomy condition cannot be recovered from the Lagrangian formalism when dealing with singular Lagrangians. See \cite{Mun1992,Ski1983}.

	\chapter{Examples in mechanics}
	\label{ch:contact-examples}
		

This last chapter of the first part is devoted to study some examples of contact mechanical systems. We will analyze them with different levels of detail.

The first example \ref{sec:damped-harmonic-oscillator} is the \textbf{damped harmonic oscillator}. We will consider the Lagrangian function of the harmonic oscillator and add to it a holonomic dissipation term. In this way we can obtain the equation of a damped harmonic oscillator. In this first example, we will give a complete description of the Lagrangian, Hamiltonian and Skinner--Rusk formulations. We will also see the energy dissipation law for this system.

In the second example \ref{sec:motion-constant-gravitational-field-friction} we will describe the \textbf{motion of a particle in a constant gravitational field with friction}. We will develop the Lagrangian formalism of this system give its energy dissipation law. We will also find a contact symmetry which will allow us to obtained its associated dissipated quantity. With these two dissipated quantities, we will find a conserved quantity.

The third example \ref{sec:parachute-equation} describes the fall of a \textbf{parachute}. This example is a particularly interesting one, because in it we consider a Lagrangian function which is {\it not} a Lagrangian with holonomic dissipation term. We see that when we make the friction go to zero, we recover the Lagrangian used in the previous example. We develop the Lagrangian formalism for this system and give its energy dissipation law.

In the fourth example \ref{sec:Lagrangian-holonomic-dissipation-term} we give a complete description of the Skinner--Rusk formalism for \textbf{Lagrangians with holonomic dissipation term}, including the constraint algorithm. These Lagrangians were introduced in \cite{Cia2018} and \cite{Gas2019} and are very common.

The fifth example \ref{sec:central-force-dissipation} develops the Skinner--Rusk formalism of a system consisting of a particle submitted to a \textbf{central potential with friction}. The corresponding Lagrangian function is regular and thus the constraint algorithm finishes in one step.

In the sixth example \ref{sec:Lagrange-multipliers} we deal with a \textbf{damped simple pendulum}. We use the \textbf{method of Lagrange multipliers} to obtain a Lagrangian function describing the behaviour of the pendulum restricted to the circumference $r = \ell$. The Lagrangians obtained this way are always singular because the velocities associated with the Lagrange multiplier do not appear in the Lagrangian. We develop the Skinner--Rusk formalism for this system.

Finally, Example \ref{sec:Cawleys-Lagrangian} develops the Skinner--Rusk formalism for the \textbf{Cawley's Lagrangian} \cite{Caw1979} modified by adding a dissipation term. This is a singular Lagrangian.

\section{The damped harmonic oscillator}
\label{sec:damped-harmonic-oscillator}

	Consider the configuration space $Q = \R$. The Lagrangian description of the one-dimensional harmonic oscillator is given by the Lagrangian function $L\colon\T Q\to\R$, where
	$$ L(q,v) = \frac{1}{2}mv^2 - \frac{1}{2}m\omega^2q^2\,. $$
	The Euler--Lagrange equation for this Lagrangian is
	$$ \ddot q + \omega^2q = 0\,, $$
	which is the equation of a harmonic oscillator.

	\subsection*{Contact Lagrangian formulation}

	Consider now the Lagrangian function $\L\colon\T Q\times\R\to\R$ given by
	\begin{equation}\label{eq:lagrangian-damped-harmonic-oscillator}
		\L(q, v, s) = L(q,v) - \gamma s = \frac{1}{2}mv^2 - \frac{1}{2}m\omega^2q^2 - \gamma s\,.
	\end{equation}
	Since
	$$ \Delta = v\parder{}{v}\ ,\quad J = \parder{}{v}\otimes\d q\,, $$
	we have that
	\begin{align*}
		\d\L &= mv\d v - m\omega^2 q\d q - \gamma \d s\,,\\
		E_\L &= \Delta(\L)-\L = \frac{1}{2}mv^2 + \frac{1}{2}m\omega^2 q^2 + \gamma s\,,\\
		\d E_\L &= mv\d v + m\omega^2 q\d q + \gamma \d s\,,\\
		\theta_\L &= \transp{\mathcal{J}}\circ\d\L = mv\d q\,,\\
		\eta_\L &= \d s - mv\d q\,,\\
		\d\eta_\L &= m\d q\wedge\d v\,,\\
		\Reeb &= \parder{}{s}\,.
	\end{align*}
	Consider a generic vector field $X\in\X(\T Q\times\R)$ with local expression
	$$ X = f\parder{}{q} + F\parder{}{v} + g\parder{}{s}\,. $$
	The left-hand side of the first contact Lagrangian equation \eqref{eq:contact-Lagrangian-equations} is
	$$ i(X)\d\eta_\L = mf\d v - mF\d q\,, $$
	while the right-hand side is
	$$ \d E_\L - \Reeb(E_\L)\eta_\L = mv\d v + m\omega^2 q\d q + mv\gamma\d q\,. $$
	Equating both expressions, we get the conditions
	\begin{equation*}
		\begin{dcases}
			f = v\,,\\
			F = -\omega^2 q - \gamma v\,.
		\end{dcases}
	\end{equation*}
	On the other hand, the second equation in \eqref{eq:contact-Lagrangian-equations} yields
	$$ g = \frac{1}{2}mv^2 - \frac{1}{2}m\omega^2q^2 - \gamma s = \L\,. $$
	Then, if the vector field $X$ is a solution to the contact Lagrangian equations \eqref{eq:contact-Lagrangian-equations}, it has local expression
	$$ X = v\parder{}{q} + (-\omega^2 q - \gamma v)\parder{}{v} + \L\parder{}{s}\,. $$
	Notice that the {\sc sode} condition for $X$ is automatically satisfied as the Lagrangian $\L$ is regular.

	Let $\sigma(t) = (q(t),v(t),s(t))$ be an integral curve of the vector field $X$. Then, it satisfies the following system of differential equations:
	$$
		\begin{dcases}
			\dot q = v\,,\\
			\dot v = -\omega^2 q - \gamma v\,,\\
			\dot s = \L\,.
		\end{dcases}
	$$
	The first two equations of this system can be combined to obtain the second-order differential equation
	$$ \ddot q + \gamma \dot q + \omega^2 q = 0\,, $$
	which corresponds to a damped harmonic oscillator.

	The dissipation of the energy is given by Theorem \ref{thm:energy-dissipation-contact}:
	$$ \Lie_X E_\L = -\gamma E_\L\,. $$

	\subsection*{Contact Hamiltonian formulation}

	Consider the contact manifold $(\cT Q\times\R,\eta)$ with natural coordinates $(q, p, s)$, where $\eta = \d s - p\d q$. The Reeb vector field is $\Reeb = \tparder{}{s}$.

	The Legendre map associated to the Lagrangian function $\L$ given in \eqref{eq:lagrangian-damped-harmonic-oscillator} is the map $\F\L\colon\T Q\times\R\to\cT Q\times\R$ given by
	$$ \F\L(q,v,s) = (q, p=mv, s)\in\cT Q\times\R\,. $$
	The Hamiltonian function defined by $\F\L^\ast H = E_\L$ is
	$$ H = \frac{1}{2m}p^2 + \frac{1}{2}m\omega^2 q^2 + \gamma s\,. $$
	Its differential is
	$$ \d H = \frac{p}{m}\d p + m\omega^2 q\d q + \gamma \d s\,. $$
	Consider the vector field $Y\in\X(\cT Q\times\R)$ with local expression
	$$ Y = f\parder{}{q} + G\parder{}{p} + g\parder{}{s}\,. $$
	The left-hand side of the first contact Hamiltonian equation \eqref{eq:contact-hamilton-equations-fields} reads
	$$ i(Y)\d\eta = f\d p - G\d q\,, $$
	while the right-hand side is
	$$ \d H - \Reeb(H)\eta = \frac{p}{m}\d p + \gamma p\d q + m\omega^2 q\d q\,, $$
	and equating them we obtain the conditions
	$$
		\begin{dcases}
			f = \frac{p}{m}\,,\\
			G = -m\omega^2 q - \gamma p\,.
		\end{dcases}
	$$
	On the other hand, the second equation in \eqref{eq:contact-hamilton-equations-fields} gives
	$$ g = \frac{1}{2m}p^2 - \frac{1}{2}m\omega^2 q^2 - \gamma s\,. $$
	Hence, the vector field $Y$ has local expression
	$$ Y = \frac{p}{m}\parder{}{q} + (-m\omega^2 q - \gamma p)\parder{}{p} + \left(\frac{1}{2m}p^2 - \frac{1}{2}m\omega^2 q^2 - \gamma s\right)\parder{}{s}\,. $$
	An integral curve $\sigma(t) = (q(t),p(t),s(t))$ of the vector field $Y$ satisfies the system of differential equations
	$$
		\begin{dcases}
			\dot q = \frac{p}{m}\,,\\
			\dot p = -m\omega^2 q - \gamma p\,,\\
			\dot s = \frac{1}{2m}p^2 - \frac{1}{2}m\omega^2 q^2 - \gamma s
		\end{dcases}
	$$
	Combining the first two equations in of the system above, we obtain the second-order differential equation
	$$ \ddot q + \gamma \dot q + \omega^2 q = 0\,, $$
	which is the equation of a damped harmonic oscillator. The dissipation of the Hamiltonian function is given by Theorem \ref{thm:energy-dissipation-contact}:
	$$ \Lie_Y H = -\gamma H\,. $$

	\subsection*{Contact Skinner--Rusk formulation}

	We have already seen the contact Lagrangian and Hamiltonian formulations of the damped harmonic oscillator. Now, we are going to state the Skinner--Rusk formulation and we will also recover the Lagrangian and Hamiltonian cases from it.

	Consider the extended Pontryagin bundle
	$$ \W = \T Q\times_Q\cT Q\times\R\,, $$
	with natural coordinates $(q, v, p, s)$. The coupling function is $\C = pv$. The 1-form $\eta = \d s - p\d q$ defines a precontact structure on $\W$ with Reeb vector field $\Reeb = \tparder{}{s}$. We have that $\d\eta = \d q\wedge\d p$. The Hamiltonian function associated to the Lagrangian $\L$ given in \eqref{eq:lagrangian-damped-harmonic-oscillator} is the function
	$$ \H = \C - \L = pv - \frac{1}{2}mv^2 + \frac{1}{2}m\omega^2 q^2 + \gamma s\in\Cinfty(\W)\,. $$
	We have
	$$ \d\H = p\d v + v\d p - mv\d v + m\omega^2 q\d q + \gamma \d s\,, $$
	and hence
	$$ \d\H - \Reeb(\H)\eta = (p-mv)\d v + v\d p + (m\omega^2 q + \gamma p)\d q\,. $$
	Given a vector field $Z\in\X(\W)$ with coordinate expression
	$$ Z = f\parder{}{q} + F\parder{}{v} + G\parder{}{p} + g\parder{}{s}\,, $$
	equations \eqref{eq:SR-contact-hamilton-equations-fields} give the conditions
	\begin{align*}
		G &= -m\omega^2 q - \gamma p\,, & f = v\,,\\
		p &= mv\,,&  g = \L\,.
	\end{align*}
	Hence, the vector field $Z$ is a {\sc sode} and has local expression
	$$ Z = v\parder{}{q} + F\parder{}{v} + (-m\omega^2 q - \gamma p)\parder{}{p} + \L\parder{}{s}\,, $$
	and we have the constraint function
	$$ \xi_1 = p - mv = 0 $$
	defining the first constraint submanifold $\W_1\hookrightarrow\W$. Now, we have to impose the tangency of the vector field $Z$ to the submanifold $\W_1$:
	$$ 0 = Z(\xi_1) = -m\omega^2 q - \gamma p - mF\,, $$
	we get the condition
	$$ F = -\omega^2 q - \gamma v $$
	and no new constraints appear. Then, we have the unique solution
	$$ Z = v\parder{}{q} + (-\omega^2 q - \gamma v)\parder{}{v} + (-m\omega^2 q - \gamma p)\parder{}{p} + \L\parder{}{s}\,. $$
	Projecting onto each factor of $\W = \T Q\times_Q\cT Q\times\R$ using the projections $\rho_1,\rho_2$, we recover the Lagrangian and Hamiltonian formalisms described above. In the Lagrangian formalism we obtain the holonomic vector field $X\in\X(\T Q\times\R)$ given by
	$$ X = v\parder{}{q} + (-\omega^2 q - \gamma v)\parder{}{v} + \L\parder{}{s}\,, $$
	while in the Hamiltonian formalism we get the vector field $Y\in\X(\cT Q\times\R)$ given by
	$$ Y = \frac{p}{m}\parder{}{q} + (-m\omega^2 q - \gamma p)\parder{}{p} + \left(\frac{1}{2m}p^2 - \frac{1}{2}m\omega^2 q^2 - \gamma s\right)\parder{}{s}\,. $$

\section{Motion in a constant gravitational field with friction}
\label{sec:motion-constant-gravitational-field-friction}

	Consider the motion of a particle in a vertical plane under the action of constant gravity. In this case, $Q = \R^2$ with coordinates $(x,y)$. This motion can be described by the Lagrangian function
	\begin{equation}\label{eq:lagrangian-falling-particle-without-friction}
		L = \frac{1}{2}mv^2 - mgy\,,
	\end{equation}
	where $v^2 = v_x^2 + v_y^2$ in the fiber bundle $\T Q$ with coordinates $(x, y, v_x, v_y)$.

	In order to introduce air friction, we consider the Lagrangian with holonomic dissipation term $\L = L-\gamma s$ in $M = \T Q\times\R$ endowed with coordinates $(x, y, v_x, v_y, s)$. In this case, we have the differential forms
	\begin{align*}
		\theta_\L &= mv_x\d x + mv_y\d y\,,\\
		\eta_\L &= \d s - \theta_\L = \d s - mv_x\d x - mv_y\d y\,,\\
		\d \eta_\L &= m\d x\wedge\d v_x + m\d y\wedge\d v_y\,.\\
	\end{align*}
	The Reeb vector field is $\Reeb_\L = \tparder{}{s}$ and the Lagrangian energy is
	$$
		E_\L = \frac{1}{2}mv^2 + mgy + \gamma s\,.
	$$
	The dynamical equations for a vector field $X\in\X(M)$ with coordinate expression
	$$ X = a\parder{}{s} + b\parder{}{x} + c\parder{}{y} + d\parder{}{v_x} + e\parder{}{v_y} $$
	are
	\begin{equation*}
		\begin{cases}
			i(X)\d\eta_\L = \d E_\L - \Reeb_\L(E_\L)\eta_\L\,,\\
			i(X)\eta_\L = -E_\L\,.
		\end{cases}
	\end{equation*}
	Using the fact that
	$$ \d E_\L = mv_x\d v_x + mv_y\d v_y + mg\d y + \gamma\d s $$
	and that
	$$ \Reeb_\L(E_\L)\eta_\L = \gamma\d s - \gamma mv_x\d x - \gamma mv_y\d y\,, $$
	we obtain the relations
	\begin{equation*}
		\begin{cases}
			a = bv_x + cv_y - E_\L\,,\\
			b = v_x\,,\\
			c = v_y\,,\\
			d = -\gamma v_x\,,\\
			e = -\gamma v_y - g\,.
		\end{cases}
	\end{equation*}
	The second and third conditions imply that $a = \L$. Hence, the contact Lagrangian vector field is
	$$ \Gamma_\L = \L\parder{}{s} + v_x\parder{}{x} + v_y\parder{}{y} - \gamma v_x\parder{}{v_x} - (g + \gamma v_y)\parder{}{v_y}\,. $$
	This gives the following system of differential equations:
	\begin{equation*}
		\begin{cases}
			\ddot x + \gamma\dot x = 0\,,\\
			\ddot y + \gamma\dot y + g = 0\,,\\
			\dot s = \L\,.
		\end{cases}
	\end{equation*}
	As in the previous example, the energy dissipation is given by Theorem \ref{thm:energy-dissipation-contact}:
	$$ \Lie_{\Gamma_\L}E_\L = -\gamma E_\L\,. $$
	Notice that $\tparder{\L}{x} = 0$. Thus, it is immediate to check that $\tparder{}{x}$ is a contact symmetry. Its associated dissipated quantity is its corresponding momentum:
	$$ p^x = \parder{\L}{x} = mv_x\,. $$
	The dissipation of this quantity is given by Theorem \ref{thm:momentum-dissipation-contact}:
	$$ \Lie_{\Gamma_\L}p^x = -\gamma p^x\,. $$
	Now, taking into account Proposition \ref{prop:conserved-dissipated-contact}, as we have two dissipated quantities, we can obtain the conserved quantity
	$$ G = \frac{E_\L}{p^x} = \frac{\frac{1}{2}mv^2 + mgy + \gamma s}{mv_x}\,. $$

\section{The parachute equation}
\label{sec:parachute-equation}

	In this example we are going to consider a contact Lagrangian function which is {\it not} a Lagrangian with holonomic dissipation term.

	Consider the vertical motion of a particle falling in a fluid under the action of constant gravity. If the friction is modeled by the drag equation, the friction force is proportional to the square of the velocity. This motion can be described as the contact Lagrangian system $(M,\L)$, where $M=\T\R\times\R$ endowed with coordinates $(y, v, s)$ and
	$$ \L = \frac{1}{2}mv^2 - \frac{mg}{2\gamma}(e^{2\gamma y} - 1) + 2\gamma v s\,, $$
	where $\gamma$ is the friction coefficient, which depends on the density of the air, the shape of the object, etc.

	\begin{rmrk}\rm
		Notice that
		$$ \lim_{\gamma\to0}\L = \frac{1}{2}mv^2 - mgy\,, $$
		which is the mechanical Lagrangian \eqref{eq:lagrangian-falling-particle-without-friction} considered at the beginning of the previous example.
	\end{rmrk}

	In this case, we have
	\begin{align*}
		\theta_\L &= (mv + 2\gamma s)\d y\,,\\
		\eta_\L &= \d s - \theta_\L = \d s - (mv + 2\gamma s)\d y\,,\\
		\d\eta_\L &= m\d y\wedge\d v + 2\gamma\d y\wedge\d s\,,\\
		E_\L &= \frac{1}{2}mv^2 + \frac{mg}{2\gamma}(e^{2\gamma y}-1)\,,\\
		\Reeb_\L &= \parder{}{s} - \frac{2\gamma}{m}\parder{}{v}\,.
	\end{align*}
	Consider the vector field
	$$ X = a\parder{}{s} + b\parder{}{y} + c\parder{}{v}\in\X(\T\R\times\R)\,. $$
	Taking into account that
	$$ \d E_\L = mv\d v + mge^{2\gamma y}\d y $$
	and that
	$$ \Reeb_\L(E_\L)\eta_\L = -2\gamma v(\d s - (mv + 2\gamma s)\d y)\,, $$
	we obtain the conditions
	\begin{equation*}
		\begin{dcases}
			a = (mv + 2\gamma)b - E_\L\,,\\
			b = v\,,\\
			c = -ge^{2\gamma y} - \frac{2\gamma}{m}a + 2\gamma v^2 + \frac{4\gamma^2}{m}vs\,.
		\end{dcases}
	\end{equation*}
	The first two condition imply that $a = \L$. Using this fact, the third equation becomes $c = -g + \gamma v^2$. Summing up, we get that the contact Lagrangian vector field is
	$$ \Gamma_\L = v\parder{}{y} + (\gamma v^2 - g)\parder{}{v} + \L\parder{}{s}\,. $$
	Hence, we have the following system of differential equations:
	\begin{equation}\label{eq:ode-system-parachute}
		\begin{cases}
			\ddot y - \gamma\dot y^2 + g = 0\,,\\
			\dot s = \L\,.
		\end{cases}
	\end{equation}

	The energy dissipation is given by Theorem \ref{thm:energy-dissipation-contact},
	$$ \Lie_{\Gamma_\L}E_\L = 2\gamma v E_\L\,. $$
	\begin{rmrk}\rm
		Equation \eqref{eq:ode-system-parachute} describes an object falling ($\dot y < 0$). To describe an object ascending it is enough to change $\gamma$ for $-\gamma$ in the Lagrangian function.
	\end{rmrk}

\section{Lagrangian with holonomic dissipation term}
\label{sec:Lagrangian-holonomic-dissipation-term}

	Let $Q$ be a smooth manifold of dimension $n$. Consider a Lagrangian function $L_\circ\in\Cinfty(\T Q)$ either regular or singular and $\gamma\in\R$. Let $\L = \tau_1^\ast L_\circ - \gamma s\in\Cinfty(\T Q\times\R)$ be a Lagrangian with holonomic dissipation term \cite{Cia2018,Gas2019}. Let $\W = \T Q\times_Q\cT Q\times\R$ be the Pontryagin bundle with coordinates $(q^i, v^i, p_i, s)$. Denote $\L = \rho_1^\ast L\in\Cinfty(\W)$, which is regular or singular Lagrangian depending on the regularity of $L_\circ$. If the Lagrangian is singular, we will assume it is almost-regular. Then,
	$$ \H = p_iv^i - L_\circ(q^i,v^i) + \gamma s\in\Cinfty(\W)\,, $$
	and
	$$ \d\H = v^i\d p_i + \left( p_i - \parder{L_\circ}{v^i} \right)\d v^i - \parder{L_\circ}{q^i}\d q^i + \gamma\d s\,. $$
	Consider a vector field $X_\H\in\X(\W)$ with local expression $X_\H = f^i\parder{}{q^i} + F^i\parder{}{v^i} + G_i\parder{}{p_i} + f\parder{}{s}$. Then, equations \eqref{eq:SR-contact-hamilton-equations-fields} give the conditions
	\begin{equation*}
		\begin{dcases}
			f^i = v^i\,,\\
			f = (f^i - v^i)p_i + \L = \L\,,\\
			p_i = \parder{L_\circ}{v^i}\,,\\
			G_i = \parder{L_\circ}{q^i} - \gamma p_i\,.
		\end{dcases}
	\end{equation*}
	We have the submanifold $\W_1 = \graph\F L\hookrightarrow\W$, and
	$$ \restr{X_\H}{\W_1} = v^i\parder{}{q^i} + F^i\parder{}{v^i} + \left( \parder{L_\circ}{q^i} - \gamma p_i \right)\parder{}{p_i} + (L_\circ - \gamma s)\parder{}{s}\,. $$
	Imposing the tangency of $X_\H$ to $\W_1$, we get
	$$ X_\H\left( p_j - \parder{L_\circ}{v^j} \right) = -\parderr{L_\circ}{q^i}{v^j}v^i - \parderr{L_\circ}{v^i}{v^j}F^i + \parder{L_\circ}{q^j} - \gamma p_j = 0\quad\mbox{(on $\W_1$)}\,. $$
	In Section \ref{sec:contact-dynamical-equations} we pointed out that if the Lagrangian is regular, the tangency condition allows us to determine all the coefficients $F^i$ and we have a unique solution. On the other hand, if the Lagrangian is singular, the tangency condition establishes some relations between the functions $F^i$. Also, new constraints may appear, defining a new constraint submanifold $\W_2\hookrightarrow\W_1\hookrightarrow\W$. The constraint algorithm continues by imposing tangency to this new submanifold and so on. Eventually, we may find (if it exists) a final constraint submanifold $\W_f$ where there exist tangent solutions $X_\H$.

	Let $\sigma(t) = (q^i(t), v^i(t), p_i(t), s(t))$ be an integral curve of a solution $X_\H\in\X(\W)$ tangent to $\W_f$. Then, equations \eqref{eq:SR-contact-hamilton-equations-curves}, on $\W_f$, are
	\begin{equation*}
		\begin{dcases}
			\dot s = L_\circ - \gamma s\,,\\
			\dot q^i = v^i\,,\\
			\dot p_i = \frac{\d}{\d t}\left( \parder{L}{v^i} \right) = \parder{L_\circ}{q^i} - \gamma p_i = \parder{L_\circ}{q^i} - \gamma\parder{L_\circ}{v^i}\,.
		\end{dcases}
	\end{equation*}

	The next three examples are of this kind: one regular system and two singular systems.

\section{Central force with dissipation}
\label{sec:central-force-dissipation}

	Consider a particle of mass $m$ in $\R^3$ submitted to a central potential with dissipation. Taking $Q = \R^3 - \{(0,0,0\}$ with coordinates $(q^i)$, the Lagrangian describing the dynamics of the system is
	$$ L = \frac{1}{2}mv_iv^i - U(r) - \gamma s\in\Cinfty(\T Q\times\R)\,, $$
	where $v_i = g_{ij}v^j$, $g_{ij}$ is the natural extension of the Euclidean metric of $\R^3$ to the extended Pontryagin bundle $\W=\T Q\times_Q\cT Q\times\R$, and $r = \sqrt{q_iq^i}$. In $\W$, with local coordinates $(q^i,v^i,p_i,s)$, we denote $\L = \rho_1^\ast L\in\Cinfty(\W)$, which has the same coordinate expression as $L$ and is a hyperregular Lagrangian. Hence,
	$$ \H = p_iv^i-\frac{1}{2}mv_iv^i + U(r) + \gamma s\in\Cinfty(\W)\,, $$
	and
	$$ \d\H = v^i\d p_i + (p_i - mv_i)\d v^i + \frac{U'(r)}{r}q_i\d q^i + \gamma\d s\,. $$
	Consider a vector field $X_\H\in\X(\W)$ with local expression $X_\H = f^i\parder{}{q^i} + F^i\parder{}{v^i} + G_i\parder{}{p_i} + f\parder{}{s}$. Then, equations \eqref{eq:SR-contact-hamilton-equations-fields} give
	\begin{equation*}
		\begin{dcases}
			f^i = v^i\,,\\
			f = (f^i - v^i)p_i + \L = \L\,,\\
			p_i = mv_i\,,\\
			G_i = -\frac{U'(r)}{r}q_i - \gamma p_i\,.
		\end{dcases}
	\end{equation*}
	The first constraint submanifold $\W_1\hookrightarrow\W$ is
	$$ \W_1 = \{(q^i,v^i,p_i,s)\in\W\ \vert\ p_i-mv_i = 0\} = \graph\F L\,, $$
	and
	$$ \restr{X_\H}{\W_1} = v^i\parder{}{q^i} + F^i\parder{}{v^i} - \left( \gamma p_i + \frac{U'(r)}{r}q_i \right)\parder{}{p_i} + \left( \frac{1}{2}mv_iv^i - U(r) - \gamma s \right)\parder{}{s}\,. $$
	The tangency condition of $X_\H$ to the first constraint submanifold $\W_1$ reads
	$$ X_\H(p_i - mv_i) = -\gamma p_i -\frac{U'(r)}{r}q_i - mF_i = 0\Longleftrightarrow F^i = -\frac{1}{m}\left( \gamma p^i + \frac{U'(r)}{r}q^i \right)\quad\mbox{(on $\W_1$)}\,, $$
	and the algorithm finishes with the unique solution
	$$ \restr{X_\H}{\W_1} = v^i\parder{}{q^i} -\frac{1}{m}\left( \gamma p^i + \frac{U'(r)}{r}q^i \right)\parder{}{v^i} - \left( \gamma p_i + \frac{U'(r)}{r}q_i \right)\parder{}{p_i} + \L\parder{}{s}\,. $$
	Hence, if $\sigma(t) = (q^i(t),v^i(t),p_i(t),s(t))$ is an integral curve of $X_\H$, equations \eqref{eq:SR-contact-hamilton-equations-curves}, on $\W_1$, read
	$$
		\begin{dcases}
			\dot s = \L\,,\\
			\dot q^i = v^i\,,\\
			\frac{1}{m}\dot p^i = \dot v^i = \ddot q^i = -\gamma\dot q^i - \frac{U'(r)}{mr}q^i\,,
		\end{dcases}
	$$
	which are the Euler--Lagrange equations for the motion of a particle in a central potential with friction.

	We are now going to recover the Lagrangian and Hamiltonian formalisms from the Skinner--Rusk formalism, as stated in Section \ref{sec:SR-contact-recovering}, by projecting onto each factor of the Pontryagin bundle $\W$. As $L$ is hyperregular, the Legendre map $\F L\colon\T Q\times\R\to\cT Q\times\R$ is a global diffeomorphism, and the constraint algorithm finishes with the first constraint submanifold $\W_1$. In the Lagrangian formalism, the contact Lagrangian vector field is the {\sc sopde}
	$$ X_L = v^i\parder{}{q^i} - \left( \gamma v^i + \frac{U'(r)}{mr}q^i \right)\parder{}{v^i} + \left( \frac{1}{2}mv_iv^i - U(r) - \gamma s \right)\parder{}{s}\in\X(\T Q\times\R)\,, $$
	and, in the Hamiltonian formalism, we obtain the contact Hamiltonian vector field
	$$ X_H = \frac{p_i}{m}\parder{}{q^i} - \left(\gamma p_i + \frac{U'(r)}{r}q_i\right)\parder{}{p_i} + \left( \frac{p_ip^i}{2m} - U(r) - \gamma s \right)\parder{}{s}\in\X(\cT Q\times\R)\,. $$

\section[Lagrange multipliers. The damped simple pendulum]{Lagrange multipliers.\\The damped simple pendulum}
\label{sec:Lagrange-multipliers}

	The \textit{method of Lagrange multipliers} is used to incorporate constraints to a system. This leads to singular Lagrangians in a very natural way, since the velocities associated to the Lagrange multipliers do not appear in the Lagrangian. We will use a simple case, the pendulum with friction, to explain how to apply the Skinner--Rusk formalism to these systems.

	Consider a damped pendulum of length $\ell$ and mass $m$. Its position in the plane can be described using polar coordinates $(r,\theta)$, where $\theta = 0$ is the position at rest. The motion of the pendulum is restricted to the circumference $r = \ell$. Hence, the corresponding Lagrangian is
	$$ L = \frac{1}{2}m(v_r^2 + r^2v_\theta^2) - mgr(1-\cos\theta) + \lambda(r - \ell) - \gamma s\in\Cinfty(\T\R^3\times\R)\,, $$
	where $\lambda$ is the Lagrange multiplier. It is a singular Lagrangian since the velocity $v_\lambda$ does not appear in the Lagrangian function. In the Pontryagin bundle $\W = \T\R^3\times_{\R^3}\cT\R^3\times\R$, with local coordinates $(r,\theta,\lambda,v_r,v_\theta,v_\lambda,p_r,p_\theta,p_\lambda,s)$, we have $\L = \rho_1^\ast L\in\Cinfty(\W)$. Then,
	$$ \H = p_rv_r + p_\theta v_\theta + p_\lambda v_\lambda - \frac{1}{2}m(v_r^2 + r^2v_\theta^2) + mgr(1 - \cos\theta) + \gamma s - \lambda(r - \ell)\in\Cinfty(\W)\,. $$
	Consider a vector field $X_\H\in\X(\W)$ with local expression
	$$ X_\H = f_r\parder{}{r} + f_\theta\parder{}{\theta} + f_\lambda\parder{}{\lambda} + F_r\parder{}{v_r} + F_\theta\parder{}{v_\theta} + F_\lambda\parder{}{v_\lambda} + G_r\parder{}{p_r} + G_\theta\parder{}{p_\theta} + G_\lambda\parder{}{p_\lambda} + f\parder{}{s}\,. $$
	Then, equations \eqref{eq:SR-contact-hamilton-equations-fields} give the conditions
	$$
		\begin{dcases}
			f = \L\,,\\
			f_r = v_r\,,\\
			f_\theta = v_\theta\,,\\
			f_\lambda = v_\lambda\,,\\
			p_r = mv_r\,,\\
			p_\theta = r^2mv_\theta\,,\\
			p_\lambda = 0\,,\\
			G_r = mrv_\theta^2 - mg(1-\cos\theta) + \lambda - \gamma p_r\,,\\
			G_\theta = -mgr\sin\theta - \gamma p_\theta\,,\\
			G_\lambda = r - \ell - \gamma p_\lambda\,.
		\end{dcases}
	$$
	The first constraint submanifold is $\W_1\hookrightarrow\W$ given by
	$$ \W_1 = \{(r,\theta,\lambda,v_r,v_\theta,v_\lambda,p_r,p_\theta,p_\lambda,s)\ \vert\ p_r = mv_r,\ p_\theta = mr^2v_\theta,\ p_\lambda = 0\} = \graph\F L\,, $$
	and the vector field $X_\H$ is
	\begin{align*}
		\restr{X_\H}{\W_1} =&\ \L\parder{}{s} + v_r\parder{}{r} + v_\theta\parder{}{\theta} + v_\lambda\parder{}{\lambda} + F_r\parder{}{v_r} + F_\theta\parder{}{v_\theta} + F_\lambda\parder{}{v_\lambda} \\
		&+ (mrv_\theta^2 - mg(1-\cos\theta) + \lambda - \gamma p_r)\parder{}{p_r} \\
		&- (mgr\sin\theta + \gamma p_\theta)\parder{}{p_\theta} + (r - \ell - \gamma p_\lambda)\parder{}{p_\lambda}\,.
	\end{align*}
	Imposing tangency of $X_\H$ to $\W_1$, we obtain the following conditions (on $\W_1$):
	\begin{equation}\label{eq:damped-pendulum-tangency-W1}
		\begin{dcases}
			F_r = rv_\theta^2 - g(1-\cos\theta) + \frac{\lambda}{m} - \gamma v_r\,,\\
			2v_rv_\theta + r F_\theta = -g\sin\theta - \gamma rv_\theta\,,\\
			r = \ell\,.
		\end{dcases}
	\end{equation}
	Notice that we have dinamically recovered the constraint $r = \ell$, defining a new constraint submanifold $\W_2\hookrightarrow\W_1\hookrightarrow\W$. The tangency condition to $\W_2$ gives
	$$ v_r = 0\quad \mbox{(on $\W_2$)}\,, $$
	defining a new constraint submanifold $\W_3$. Imposing tangency of $X_\H$ to $\W_3$ we obtain the equation
	$$ F_r = 0\,, $$
	allowing us to compute the Lagrange multiplier $\lambda$:
	$$ \lambda = mg(1-\cos\theta) - m\ell v_\theta^2\quad\mbox{(on $\W_3$)}\,. $$
	This is a new constraint, which defines a new constraint submanifold $\W_4$. The tangency condition to $\W_4$ gives a last constraint
	$$ v_\lambda = m(3gv_\theta\sin\theta + 2\ell\gamma v_\theta^2)\quad\mbox{(on $\W_4$)}\,. $$
	Finally, imposing the tangency condition to this las constraint, we determine the coefficient $F_\lambda:$
	$$ F_\lambda = mg\left(3v_\theta\cos\theta - 3\frac{g}{\ell}\sin^2\theta - 5\gamma v_\theta\sin\theta - 2\ell gv_\theta^2\right) \quad\mbox{(on $\W_4$)}\,, $$
	and no new constraints appear. Hence, the constraint algorithm ends with the final constraint submanifold $\W_f = \W_4$, which is defined as
	\begin{align*}
		\W_f = \{ & (r,\theta,\lambda,v_r,v_\theta,v_\lambda,p_r,p_\theta,p_\lambda, s)\ \vert\ p_r = mv_r,\ p_\theta = mr^2v_\theta,\ p_\lambda = 0,\ r = \ell,\\
		& v_r = 0,\ \lambda = mg(1-\cos\theta)-m\ell v_\theta^2,\ v_\lambda = m(3gv_\theta\sin\theta + 2\ell\gamma v_\theta^2) \}\,,
	\end{align*}
	and the unique solution
	\begin{align*}
		\restr{X_\H}{\W_f} =&\ v_\theta\parder{}{\theta} + m\left(3gv_\theta\sin\theta + 2\ell\gamma v_\theta^2\right)\parder{}{\lambda} - \left( \frac{g}{\ell}\sin\theta + \gamma v_\theta \right)\parder{}{v_\theta} + \\
		&\ mg\left( 3v_\theta\cos\theta - 3\frac{g}{\ell}\sin^2\theta - 5\gamma v_\theta\sin\theta - 2\ell gv_\theta^2 \right)\parder{}{v_\lambda}\\
		& - m\ell(g\sin\theta + \gamma\ell v_\theta)\parder{}{p_\theta} + \left( \frac{1}{2}m\ell^2v_\theta^2 - mg\ell(1 - \cos\theta) - \gamma s \right)\parder{}{s}
	\end{align*}
	Notice that we only have three independent variables: $s$, $\theta$ and $v_\theta$. Therefore, for an integral curve of $X_\H$, the second equation in \eqref{eq:damped-pendulum-tangency-W1} gives the equation of motion
	$$ \ddot\theta = -\frac{g}{\ell}\sin\theta - \gamma\dot\theta\,, $$
	which is the equation of motion of the simple pendulum with friction.

	Following Section \ref{sec:SR-contact-recovering}, we can recover the Lagrangian and Hamiltonian formalisms by projecting onto each factor of the Pontryagin bundle $\W = \T\R^3\times_{\R^3}\cT\R^3\times\R$. In the Lagrangian formalism we have the final constraint submanifold
	\begin{align*}
		S_f =&\, \{ (r,\theta,\lambda,v_r,v_\theta,v_\lambda,s)\in\T\R^3\times\R\ \vert\ r = \ell,\ v_r = 0,\ \lambda = mg(1-\cos\theta) - m\ell v_\theta^2,\\
		& \ v_\lambda = m(3gv_\theta\sin\theta + 2\ell\gamma v_\theta^2) \}\,, 
	\end{align*}
	and the contact Lagrangian vector field is the {\sc sopde}
	\begin{align*}
		\restr{X_L}{S_f} =& \ v_\theta\parder{}{\theta} + v_\lambda\parder{}{\lambda} - \left( \frac{g}{\ell}\sin\theta + \gamma v_\theta \right)\parder{}{v_\theta} \\
		& + mg\left(3v_\theta\cos\theta - 3\frac{g}{\ell}\sin^2\theta - 5\gamma v_\theta\sin\theta - 2\ell gv_\theta^2\right)\parder{}{v_\lambda} \\
		& + \left( \frac{1}{2}m\ell^2v_\theta^2 - mg\ell(1-\cos\theta) - \gamma s \right)\parder{}{s}\in\X(\T\R^3\times\R)\,.
	\end{align*}
	In the Hamiltonian counterpart, we have
	$$ P_f = \{ (r,\theta,\lambda,p_r,p_\theta,p_\lambda,s)\in\cT\R^3\times\R\ \vert\ r = \ell,\ p_\lambda = 0,\ p_r = 0,\ \lambda = mg(1-\cos\theta) - \frac{p_\theta^2}{m\ell^3} \}\,, $$
	and the contact Hamiltonian vector field
	\begin{align*}
		\restr{X_H}{P_f} =&\ \frac{p_\theta}{m\ell^2}\parder{}{\theta} + \left( \frac{3g}{\ell^2}p_\theta\sin\theta + \frac{2\gamma}{m\ell^3}p_\theta^2 \right)\parder{}{\lambda} - (m\ell g\sin\theta + \gamma p_\theta)\parder{}{p_\theta} \\
		& + \left( \frac{p_\theta^2}{2m\ell^2} - mg\ell(1-\cos\theta) - \gamma s \right)\parder{}{s}\in\X(\cT\R^3\times\R)\,.
	\end{align*}

\section{Cawley's Lagrangian with dissipation}
\label{sec:Cawleys-Lagrangian}

	This last example is an academic model based on the Lagrangian introduced by R. Cawley to study some features of singular Lagrangians in Dirac's theory of constraint systems \cite{Caw1979}.

	Consider the manifold $\T\R^3\times\R$ with coordinates $(q^i, v^i, s)$ and the Lagrangian function
	$$ L = v^1v^3 + \frac{1}{2}q^2(q^3)^2 - \gamma s\in\Cinfty(\T\R^3\times\R)\,. $$
	In the Pontryagin bundle $\W = \T\R^3\times_{\R^3}\cT\R^3\times\R$ with local coordinates $(q^i,v^i,p_i,s)$, we denote $\L = \rho_1^\ast L\in\Cinfty(\W)$. The Lagrangians $\L$ and $L$ have the same local expression. Then,
	$$ \H = p_iv^i - v^1v^3 - \frac{1}{2}q^2(q^3)^2 + \gamma s\in\Cinfty(\W)\,. $$
	Consider now a vector field $X_\H\in\X(\W)$ with local expression
	$$ X_\H = f^i\parder{}{q^i} + F^i\parder{}{v^i} + G_i\parder{}{p_i} + f\parder{}{s}\,. $$
	Then, equations \eqref{eq:SR-contact-hamilton-equations-fields} give the conditions
	\begin{equation*}
		\begin{dcases}
			f^i = v^i\,,\\
			f = \L\,,\\
			p_1 = v^3\,,\\
			p_2 = 0\,,\\
			p_3 = v^1\,,\\
			G_1 = -\gamma p_1\,,\\
			G_2 = \frac{1}{2}q^3 - \gamma p_2\,,\\
			G_3 = q^2q^3 - \gamma p_3\,.
		\end{dcases}
	\end{equation*}
	Hence, the first constraint submanifold $\W_1\hookrightarrow\W$ is defined as
	$$ \W_1 = \{ (q^i,v^i,p_i,s)\in\W\ \vert\ p_1 = v^3,\ p_2 = 0,\ p_3 = v^1 \}\,, $$
	and the vector field $X_\H$ has local expression
	$$ \restr{X_\H}{\W_1} = v^i\parder{}{q^i} + F^i\parder{}{v^i} - \gamma p_1\parder{}{p_1} + \frac{1}{2}q^3\parder{}{p_2} + (q^2q^3 - \gamma p_3)\parder{}{p_3} + \L\parder{}{s}\,. $$
	Imposing the tangency of $X_\H$ to the first constraint submanifold $\W_1$ we get
	$$ 
		\begin{dcases}
			F^1 = q^2q^3 - \gamma p_3\,,\\
			F^3 = -\gamma p_1\,,\\
			q^3 = 0\,,
		\end{dcases}
	$$
	determining the coefficients $F^1$ and $F^3$ and adding a new constraint defining the submanifold $\W_2$. Imposing tangency of the vector field $X_\H$ to $\W_2$ we obtain
	$$ v^3 = 0\quad\mbox{(on $\W_2$)}\,, $$
	which, taking into account that $p_1 = v^3$, implies that $p_1 = 0$ on $\W_2$. Now, the tangency condition holds and gives the final constraint submanifold
	$$ \W_f = \{ (q^i,v^i,p_i,s)\in\W\ \vert\ p_1 = v^3 = 0,\ p_2 = 0,\ p_3 = v^1,\ q^3 = 0 \}\,, $$
	and the family of solutions
	$$ \restr{X_\H}{\W_f} = v^1\parder{}{q^1} + v^2\parder{}{q^2} - \gamma v^1\parder{}{v^1} + F^2\parder{}{v^2} - \gamma s\parder{}{s}\,. $$
	We can now recover the Lagrangian and Hamiltonian formalisms by projecting onto each factor of the Pontryagin bundle $\W = \T\R^3\times_{\R^3}\cT\R^3\times\R$. In the Lagrangian formalism we have the final constraint submanifold
	$$ S_f = \{ (q^i,v^i,s)\in\T\R^3\times\R \ \vert\ q^3 = 0,\ v^3 = 0 \}\,, $$
	and the contact Lagrangian vector field is the {\sc sopde}
	$$ \restr{X_L}{S_f} = v^1\parder{}{q^1} + v^2\parder{}{q^2} - \gamma v^1\parder{}{v^1} + F^2\parder{}{v^2} - \gamma s\parder{}{s}\in\X(\T\R^3\times\R)\,. $$
	On the other hand, in the Hamiltonian formalism we have the final constraint submanifold
	$$ P_f = \{ (q^i,p_i,s)\in\cT\R^3\times\R\ \vert\ p_1 = 0,\ p_2 = 0,\ q^3 = 0 \}\,, $$
	and the unique contact Hamiltonian vector field is
	$$ \restr{X_H}{P_f} = p_3\parder{}{q^1} + v^2\parder{}{q^2} - \gamma p_1\parder{}{p_1} - \gamma s\parder{}{s}\,. $$
	Notice that $\ker\F L = \left\langle \dparder{}{v^2} \right\rangle$.

\part{Field theory}
\label{pt:field-theory}

	\chapter[Review on \texorpdfstring{$k$}--symplectic and \texorpdfstring{$k$}--cosymplectic formalisms]{Review on \texorpdfstring{$k$}--symplectic and \phantom{n} \texorpdfstring{$k$}--cosymplectic formalisms}
	\label{ch:k-symplectic-k-cosymplectic-formalisms}


This chapter is devoted to review both the Hamiltonian and Lagrangian formalisms of autonomous and nonautonomous field theories. In Section \ref{sec:k-vector-fields-integral-sections}, we define the notion of $k$-vector field and integral section, which will be of great interest when developing the Hamiltonian and Lagrangian formulation of field theories. We also stablish the conditions for a $k$-vector field to be integrable. Section \ref{sec:k-symplectic-geometry} is devoted to present to framework of $k$-symplectic geometry. The notion of $k$-symplectic manifold is introduced and some of its most relevant properties are stated. In particular, we give a proof of the Darboux theorem for $k$-symplectic manifolds different from the one given in \cite{Awa1992}. In Section \ref{sec:k-symplectic-Hamiltonian-systems} we develop the $k$-symplectic Hamiltonian formalism, which is the natural formalism to deal with autonomous Hamiltonian field theories, and obtain the $k$-symplectic Hamilton--De Donder--Weyl equations. Section \ref{sec:k-symplectic-Lagrangian-systems} begins by presenting the canonical geometric structures of the tangent bundle of $k$-velocities $\oplus^k\T Q$: the vertical lifts, the canonical $k$-tangent structure and the Liouville vector field. The notions of second-order partial differential equation and holonomic map are introduced and we establish the relation between them. With these geometric tools, we can define the Lagrangian energy, the Cartan forms and the Legendre map associated with a Lagrangian function. Then, we develop the Lagrangian formalism for autonomous field theories and obtain the $k$-symplectic Euler--Lagrange equations. In Section \ref{sec:k-cosymplectic-geometry} we offer a review on $k$-cosymplectic geometry, which is the natural framework when dealing with nonautonomous field theories. We define the notions of $k$-cosymplectic manifold, Reeb vector fields and state the Darboux theorem for $k$-cosymplectic manifolds. We include a detailed study of the canonical $k$-cosymplectic manifold $\R^k\times\oplus^k\cT Q$ and show that its natural coordinates are, in fact, Darboux coordinates. Finally, in Sections \ref{sec:k-cosymplectic-Hamiltonian-systems} and \ref{sec:k-cosymplectic-Lagrangian-systems} we give a complete description of the Hamiltonian and Lagrangian formalisms for nonautonomous field theories. Some references on these topics are \cite{Awa1992,DeLeo1998,DeLeo2001,DeLeo2015,Gun1987,Mun2010,Rom2011}.

\section{\texorpdfstring{$k$}--vector fields and integral sections}
\label{sec:k-vector-fields-integral-sections}

The notion of $k$-vector field is of great use in the geometric study of partial differential equations. See, for instance, \cite{DeLeo2015}.

Let $M$ be a smooth $n$-dimensional manifold. Consider the direct sum of $k$ copies of its tangent bundle: $\oplus^k\T M$. We have the natural projections
\begin{equation*}
	\tau^\alpha\colon\oplus^k\T M\to\T M\ , \qquad \tau^1_M\colon\oplus^k\T M\to M\,.
\end{equation*}

\begin{dfn}
	A \textbf{$k$-vector field} on a manifold $M$ is a section
	$$ \bfX\colon M\to\oplus^k\T M $$
	of the natural projection $\tau^1_M$ defined above. We will denote by $\X^k(M)$ the set of all $k$-vector fields on $M$.
\end{dfn}
$$
	\xymatrix{
		& & \oplus^k\T M \ar[dd]^{\tau^\alpha} \\\\
		M \ar[uurr]^{\bf X} \ar[rr]^{X_\alpha} & & \T M
	}
$$
Taking into account the diagram above, a $k$-vector field $\bfX\in\X^k(M)$ can be given by $k$ vector fields $X_1,\dotsc,X_k\in\X(M)$, obtained as $X_\alpha = \tau^\alpha\circ\bfX$. With this in mind, we can denote $\bfX = (X_1,\dotsc,X_k)$. A $k$-vector field $\bfX = (X_1,\dotsc,X_k)$ induces a decomposable contravariant skew-symmetric tensor field, $X_1\wedge\dotsb\wedge X_k$, which is a section of the bundle $\bigwedge^k\T M\to M$. This also induces a tangent distribution on $M$.

\begin{dfn}\label{dfn:first-prolongation-k-tangent-bundle}
	Given a map $\phi\colon U\subset\R^k\to M$, we define its \textbf{first prolongation} to $\oplus^k\T M$ as the map
	$$ \phi'\colon U\subset\R^k\to\oplus^k\T M\,, $$
	defined by
	$$ \phi'(t) = \left( \phi(t); \T\phi\left( \parder{}{t^1}\bigg\vert_t \right),\dotsc,\T\phi\left( \parder{}{t^k}\bigg\vert_t \right) \right) \equiv (\phi(t); \phi'_\alpha(t))\,, $$
	where $t = (t^1,\dotsc,t^k)$ are the canonical coordinates of $\R^k$.
\end{dfn}

In the same way as we have integral curves of vector fields, we can define de notion of integral section of a $k$-vector field:

\begin{dfn}
	Let $\bfX = (X_1,\dotsc,X_k)\in\X^k(M)$ be a $k$-vector field. An \textbf{integral section} of $\bfX$ is a map $\phi\colon U\subset\R^k\to M$ such that
	$$ \phi' = \bfX\circ\phi\,, $$
	that is, $\T\phi\circ\dparder{}{t^\alpha} = X_\alpha\circ\phi$ for every $\alpha$.

	We say that a $k$-vector field $\bfX\in\X^k(M)$ is \textbf{integrable} if every point of $M$ is in the image of an integral section of $\bfX$.
\end{dfn}

Consider a $k$-vector field $\bfX = (X_\alpha)$ with local expression
$$ X_\alpha = X_\alpha^i\parder{}{x^i}\,. $$
Then, $\phi\colon U\subset\R^k\to M$ is an integral section of $\bfX$ if, and only if, it is a solution of the system of partial differential equations
$$ \parder{\phi^i}{t^\alpha} = X_\alpha^i(\phi)\,. $$

Let $\bfX = (X_1,\dotsc,X_k)$ be a $k$-vector field on $M$. Then, $\bfX$ is integrable if, and only if, $[X_\alpha,X_\beta] = 0$ for every $\alpha,\beta$. These are precisely the necessary and sufficient conditions for the integrability of the above systems of partial differential equations \cite{Lee2013}.

\section{\texorpdfstring{$k$}--symplectic geometry}
\label{sec:k-symplectic-geometry}

In this section we will review the concepts of $k$-symplectic geometry which we will be using in the following sections to develop both the Hamiltonian and Lagrangian formalisms of autonomous field theories. See \cite{Awa1992,Gun1987,Mun2010,Rom2011}. We will begin by defining the notion of $k$-symplectic manifold and we will also prove the Darboux theorem for $k$-symplectic manifolds, which states that every $k$-symplectic manifold is locally diffeomorphic to $\oplus^k\cT Q$.

\begin{dfn}\label{dfn:k-symplectic manifold}
	Let $M$ be a manifold of dimension $m = n + kn$. A \textbf{$k$-symplectic structure} on $M$ is family $(\omega^1,\dotsc,\omega^k,V)$, where $\omega^\alpha\in\Omega^2(M)$ are closed and $V$ is an integrable $nk$-dimensional tangent distribution on $M$ such that
	\begin{enumerate}[{\rm(1)}]
		\item $\restr{\omega^\alpha}{V\times V} = 0$ for every $\alpha = 1,\dotsc,k$,
		\item $\bigcap_{\alpha = 1}^k\ker\omega^\alpha = \{0\}$.
	\end{enumerate}
	We say that $(M,\omega^\alpha,V)$ is a \textbf{$k$-symplectic manifold}.
\end{dfn}

\begin{thm}[Darboux theorem for $k$-symplectic manifolds]\label{thm:k-symplectic-Darboux}
	Let $(M,\omega^\alpha,V)$ be a $k$-symplectic manifold. Then, for every point of $p\in M$, there exists a local chart $(U;q^i,p_i^\alpha)$, $p\in U$, such that
	$$ \restr{\omega^\alpha}{U} = \d q^i\wedge \d p_i^\alpha\ ,\quad V = \left\langle \parder{}{p_i^\alpha} \right\rangle\,. $$
	These coordinates are called \textbf{Darboux} or \textbf{canonical} coordinates of the $k$-symplectic manifold.
\end{thm}
\begin{proof}
	Let $p$ be a point of $M$. We can find a local chart $(U_p;x^1,\dotsc,x^n,y_1,\dotsc,y_{nk})$ around $p$ such that $V = \ds\left\langle \parder{}{y_i} \right\rangle$. By Poincaré's Lemma, as the 2-forms $\omega^\alpha$ are closed, they are locally exact. Hence, we can say that $\restr{\omega^\alpha}{U_p} = \restr{\d\theta^\alpha}{U_p}$. Using this, we have that, in $U_p$,
	\begin{align*}
		\omega^\alpha &= \d\left( f_i^\alpha\d x^i + g^{\alpha r}\d y_r \right)\\
		&= -\parder{f^\alpha_i}{x^j}\d x^i\wedge\d x^j - \parder{f^\alpha_i}{y_r}\d x^i\wedge\d y_r +  \parder{g^{\alpha r}}{x^i}\d x^i\wedge\d y_r - \parder{g^{\alpha r}}{y_s}\d y_r\wedge\d y_s\\
		&= \frac{1}{2}\left( \parder{f^\alpha_j}{x^i} - \parder{f^\alpha_i}{x^j} \right)\d x^i\wedge\d x^j + \left( \parder{g^{\alpha r}}{x^i} - \parder{f^\alpha_i}{y_r} \right)\d x^i\wedge\d y_r \\
		& \quad + \frac{1}{2}\left( \parder{g^{\alpha s}}{y^r} - \parder{g^{\alpha r}}{y^s} \right)\d y_r\wedge\d y_s\,.\\
	\end{align*}
	It is clear that
	$$ \parder{g^{\alpha s}}{y^r} = \parder{g^{\alpha r}}{y^s}\,, $$
	because $\restr{\omega^\alpha}{V\times V} = 0$. We want to prove that $\omega^\alpha = \d x^i\d p_i^\alpha$ for some functions $p_i^\alpha(x^j,y_r)$.
	\begin{align*}
		\d x^i\wedge\d p^\alpha_i &= \d x^i\wedge\left( \parder{p_i^\alpha}{x^j}\d x^j + \parder{p_i^\alpha}{y_r}\d y_r \right)\\
		&= \parder{p^\alpha_i}{x^j}\d x^i\wedge\d x^j + \parder{p_i^\alpha}{y_r}\d x^i\wedge\d y_r\\
		&= \frac{1}{2}\left( \parder{p_i^\alpha}{x^j} - \parder{p_j^\alpha}{x^i} \right)\d x^i\wedge\d x^j + \parder{p_i^\alpha}{y_r}\d x^i\wedge\d y_r\,.
	\end{align*}
	Hence, $(x^i,p_i^\alpha)$ are Darboux coordinates if, and only if, the following conditions hold:
	\begin{equation}\label{eq:Darboux-k-symp-conditions}
		\begin{split}
			\parder{f^\alpha_j}{x^i} - \parder{f^\alpha_i}{x^j} = \parder{p^\alpha_i}{x^j} - \parder{p^\alpha_j}{x^i}\,,\\
			\parder{g^{\alpha r}}{x^i} - \parder{f^\alpha_i}{y_r} = \parder{p_i^\alpha}{y_r}\,.
		\end{split}
	\end{equation}
	We will prove that we can define $p_i^\alpha$ as
	$$ p_i^\alpha = -f_i^\alpha + \int_0^1\sum_r\parder{g^{\alpha r}}{x^i}(x^1,\dotsc,x^n,ty_1,\dotsc,ty_{nk})y_r\d t\,, $$
	fulfilling conditions \eqref{eq:Darboux-k-symp-conditions}. First of all, we have that
	$$ \parder{p^\alpha_i}{x^j} = -\parder{f^\alpha_i}{x^j} + \int_0^1\sum_r\parderr{g^{\alpha r}}{x^i}{x^j}(x, ty)y_r\d t\,, $$
	and hence,
	$$ \parder{p^\alpha_i}{x^j} - \parder{p^\alpha_j}{x^i} = -\parder{f^\alpha_i}{x^j} + \parder{f^\alpha_j}{x^i} + \int_0^1\sum_r\left(\parderr{g^{\alpha r}}{x^i}{x^j} - \parderr{g^{\alpha r}}{x^j}{x^i}\right)(x, ty)y_r\d t = \parder{f^\alpha_j}{x^i} - \parder{f^\alpha_i}{x^j} \,. $$
	On the other hand,
	\begin{align*}
		\parder{p_i^\alpha}{y_r} &= -\parder{f^\alpha_i}{y_r} + \int_0^1\parder{g^{\alpha r}}{x^i}(x, ty)\d t + \int_0^1\sum_{r,s}\parderr{g^{\alpha r}}{x^i}{y_s}(x,ty)ty_r\d t\\
		&= -\parder{f^\alpha_i}{y_r} + \parder{}{x^i}\int_0^1\left( g^{\alpha r}(x, ty) + \sum_{r,s}\parder{g^{\alpha r}}{y_s}(x, ty)ty_r \right)\d t\\
		&= -\parder{f^\alpha_i}{y_r} + \parder{}{x^i}\int_0^1\frac{\d}{\d t}\big(g^{\alpha r}(x, ty)t\big)\d t\\
		&= -\parder{f^\alpha_i}{y_r} + \parder{}{x^i}\Big[ g^{\alpha r}(x, ty)t \Big]^{t = 1}_{t = 0}\\
		&= -\parder{f^\alpha_i}{y_r} + \parder{g^{\alpha r}}{x^i}\,.
	\end{align*}
\end{proof}

An alternative proof of this Theorem can be found in \cite{Awa1992}.

\begin{exmpl}[Canonical model for $k$-symplectic manifolds]\label{ex:canonical-model-k-symplectic}\rm
	Let $Q$ be a smooth $n$-manifold with local coordinates $(q^i)$ and consider the direct sum
	$$ \oplus^k\cT Q = \cT Q\oplus_Q\overset{(k)}{\dotsb}\oplus_Q \cT Q\,, $$
	with natural projections
	$$ \pi^\alpha\colon\oplus^k\cT Q\to\cT Q\ ,\quad \pi_Q^1\colon\oplus^k\cT Q\to Q\,. $$
	In the same way as in the case of the contangent bundle, a local chart $(U;q^i)$ in $Q$ induces a natural chart $\left((\pi^1_Q)^{-1}(U);q^i,p_i^\alpha\right)$ in $\oplus^k\cT Q$.

	Consider the canonical forms in the contangent bundle $\theta\in\Omega^1(\cT Q)$ and $\omega = -\d\theta\in\Omega^2(\cT Q)$. Hence, the direct sum $\oplus^k\cT Q$ has the canonical forms
	$$ \theta^\alpha = (\pi^\alpha)^\ast\theta\ ,\quad \omega^\alpha = (\pi^\alpha)^\ast\omega = -(\pi^\alpha)^\ast\d\theta = -\d\theta^\alpha\,, $$
	which in natural coordinates read
	$$ \theta^\alpha = p_i^\alpha\d q^i\ ,\quad\omega^\alpha = \d q^i\wedge\d p_i^\alpha\,. $$
	Taking all this into account, the triple $(\oplus^k\cT Q,\omega^\alpha,V)$, whith $V = \ker\T\pi_Q^1$, is a $k$-symplectic manifold. Notice that the natural coordinates $(q^i,p_i^\alpha)$ in $\oplus^k\cT Q$ are Darboux coordinates.
\end{exmpl}

\section{\texorpdfstring{$k$}--symplectic Hamiltonian systems}
\label{sec:k-symplectic-Hamiltonian-systems}

\begin{dfn}
	A \textbf{$k$-symplectic Hamiltonian system} is a family $(M,\omega^\alpha,V,H)$, where $(M,\omega^\alpha,V)$ is a $k$-symplectic manifold and $H\in\Cinfty(M)$ is a function called the \textbf{Hamiltonian function}.
\end{dfn}

Given a $k$-symplectic Hamiltonian system $(M,\omega^\alpha,V,H)$, we can define the vector bundle morphism $\flat\colon\oplus^k\T M\to \cT M$ as
\begin{equation}\label{eq:k-symplectic-flat}
	\flat(v_1,\dotsc,v_k) = i(v_\alpha)\omega^\alpha\,.
\end{equation}
This morphism induces a morphism of $\Cinfty(M)$-modules $\flat\colon\X^k(M)\to\Omega^1(M)$.

\begin{rmrk}\rm
	The morphism $\flat$ is surjective.
\end{rmrk}

\begin{dfn}
	A $k$-vector field $\bfX = (X_1,\dotsc,X_k)\in\X^k(M)$ is a \textbf{$k$-symplectic Hamiltonian $k$-vector field} of a $k$-symplectic Hamiltonian system $(M, \omega^\alpha, V, H)$ if it is a solution to
	\begin{equation}\label{eq:k-symplectic-Hamiltonian-equation}
		\flat(\mathbf{X}) = i(X_\alpha)\omega^\alpha = \d H\,,
	\end{equation}
	called the \textbf{$k$-symplectic Hamiltonian equation}. The set of $k$-symplectic Hamiltonian $k$-vector fields is denoted by $\X^k_H(M)$.
\end{dfn}

Notice that the surjectivity of the morphism $\flat$ ensures the existence of solutions to equation \eqref{eq:k-symplectic-Hamiltonian-equation}. However, in general, we do not have uniqueness of solutions. In fact, if $\bfX$ is a solution to \eqref{eq:k-symplectic-Hamiltonian-equation}, any element of the set $\bfX + \ker\flat$ is also a solution.

Consider a $k$-symplectic Hamiltonian system $(M,\omega^\alpha,V,H)$ with Darboux coordinates $(q^i, p_i^\alpha)$ and a $k$-vector field $\bfX = (X_1,\dotsc,X_k)$ locally given by
$$ X_\alpha = (X_\alpha)^i\parder{}{q^i} + (X_\alpha)^\beta_i\parder{}{p^\beta_i}\,. $$
Then, equation \eqref{eq:k-symplectic-Hamiltonian-equation} reads
\begin{equation}
	\begin{dcases}
		\parder{H}{q^i} = -\sum_{\beta = 1}^{k}(X_\beta)^\beta_i\,,\\
		\parder{H}{p_i^\alpha} = (X_\alpha)^i\,.
	\end{dcases}
\end{equation}

\begin{thm}
	Let $(M,\omega^\alpha,V,H)$ be a $k$-symplectic Hamiltonian system and $\bfX = (X_1,\dotsc,X_k)\in\X^k_H(M)$ an integrable vector field solution to \eqref{eq:k-symplectic-Hamiltonian-equation}. If $\phi\colon\R^k\to M$ is an integral section of $\bfX$, with local expression $\phi(t) = (\phi^i(t), \phi^\alpha_i(t))$, $t\in\R^k$, then $\phi$ is a solution to
	\begin{equation}\label{eq:k-symp-HDW}
		i(\phi'_\alpha)\omega^\alpha = \d H\circ\phi\,,
	\end{equation}
	called \textbf{Hamilton--De Donder--Weyl equation}.
\end{thm}

In local coordinates, equation \eqref{eq:k-symp-HDW} is equivalent to the system of partial differential equations
\begin{equation*}
	\begin{dcases}
		\restr{\parder{H}{q^i}}{\phi(t)} = -\sum_{\beta = 1}^k \restr{\parder{\phi^\beta_i}{t^\beta}}{t}\,,\\
		\restr{\parder{H}{p_i^\alpha}}{\phi(t)} = \restr{\parder{\phi^i}{t^\alpha}}{t}\,.
	\end{dcases}
\end{equation*}

\section[Lagrangian formalism for autonomous field theories]{Lagrangian formalism for autonomous field \phantom{.} theories}
\label{sec:k-symplectic-Lagrangian-systems}

Let $Q$ be a smooth $n$-dimensional manifold with coordinates $(q^i)$. Consider the tangent bundle of $k$-velocities $\oplus^k\T Q$ with natural coordinates $(q^i,v^\alpha_i)$.

\begin{dfn}\label{dfn:k-symplectic-vertical-lifts}
	The \textbf{vertical $\alpha$-th lift} $(u_q)^\alpha$ of a vector $u_q\in\T_qQ$ is defined as
	$$ (u_q)^\alpha(v_{1q},\dotsc,v_{kq}) = \restr{\frac{\d}{\d s}\left( v_{1q},\dotsc,v_{\alpha-1q},v_{\alpha q} + s u_q,v_{\alpha+1q},\dotsc,v_{kq} \right)}{s = 0}\,, $$
	where $v = (v_{1q},\dotsc,v_{kq})\in\oplus^k\T Q$.
\end{dfn}
In local coordinates, if $v_q = a^i\restr{\dparder{}{q^i}}{q}$, then $(u_q)^\alpha  = a^i\restr{\dparder{}{v^i_\alpha}}{v}$.

\begin{dfn}\label{dfn:k-symplectic-canonical-tangent-structure}
	The \textbf{canonical $k$-tangent structure} on $\oplus^k\T Q$ is the set $(J^1,\dotsc,J^k)$ of (1,1)-tensors in $\oplus^k\T Q$ given by
	$$ J^\alpha(v)(Z_v) = (\T_v\tau_Q(Z_v))^\alpha\,, $$
	where $Z_v\in\T_v(\oplus^k\T Q)$ and $v = (v_{1q},\dotsc,v_{kq})\in\oplus^k\T Q$.
\end{dfn}
In local coordinates, we have
$$ J^\alpha = \parder{}{v^i_\alpha}\otimes\d q^i\,. $$
\begin{dfn}\label{dfn:k-symplectic-Liouville-vector-field}
	Consider the vector fields $\Delta_\alpha\in\X(\oplus^k\T Q)$ infinitesimal generators of the flows
	$$
	\begin{array}{ccl}
	\R \times \oplus^k\T Q & \longrightarrow &
	 \oplus^k\T Q  \\
	\noalign{\medskip} (s,({v_1}_q,\dotsc , {v_k}_q)) &
	\longmapsto & ({v_1}_q,\dotsc,{v_{\alpha-1}}_q, e^s \,
	{v_\alpha}_q,{v_{\alpha+1}}_q, \ldots,{v_k}_q)\, .
	\end{array}
	$$
	The \textbf{Liouville vector field} is $\Delta = \sum_\alpha\Delta_\alpha$.
\end{dfn}
In local coordinates,
$$ \Delta_\alpha = \sum_i v_\alpha^i\parder{}{v_\alpha^i}\ ,\qquad \Delta = v_\alpha^i\parder{}{v_\alpha^i}\,. $$
\begin{dfn}
	A $k$-vector field $\bfX = (X_1\dotsc,X_k)\in\X^k(\oplus^k\T Q)$ is a \textbf{second order partial differential equation} ({\sc sopde} for short) if $J^\alpha(X_\alpha) = \Delta$.
\end{dfn}
A \textsc{sopde} $\bfX = (X_1\dotsc,X_k)$ is locally given by
$$ X_\alpha = v_\alpha^i\parder{}{q^i} + (X_\alpha)^i_\beta\parder{}{v^i_\beta}\,. $$

\begin{dfn}
	Consider a map $\phi\colon\R^k\to Q$ and let $\phi'$ be its first prolongation to $\oplus^k\T Q$. The map $\phi'$ is said to be \textbf{holonomic}.
\end{dfn}
In local coordinates,
$$ \phi'(t) = \left( \phi^i(t),\parder{\phi^i}{t^\alpha}(t) \right)\,. $$


\begin{prop}
	Let $\bfX = (X_1,\dotsc,X_k)$ be an integrable {\sc sopde} with local expression
	$$ X_\alpha = v_\alpha^i\parder{}{q^i} + (X_\alpha)^i_\beta\parder{}{v^i_\beta}\,. $$

	If $\psi\colon\R^k\to\oplus^k\T Q$ is an integral section of $\bfX$, then $\psi = \phi'$, where $\phi$ is the first prolongation of the map $\phi = \tau^1_Q\circ\psi\colon \R^k\to Q$, $\tau^1_Q\colon\oplus^k\T Q\to Q$ is the canonical projection, and $\phi$ is a solution of the system of second order partial differential equations
	\begin{equation}\label{eq:k-symp-sopde-condition-coord}
		\parderr{\psi^i}{t^\alpha}{t^\beta}(t) = (X_\alpha)^i_\beta(\psi(t))\,.
	\end{equation}
	Conversely, if $\phi\colon\R^k\to Q$ is a map satisfying conditions \eqref{eq:k-symp-sopde-condition-coord}, then $\phi'$ is an integral section of $\bfX = (X_1,\dotsc,X_k)$.

\end{prop}

\begin{rmrk}\rm
	If $\bfX = (X_\alpha)$ is an integrable {\sc sopde}, the proposition above implies that $(X_\alpha)^i_\beta = (X_\beta)^i_\alpha$.
\end{rmrk}

\begin{dfn}\label{dfn:k-symplectic-Cartan-forms}
	A \textbf{Lagrangian function} is a function $L\colon\oplus^k\T Q\to\R$. The \textbf{Lagrangian energy} associated to $L$ is $E_L = \Delta(L) - L\in\Cinfty(\oplus^k\T Q)$.

	The \textbf{Cartan forms} associated to $L$ are
	$$ \theta^\alpha_L = \transp{J^\alpha}\circ\d L\in\Omega^1(\oplus^k\T Q)\ ,\quad \omega^\alpha_L = -\d\theta^\alpha_L\in\Omega^2(\oplus^k\T Q)\,. $$
\end{dfn}
Taking coordinates $(q^i,v^i_\alpha)$ in $\oplus^k\T Q$, we have
\begin{align*}
	E_L &= v^i_\alpha\parder{L}{v^i_\alpha} - L\,,\\
	\theta^\alpha_L &= \parder{L}{v^i_\alpha}\d q^i\,,\\
	\omega^\alpha_L &= -\parderr{L}{q^j}{v^i_\alpha}\d q^j\wedge\d q^i - \parderr{L}{v^j_\beta}{v^i_\alpha}\d v^j_\beta\wedge\d q^i\,.
\end{align*}

Now, taking into account Definition \ref{dfn:fibre-derivative}, we can define
\begin{dfn}
	The \textbf{Legendre map} of $L$ is its fibre derivative $\F L\colon\oplus^k\T Q\to\oplus^k\cT Q$.
\end{dfn}

In local coordinates,
$$ \F L(q^i,v^i_\alpha) = \left( q^i,\parder{L}{v^i_\alpha} \right)\,. $$

\begin{dfn}
	A Lagrangian function $L\in\Cinfty(\oplus^k\T Q)$ is said to be \textbf{regular} if the following equivalent conditions hold:
	\begin{enumerate}[{\rm(1)}]
		\item The matrix $\left( \dparderr{L}{v^i_\alpha}{v^j_\beta} \right)$ is everywhere nonsingular.
		\item The second fibre derivative $\F^2 L\colon\oplus^k\T Q\to \oplus^k\cT Q\otimes\oplus^k\cT Q$ is everywhere nonsingular.
		\item The Legendre map $\F L$ is a local diffeomorphism.
		\item The family $(\oplus^k\T Q, \omega_L^\alpha, V = \ker\T\tau^1_Q)$ is a $k$-symplectic manifold.
	\end{enumerate}
	Otherwise, the Lagrangian $L$ is \textbf{singular}. The Lagrangian $L$ is \textbf{hyperregular} if the Legendre map $\F L$ is a global diffeomorphism.
\end{dfn}

\begin{dfn}
	A singular Lagrangian $L$ is \textbf{almost-regular} if
	\begin{enumerate}[{\rm(1)}]
		\item $\P = \F L(\oplus^k\T Q)\subseteq\oplus^k\cT Q$ is a closed submanifold.
		\item The Legendre map $\F L$ is a submersion onto its image.
		\item The fibres $\F L^{-1}(p)$ are connected submanifolds of $\oplus^k\T Q$, for every $p\in\P$.
	\end{enumerate}
\end{dfn}

Then, $(\oplus^k\T Q, \omega^\alpha_L, E_L)$ is a $k$-symplectic or $k$-presymplectic Hamiltonian system, depending on the regularity of the Lagrangian function $L$.

\begin{dfn}
	Consider the Lagrangian function $L\in\Cinfty(\oplus^k\T Q)$ and its associated Hamiltonian system $(\oplus^k\T Q, \omega^\alpha_L, E_L)$. A $k$-vector field $\bfX=(X_1,\dotsc,X_k)\in\X^k(\oplus^k\T Q)$ is a \textbf{$k$-symplectic Lagrangian $k$-vector field} if it is a solution to the \textbf{$k$-symplectic Lagrangian equation}
	\begin{equation}\label{eq:k-symplectic-Lagrangian-equation}
		i(X_\alpha)\omega^\alpha_L = \d E_L\,.
	\end{equation}
	We will denote by $\X_L^k(\oplus^k\T Q)$ the set of $k$-symplectic Lagrangian $k$-vector fields.
\end{dfn}

Consider a $k$-vector field $\bfX = (X_\alpha)\in\X^k(\oplus^k\T Q)$ with local expression
$$ X_\alpha = (X_\alpha)^i\parder{}{q^i} + (X_\alpha)^i_\beta\parder{}{v^i_\beta}\,. $$
Then, equation \eqref{eq:k-symplectic-Lagrangian-equation} reads
\begin{align*}
	\left( \parderr{L}{q^i}{v^j_\alpha} - \parderr{L}{q^j}{v^i_\alpha} \right)(X_\alpha)^j - \parderr{L}{v^i_\alpha}{v^j_\beta}(X_\alpha)^j_\beta &= v_\alpha^j\parderr{L}{q^i}{v^j_\alpha} - \parder{L}{q^i}\,,\\
	\parderr{L}{v^j_\beta}{v^i_\alpha}(X_\alpha)^i &= \parderr{L}{v^j_\beta}{v^i_\alpha}v^i_\alpha\,.
\end{align*}
If the Lagrangian $L$ is regular, these equations become
\begin{align*}
	\parderr{L}{q^j}{v^i_\alpha}v^j_\alpha + \parderr{L}{v^i_\alpha}{v^j_\beta}(X_\alpha)^j_\beta &= \parder{L}{q^i}\,,\\
	(X_\alpha)^i &= v^i_\alpha\,.
\end{align*}
Hence, we can state the following theorem:
\begin{thm}
	Consider a Lagrangian $L\in\Cinfty(\oplus^k\T Q)$ and let $\bfX = (X_1,\dotsc,X_k)\in\X^k_L(\oplus^k\T Q)$. Then,
	\begin{enumerate}[{\rm(1)}]
		\item If the Lagrangian $L$ is regular, $\bfX$ is a {\sc sopde}. Moreover, if $\psi\colon\R^k\to\oplus^k\T Q$ is an integral section of $\bfX$, the map $\phi = \tau^1_Q\circ\psi\colon\R^k\to Q$ is a solution to the \textbf{Euler--Lagrange field equations}
		\begin{equation}\label{eq:k-symp-Euler-Lagrange-field-equations}
			\restr{\parder{}{t^\alpha}}{t}\left(\restr{\parder{L}{v^i_\alpha}}{\phi'(t)}\right) = \restr{\parder{L}{q^i}}{\phi'(t)}\,.
		\end{equation}
		\item If $\bfX = (X_1,\dotsc,X_k)$ is integrable and $\phi'\colon\R^k\to\oplus^k\T Q$ is an integral section of $\bfX$, then $\phi\colon\R^k\to Q$ is a solution to the Euler--Lagrange field equations \eqref{eq:k-symp-Euler-Lagrange-field-equations}.
	\end{enumerate}
\end{thm}

\section{\texorpdfstring{$k$}--cosymplectic geometry}
\label{sec:k-cosymplectic-geometry}

$k$-cosymplectic geometry is the natural geometric framework to deal with nonautonomous Hamiltonian and Lagrangian field theories. In this section we will define the concept of $k$-cosymplectic manifold and Reeb vector fields. We will also state the Darboux theorem for $k$-cosymplectic manifolds, which says that every $k$-cosymplectic manifold is locally diffeomorphic to $\R^k\times\oplus^k\cT Q$. See \cite{DeLeo1998,DeLeo2015}.

\begin{dfn}\label{dfn:k-cosymplectic-manifold}
	Consider a smooth manifold $M$ of dimension $m = k(n+1) + n$. A \textbf{$k$-cosymplectic structure} on $M$ is a family $(\eta^\alpha, \omega^\alpha, V)$, where $\alpha = 1,\dotsc,k$, $\eta^\alpha\in\Omega^1(M)$ are closed 1-forms, $\omega^\alpha\in\Omega^2(M)$ are closed 2-forms and $V$ is an integrable $nk$-dimensional distribution on $M$ satisfying
	\begin{enumerate}[{\rm(1)}]
		\item $\eta^1\wedge\dotsb\wedge\eta^k\neq 0$, $\restr{\eta^\alpha}{V} = 0$, $\restr{\omega^\alpha}{V\times V} = 0$,
		\item $\left(\bigcap_\alpha\ker\eta^\alpha\right)\cap\left(\bigcap_\alpha\ker\omega^\alpha\right) = \{ 0 \}$,
		\item $\dim\left(\bigcap_\alpha\ker\omega^\alpha\right) = k$.
	\end{enumerate}
	Under these hypotheses, $(M,\eta^\alpha,\omega^\alpha,V)$ is a \textbf{$k$-cosymplectic manifold}.
\end{dfn}

\begin{rmrk}\rm
	In particular, if $k=1$, then $\dim M = 2n+1$ and $(M,\eta^1,\omega^1)$ is a cosymplectic manifold.
\end{rmrk}

\begin{prop}\label{prop:k-cosymplectic-Reeb}
	Let $(M,\eta^\alpha,\omega^\alpha,V)$ be a $k$-cosymplectic manifold. There exist a unique family of $k$ vector fields $R_\alpha\in\X(M)$ such that
	\begin{equation*}
		\begin{cases}
			i(R_\alpha)\eta^\beta = \delta_\alpha^\beta\,,\\
			i(R_\alpha)\omega^\beta = 0\,.
		\end{cases}
	\end{equation*}
	These vector fields are called \textbf{Reeb vector fields}.
\end{prop}

The following theorem is the $k$-cosymplectic counterpart of Theorem \ref{thm:k-symplectic-Darboux}. A proof of this theorem can be found in \cite{DeLeo1998}.

\begin{thm}[Darboux Theorem for $k$-cosymplectic manifolds]
	Consider a $k$-cosymplectic manifold $(M,\eta^\alpha,\omega^\alpha,V)$ of dimension $m = k(n+1) + n$. Then, around every point of $M$, there exists local coordinates $(t^\alpha, q^i, p_i^\alpha)$, with $1 \leq\alpha\leq k$, $1\leq i \leq n$, such that
	$$ \eta^\alpha = \d t^\alpha\ ,\quad\omega^\alpha = \d q^i\wedge\d p_i^\alpha\ ,\quad V = \left\langle \parder{}{p_i^\alpha}\right\rangle\ . $$
	These coordinates are called \textbf{Darboux} or \textbf{canonical coordinates} of the $k$-cosymplectic manifold $M$.
\end{thm}

Taking Darboux coordinates, the Reeb vector fields are
$$ R_\alpha = \parder{}{t^\alpha}\,. $$

Let $(M,\eta^\alpha,\omega^\alpha,V)$ be a $k$-cosymplectic manifold. We can define the vector bundle morphisms
$$ \tilde\flat\colon\oplus^k\T M\to\oplus^k\cT M$$
given by $\tilde\flat(v) = \left( i(v_1)\omega^1 + (i(v_1)\eta^1)\eta^1,\dotsc, i(v_k)\omega^k + (i(v_k)\eta^k)\eta^k \right)$, and
\begin{equation}\label{eq:k-cosymplectic-flat}
	\flat\colon\oplus^k\T M\to\cT M
\end{equation}
defined by $\flat(v) = \sum_\alpha\left(i(v_\alpha)\omega^\alpha + (i(v_\alpha)\eta^\alpha)\eta^\alpha\right)$, where $v = (v_1,\dotsc,v_k)\in\oplus^k\T M$.

\begin{rmrk}\rm
	Notice that $\flat = \tr(\tilde\flat)$, and hence in the case $k=1$, we have that $\flat = \tilde\flat$, which is the $\flat$ morphism defined for cosymplectic manifolds.
\end{rmrk}

\subsection*{Trivial \texorpdfstring{$k$}--cosymplectic manifolds}

Let $(N, \varpi^1,\dotsc,\varpi^k, {\cal V})$ be a $k$-symplectic manifold of dimension $n(k+1)$. Consider now the product manifold $M = \R^k\times N$ and the canonical projections
$$ \pi_{\R^k}\colon\R^k\times N\to\R^k\ ,\qquad \pi_N\colon\R^k\times N\to N\ . $$
We can define the differential forms
$$ \eta^\alpha = \pi_{\R^k}^\ast(\d t^\alpha)\in\Omega^1(M)\ ,\quad \omega^\alpha = \pi_N^\ast\varpi^\alpha\in\Omega^2(M)\,, $$
where $(t^\alpha)$ are the canonical coordinates in $\R^k$. On the other hand, the distribution ${\cal V}$ in $N$ defines a distribution $V$ in $M = \R^k\times N$ in a natural way. Notice that all conditions in Definition \ref{dfn:k-cosymplectic-manifold} are fulfilled and hence $(M,\eta^\alpha,\omega^\alpha,V)$ is a $k$-cosymplectic manifold.

Taking into account the previous example, the simplest model of $k$-cosymplectic manifold is the so called \textbf{stable contangent bundle of $k^1$-covelocities} of an $n$-dimensional smooth manifold $Q$, denoted by $\R^k\times\oplus^k\cT Q$, where $\oplus^k\cT Q$ is the Whitney sum of $k$ copies of the cotangent bundle of $Q$, i.e. $\oplus^k\cT Q = \cT Q\oplus_Q\overset{(k)}{\dotsb}\oplus_Q\cT Q$. Thus, the elements of $\R^k\times\oplus^k\cT Q$ are of the form $(t, {\nu_1}_q, \dotsc,{\nu_k}_q)$ where $t\in\R^k$, $q\in Q$ and ${\nu_\alpha}_q\in\T_q^\ast Q$.

The following diagram summarizes the projections we will use from now on:
\begin{equation*}
	\xymatrix{ &&&& \R^k\times (T^1_k)^\ast Q \ar[rr]^{\overline{\pi}_2} \ar[ddrr]^{(\pi_Q)_1} \ar[dd]^{(\pi_Q)_{1,0}} \ar[ddll]_{\overline{\pi}_1} \ar@/_2pc/[ddllll]_{\pi_1^\alpha} \ar@/^2.5pc/[rrrr]^{\pi_2^\alpha} && (T^1_k)^\ast Q \ar[rr]^{\pi^{k,\alpha}} \ar[dd]^{\pi^k} && T^\ast Q \ar[ddll]_{\pi} \\\\
	\R && \R^k \ar[ll]_{\pi^\alpha} && \R^k\times Q \ar[ll]_{\pi_{\R^k}} \ar[rr]^{\pi_Q} && Q}
\end{equation*}
If $(q^i)$, $1\leq i\leq n$, is a local coordinate system defined on an open subset $U\subset Q$, the induced local coordinates $(t^\alpha,q^i,p_i^\alpha)$, $1\leq\alpha\leq k$, on $\R^k\times\oplus^k\cT U = ((\pi_Q)_1)^{-1}(U)$ given by
\begin{align*}
	t^\alpha(t, {\nu_1}_q, \dotsc,{\nu_k}_q) &= t^\alpha(t) = t^\alpha\,,\\
	q^i(t, {\nu_1}_q, \dotsc,{\nu_k}_q) &= q^i(q)\,,\\
	p_i^\alpha(t, {\nu_1}_q, \dotsc,{\nu_k}_q) &= {\nu_\alpha}_q\left( \restr{\parder{}{q^i}}{q} \right)\,.
\end{align*}
Hence, $\R^k\times\oplus^k\cT Q$ is endowed with a $k$-cosymplectic structure and thus it is a $k$-cosymplectic manifold of dimension $k + n(k+1)$. This manifold has the structure of a vector bundle over $Q$ with the projection $(\pi_Q)_1$.

On $\R^k\times\oplus^k\cT Q$ we can define the family of canonical forms
$$ \eta^\alpha = (\pi^\alpha_1)^\ast \d t\ ,\quad\theta^\alpha = (\pi_2^\alpha)^\ast\theta\ ,\quad \omega^\alpha = (\pi_2^\alpha)^\ast\omega\ , $$
with $1\leq\alpha\leq k$, being $\pi_1^\alpha\colon\R^k\times\oplus^k\cT Q\to\R$ and $\pi^\alpha_2\colon\R^k\times\oplus^k\cT Q\to \cT Q$ the projections given by
$$ \pi_1^\alpha(t, {\nu_1}_q, \dotsc,{\nu_k}_q) = t^\alpha\ ,\quad \pi^\alpha_2(t, {\nu_1}_q, \dotsc,{\nu_k}_q) = {\nu_\alpha}_q\,, $$
and $\theta$ and $\omega$ are the canonical Liouville and symplectic forms on $\cT Q$ respectively. Notice that, since $\omega = -\d\theta$, we have that $\omega^\alpha = -\d\theta^\alpha$.

Considering the local coordinate system $(t^\alpha,q^i,p_i^\alpha)$ on $\R^k\times\oplus^k\cT Q$, the canonical forms $\eta^\alpha$, $\theta^\alpha$ and $\omega^\alpha$ have local expressions
$$ \eta^\alpha = \d t^\alpha\ ,\quad \theta^\alpha = p_i^\alpha\d q^i\ ,\quad \omega^\alpha = \d q^i\wedge\d p_i^\alpha\,. $$
In addition, consider the distribution $V = \ker\T(\pi_Q)_{1,0}$. In local coordinates, the forms $\eta^\alpha$ and $\omega^\alpha$ are closed and satisfy the relations
\begin{enumerate}[{\rm(1)}]
	\item $\d t^1\wedge\dotsb\wedge\d t^k\neq 0$, $\restr{\d t^\alpha}{V} = 0$, $\restr{\omega^\alpha}{V\times V} = 0$,
	\item $\left(\bigcap_\alpha\ker\d t^\alpha\right)\cap\left(\bigcap_\alpha\ker\omega^\alpha\right) = \{ 0 \}$,
	\item $\dim\left(\bigcap_\alpha\ker\omega^\alpha\right) = k$.
\end{enumerate}
\begin{rmrk}\rm
	Notice that the canonical forms on $\oplus^k\cT Q$ and $\R^k\times\oplus^k\cT Q$ are $(\bar\pi_2)^\ast$-related.
\end{rmrk}

\section{\texorpdfstring{$k$}--cosymplectic Hamiltonian systems}
\label{sec:k-cosymplectic-Hamiltonian-systems}

\begin{dfn}
	Consider a $k$-cosymplectic manifold $(M,\eta^\alpha,\omega^\alpha, V)$ and let $\gamma\in\Omega^1(M)$ be a closed 1-form on $M$, which will be called the \textbf{Hamiltonian 1-form}. The family $(M,\eta^\alpha,\omega^\alpha,V,\gamma)$ is a \textbf{$k$-cosymplectic Hamiltonian system}.

	A $k$-vector field $\bfX = (X_1,\dotsc,X_k)\in\X^k(M)$ is a \textbf{$k$-cosymplectic Hamiltonian $k$-vector field} if it is a solution to the system of equations
	\begin{equation}\label{eq:k-cosymplectic-Hamilton-equations-fields}
		\begin{cases}
			i(X_\alpha)\omega^\alpha = \gamma - \gamma(R_\alpha)\eta^\alpha\,,\\
			i(X_\alpha)\eta^\beta = \delta_\alpha^\beta\,.
		\end{cases}
	\end{equation}
	We will denote by $\X_H^k(M)$ the set of $k$-cosymplectic Hamiltonian $k$-vector fields on $M$. 
\end{dfn}

\begin{rmrk}\rm
	Notice that in the case $k=1$ we recover the equation of motion for a cosymplectic Hamiltonian system \cite{Chi1994,DeLeo2002}.
\end{rmrk}

As the Hamiltonian 1-form $\gamma$ is closed, by Poincaré's Lemma, there exists a local function $H$ such that $\gamma = \d H$. Using the $\flat$ morphism defined in \eqref{eq:k-cosymplectic-flat}, we can rewrite equations \eqref{eq:k-cosymplectic-Hamilton-equations-fields} as
\begin{equation*}
	\begin{cases}
		\flat(\bfX) = \gamma + (1 - \gamma(R_\alpha))\eta^\alpha\,,\\
		i(X_\alpha)\eta^\beta = \delta_\alpha^\beta
	\end{cases}
\end{equation*}
for a $k$-vector field $\bfX = (X_1,\dotsc,X_k)\in\X^k(M)$. Consider now an arbitrary $k$-vector field $\bfX = (X_1,\dotsc,X_k)$ with local expression
$$ X_\alpha = (X_\alpha)^\beta\parder{}{t^\beta} + (X_\alpha)^i\parder{}{q^i} + (X_\alpha)^\beta_i\parder{}{p_i^\beta}\,. $$
Imposing equations \eqref{eq:k-cosymplectic-Hamilton-equations-fields}, we obtain the relations
\begin{equation}\label{eq:k-cosymplectic-Hamilton-equations-fields-coordinates}
	\begin{dcases}
		(X_\alpha)^\beta = \delta_\alpha^\beta\,,\\
		\parder{H}{p_i^\alpha} = (X_\alpha)^i\,,\\
		\parder{H}{q^i} = -\sum_{\alpha=1}^k(X_\alpha)^\alpha_i\,.
	\end{dcases}
\end{equation}
Notice that these conditions do not depend on the Reeb vector fields. However, we need Reeb vector fields to write the equations.

Consider the map $\R^k\to\R^k\times\oplus^k\cT Q$ given by $\psi(t) = (\psi^\alpha(t),\psi^i(t),\psi^\alpha_i(t))$. If $\psi$ is an integral section of the $k$-vector field $\bfX$, from \eqref{eq:k-cosymplectic-Hamilton-equations-fields-coordinates}, we have that $\psi$ has to be a solution to the so called \textbf{$k$-cosymplectic Hamiltonian field equations}
\begin{equation*}
	\begin{dcases}
		\parder{H}{q^i} = -\sum_{\alpha=1}^k\parder{\psi_i^\alpha}{t^\alpha}\,,\\
		\parder{H}{p_i^\alpha} = \parder{\psi^i}{t^\alpha}\,.
	\end{dcases}
\end{equation*}

\section{Lagrangian formalism for nonautonomous field theories}
\label{sec:k-cosymplectic-Lagrangian-systems}

Consider the phase space $\R^k\times\oplus^k\T Q$ endowed with canonical coordinates $(t^\alpha,q^i, v^i_\alpha)$. The canonical structures $J^\alpha$ and the Liouville vector field $\Delta$ introduced in Definitions \ref{dfn:k-symplectic-canonical-tangent-structure} and \ref{dfn:k-symplectic-Liouville-vector-field} can be trivially extended from $\oplus^k\T Q$ to $\R^k\times\oplus^k\T Q$. Their local expressions remain the same:
$$ J^\alpha = \parder{}{v^i_\alpha}\ ,\quad \Delta = v^i_\alpha\parder{}{v^i_\alpha}\,. $$
Using these structures, we can define
\begin{dfn}
	A $k$-vector field $\bfX = (X_1,\dotsc,X_k)\in\X^k(\R^k\times\oplus^k\T Q)$ is a \textbf{second order partial differential equation} ({\sc sopde}) if the following conditions hold:
	\begin{enumerate}[{\rm(1)}]
		\item $J^\alpha(X_\alpha) = \Delta$,
		\item $i(X_\alpha)\d t^\beta = \delta_\alpha^\beta$.
	\end{enumerate}
\end{dfn}
If $\bfX=(X_1,\dotsc,X_k)$ is a {\sc sopde}, then it has local expression
$$ X_\alpha = \parder{}{t^\alpha} + v^i_\alpha\parder{}{q^i} + (X_\alpha)^i_\beta\parder{}{v^i_\beta}\,. $$
\begin{dfn}
	Consider the map $\phi\colon\R^k\to Q$. Its \textbf{first prolongation} is the map $\psi'\colon\R^k\to\R^k\times\oplus^k\T Q$ given by
	$$ \phi'(t) = \left(t, j_0^1\phi_t\right) \equiv \left( t, \T_t\phi\left( \restr{\parder{}{t^1}}{t} \right),\dotsc,\T_t\phi\left(\restr{\parder{}{t^k}}{t} \right) \right)\,. $$
	The map $\phi'$ is said to be \textbf{holonomic} and its local expression is
	$$ \phi'(t^1,\dotsc,t^k) = \left( t^1,\dotsc,t^k,\phi^i(t),\parder{\phi^i}{t^\alpha}(t) \right) $$
\end{dfn}

\begin{prop}
	Let $\bfX = (X_1,\dotsc,X_k)\in\X^k(M)$ be an integrable {\sc sopde} and consider the map $\psi\colon\R^k\to\R^k\times\oplus^k\T Q$ given by $\psi(t) = (\psi^\alpha(t),\psi^i(t),\psi^i_\alpha(t))$. Then, $\psi$ is an integral section of $\bfX$ if, and only if, the following conditions hold:
	\begin{equation*}
		\begin{dcases}
			\psi^\alpha(t) = t^\alpha\,,\\
			\psi^i_\alpha(t) = \parder{\psi^i}{t^\alpha}(t)\,,\\
			\parderr{\psi^i}{t^\alpha}{t^\beta}(t) = (X_\alpha)^i_\beta(\psi(t))\,.
		\end{dcases}
	\end{equation*}
	In this case, $\psi$ is a holonomic section of $\bfX$.
\end{prop}
Notice that if $\bfX = (X_1,\dotsc,X_k)$ is integrable, from the previous proposition, we deduce that $(X_\alpha)^i_\beta = (X_\beta)^i_\alpha$.

\begin{dfn}
	A \textbf{Lagrangian function} on $\R^k\times\oplus^k\T Q$ is a function $L\in\Cinfty(\R^k\times\oplus^k\T Q)$. Its associated \textbf{Lagrangian energy} is the function $E_L = \Delta(L)-L$. The \textbf{Cartan forms} associated to the Lagrangian $L$ are
	\begin{align*}
		\theta^\alpha_L &= \transp{J^\alpha}\circ\d L\in\Omega^1(\R^k\times\oplus^k\T Q)\,,\\
		\omega^\alpha_L &= -\d\theta^\alpha_L\in\Omega^2(\R^k\times\oplus^k\T Q)\,.
	\end{align*}
\end{dfn}
Taking coordinates $(t^\alpha,q^i,v^i_\alpha)$ in $\R^k,\oplus^k\T Q$, we have
\begin{align*}
	E_L &= v^i_\alpha\parder{L}{v^i_\alpha} - L\,,\\
	\theta^\alpha_L &= \parder{L}{v^i_\alpha}\d q^i\,,\\
	\omega^\alpha_L &= -\parderr{L}{t^\beta}{v^i_\alpha}\d t^\beta\wedge\d q^i-\parderr{L}{q^j}{v^i_\alpha}\d q^j\wedge\d q^i - \parderr{L}{v^j_\beta}{v^i_\alpha}\d v^j_\beta\wedge\d q^i\,.
\end{align*}

\begin{dfn}
	A Lagrangian function $L\in\Cinfty(\R^k\times\oplus^k\T Q)$ is \textbf{regular} if the matrix
	$$ \left( \parderr{L}{v^i_\alpha}{v^j_\beta} \right) $$
	is invertible. Otherwise, the Lagrangian is \textbf{singular}.
\end{dfn}
\begin{prop}
	Consider a Lagrangian $L\in\Cinfty(\R^k\times\oplus^k\T Q)$. Then, $L$ is regular if, and only if $(\d t^\alpha,\omega^\alpha_L,V = \ker\T(\pi_{\R^k})_{1,0})$ is a $k$-cosymplectic structure on $\R^k\times\oplus^k\T Q$.
\end{prop}

\begin{dfn}
	A $k$-vector field $\bfX = (X_1,\dotsc,X_k)\in\X^k(\R^k\times\oplus^k\T Q)$ is called a \textbf{$k$-cosymplectic Lagrangian $k$-vector field} if it is a solution to the \textbf{$k$-cosymplectic Lagrangian equations}
	\begin{equation}\label{eq:k-cosymplectic-Lagrange-equations-fields}
		\begin{dcases}
			i(X_\alpha)\omega^\alpha_L = \d E_L + \parder{L}{t^\alpha}\d t^\alpha\,,\\
			i(X_\alpha)\d t^\beta = \delta_\alpha^\beta\,.
		\end{dcases}
	\end{equation}
	We will denote by $\X^k_L(\R^k\times\oplus^k\T Q)$ the set of $k$-cosymplectic Lagrangian $k$-vector fields.
\end{dfn}

\begin{rmrk}\rm
	Notice that if the Lagrangian $L$ is regular, $(\R^k\times\oplus^k\T Q,\d t^\alpha,\omega^\alpha_L,V)$ is a $k$-cosymplectic manifold. Denote by $R_\alpha^L$ its corresponding Reeb vector fields. Hence, we can rewrite equations \eqref{eq:k-cosymplectic-Lagrange-equations-fields} as
	\begin{equation}\label{eq:k-cosymplectic-Lagrange-equations-fields-Reeb}
		\begin{dcases}
			i(X_\alpha)\omega^\alpha_L = \d E_L + R_\alpha^L(L)\d t^\alpha\,,\\
			i(X_\alpha)\d t^\beta = \delta_\alpha^\beta\,.
		\end{dcases}
	\end{equation}
\end{rmrk}

If $\bfX = (X_1,\dotsc,X_k)\in\X^k(\R^k\times\oplus^k\T Q)$ is an integrable {\sc sopde} solution to \eqref{eq:k-cosymplectic-Lagrange-equations-fields-Reeb}, its integral sections are solutions to the \textbf{Euler--Lagrange equations} for the Lagrangian $L$:
\begin{equation*}
	\parderr{L}{t^\alpha}{v^i_\alpha} + \parderr{L}{q^j}{v^i_\alpha}\parder{\psi^j}{t^\alpha} + \parderr{L}{v^j_\beta}{v^i_\alpha}\parderr{\psi^j}{t^\alpha}{t^\beta} = \parder{L}{q^i}\,.
\end{equation*}

\begin{rmrk}\rm
	In the Hamiltonian framework, we saw that the Reeb vector fields appear in the equations but not in the solutions. In the Lagrangian framework we can go one step further and write equations \eqref{eq:k-cosymplectic-Lagrange-equations-fields-Reeb} without using the Reeb vector fields \cite{Bua2015}. Consider the \textbf{Poincaré-Cartan 1-forms}:
	$$ \Theta^\alpha_L = \theta^\alpha_L + (\delta^\alpha_\beta L - \Delta^\alpha_\beta(L))\d t^\beta\,, $$
	where $\Delta^\alpha_\beta = v^i_\beta\dparder{}{v^i_\alpha}$. Defining $\Omega^\alpha = -\d\Theta^\alpha$, equations \eqref{eq:k-cosymplectic-Lagrange-equations-fields-Reeb} become
	\begin{equation*}
		\begin{dcases}
			i(X_\alpha)\Omega^\alpha_L = (k-1)\d L\,,\\
			i(X_\alpha)\d t^\beta = \delta^\beta_\alpha\,,
		\end{dcases}
	\end{equation*}
	which are equivalent to \eqref{eq:k-cosymplectic-Lagrange-equations-fields}.
\end{rmrk}

\begin{dfn}
	The \textbf{Legendre map} of a Lagrangian function $L\in\Cinfty(\R^k\times\oplus^k\T Q)$ is its fibre derivative $\F L\colon\R^k\times\oplus^k\T Q\to\R^k\times\oplus^k\cT Q$.
\end{dfn}

Taking coordinates $(t^\alpha,q^i,v^i_\alpha)$ in $\R^k\times\oplus^k\T Q$,
$$ \F L(t^\alpha,q^i,v^i_\alpha) = \left( t^\alpha,q^i,\parder{L}{v^i_\alpha} \right)\,. $$
Notice that using the Legendre map $\F L$, we can redefine the Cartan forms as $\theta^\alpha_L = \F L^\ast\theta^\alpha$ and $\omega^\alpha_L = \F L^\ast\omega^\alpha$.
\begin{prop}
	The Lagrangian $L$ is regular if, and only if, the Legendre map $\F L$ is a local diffeomorphism.
\end{prop}
\begin{dfn}
	If the Legendre map $\F L$ is a global diffeomorphism, the Lagragian function $L$ is said to be \textbf{hyperregular}. A singular Lagrangian is \textbf{almost-regular} if the following conditions are satisfied:
	\begin{enumerate}[{\rm(1)}]
		\item $\P = \F L(\R^k\times\oplus^k\T Q)\subset\R^k\times\oplus^k\cT Q$ is a closed submanifold.
		\item The Legendre map $\F L$ is a submersion onto its image.
		\item The fibres $\F L^{-1}(\F L(v))\subset\R^k\times\oplus^k\T Q$ are connected submanifolds for every $v\in\R^k\times\oplus^k\T Q$.
	\end{enumerate}
\end{dfn}

	\chapter[Constraint algorithms for singular field theories]{Constraint algorithms for \phantom{mm} singular field theories}
	\label{ch:constraint-algorithms}


This chapter is devoted to the study of constraint algorithms to deal with field theories described by singular Lagrangian functions. In Section \ref{sec:k-presymplectic-constraint-algorithm} we review the constraint algorithm for singular autonomous field theories introduced in \cite{Gra2009}. We begin by defining the notion of $k$-presymplectic manifold, which is a weakened version of the definition of $k$-symplectic manifold. We also prove a Darboux theorem for $k$-presymplectic manifolds. Then the constraint algorithm for singular autonomous field theories is summarized. In Section \ref{sec:k-precosymplectic-geometry} we introduce the concept of $k$-precosymplectic manifold as a weakened version of the notion of $k$-cosymplectic manifold in order to deal with nonautonomous field theories described by singular Lagrangian functions. We also prove the existence of a family of global Reeb vector fields for $k$-precosymplectic manifolds, although this family will not be unique. In Section \ref{sec:constraint-algorithm-k-precosymplectic} we generalize the constraint algorithm described in \cite{Gra2009} in order to deal with nonautonomous field theories. Finally, in Section \ref{sec:constraint-algorithm-examples} we analyze several examples of field theories described by singular Lagrangians and apply the constraint algorithm to both their Lagrangian and Hamiltonian formulations. See \cite{Gra2020}.

\section{The constraint algorithm for autonomous singular field theories}
\label{sec:k-presymplectic-constraint-algorithm}

In this section we will review the constraint algorithm for $k$-presymplectic Hamiltonian systems developed in \cite{Gra2009}. When dealing with field theories defined by singular Lagrangians, we have to weaken some condition in Definition \ref{dfn:k-symplectic manifold}, because the Cartan forms defined in \ref{dfn:k-symplectic-Cartan-forms} do not constitute a $k$-symplectic structure in $\oplus^k\T Q$. This motivates the following definition:

\begin{dfn}
	A family $(\omega^1,\dotsc,\omega^k)$, of closed 2-forms in a manifold $M$ is a \textbf{$k$-presymplectic structure} on $M$. Then $(M,\omega^1,\dotsc,\omega^k)$ is a \textbf{$k$-presymplectic manifold}.
\end{dfn}

\begin{rmrk}\rm
	Notice that in the particular case $k=1$ we recover the definition of presymplectic manifold. See \cite{DeLeo1989,Got1979,Got1980,Lib1987} for more details in presymplectic mechanics.
\end{rmrk}

The following theorem states that, in certain cases, we have Darboux-type coordinates in a $k$-presymplectic manifold.

\begin{thm}[Darboux Theorem for $k$-presymplectic manifold]\label{thm:k-presymplectic-Darboux-theorem}
	Consider a $k$-presymplectic manifold $(M,\omega^\alpha)$ with $\rk\omega^\alpha = 2r_\alpha$ and $\dim M = m = nk-\sum_\alpha r_\alpha - d$ equipped with a $nk$-dimensional integrable distribution $V$ such that
	$$ \restr{\omega^\alpha}{V\times V} = 0\,. $$
	Then, around every point $p\in M$, there exist a local chart $(U,q^i,p_{i_\alpha}^\alpha, z^j)$ such that
	$$ \restr{\omega^\alpha}{U} = \d q^{i_\alpha}\wedge\d p_{i_\alpha}^\alpha\ ,\quad \restr{V}{U} = \left\langle \parder{}{p_{i_\alpha}^\alpha},\parder{}{z^j} \right\rangle\ ,\quad \restr{\left( \bigcap_{\alpha=1}^k\ker\omega^\alpha \right)}{U} = \left\langle \parder{}{z^j} \right\rangle\,. $$
\end{thm}
\begin{proof}
	First, notice that, denoting by $K$ the distribution generated by $\bigcap_{\alpha=1}^k\ker\omega^\alpha$, we have that $K\subset V$. Let $d = \rk K$. Let us show that this distribution $K$ is involutive. Consider $X,Y\in K$. Then,
	$$ i([X,Y])\omega^\alpha = (\Lie_X\circ i_Y - i_Y\circ\Lie_X)\omega^\alpha = -i_Y\Lie_X\omega^\alpha = -i_Y(i_X\d\omega^\alpha + \d i_X\omega^\alpha) = 0\,. $$
	As the distribution $K$ is involutive, its integral submanifolds give a foliation in $M$. Then, for every $p\in M$, we can take a local chart of coordinates adapted to this foliation
	$$ (U,u^\ell,z^j)\ ,\quad 1\leq\ell\leq n + \sum_\alpha r_\alpha = n(k+1) - m - d\,, $$
	that is, $(z^j)$ are local coordinates in the leaves, and such that
	\begin{equation}\label{eq:proof-k-presymp-darboux-thm-K}
		\restr{\left( \bigcap_{\alpha=1}^k\ker\omega^\alpha \right)}{U} = \left\langle \parder{}{z^j} \right\rangle\,.
	\end{equation}
	Now, consider the quotient $\widetilde M = M/K$, which we assume to be a smooth manifold (if this hypothesis does not hold, we can do a local reasoning on the local chart $U$, taking the distribution given by \eqref{eq:proof-k-presymp-darboux-thm-K} as $K$) and the natural projection $\tau\colon M\to\widetilde M$. As $K\subset V$ and $V$ is involutive, the closed 2-forms $\omega^\alpha$ are $\tau$-projectable to closed forms $\widetilde\omega^\alpha\in\Omega^2(\widetilde M)$, and the distribution $V$ restricts to a $r$-dimensional integrable distribution $\widetilde V$ in $\widetilde M$, where $r = \sum_\alpha r_\alpha$, in such a way that,
	$$ \restr{\widetilde\omega^\alpha}{\widetilde V\times\widetilde V} = 0\ , \quad \bigcap_{\alpha = 1}^k\ker\widetilde\omega^\alpha = \{ 0 \}\,. $$
	At this point, we can adapt the proof of Theorem \ref{thm:k-symplectic-Darboux} to this situation, with the only difference that, now, the forms $\widetilde\omega^\alpha$ do not have the same rank (equal to $2n$). The final consequence is that, for $\tau(p) = \widetilde p$ we have a local chart of coordinates
	$$ (\widetilde U,\widetilde q^{i_\alpha},\widetilde p_{i_\alpha}^\alpha)\ ,\quad i_\alpha \in I_\alpha \subseteq\{ 1,\dotsc,n \}\ ,\quad \vert I_\alpha\vert = r_\alpha\ ,\quad 1\leq\alpha\leq n\,, $$
	where $\widetilde U = \tau (U)$, such that
	$$ \restr{\widetilde\omega^\alpha}{\widetilde U} = \d \widetilde q^{i_\alpha}\wedge\d \widetilde p_{i_\alpha}^\alpha\ ,\quad \restr{\widetilde V}{\widetilde U} = \left\langle  \parder{}{\widetilde p_{i_\alpha}^\alpha}\right\rangle\,. $$
	Therefore, in $U$ we can take coordinates $(u^\ell) = (q^i,p_{i_\alpha}^\alpha)$, with $q^i = \widetilde q^i\circ\tau$ and $p_{i_\alpha}^\alpha = \widetilde p_{i_\alpha}^\alpha\circ\tau$, and the chart $(U,q^i,p_{i_\alpha}^\alpha,z^j)$ verifies the conditions given in the statement of the theorem.
\end{proof}

\begin{dfn}
	A \textbf{$k$-presymplectic Hamiltonian system} is a family $(M,\omega^\alpha,H)$ where $(M,\omega^\alpha)$ is a $k$-presymplectic manifold and $H\in\Cinfty(M)$ is the \textbf{Hamiltonian function}.
\end{dfn}

\begin{dfn}
	Consider a $k$-presymplectic manifold $(M,\omega^\alpha,H)$. Let $\bfX = (X_\alpha)\in\X^k(M)$ be a $k$-vector field in $M$. $\bfX$ is said to be a \textbf{$k$-presymplectic Hamiltonian $k$-vector field} if it is a solution to the geometric field equation
	\begin{equation}\label{eq:k-presymplectic-Hamilton-equations-fields}
			i(X_\alpha)\omega^\alpha = \d H\,.
	\end{equation}
\end{dfn}

Notice that in the $k$-presymplectic setting the existence of solutions to equation \eqref{eq:k-presymplectic-Hamilton-equations-fields} is not assured everywhere in $M$. In what follows, we will see how to obtain a submanifold of $M$ where we can ensure the existence of solutions tangent to this submanifold.

Given a $k$-presymplectic Hamiltonian system, we want to find a submanifold $C$ of $M$ and integrable $k$-vector fields $\bfX = (X_1,\dotsc,X_k)\in\X^k(M)$ such that
$$ i(X_\alpha)\omega^\alpha = \d H\ ,\quad \mbox{(on $C$)} $$
and such that the $k$-vector field $\bfX$ is tangent to the submanifold $C$ (i.e., the vector fields $X_\alpha$ are tangent to the submanifold $C$).

Given a submanifold $C\subset M$, with natural embedding $j_C\colon C\hookrightarrow M$, consider the natural extension of $j_C$ to the $k$-tangent bundles,
$$ T^k j_C\colon\oplus^k\T C\to\oplus^k\T M\,, $$
and denote by $\underline{\oplus^k\T C} = \T^kj_C(\oplus^k\T C)$.

\begin{dfn}
	The \textbf{k-presymplectic orthogonal complement} of $\underline{\oplus^k\T C}$ in $\oplus^k\T M$ is the anihilator of the image of $\underline{\oplus^k\T C}$ by the flat morphism defined in \eqref{eq:k-symplectic-flat}:
	$$ (\T C)^\perp = \left[ \flat(\oplus^k\T C) \right]^0 = \left\{ u_p\in\T M\ \vert\ \forall({v_1}_p, \dotsc, {v_k}_p)\in\oplus^k\T C\,,\ \left\langle i({v_\alpha}_p)\omega^\alpha_p,u_p \right\rangle \right\}\,. $$
\end{dfn}

\begin{rmrk}\rm
	In the particular case $C = M$, we have that
	$$ (\T M)^\perp = \left\{ u_p\in\T M\ \vert\ u_p\in\cap_\alpha\ker\omega^\alpha_p \right\}\,.$$
\end{rmrk}

The following theorem is the main result that we need to define the constraint algorithm:
\begin{thm}\label{thm:k-presymplectic-constraint-algorithm}
	Consider a submanifold $C\subset M$. The following two conditions are equivalent:
	\begin{itemize}
		\item there exists a $k$-vector field $\bfX = (X_1,\dotsc,X_k)\in\X^k(M)$, tangent to $C$, such that equation \eqref{eq:k-presymplectic-Hamilton-equations-fields} holds.
		\item $i(Y_p)\d_p H = 0$ for every $p\in C$ and $Y_p\in(\T_p C)^\perp$.
	\end{itemize}
\end{thm}

This last theorem allows us to define an algorithmic procedure which gives a sequence of subsets
$$ \dotsb\subset C_j\subset\dotsb\subset C_2\subset C_1\subset M. $$
We will assume that every subset $C_j$ in the above sequence is a regular submanifold of $M$. We begin by defining the submanifold $C_1$ as the submanifold of $M$ where equation \eqref{eq:k-presymplectic-Hamilton-equations-fields} is consistent:
$$ C_1 = \{ p\in M\ \vert\ \exists \bfX_p \mbox{ such that } i({X_\alpha}_p)\omega^\alpha_p = \d_p H \}\,. $$
Hence, there exist $k$-vector field $\bfX$ on the submanifold $C_1$ which are solutions to equation \eqref{eq:k-presymplectic-Hamilton-equations-fields} on $C_1$. Nevertheless, in general, these $k$-vector fields may not be tangent to $C_1$. Hence, we need to consider the submanifold
$$ C_2 = \{ p\in C_1\ \vert\ \exists \bfX_p\in\oplus^k\T C \mbox{ such that } i({X_\alpha}_p)\omega^\alpha_p = \d_p H \}\,, $$
and so on. Following this method, we obtain a sequence of constraint submanifolds
$$ \dotsb\hookrightarrow C_j\hookrightarrow\dotsb\hookrightarrow C_2\hookrightarrow C_1\hookrightarrow M. $$
Now, considering Theorem \ref{thm:k-presymplectic-constraint-algorithm}, each constraint submanifold $C_j$, called the \textbf{$j$-th constraint submanifold}, can be defined as
$$ C_j = \{ p\in C_{j-1}\ \vert\ i(Y_p)\d_p H = 0 \mbox{ for every } Y_p\in(\T_pC_{j-1})^\perp \}\,. $$
Denoting by $\X(C_j)^\perp$ the set of vector fields $Y\in\X(M)$ such that $Y_p\in(\T_p C_j)^\perp$, one can obtain constraint function $\xi_\mu$ defining each $C_j$ from a local basis $\{Z_1,\dotsc,Z_r\}$ of vector fields of $\X(C_{j-1})^\perp$ by setting $\xi_\mu = i(Z_\mu)\d H$. This procedure, known as the \textbf{$k$-presymplectic constraint algorithm}, can be summarized as follows:
\begin{enumerate}[{\rm(1)}]
	\item obtain a local basis $\{Z_1,\dotsc,Z_r\}$ of vector fields of $\cap_\alpha\ker\omega^\alpha$\,,
	\item apply Theorem \ref{thm:k-presymplectic-constraint-algorithm} to obtain a set of independent constraint functions $\xi_\mu = i(Z_\mu)\d H$ defining the first constraint submanifold $C_1\hookrightarrow M$,
	\item compute the solutions $\bfX = (X_\alpha)$ to equation \eqref{eq:k-presymplectic-Hamilton-equations-fields} on the submanifold $C_1$,
	\item impose the tangency condition of the vector fields $X_1,\dotsc,X_k$ to the submanifold $C_1$, i.e. $X_\alpha(\xi_\mu) = 0$,
	\item iterate the previous step until no new constraints appear.
\end{enumerate}
When the previous algorithm finishes, we have two possibilities:
\begin{itemize}
	\item The algorithm finishes with a submanifold $C_j$, with $j>0$, such that $C_{j+1} = C_j \equiv C_f$, called the \textbf{final constraint submanifold}, with $\dim C_j > 0$. Then, there exists a family of $k$-vector fields $\bfX = (X_\alpha)$ in $M$, tangent to $C_f$, such that equation \eqref{eq:k-presymplectic-Hamilton-equations-fields} holds on $C_f$. This is the interesting case.
	\item The algorithm finishes with an empty set or a set of isolated points. This means that equation \eqref{eq:k-presymplectic-Hamilton-equations-fields} has no solution on a submanifold of $M$.
\end{itemize}

\begin{rmrk}\rm
	Notice that the $k$-presymplectic constraint algorithm described above does not include the {\sc sopde} condition. If we want our solutions to be {\sc sopde}s, we need to impose it as an extra requirement.
\end{rmrk}

\section{\texorpdfstring{$k$}--precosymplectic geometry}
\label{sec:k-precosymplectic-geometry}

In the same way as $k$-presymplectic manifolds are a generalization of $k$-symplectic manifolds that allows us to work with field theories described by singular Lagrangians, we are now going to define the concept of $k$-precosymplectic manifold \cite{Gra2020}:
\begin{dfn}\label{dfn:k-precosymplectic-manifold}
	Consider a smooth manifold $M$ of dimension $m = k(n+1) + n - \ell$ (with $1\leq\ell\leq nk$). A \textbf{$k$-precosymplectic structure} in $M$ is a family $(\eta^\alpha,\omega^\alpha,V)$, $1\leq \alpha\leq k$, where $\eta^\alpha\in\Omega^1(M)$ are closed 1-forms, $\omega^\alpha\in\Omega^2(M)$ are closed 2-forms such that $\rk\omega^\alpha = 2 r_\alpha$, with $1\leq r_\alpha\leq n$, and $V$ is an integrable $nk$-dimensional distribution in $M$ such that
	\begin{enumerate}[{\rm(1)}]
		\item $\eta^1\wedge\dotsb\wedge\eta^k\neq 0$, $\restr{\eta^\alpha}{V} = 0$, $\restr{\omega^\alpha}{V\times V} = 0$,
		\item $\dim\left( \bigcap_\alpha\ker\omega^\alpha_p \right) \geq k$ for every $p\in M$.
	\end{enumerate}
	A manifold $M$ equipped with a $k$-precosymplectic structure is called a \textbf{$k$-precosymplec\-tic manifold}.
\end{dfn}
In particular, in the case $k=1$, we have that $\dim M = 2n + 1 - \ell$ and $(M,\eta^1,\omega^1)$ is a precosymplectic manifold as is defined in \cite{Chi1994,Ibo1992}, and the so-called gauge distribution is $\ker\eta^1\cap\ker\omega^1$.

\begin{exmpl}\rm
	Consider a $k$-presymplectic manifold $(P,\varpi^\alpha,{\cal V})$. Then, the product manifold $\R^k\times P$ equipped with the 1-forms $\eta^\alpha = \tau^\ast\d t^\alpha$, where $(t^\alpha)$ are the canonical coordinates in $\R^k$ and $\tau$ is the natural projection $\R^k\times P\overset{\tau}{\longrightarrow}\R^k$, and the 2-forms $\omega^\alpha = \pi^\ast\varpi^\alpha$, where $\pi$ is the natural projection $\R^k\times P\overset{\pi}{\longrightarrow}P$. When describing the constraint algorithm, we will ask our manifolds to be of this type in order to have the problem well defined.
\end{exmpl}

In Definition \ref{dfn:k-precosymplectic-manifold} we have imposed the existence of a distribution $V$ because it is this condition what ensures the existence of Darboux coordinates in the regular case. It is still an open problem to fully characterize the conditions for the existence of Darboux-type coordinates in the singular case. From now on, we will assume the existence of Darboux coordinates around every point. For instance, the manifolds of the form $M = \R^k\times P$, with $P$ a $k$-presymplectic manifold, satisfies our requirements (see previous example). In more detail, consider $k$-precosymplectic manifold $M$ such that $\rk\omega^\alpha = 2r_\alpha$, with $1\leq r_\alpha\leq n$ and define $d = kn-\sum_{\alpha=1}^k r_\alpha - \ell$. We assume the existence around every point $p\in M$ of a local chart $(U_p; t^\alpha, q^i, p_{i_\alpha}^\alpha, z^j)$, with $1\leq\alpha\leq k$, $1\leq i\leq n$, $i_\alpha\in I_\alpha\subseteq\{1,\dotsc,n\}$, $\vert I_\alpha\vert = r_\alpha$ and $1\leq j\leq d$, such that
\begin{align*}
	\restr{\eta^\alpha}{U_p} &= \d t^\alpha\,,\\
	\restr{\omega^\alpha}{U_p} &= \d q^{i_\alpha}\wedge\d p_{i_\alpha}^\alpha\,,\\
	\restr{V}{U_p} &= \left\langle \parder{}{p_{i_\alpha}^\alpha}, \parder{}{z^j}\right\rangle\,,
\end{align*}
and
$$
	\restr{\left[\left(\bigcap_{\alpha=1}^k\ker\eta^\alpha\right)\cap\left(\bigcap_{\alpha=1}^k\ker\omega^\alpha\right)\right]}{U_p} = \left\langle \parder{}{z^j} \right\rangle\,.
$$

In order to work with Hamilton's equation, we need the Reeb vector fields $R_\alpha$ defined in Proposition \ref{prop:k-cosymplectic-Reeb}. In that proposition we already mentioned its existence and uniqueness. In the following proposition we will prove their existence in the $k$-precosymplectic case. However, we will see afterwards that they are not unique.

\begin{prop}\label{prop:Reeb-k-precosymplectic}
	Let $(M,\eta^\alpha,\omega^\alpha,V)$ be a $k$-precosymplectic manifold with Darboux charts. Then, there exists a family of vector fields $Y_1,\dotsc,Y_k\in\X(M)$ such that
	\begin{equation*}
		\begin{dcases}
			i(Y_\alpha)\omega^\beta = 0\,,\\
			i(Y_\alpha)\eta^\beta = \delta_\alpha^\beta\,.
		\end{dcases}
	\end{equation*}
\end{prop}
\begin{proof}
	Consider a partition of unity $\{(U_\lambda,\psi_\lambda)\}_{\lambda\in\Lambda}$ such that we have a Darboux chart $(U_\lambda;t^\alpha_\lambda,q^i_\lambda,p^\alpha_{i_\alpha,\lambda},z^j_\lambda)$ for every $\lambda\in\Lambda$. Consider now the vector fields
	$$ Y_\alpha^\lambda = \parder{}{t^\alpha_\lambda}\in\X(U_\lambda)\,. $$
	These vector fields clearly satisfy
	\begin{equation*}
		\begin{cases}
			i(Y_\alpha^\lambda)\omega^\beta = 0\,,\\
			i(Y_\alpha^\lambda)\eta^\beta = \delta_\alpha^\beta
		\end{cases}
	\end{equation*}
	on $U_\lambda$. Using the partition of unity introduced above, we can define global vector fields
	\begin{equation*}
		\widetilde Y_\alpha^\lambda(p) = \begin{cases}
			\psi_\lambda(p)Y_\alpha^\lambda(p) & \mbox{if }p\in U_\lambda\,,\\
			0 & \mbox{if }p\notin U_\lambda\,.\\
		\end{cases}
	\end{equation*}
	Now we can construct global vector fields $Y_\alpha = \sum_{\lambda\in\Lambda}\widetilde Y_\alpha^\lambda$ satisfying
	\begin{equation*}
		\begin{dcases}
			i(Y_\alpha)\omega^\beta = 0\,,\\
			i(Y_\alpha)\eta^\beta = \delta_\alpha^\beta\,,
		\end{dcases}
	\end{equation*}
	for every $\alpha,\beta = 1,\dotsc,k$.
\end{proof}
These Reeb vector fields are not necessarily unique. In fact, the Reeb vector fields can be written in Darboux coordinates as
$$ R_\alpha = \parder{}{t^\alpha} + D^j_\alpha\parder{}{z^j} $$
for arbitrary functions $D_\alpha^j$.
\begin{rmrk}\rm
	Nevertheless, sometimes one can impose some extra conditions that determine the Reeb vector fields uniquely. Consider for instance the $k$-precosymplectic manifold $M=\R^k\times P$, where $P$ is a $k$-presymplectic manifold. In this situation, the canonical vector fields $\tparder{}{t^\alpha}$ of $\R^k$ can be canonically lifted to the product manifold $\R^k \times P$. These vector fields are also denoted by $\tparder{}{t^\alpha}$ and are a family of Reeb vector fields of the $k$-precosymplectic manifold $\R^k \times M$.
\end{rmrk}

\section{Constraint algorithm for \texorpdfstring{$k$}--precosymplectic field theories}
\label{sec:constraint-algorithm-k-precosymplectic}

This section is devoted to generalize the constraint algorithm for $k$-presymplectic field theories developed in \cite{Gra2009} to the case of singular $k$-cosymplectic field theories \cite{Gra2020}. Throughout this section we are considering $k$-precosymplectic manifolds of the form $M = \R^k\times P$, where $P$ is a $k$-presymplectic manifold. As we pointed out in the previous section, these manifolds have Darboux-type coordinates and have a uniquely defined collection of Reeb vector fields $R_1,\dotsc,R_k$. This particular case is in fact the most common case that arises when studying classical field theories and applied mathematics. We will begin by defining the notion of $k$-precosymplectic Hamiltonian system:

\begin{dfn}
	Let $(M,\eta^\alpha,\omega^\alpha,V)$ be a $k$-precosymplectic manifold of the form $M=\R^k\times P$, with $P$ a $k$-presymplectic manifold, and $\gamma\in\Omega^1(M)$ a closed 1-form. Then the family $(M,\eta^\alpha,\omega^\alpha,V,\gamma)$ is called a \textbf{$k$-precosymplectic Hamiltonian system} and $\gamma$ is called the \textbf{Hamiltonian 1-form}. Since $\gamma$ is closed, by Poincaré's Lemma, $\gamma = \d H$ locally for some $H\in\Cinfty(U)$, $U\subset M$, called a \textbf{local Hamiltonian function}.
\end{dfn}

\begin{dfn}
	Let $\bfX = (X_1,\dotsc,X_k)\in\X^k(M)$ be a $k$-vector field in $M$. $\bfX$ is said to be a \textbf{$k$-precosymplectic Hamiltonian $k$-vector field} if it is a solution to the system
	\begin{equation}\label{eq:k-precosymplectic-Hamilton-equations-fields}
		\begin{cases}
			i(X_\alpha)\omega^\alpha = \gamma - \gamma(R_\alpha)\eta^\alpha\,,\\
			i(X_\alpha)\eta^\beta = \delta_\alpha^\beta\,.
		\end{cases}
	\end{equation}
\end{dfn}
The solutions to the field equations defined by the $k$-precosymplectic Hamiltonian system $(M,\eta^\alpha,\omega^\alpha,V,\gamma)$ are the integral sections of these $k$-precosymplectic Hamiltonian $k$-vector fields.

\begin{rmrk}\rm
	In the particular case $k=1$ we recover the case of singular nonautonomous mechanics studied in \cite{Chi1994}. In this case, the Poincaré--Cartan 2-form $\Omega = \omega + \gamma\wedge\eta$ is used in order to write Hamilton's equation without using the Reeb vector field.
\end{rmrk}

We want to develop an algorithm that allows us to find a submanifold $N\hookrightarrow M$ where the system of equations \eqref{eq:k-precosymplectic-Hamilton-equations-fields} has solutions tangent to $N$. In order to find this submanifold $N$ (if it exists) we need to develop an algorithm that introduces some constraint in every step and so providing a sequence of submanifolds
$$ \dotsb\hookrightarrow M_j\hookrightarrow\dotsb\hookrightarrow M_2\hookrightarrow M_1\hookrightarrow M $$
which, in some favorable cases, will end with a \textbf{final constraint submanifold $M_f$} with $\dim M_f\geq 1$. The cases where $M_f$ is empty or a union of isolated points have no interest to us.

\begin{thm}\label{thm:k-cosymplec-constraint-algorithm}
	Consider a $k$-precosymplectic Hamiltonian system $(M,\eta^\alpha,\omega^\alpha,V,\gamma)$, a submanifold $C\hookrightarrow M$ and a $k$-vector field $\bfX = (X_1,\dotsc,X_k)\in\X^k(C)$. Under these hypotheses, the following are equivalent:
	\begin{enumerate}[{\rm(1)}]
		\item There exists a $k$-vector field $\bfX = (X_\alpha)\in\X^k(C)$ such that the system of equations
		\begin{equation}\label{eq:k-cosymplectic-Hamilton-equations-thm-constraint}
			\begin{cases}
				i(X_\alpha)\omega^\alpha = \gamma - \gamma(R_\alpha)\eta^\alpha\,,\\
				i(X_\alpha)\eta^\beta = \delta_\alpha^\beta
			\end{cases}
		\end{equation}
		holds on $C$.
		\item For every $p\in C$, there exists $\bfZ_p = (Z_\alpha)_p\in\oplus^k\T_pC$ such that, if $\widetilde\gamma_p = \gamma_p - \gamma_p({R_\alpha}_p)\eta^\alpha_p$, then
		$$ i({Z_\alpha}_p)\eta^\beta_p = \delta_\alpha^\beta\ ,\qquad \sum_\alpha\eta^\alpha_p + \widetilde\gamma_p = \flat(\bfZ_p)\,. $$
	\end{enumerate}
\end{thm}
\begin{proof}
	Consider the $k$-vector field $\bfZ = \bfX$. It is clear that $i({Z_\alpha}_p)\eta^\beta_p = \delta_\alpha^\beta$ for every $p\in C$ and that
	$$ \flat(\bfZ_p) = i({Z_\alpha}_p)\omega^\alpha_p + (i({Z_\alpha}_p)\eta^\alpha_p)\eta^\alpha_p = \widetilde\gamma_p + \sum_\alpha\eta^\alpha_p\,. $$

	Conversely, suppose that for every $p\in C$, there exists $\bfZ_p\in\oplus^k\T_pC$ satisfying
	$$ i({Z_\alpha}_p)\eta^\beta_p = \delta_\alpha^\beta\ ,\qquad \sum_\alpha\eta^\alpha_p + \widetilde\gamma_p = \flat(\bfZ_p)\,. $$
	Consider $p\in C$. Taking a Darboux chart $(U;t^\alpha,q^i,p_{i_\alpha}^\alpha,z^j)$ around $p$, we have
	\begin{align*}
		\eta^\alpha &= \d t^\alpha\,,\\
		\omega^\alpha &= \sum_{i\in I_\alpha}\d q^i\wedge \d p_i^\alpha\,,\\
		\gamma &= \parder{H}{t^\alpha}\d t^\alpha + \parder{H}{q^i}\d q^i + \parder{H}{p_{i_\alpha}^\alpha}\d p_{i_\alpha}^\alpha + \parder{H}{z^j}\d z^j\,.
	\end{align*}
	With this in mind, the local expression of $\widetilde\gamma = \gamma - \gamma(R_\alpha)\eta^\alpha$ is
	$$ \widetilde\gamma = \parder{H}{q^i}\d q^i + \parder{H}{p_{i_\alpha}^\alpha}\d p_{i_\alpha}^\alpha + \parder{H}{z^j}{\d z^j}\,. $$
	In what follows, in order to keep the notation as simple as possible, we will ommit the point $p$. Consider the $k$-vector field $\bfZ = (Z_\alpha)$ with local expression
	$$ Z_\alpha = A_\alpha^\beta\parder{}{t^\beta} + B_\alpha^i\parder{}{q^i} + C_{\alpha,i_\beta}^\beta\parder{}{p_{i_\beta}^\beta} + D_\alpha^j\parder{}{z^j}\,. $$
	Computing its image by the morphism $\flat$,
	\begin{align*}
		\flat(\bfZ) &= \sum_\alpha \left(i(Z_\alpha)\omega^\alpha + (i(Z_\alpha)\eta^\alpha)\eta^\alpha\right)\\
		&= \sum_\alpha\sum_{i\in I_\alpha}i(Z_\alpha)(\d q^i\wedge \d p_i^\alpha) + \sum_\alpha (i(Z_\alpha)\d t^\alpha)\d t^\alpha\\
		&= \sum_\alpha\sum_{i\in I_\alpha}(i(Z_\alpha)\d q^i)\d p_i^\alpha - \sum_\alpha\sum_{i\in I_\alpha} (i(Z_\alpha)\d p^\alpha_i) \d q^i + \sum_\alpha (i(Z_\alpha)\d t^\alpha)\d t^\alpha\\
		&= \sum_\alpha\sum_{i\in I_\alpha}B_\alpha^i\d p_i^\alpha - \sum_\alpha\sum_{i\in I_\alpha}C_{\alpha,i}^\alpha\d q^i + \sum_\alpha A_\alpha^\alpha\d t^\alpha\,.
	\end{align*}
	Comparing this last expression with
	$$ \sum_\alpha\eta^\alpha + \widetilde\gamma = \sum_\alpha\d t^\alpha + \parder{H}{q^i}\d q^i + \parder{H}{p_{i_\alpha}^\alpha}\d p_{i_\alpha}^\alpha + \parder{H}{z^j}{\d z^j}\,, $$
	we obtain the relations
	\begin{equation*}
		\begin{dcases}
			A_\alpha^\alpha = 1\,,\\
			\parder{H}{z^j} = 0\,,\\
			\parder{H}{q^i} = -\sum_{\substack{\alpha\mbox{ \scriptsize such} \\ \mbox{\scriptsize that } i\in I_\alpha}}C^\alpha_{\alpha,i}\,,\\
			\parder{H}{p_{i_\alpha}^\alpha} = B_\alpha^{i_\alpha}\,.
		\end{dcases}
	\end{equation*}
	Moreover, we know by hypothesis that $A_\alpha^\beta = \delta_\alpha^\beta$. The second condition $\tparder{H}{z^j}=0$ is a compatibility condition of the Hamilton equations in the $k$-precosymplectic case. We can state this condition as: the Hamiltonian function $H$ cannot depend on the so-called {\it gauge variables} $z^j$. Finally, the third and fourth equations, along with $A_\alpha^\beta = \delta_\alpha^\beta$, are equivalent to \eqref{eq:k-cosymplectic-Hamilton-equations-thm-constraint} when written in coordinates (see \eqref{eq:k-cosymplectic-Hamilton-equations-fields-coordinates}).
\end{proof}

We can use the previous theorem to give a description of the constraint algorithm. First, we must restrict ourselves to the points satisfying the condition
$$ \gamma\left(\parder{}{z^j}\right) = 0\,, $$
because we have already seen that it is a compatibility condition of the system. We define the $j$-th constraint submanifold $M_j\subset M_{j-1}$ as
\begin{multline*}
	M_j = \Big\{p\in M_{j-1}\ \vert\ \exists\bfZ = (Z_\alpha)\in\X^k(M_{j-1}) \mbox{ such that } \\
	\flat(\bfZ) = \widetilde\gamma + \sum_\alpha\eta^\alpha \mbox{ and } i(Z_\alpha)\eta^\beta = \delta_\alpha^\beta \Big\}\,,
\end{multline*}
where $M_0 = M$.

\begin{dfn}
	Consider a submanifold $C\hookrightarrow M$ of a $k$-precosymplectic manifold $M$. We define the \textbf{$k$-precosymplectic orthogonal complement} of $C$ as
	$$ \T C^\perp = \left( \flat(\oplus^k\T C\cap D_C) \right)^\circ\,, $$
	where $D_C$ is the set of all $k$-vectors $\bfZ_p = (Z_\alpha)_p$ on $C$ such that $i({Z_\alpha}_p)\eta^\beta_p = \delta_\alpha^\beta$.
\end{dfn}
Taking into account the previous definition and Theorem \ref{thm:k-cosymplec-constraint-algorithm}, we can redefine the constraint submanifolds as
$$ M_j = \Big\{ p\in M_{j-1}\ \vert\ \widetilde\gamma_p + \sum_\alpha\eta^\alpha_p\in((\T M_{j-1})^\perp_p)^\circ \Big\}\,. $$
This last characterization allows us to effectively compute the constraints at every step of the constraint algorithm. However, there exists an alternative and equivalent way to compute the constraint submanifolds of the algorithm, which is much more operational:
\begin{enumerate}[{\rm(1)}]
	\item Obtain a local basis $\{Z_1,\dots,Z_k\}$ of $(\T M)^\perp$.
	\item Use Theorem \ref{thm:k-cosymplec-constraint-algorithm} to obtain a set of independent constraint functions
	$$ f_\mu = i(Z_\mu)\bigg(\widetilde\gamma + \sum_\alpha\eta^\alpha\bigg)\,, $$
	defining a submanifold $M_1\hookrightarrow M$.
	\item Compute solutions $\bfX = (X_\alpha)$ to equations \eqref{eq:k-precosymplectic-Hamilton-equations-fields}.
	\item Impose tangency of $X_1,\dotsc,X_k$ to $M_1$.
	\item Iterate item (4) until no new constraints appear.
\end{enumerate}
If this iterative procedure ends with a submanifold $M_f\hookrightarrow M$ with $\dim M_f\geq 1$, we can ensure the existence of solutions to \eqref{eq:k-precosymplectic-Hamilton-equations-fields} in $M_f$.

\begin{rmrk}\rm
	In particular, the constraint algorithm described above works for a singular Lagrangian field theory $(\R^k\times\oplus^k\T Q, \d t^\alpha, \omega^\alpha_L, L)$ and for its associated Hamiltonian formalism on $\P$. However, in the Lagrangian formalism, the problem of finding {\sc sopdes} solving the field equations is not considered in the algorithm above. In general, imposing the {\sc sopde} condition leads to new constraints and, in favorable cases, we get a new final constraint submanifold $S_f\hookrightarrow M_f$ where we have {\sc sopde} $k$-vector fields solutions tangent to $S_f$. In the examples analyzed in the next section we give some insights on how to proceed in these cases. Nevertheless, it is still an open problem to find a rigorous characterization of all these constraints arising from the {\sc sopde} condition. See \cite{DeLeo2002} for a deeper study on this topics in the case of singular Lagrangian mechanics.
\end{rmrk}

It is important to point out that we can treat singular $k$-symplectic field theories as a particular case of $k$-precosymplectic field theories. In this case, we do not have the 1-forms $\eta^\alpha$ and the $k$-presymplectic algorithm described in Section \ref{sec:k-presymplectic-constraint-algorithm} is recovered. See \cite{Gra2009} for details.

\section{Examples}
\label{sec:constraint-algorithm-examples}

	In this last section of the chapter, we are going to analyze several examples of field theories described by singular Lagrangian functions and how to apply the constraint algorithm developed above to each of them. We will apply the constraint algorithm to both the Lagrangian and Hamiltonian formulations of every example.

	The first example analyzes the systems defined by \textbf{Lagrangians which are affine in the velocities}. Such Lagrangians are of great interest in many areas of theoretical physics, such as gravitation or quantum field theory. The second example studies in detail a \textbf{particular example of affine Lagrangian} in order to see how does the constraint algorithm works. The last example deals with a \textbf{singular quadratic Lagrangian}.

	\subsection*{Affine Lagrangians}\label{sub:affine-Lagrangians}

	Affine Lagrangians are very important in physics. For instance, the so-called Einstein--Palatini (or {\it metric-affine}) approach to gravitation and Dirac fermion fields \cite{Gia1997} among others, are examples of theories described by affine Lagrangians.

	Consider a field theory with configuration manifold $Q$ with coordinates $(q^i)$. The bundle
	$$ \bar\tau_1\colon\R^k\times\oplus^k\T Q\to\R^k $$
	is its nonautonomous phase space of $k^1$-velocities and has coordinates $(t^\alpha,q^i,v^i_\alpha)$. Consider an affine Lagrangian $L\colon\R^k\times\oplus^k\T Q\to\R$ of the form
	\begin{equation}\label{eq:affine-lagrangian}
		L(t^\alpha,q^i,v^i_\alpha) = f^\mu_j(q^i)v^j_\mu + g(t^\alpha,q^i)\,.
	\end{equation}
	Such a Lagrangian is the sum of the pullbacks to $\R^k\times\oplus^k\T Q$ of two functions:
	\begin{itemize}
		\item a linear function $\oplus^k\T Q\to\R$ on the fibers of the bundle $\oplus^k\T Q\to Q$\,,
		\item an arbitrary function $\R^k\times Q\to\R$\,.
	\end{itemize}

	\subsubsection*{Lagrangian formalism}

	Consider a Lagrangian function $L\in\Cinfty(\R^k\times\oplus^k\T Q$ of the form \eqref{eq:affine-lagrangian}. Associated to this affine Lagrangian we have
	\begin{align*}
		E_L &= \Delta(L)-L = -g(t^\alpha,q^i)\in\Cinfty(\R^k\times\oplus^k\T Q)\,,\\
		\omega^\alpha_L &= -\parder{f^\alpha_i}{q^j}\d q^j\wedge\d q^k\in\Omega^2(\R^k\times\oplus^k\T Q)\,,
	\end{align*}
	defining a $k$-precosymplectic structure $(\d t^\alpha,\omega^\alpha_L,V)$, with $V = \left\langle\tparder{}{v^j_\mu}\right\rangle$, in $\R^k\times\oplus^k\T Q$. We can take $R_\alpha = \tparder{}{t^\alpha}$ as Reeb vector fields. Consider now a $k$-vector field $\bfX = (X_1,\dotsc,X_k)\in\X^k(\R^k\times\oplus^k\T Q)$ with local expression
	$$ X_\alpha = \parder{}{t^\alpha} + F_\alpha^\ell\parder{}{q^\ell} + G_{\alpha\nu}^\ell\parder{}{v^\ell_\nu}\,. $$
	The $k$-cosymplectic Lagrangian equation
	$$ i(X_\alpha)\omega^\alpha_L = \d E_L + \parder{L}{t^\mu}\d t^\mu $$
	for this $k$-vector field $\bfX$ gives
	\begin{equation}\label{eq:affine-lagrangians-first}
		\parder{g}{q^j} + F^\ell_\alpha\left( \parder{f^\alpha_\ell}{q^j} - \parder{f^\alpha_j}{q^\ell} \right) = 0\,.
	\end{equation}
	This is a system of (linear) equations for the functions $F_\alpha^\ell$ which allows us to partially determine these functions. Eventually, new constraints may appear depending on the ranks of the matrices involved. In this case, the constraint algorithm goes on by demanding the tangency of the vector fields $X_\alpha$ to the new manifold described by the constraints. Notice that, in any case, the functions $G_{\alpha\nu}^i$ remain undetermined.

	If we impose the {\sc sopde} condition to the solutions, that is, $F^k_\nu = v_\nu^k$, equations \eqref{eq:affine-lagrangians-first} read
	\begin{equation*}
		\parder{g}{q^j} + v^\ell_\alpha\left( \parder{f^\alpha_\ell}{q^j} - \parder{f^\alpha_j}{q^\ell} \right) = 0\,,
	\end{equation*}
	which are new constraints. Imposing the tangency condition for the vector fields
	$$ X_\nu = \parder{}{t^\nu} + v_\nu^\ell\parder{}{q^\ell} + G_{\nu\alpha}^\ell\parder{}{v^\ell_\alpha} $$
	we obtain the relations
	\begin{equation*}
		\parder{g}{q^j} + G_{\nu\alpha}^\ell\left( \parder{f^\alpha_\ell}{q^j} - \parder{f^\alpha_j}{q^\ell} \right) = 0\,,
	\end{equation*}
	which allows us to partially determine the coefficients $G_{\nu\alpha}^\ell$. In addition, new constraints may appear depending on the rank of the matrix $\left( \dparder{f^\alpha_\ell}{q^j} - \dparder{f^\alpha_j}{q^\ell} \right)$. If this is the case, the algorithm continues by imposing again the tangency condition.

	\subsubsection*{Hamiltonian formalism}

	In the Hamiltonian formalism, the phase space of $k^1$-momenta is the bundle $\bar\pi_1\colon\R^k\times\oplus^k\cT Q\to\R^k$. The Legendre map associated to the Lagrangian $L$ given in \eqref{eq:affine-lagrangian} is
	$$ t^\mu\circ\F L = t^\mu\ ,\quad q^i\circ\F L = q^i\ ,\quad p_i^\mu\circ\F L = \parder{L}{v^i_\mu} = f^\mu_i(q^j)\,. $$
	Notice that the submanifold $\P = \F L(\R^k\times\oplus^k\T Q)$ is given by the constraints $p_i^\mu = f_i^\mu(q^j)$. Thus, it is the image of a section $\xi\colon\R^k\times Q\to \R^k\times\oplus^k\cT Q$ of the projection $(\pi_Q)_{1,0}\colon\R^k\times\oplus^k\cT Q\to\R^k\times Q$, and can be identified in a natural way with $\R^k\times Q$. Hence, as $\xi\circ\tau_1$ is a surjective submersion and its fibers are connected, then so is the restriction of the Legendre map $\F L$ onto its image $\P$, $\F L_0\colon\R^k\times\oplus^k\T Q\to\P$, since $\F L_0 = \xi\circ\tau_1$.

	Summing up, affine Lagrangians are almost-regular Lagrangians and then $\P$ is an embedded submanifold of $\R^k\times\oplus^k\cT Q$, which is diffeomorphic to $\R^k\times Q$.

	Hence, we can introduce
	\begin{align*}
		H &= -g(t^\alpha, q^i)\in\Cinfty(\P)\,,\\
		\omega^\alpha &= -\parder{f^\alpha_k}{q^j}\d q^j\wedge\d q^k\in\Omega^2(\P)\,,
	\end{align*}
	such that
	\begin{align*}
		\F L_0^\ast E_L = H\,,\\
		\F L_0^\ast \omega_L^\alpha = \omega^\alpha\,.
	\end{align*}
	As above, we have
	$$ \eta^\alpha = \d t^\alpha\ ,\quad R_\alpha = \parder{}{t^\alpha}\,. $$
	For a $k$-vector field $\bfX = (X_1,\dotsc,X_k)\in\X^k(\P)$ with local expression
	$$ X_\alpha = \parder{}{t^\alpha} + F^\ell_\alpha\parder{}{q^\ell}\,, $$
	the Hamilton equation
	$$ i(X_\alpha)\omega^\alpha = \d H - R_\alpha(H)\d t^\alpha $$
	yields the conditions
	\begin{equation}\label{eq:affine-lagrangians-second}
		\parder{g}{q^j} + F^\ell_\alpha\left( \parder{f^\alpha_\ell}{q^j} - \parder{f^\alpha_j}{q^\ell} \right) = 0\,.
	\end{equation}
	As in the Lagrangian formalism, this system of linear equations allows us to partially determine the functions $F_\alpha^\ell$. In addition, new constraints may appear depending on the rank of the matrices involved. If new constraints appear, the constraint algorithm continues by demanding the tangency condition to the new constraints.

	\subsection*{A simple affine Lagrangian model}

	In this example we are going to consider a particular case of affine Lagrangian in order to clearly see how does the constraint algorithm work in a concrete example.

	\subsubsection*{Lagrangian formalism}

	Consider the configuration manifold $\R^2\times Q = \R^2\times\R^2$ with coordinates $(t^1,t^2; q^1, q^2)$. The Lagrangian formalism takes place in the bundle $\R^2\times\oplus^2\T Q$, equipped with natural coordinates $(t^1,t^2; q^1, q^2, v^1_1, v^1_2, v^2_1, v^2_2)$. Consider the Lagrangian function
	\begin{equation}\label{eq:affine-lagrangian-particular}
		L = q^2v^1_1 - q^1v^2_2 + q^1q^2t^1\in\Cinfty(\R^2\times\oplus^2\T Q)\,.
	\end{equation}
	Hence, the functions in \eqref{eq:affine-lagrangian} are
	$$ f^1_1 = q^2\ ,\quad f^2_1 = 0\ ,\quad f^1_2 = 0\ ,\quad f^2_2 = -q^1\ ,\quad g = q^1q^2t^1\,. $$
	We have the forms
	\begin{align*}
		\eta^1 &= \d t^1\,,\\
		\eta^2 &= \d t^2\,,\\
		\omega^1_L &= \d q^1\wedge\d q^2\,,\\
		\omega^2_L &= \d q^1\wedge\d q^2\,.
	\end{align*}
	The Reeb vector fields are
	$$ R_1^L = \parder{}{t^\alpha}\ ,\quad R_2^L = \parder{}{t^2}\,. $$
	The Lagrangian energy is $E_L = -q^1q^2t^1\in\Cinfty(\R^2\times\oplus^2\T Q)$. Consider the 2-vector field $\bfX = (X_1,X_2)\in\X^2(\R^2\times\oplus^2\T Q)$ with local expression
	\begin{align*}
		X_1 &= \parder{}{t^1} + F_1^1\parder{}{q^1} + F_1^2\parder{}{q^2} + G_{11}^1\parder{}{v^1_1} + G_{12}^1\parder{}{v^1_2} + G_{11}^2\parder{}{v^2_1} + G_{12}^2\parder{}{v^2_2}\,,\\
		X_2 &= \parder{}{t^2} + F_2^1\parder{}{q^1} + F_2^2\parder{}{q^2} + G_{21}^1\parder{}{v^1_1} + G_{22}^1\parder{}{v^1_2} + G_{21}^2\parder{}{v^2_1} + G_{22}^2\parder{}{v^2_2}\,.
	\end{align*}
	The Lagrangian equation $i(X_\alpha)\omega^\alpha_L = \d E_L - R_\alpha^L(E_L)\d t^\alpha$ for the 2-vector field $\bfX$ reads
	$$ F_1^1\d q^2 - F_1^2\d q^1 + F_2^1\d q^2 - F_2^2\d q^1 = -q^2t^1\d q^1 - q^1t^1\d q^2\,, $$
	and conditions \eqref{eq:affine-lagrangians-first} give
	$$ F_1^2 + F_2^2 = q^2t^1\ ,\quad F_1^1 + F_2^1 = -q^1t^1\,, $$
	which can be written as
	\begin{equation*}
		\begin{pmatrix}
			0 & 1 & 0 & 1\\
			-1 & 0 & -1 & 0
		\end{pmatrix}
		\begin{pmatrix}
			F_1^1\\
			F_1^2\\
			F_2^1\\
			F_2^2
		\end{pmatrix}
		= \begin{pmatrix}
			q^2x^1\\
			q^1x^1
		\end{pmatrix}\ .
	\end{equation*}
	If we impose the {\sc sopde} condition, i.e. $F_\mu^\ell = v_\mu^\ell$, the 2-vector field $\bfX$ becomes
	\begin{align*}
		X_1 &= \parder{}{t^1} + v^1_1\parder{}{q^1} + v_1^2\parder{}{q^2} + G_{1\nu}^\ell\parder{}{v^\ell_\nu}\,,\\
		X_2 &= \parder{}{t^2} + v^1_2\parder{}{q^1} + v_2^2\parder{}{q^2} + G_{2\nu}^\ell\parder{}{v^\ell_\nu}
	\end{align*}
	and we obtain two new constraints
	\begin{equation*}
		\begin{dcases}
			\zeta_1 = v_1^2 + v_2^2 - q^2t^1 = 0\\
			\zeta_2 = v_1^1 + v_2^1 + q^1t^1 = 0
		\end{dcases}
	\end{equation*}
	These two constraints $\zeta_1,\zeta_2$ define the submanifold $S_1\hookrightarrow\R^2\times\oplus^2\T Q$. Imposing the tangency condition to this new manifold, we get
	\begin{equation*}
		\begin{cases}
			X_1(\zeta_1) = -q^2 + G_{11}^2 + G_{12}^2 - t^1v_1^2 = 0\,,\\
			X_1(\zeta_2) = q^1 + t^1v_1^1 + G_{11}^1 + G_{12}^1 = 0\,,\\
			X_2(\zeta_1) = -t^1v_2^2 + G_{21}^2 + G_{22}^2 = 0\,,\\
			X_2(\zeta_2) = t^1v_2^1 + G_{21}^1 + G_{22}^1 = 0\,,
		\end{cases}
	\end{equation*}
	which, written in matrix forms, becomes
	\begin{equation*}
		\begin{pmatrix}
			0 & 1 & 0 & 1 & 0 & 0 & 0 & 0\\
			1 & 0 & 1 & 0 & 0 & 0 & 0 & 0\\
			0 & 0 & 0 & 0 & 0 & 1 & 0 & 1\\
			0 & 0 & 0 & 0 & 1 & 0 & 1 & 0\\
		\end{pmatrix}
		\begin{pmatrix}
			G_{11}^1\\
			G_{11}^2\\
			G_{12}^1\\
			G_{12}^2\\
			G_{21}^1\\
			G_{21}^2\\
			G_{22}^1\\
			G_{22}^2\\
		\end{pmatrix}
		= \begin{pmatrix}
			q^2 + t^1v_1^2\\
			-q^1 - t^1v_1^1\\
			t^1v_2^2\\
			-t^1v_2^1
		\end{pmatrix}\,,
	\end{equation*}
	which allows us to partially determine the functions $G_{\alpha\nu}^\ell$. Notice that no new constraints appear, and hence the final constraint submanifold is $S_1$.

	\subsubsection*{Hamiltonian formalism}

	The Hamiltonian formalism takes place in the bundle $(\R^2\times\oplus^2\cT Q$, endowed with natural coordinates $(t^1,t^2; q^1,q^2,p_1^1,p_1^2,p_2^1,p_2^2)$. The Legendre map $\F L$ given by the Lagrangian defined in \eqref{eq:affine-lagrangian-particular} is the map
	$$ \F L\colon \R^2\times\oplus^2\T Q\to\R^2\times\oplus^2\cT Q\,, $$
	given by
	$$ (t^1,t^2; q^1,q^2,p_1^1,p_1^2,p_2^1,p_2^2) = \F L(t^1,t^2; q^1,q^2,v_1^1,v_2^1,v_1^2,v_2^2) = (t^1,t^2; q^1,q^2,q^2,0,0,-q^1)\,. $$
	Hence, its image $\P = \F L(\R^2\times\oplus^2\T Q)$ is given by the primary constraints
	$$ p_1^1 = q^2\ ,\quad p_1^2 = 0\ ,\quad p_2^1 = 0\ ,\quad p_2^2 = -q^1\,. $$
	With this in mind, it is clear that we can describe the manifold $\P$ with coordinates $(t^1,t^2, q^1, q^2)$. In $\P$, we have the forms
	\begin{equation*}
		\eta^1 = \d t^1\ ,\quad
		\eta^2 = \d t^2\ ,\quad
		\omega^1 = \d q^1\wedge\d q^2\ ,\quad
		\omega^2 = \d q^1\wedge\d q^2\,.
	\end{equation*}
	The Reeb vector fields are $R_1 = \tparder{}{t^1}, R_2 = \tparder{}{t^2}$. The Hamiltonian function is $H = -q^1q^2t^1$. Consider a generic 2-vector field in $\P$ $\bfX = (X_1,X_2)\in\X^2(\P)$ with local expression
	\begin{align*}
		X_1 = \parder{}{t^1} + B_1^1\parder{}{q^1} + B_1^2\parder{}{q^2}\,,\\
		X_2 = \parder{}{t^2} + B_2^1\parder{}{q^1} + B_2^2\parder{}{q^2}\,.
	\end{align*}
	The Hamiltonian equation $i(X_\alpha)\omega^\alpha = \d H - R_\alpha(H)\d t^\alpha$ for the 2-vector field $\bfX$ gives
	$$ B_1^1\d q^2 - B_1^2\d q^1 + B_2^1\d q^2 - B_2^2\d q^1 = -t^1q^2\d q^1 - t^1q^1\d q^2\,. $$
	Now, conditions \eqref{eq:affine-lagrangians-second} read
	$$ B_1^2 + B_2^2 = t^1q^2\ ,\quad B_1^1 + B_2^1 = -t^1q^1\,. $$
	These two relations allow us to partially determine the functions $B_\alpha^j$. In this case, no new constraints appear.

	\subsection*{A singular quadratic Lagrangian}

	\subsubsection*{Lagrangian formalism}

	Consider the configuration manifold $Q = \R\times\R^+$ equipped with coordinates $(q,e)$, and two independent variables $(t, s)\in\R^2$. The corresponding phase space is the bundle $\R^2\times\oplus^2\T Q$ with natural coordinates $(t, s; q, e, q_t, q_s, e_t, e_s)$. Consider the Lagrangian function $L\colon \R^2\times\oplus^2\T Q\to\R$ given by
	\begin{equation}\label{eq:singular-quadratic-lagrangian}
		L = \frac{1}{2e}q_t^2 + \frac{1}{2}\sigma^2 e - \frac{1}{2}\tau q_s^2\,,
	\end{equation}
	where $\tau\in\R$ is a constant parameter and $\sigma = \sigma(t, s)\in\Cinfty(\R^2)$ is a given function. This Lagrangian is quite similar to one introduced in \cite{Gra2009}, but considering one of its parameters as a function of $(t,s)$ in order to illustrate the nonautonomous setting.

	We begin by computing the canonical structure and the Liouville vector field of the bundle $\R^2\times\oplus^2\T Q$:
	\begin{align*}
		J^t &= \parder{}{q_t}\otimes\d q + \parder{}{e_t}\otimes\d e\,,\\
		J^s &= \parder{}{q_s}\otimes\d q + \parder{}{e_s}\otimes\d e\,,\\
		\Delta_t &= q_t\parder{}{q_t} + e_t\parder{}{e_t}\,,\\
		\Delta_s &= q_s\parder{}{q_s} + e_s\parder{}{e_s}\,,\\
		\Delta &= \Delta_t + \Delta_s\,.
	\end{align*}
	The Cartan 1 and 2-forms are
	\begin{align*}
		\theta^t_L &= \transp{J^t}\circ\d L = \frac{1}{2}q_t\d q\,,\\
		\theta^s_L &= \transp{J^s}\circ\d L = -\tau q_s\d q\,,\\
		\omega^t_L &= -\d\theta^t_L = \frac{q_t}{e^2}\d e\wedge\d q - \frac{1}{e}\d q_t\wedge\d q\,,\\
		\omega^s_L &= -\d\theta^s_L = \tau\d q_s\wedge\d q\,.
	\end{align*}
	The Lagrangian energy $E_L$ is
	$$ E_L = \Delta(L)-L = \frac{1}{2e}q_t^2 - \frac{1}{2}\sigma^2 e - \frac{1}{2}\tau q_s^2\,. $$
	Consider now a 2-vector field $\bfX = (X_t,X_s)\in\X^2(\R^2\times\oplus^2\T Q)$ and consider the $k$-cosymplectic Lagrangian equations \eqref{eq:k-cosymplectic-Lagrange-equations-fields} for it. Applying the second group of equations, namely $i(X_\alpha)\d t^\beta = \delta_\alpha^\beta$, the 2-vector field $\bfX$ becomes
	\begin{align*}
		X_t = \parder{}{t} + B_t^q\parder{}{q} + B_t^e\parder{}{e} + C_t^{q_t}\parder{}{q_t} + C_t^{q_s}\parder{}{q_s} + C_t^{e_t}\parder{}{e_t} + C_t^{e_s}\parder{}{e_s}\,,\\
		X_s = \parder{}{s} + B_s^q\parder{}{q} + B_s^e\parder{}{e} + C_s^{q_t}\parder{}{q_t} + C_s^{q_s}\parder{}{q_s} + C_s^{e_t}\parder{}{e_t} + C_s^{e_s}\parder{}{e_s}\,.
	\end{align*}
	Applying now the first $k$-cosymplectic Lagrangian equation \eqref{eq:k-cosymplectic-Lagrange-equations-fields}, $i(X_\alpha)\omega^\alpha_L = \d E_L + \parder{L}{t^\alpha}\d t^\alpha$, and equating the coefficients, we obtain the relations
	\begin{equation*}
		\begin{dcases}
			B_t^e = \frac{e^2}{q_t}\left( \frac{1}{e}C_t^{q_t} - \tau C_s^{q_s} \right)\,,\\
			B_t^q = q_t\,,\\
			B_s^q = q_s\,,\\
			\frac{q_t^2}{e^2} = \sigma^2\,.
		\end{dcases}
	\end{equation*}
	These equations determine the coefficients $B_t^e$, $B_t^q$ and $B_s^q$ of the 2-vector field $\bfX$ in terms of the variables and the other coefficients. The last equation is a constraint,
	$$ \zeta_1 = \frac{1}{2}\left( \frac{q_t^2}{e^2} - \sigma^2 \right)\,, $$
	which defines the submanifold $S_1\hookrightarrow\R^2\times\oplus^2\T Q$. At this point, $\bfX$ has nine undetermined coefficients. Imposing the tangency of the 2-vector field $\bfX = (X_t,X_s)$ to the submanifold $S_1$, that is, imposing
	$$ \restr{X_t(\zeta_1)}{S_1} = 0\ ,\quad \restr{X_s(\zeta_1)}{S_1} = 0\,, $$
	we obtain two more relations between the undetermined coefficients (on $S_1$) and no more constraints.

	In order to complete our analysis, we are going to impose the {\sc sopde} condition. Then, the generic expression of $\bfX$ is
	\begin{align*}
		X_t = \parder{}{t} + q_t\parder{}{q} + e_t\parder{}{e} + C_t^{q_t}\parder{}{q_t} + C_t^{q_s}\parder{}{q_s} + C_t^{e_t}\parder{}{e_t} + C_t^{e_s}\parder{}{e_s}\,,\\
		X_s = \parder{}{s} + q_s\parder{}{q} + e_s\parder{}{e} + C_s^{q_t}\parder{}{q_t} + C_s^{q_s}\parder{}{q_s} + C_s^{e_t}\parder{}{e_t} + C_s^{e_s}\parder{}{e_s}\,.
	\end{align*}
	Now, the first equation in \eqref{eq:k-cosymplectic-Lagrange-equations-fields} gives the relation
	$$ e_t\frac{q_t}{e^2} - \frac{1}{e}C_t^{q_t} + \tau C_s^{q_s} = 0\,, $$
	and the same constraint $\zeta_1$ obtained before. Imposing now the tangency to the submanifold $S_1$ determines the functions $C_t^{q_t}$ and $C_s^{q_t}$ (on $S_1$) and no new constraints appear. In conclusion, the 2-vector field $\bfX = (X_t,X_s)$ has five undetermined coefficients.

	\subsubsection*{Hamiltonian formalism}

	The Hamiltonian counterpart takes place in the bundle $\R^2\times\oplus^2\cT Q$, equipped with natural coordinates $(t,s;q, e, p^t, p^s, \pi^t, \pi^s)$. The Legendre map associated to the Lagrangian function \eqref{eq:singular-quadratic-lagrangian} is the map
	$$ \F L\colon\R^2\times\oplus^2\T Q\to\R^2\times\oplus^2\cT Q$$
	given by
	$$ \F L(t, s; q, e, q_t, q_s, e_t, e_s) = \left( t, s; q, e, \frac{1}{e}q_t, -\tau q_s, 0, 0 \right)\,. $$
	The primary Hamiltonian constraint submanifold
	$$ \P = \F L(\R^2\times\oplus^2\T Q)\hookrightarrow\R^2\times\oplus^2\cT Q$$
	is defined by the constraints
	$$ \pi^t = 0\ ,\quad\pi^s = 0\,. $$
	Taking coordinates $(t, s; q, e, p^t, p^s)$ as coordinates on $\P$, its 2-precosymplectic structure is
	\begin{align*}
		\eta^t &= \d t\,,\\
		\eta^s &= \d s\,,\\
		\omega^t &= \d q\wedge\d p^t\,,\\
		\omega^s &= \d q\wedge\d p^s\,.
	\end{align*}
	We have that
	$$ \ker\eta^t\cap\ker\eta^s\cap\ker\omega^t\cap\ker\omega^s = \left\langle\parder{}{e}\right\rangle\,. $$
	The Reeb vector fields are
	$$ R_t = \parder{}{t}\ ,\quad R_s = \parder{}{s}\,. $$
	The Hamiltonian function on the submanifold $\P$ is given by
	$$ H = \frac{1}{2}e(p^t)^2 - \frac{1}{2}\sigma^2 e - \frac{1}{2\tau}(p^s)^2\,. $$
	Consider a generic 2-vector field on $\P$, $\bfX = (X_t,X_s)\in\X^2(\P)$, with local expression
	\begin{align*}
		X_t &= A_t^1\parder{}{t} + A_t^2\parder{}{s} + B_t^1\parder{}{q} + B_t^2\parder{}{e} + C_t^1\parder{}{p^t} + C_t^2\parder{}{p^s}\,,\\
		X_s &= A_s^1\parder{}{t} + A_s^2\parder{}{s} + B_s^1\parder{}{q} + B_s^2\parder{}{e} + C_s^1\parder{}{p^t} + C_s^2\parder{}{p^s}\,.
	\end{align*}
	Hamilton equations \eqref{eq:k-precosymplectic-Hamilton-equations-fields} are
	\begin{equation*}
		\begin{cases}
			i(X_t)\omega^t + i(X_s)\omega^s = \d H - \d H(R_t)\eta^t - \d H(R_s)\eta^s\,,\\
			i(X_t)\eta^t = 1\,,\\
			i(X_t)\eta^s = 0\,,\\
			i(X_s)\eta^t = 0\,,\\
			i(X_s)\eta^s = 1\,,
		\end{cases}
	\end{equation*}
	which give the relations
	\begin{equation*}
		\begin{dcases}
			B_t^1 = ep^t\,,\\
			B_s^1 = \frac{-1}{\tau}p^s\,,\\
			C_t^1 + C_s^2 = \,,\\
			A_t^1 = 1\ ,\quad A_t^2 = 0\,,\\
			A_s^1 = 0\ ,\quad A_s^2 = 1\,,
		\end{dcases}
	\end{equation*}
	which partially determine the coefficients of $\bfX$ and imposes as a consistency condition the secondary Hamiltonian constraint
	$$ \xi_1 = i\left(\parder{}{e}\right)\d H = \frac{1}{2}(p^t)^2 - \frac{1}{2}\sigma^2 = 0 \quad\mbox{(on $\P$)}\,, $$
	defining a new constraint submanifold $\P_1\hookrightarrow\P$. Imposing the tangency of the 2-vector field $\bfX$ to this new submanifold $\P_1$, $\restr{X_t(\xi_1)}{\P_1} = 0$, $\restr{X_s(\xi_1)}{\P_1} = 0$, determines the functions
	\begin{equation*}
		\restr{C_t^1}{\P_1} = \frac{1}{p^t}\sigma\parder{\sigma}{t}\ ,\quad
		\restr{C_s^1}{\P_1} = \frac{1}{p^t}\sigma\parder{\sigma}{s}\,,
	\end{equation*}
	and no new constraints appear.

	\begin{rmrk}\rm
		Notice that $\F L^\ast(\xi_1) = \zeta_1$ and $\F L(S_1) = \P_1$. Hence, as the {\sc sopde} condition does not yield new constraints in the Lagrangian formalism, there are no non-$\F L$-projectable Lagrangian constraints.
	\end{rmrk}

	\chapter[\texorpdfstring{$k$}--contact Hamiltonian systems]{\texorpdfstring{$k$}--contact Hamiltonian systems}
	\label{ch:k-contact-Hamiltonian}


In this chapter we introduce a geometric formalism for autonomous dissipative field theories. In order to do this we need to define a new geometric framework: $k$-contact geometry. The main reference on this topic is \cite{Gas2020}.

In Section \ref{sec:k-contact-geometry} we present the notion of $k$-contact manifold and we prove the existence and uniqueness of a family of Reeb vector fields in every $k$-contact manifold. We also prove the existence of two types of special coordinate systems in $k$-contact manifolds: adapted coordinates and Darboux coordinates. We also define the concept of $k$-precontact manifold, which is a weakened version of the concept of $k$-contact manifold. This will be useful when extending the Skinner--Rusk formalism for $k$-contact systems in Chapter \ref{ch:Skinner-Rusk-k-contact}. Section \ref{sec:k-contact-Hamiltonian-systems} uses the geometric framework introduced in the previous section to develop a geometric formalism, the so-called $k$-contact Hamiltonian formalism, for autonomous dissipative field theories. In Section \ref{sec:k-contact-Hamiltonian-symmetries} we generalize the different notions of symmetry presented in Chapter \ref{ch:symmetries-contact-systems} to field theories. We also prove that every Hamiltonian $k$-contact Hamiltonian symmetry is a dynamical symmetry. Finally, Section \ref{sec:k-contact-Hamiltonian-dissipation-laws} is devoted to study the dissipation laws of $k$-contact Hamiltonian systems. In particular, we prove the dissipation theorem, which states that every infinitesimal dynamical symmetry yields a dissipation law.

\section{\texorpdfstring{$k$}--contact geometry}
\label{sec:k-contact-geometry}

\begin{dfn}
	Let $M$ be an $m$-dimensional smooth manifold.
	\begin{itemize}
		\item A \textbf{generalized distribution} on $M$ is a subset $D\subset \T M$ such that, $D_x\subset\T_xM$ is a vector subspace for every $x\in M$.
		\item A distribution $D$ is said \textbf{smooth} if it can be locally spanned by a family of vector fields.
		\item A distribution $D$ is \textbf{regular} if it is smooth and of locally constant rank.
		\item A \textbf{codistribution} on $M$ is a subset $C\subset\cT M$ such that, $C_x\subset\cT_xM$ is a vector subspace for every $x\in M$.
	\end{itemize}
\end{dfn}

Given a distribution $D$, the anihilator $D^\circ$ of $D$ is a codistribution. If $D$ is not regular, $D^\circ$ may not be smooth. Using the usual identification $E^{\ast\ast} = E$ of finite-dimensional linear algebra, it is clear that $(D^\circ)^\circ = D$.

Consider a differential 1-form $\eta\in\Omega^1(M)$. Then, $\eta$ generates a smooth codistribution, denoted by $\langle\eta\rangle\subset\cT M$. This codistribution has rank 1 at every point where $\eta$ does not vanish. Its anihilator is a distribution $\langle\eta\rangle^\circ\subset\T M$ that can be described as the kernel of the linear morphism $\widehat\eta\colon\T M\to M\times\R$ defined by $\eta$. This codistribution has corank 1 at every point where $\eta$ dos not vanish.

In the same way, every 2-form $\omega\in\Omega^2(M)$ induces a linear morphism $\widehat\omega\colon\T M\to\cT M$ defined by $\widehat\omega(v) = i(v)\omega$. The kernel of this morphism $\widehat\omega$ is a distribution $\ker\widehat\omega\subset\T M$. Notice that the rank of $\widehat\omega$ is even.

Given a family of $k$ differential 1-forms $\eta^1,\dotsc,\eta^k\in\Omega^1(M)$, we will denote
\begin{itemize}
	\item $\C^\rmC = \langle\eta^1,\dotsc,\eta^k\rangle\subset\cT M$,
	\item $\D^\rmC = \left(\C^\rmC\right)^\circ = \ker\widehat{\eta^1}\cap\dotsb\cap\ker\widehat{\eta^k}\subset\T M$,
	\item $\D^\rmR = \ker\widehat{\d\eta^1}\cap\dotsb\cap\ker\widehat{\d\eta^k}\subset\T M$,
	\item $\C^\rmR = \left( \D^\rmR \right)^\circ\subset\cT M$.
\end{itemize}
With the notations we just introduced, we are ready to introduce the concept of $k$-contact manifold:
\begin{dfn}\label{dfn:k-contact-manifold}
	A \textbf{$k$-contact structure} on a manifold $M$ is a family of $k$ differential 1-forms $\eta^1,\dotsc,\eta^k\in\Omega^1(M)$ such that, with the preceding notations,
	\begin{enumerate}[{\rm(1)}]
		\item $\D^\rmC\subset\T M$ is a regular distribution of corank $k$,
		\item $\D^\rmR\subset\T M$ is a regular distribution of rank $k$,
		\item $\D^\rmC\cap\D^\rmR = \{0\}$.
	\end{enumerate}
	We call $\C^\rmC$ the \textbf{contact codistribution}, $\D^\rmC$ the \textbf{contact distribution}, $\D^\rmR$ the \textbf{Reeb distribution} and $\C^\rmR$ the \textbf{Reeb codistribution}.

	A manifold $M$ endowed with a $k$-contact structure $\eta^1,\dotsc,\eta^k\in\Omega^1(M)$ is a \textbf{$k$-contact manifold}.
\end{dfn}

\begin{rmrk}\rm
	Notice that condition (1) in Definition \ref{dfn:k-contact-manifold} is equivalent to each one of the following two conditions:
	\begin{enumerate}
		\item[(1$'$)] $\C^\rmC\subset\cT M$ is a regular codistribution of rank $k$,
		\item[(1$''$)] $\eta^1\wedge\dotsb\wedge\eta^k\neq 0$ everywhere.
	\end{enumerate}
	Condition (3) can be rewritten as
	\begin{enumerate}
		\item[(3$'$)] $\bigcap_\alpha\left( \ker\widehat{\eta^\alpha}\cap\ker\widehat{\d\eta^\alpha} \right) = \{0\}$.
	\end{enumerate}
	If conditions (1) and (2) in Definition \ref{dfn:k-contact-manifold} hold, then condition (3) is equivalent to each of the following two conditions:
	\begin{enumerate}
		\item[(3$''$)] $\T M = \D^\rmC\oplus\D^\rmR$,
		\item[(3$'''$)] $\cT M = \C^\rmC\oplus\C^\rmR$.
	\end{enumerate}
	Furthermore, using the definition of $\D^\rmC$, one can prove that, in Definition \ref{dfn:k-contact-manifold}, conditions (2) and (3) imply (1).
\end{rmrk}

\begin{rmrk}\rm
	In the particular case $k=1$, a 1-contact structure is given by a 1-form $\eta$. In this case, the conditions in Definition \ref{dfn:k-contact-manifold} mean the following: (1) $\eta\neq 0$ everywhere, (3) $\ker\widehat\eta\cap\ker\widehat{\d\eta} = \{0\}$, which implies that $\ker\widehat{\d\eta}$ has rank 0 or 1, and (2) means that $\ker\widehat{\d\eta}$ has rank 1. Hence, if conditions (1) and (3) hold, the second condition is equivalenht to saying that $\dim M$ is odd. Thus, we have recovered the notion of contact manifold introduced in Definition \ref{dfn:contact-manifold}.
\end{rmrk}

\begin{lem}\label{lem:Reeb-distribution-involutive}
	The Reeb distribution $\D^\rmR$ is involutive, and therefore it is also integrable.
\end{lem}
\begin{proof}
	Consider the relation
	$$ i([X,X']) = \Lie_Xi(X') - i(X')\Lie_X = \d i(X)i(X') + i(X)\d i(X') - i(X')\d i(X) - i(X')i(X)\d\,. $$
	Notice that if $X$ and $X'$ are sections of $\D^\rmR$ and we apply this relation to the closed 2-form $\d\eta^\alpha$, the result is zero.
\end{proof}
\begin{thm}[Reeb vector fields]\label{thm:k-contact-Reeb}
	Let $(M,\eta^\alpha)$ be a $k$-contact manifold. Then, there exists a unique family of $k$ vector fields $\Reeb_\alpha\in\X(M)$, called the \textbf{Reeb vector fields} of $M$, such that
	\begin{equation}\label{eq:k-contact-Reeb}
		\begin{dcases}
			i(\Reeb_\alpha)\eta^\beta = \delta_\alpha^\beta\,,\\
			i(\Reeb_\alpha)\d\eta^\beta = 0\,.
		\end{dcases}
	\end{equation}
	The Reeb vector fields $\Reeb_1,\dotsc,\Reeb_k$ commute:
	$$ [\Reeb_\alpha,\Reeb_\beta] = 0\,. $$
	In addition, the Reeb distribution introduced in Definition \ref{dfn:k-contact-manifold} is spanned by the Reeb vector fields,
	$$ \D^\rmR = \langle \Reeb_1,\dotsc,\Reeb_k \rangle\,, $$
	motivating its name.
\end{thm}
\begin{proof}
	Take $\cT M = \C^\rmC\oplus\C^\rmR$. The family of 1-forms $\eta^\alpha$ is a global frame of the contact codistribution $\C^\rmC$. We can find a local frame $\eta^\mu$ of the Reeb codistribution $\C^\rmR$ so that $(\eta^\alpha;\eta^\mu)$ is a local frame for $\cT M$. The corresponding dual frame for $\T M$ constituted by vector fields $(\Reeb_\beta;\Reeb_\nu)$, where the $\Reeb_\beta$ are uniquely defined by the relations
	\begin{equation*}
		\begin{dcases}
			\langle \eta^\alpha,\Reeb_\beta\rangle = \delta^\alpha_\beta\,,\\
			\langle \eta^\mu,\Reeb_\beta\rangle = 0\,.
		\end{dcases}
	\end{equation*}
	Notice that the second set of relations does not depend on the choice of the $\eta^\mu$, in fact it means that the vector fields $\Reeb_\beta$ are sections of $\left( \C^\rmR \right)^\circ = \D^\rmR$, the Reeb distribution. This is equivalent to the condition $i(\Reeb_\beta)\d\eta^\alpha = 0$ for every $\alpha$. As the 1-forms $\eta^\alpha$ are globally defined, so are the vector fields $\Reeb_\alpha$.

	To prove that the Reeb vector fields commute, notice that
	$$ i([\Reeb_\alpha,\Reeb_\beta])\eta^\gamma = 0\ ,\quad i([\Reeb_\alpha,\Reeb_\beta])\d\eta^\gamma = 0\,, $$
	which is a consequence of their definition and of the above formula for $i([X,X'])$ when applied to them.
\end{proof}

\begin{prop}\label{prop:k-contact-adapted-coordinates}
	Let $(M,\eta^\alpha)$ be a $k$-contact manifold. There exist local coordinates $(x^I, s^\alpha)$, called \textbf{adapted coordinates}, such that
	$$ \Reeb_\alpha = \parder{}{s^\alpha}\ ,\quad \eta^\alpha = \d s^\alpha - f_I^\alpha(x)\d x^I\,, $$
	where the functions $f_I^\alpha$ only depend on the $x^I$.
\end{prop}
\begin{proof}
	As the Reeb vector fields commute, there exists a set of local coordinates $(x^I,s^\alpha)$ where they can be simultaneously straightened out (see \cite[p.234]{Lee2013} for details):
	$$ \Reeb_\alpha = \parder{}{s^\alpha}\,. $$
	We are going to write the contact forms using these coordinates. The condition $i(\Reeb_\alpha)\eta^\beta = \delta_\alpha^\beta$ implies that $\eta^\alpha = \d s^\alpha - f_I^\alpha\d x^I$, where the functions $f_I^\alpha$ may depend on all the coordinates $(x^I,s^\alpha)$. Then, we have that $\d\eta^\alpha = \d x^I\wedge\d f^\alpha_I$. But the condition $i(\Reeb_\alpha)\d\eta^\beta = 0$ must be fulfilled. The only way to ensure this is that $\tparder{f^\alpha_I}{s^\beta} = 0$ and this concludes the proof.
\end{proof}

\begin{exmpl}[Canonical $k$-contact structure]\label{ex:canonical-k-contact-structure}\rm
	Let $k\geq 1$ and $Q$ a smooth manifold. The manifold $M = \oplus^k\cT Q\times\R^k$ has a canonical contact structure given by the 1-forms $\eta^1,\dotsc,\eta^k\in\Omega^1(M)$ defined as
	$$ \eta^\alpha = \d s^\alpha - \theta^\alpha\,, $$
	where $(s^1,\dotsc,s^k)$ are the canonical coordinates of $\R^k$ and $\theta^\alpha$ is the pull-back of the Liouville 1-form $\theta$ of the cotangent bundle $\cT Q$ with respect to the projection $M\to\cT Q$ to the $\alpha$-th summand.

	Take coordinates $(q^i)$ on $Q$. Then, $M$ has natural coordinates $(q^i,p_i^\alpha,s^\alpha)$. Using these coordinates, the contact forms $\eta^\alpha$ are
	$$ \eta^\alpha = \d s^\alpha - p_i^\alpha\d q^i\,. $$
	Hence, $\d\eta^\alpha = \d q^i\wedge\d p_i^\alpha$, the Reeb distribution $\D^\rmR$ is
	$$ \D^\rmR = \left\langle\parder{}{s^1},\dotsc,\parder{}{s^k}\right\rangle\,, $$
	and the Reeb vector fields are
	$$ \Reeb_\alpha = \parder{}{s^\alpha}\,. $$
\end{exmpl}

\begin{exmpl}[Contactification of a $k$-symplectic manifold]\label{ex:contactification-k-symplectic-manifold}\rm
	Consider a $k$-symplec\-tic manifold $(P,\omega^\alpha)$ such that $\omega^\alpha = -\d\theta^\alpha$ and the product manifold $M = P\times\R^k$. Let $(s^\alpha)$ be the cartesian coordinates of $\R^k$ and denote also by $\theta^\alpha$ the pull-back of $\theta^\alpha$ to the product manifold $M$. Consider the 1-forms $\eta^\alpha = \d s^\alpha - \theta^\alpha\in\Omega^1(M)$.

	Then, $(M,\eta^\alpha)$ is a $k$-contact manifold because $\C^\rmC = \langle\eta^1,\dotsc,\eta^k\rangle$ has rank $k$, $\d\eta^\alpha = -\d\theta^\alpha$, and $\D^\rmR = \bigcap_\alpha\ker\widehat{\d\theta^\alpha} = \langle\tparder{}{s^1},\dotsc,\tparder{}{s^k}\rangle$ has rank $k$ since $(P,\omega^\alpha)$ is $k$-symplectic, and the last condition is immediate.

	Notice that the so-called canonical $k$-contact structure described in the previous example is just the contactification of the $k$-symplectic manifold $P = \oplus^k\cT Q$.

	Consider now the particular case $k=1$. Let $P$ be a manifold with a 1-form and consider the product manifold $M = P\times\R$ with the 1-form $\eta = \d s - \theta\in\Omega^1(M)$. In this case, $\C^\rmC = \langle\eta\rangle$ has rank 1, $\d\eta = -\d\theta$, and $\D^\rmR = \ker\widehat{\d\theta}$ has rank 1 if, and only if, $\d\theta$ is a symplectic form on $P$. Under these hypotheses, $M$ is a 1-contact manifold.
\end{exmpl}

\begin{exmpl}\label{ex:2-contact-R6}\rm
	Consider the manifold $M = \R^6$ with coordinates $(x, y, p, q, s, t)$. Then, the 1-forms
	$$ \eta^1 = \d s - \frac{1}{2}(y\d x - x\d y)\ ,\quad\eta^2 = \d t - p\d x - q\d y $$
	define a 2-contact structure on $M$. We are going to check that the conditions in Definition \ref{dfn:k-contact-manifold} are fulfilled. In first place, it is clear that the forms $\eta^1$ and $\eta^2$ are linearly independent. Then,
	$$ \d\eta^1 = \d x\wedge\d y\ ,\quad\d\eta^2 = \d x\wedge\d p + \d y\wedge\d q\,, $$
	and hence
	$$ \D^\rmR = \left\langle\parder{}{s},\parder{}{t}\right\rangle\,, $$
	which has rank 2. It is clear that none of these vector fields belong to the kernel of the 1-forms $\eta^1,\eta^2$, which is the third condition in Definition \ref{dfn:k-contact-manifold}. The Reeb vector fields are
	$$ \Reeb_1 = \parder{}{s}\ ,\quad\Reeb_2 = \parder{}{t}\,. $$
\end{exmpl}

Now we are going to state and proof the Darboux Theorem for $k$-contact manifolds. This theorem ensures the existence of canonical coordinates for a particular kind of $k$-contact manifolds.

\begin{thm}[$k$-contact Darboux Theorem]
	Consider a $k$-contact manifold $(M,\eta^\alpha)$ of dimension $\dim M = n + kn + k$ such that there exists an integrable subdistribution $\V$ of $\D^\rmC$ with $\rk\V = nk$. Then, under these hypotheses, around every point of $M$ there exists a local chart $(U,q^i,p_i^\alpha,s^\alpha)$, $1\leq\alpha\leq k$, $1\leq i\leq n$, such that
	$$ \restr{\eta^\alpha}{U} = \d s^\alpha - p_i^\alpha\d q^i\,. $$
	Using these coordinates,
	$$ \restr{\D^\rmR}{U} = \left\langle\Reeb_\alpha = \parder{}{s^\alpha}\right\rangle\ ,\quad\restr{\V}{U} = \left\langle\parder{}{p_i^\alpha}\right\rangle\,. $$
	These coordinates are called \textbf{canonical} or \textbf{Darboux coordinates} of the $k$-contact manifold $(M,\eta^\alpha)$.
\end{thm}
\begin{proof}
	The proof of this theorem will be divided into several steps.
	\begin{enumerate}[(i)]
		\item By Proposition \ref{prop:k-contact-adapted-coordinates}, there exists a local chart $(y^I,s^\alpha)$ of adapted coordinates such that
		$$ \Reeb_\alpha = \parder{}{s^\alpha}\ ,\quad \eta^\alpha = \d s^\alpha - f_I^\alpha(y)\d y^I\,. $$
		Hence, we can locally construct the quotient manifold $\widetilde M\equiv M/\D^\rmR$, with the projection $\widetilde\tau\colon M\to\widetilde M$ and local coordinates $(\widetilde y^I)$.
		\item The distribution $\D^\rmC$, which has $\rk\D^\rmC = nk + k$, is $\widetilde\tau$-projectable because, for every $\Reeb_\alpha\in\X(\D^\rmR)$, $Z\in\X(\D^\rmC)$ and $\d\eta^\beta$, we have that
		$$ i([\Reeb_\alpha,Z])\d\eta^\beta = \Lie_{\Reeb_\alpha}i(Z)\d\eta^\beta - i(Z)\Lie_{\Reeb_\alpha}\d\eta^\beta = -i(Z)\d i(\Reeb_\alpha)\eta^\beta = -i(Z)\d\delta_\alpha^\beta = 0\,, $$
		and thus $[\Reeb_\alpha,Z]\in\X(\D^\rmR)$. Notice that this is also a consequence of condition (3) in Definition \ref{dfn:k-contact-manifold}.
		\item The forms $\d\eta^\beta$ are $\widetilde\tau$-projectable because, by Theorem \ref{thm:k-contact-Reeb}, we have that $i(\Reeb_\alpha)\d\eta^\beta = 0$ for every $\Reeb_\alpha\in\X(\D^\rmR)$, and thus
		$$ \Lie_{\Reeb_\alpha}\d\eta^\beta = \d i(\Reeb_\alpha)\eta^\beta = \d\delta_\alpha^\beta = 0\,. $$
		It is clear that the $\widetilde\tau$-projected 2-forms $\widetilde\omega^\beta\in\Omega^2(\widetilde M)$ such that $\d\eta^\beta = \widetilde\tau^\ast\widetilde\omega^\beta$ are closed. Their local expression is
		$$ \widetilde\omega^\beta = \d\widetilde f_I^\beta(\widetilde y)\wedge\d\widetilde y^I\,. $$
		Moreover, as $\V$ is involutive, we have that for every $Z,Y\in\Gamma(\V)$,
		\begin{align*}
			i(Z)i(Y)\d\eta^\beta &= i(Z)(\Lie_Y\eta^\beta - \d i(Y)\eta^\beta) \\
			&= i(Z)\Lie_Y\eta^\beta \\
			&= \Lie_Y i(Z)\eta^\beta - i([Y,Z])\eta^\beta = 0\,.
		\end{align*}
		Denote by $\widetilde\V$ the distribution in $\widetilde M$ induced by $\V$. This distribution has $\rk\widetilde\V = nk$. Then, for every $\widetilde Z,\widetilde Y\in\Gamma(\widetilde \V)$, if $Z,Y\in\Gamma(\V)$ are such that $\widetilde\tau_\ast Z = \widetilde Z$ and $\widetilde\tau_\ast Y = \widetilde Y$, we have that
		\begin{equation}\label{eq:k-contact-Darboux-1}
			0 = i(Z)i(Y)\d\eta^\beta = i(Z)i(Y)(\widetilde\tau^\ast\widetilde\omega^\beta) = \widetilde\tau^\ast i(\widetilde Z)i(\widetilde Y)\widetilde\omega^\beta\,,
		\end{equation}
		and, as $\widetilde\tau$ is a submersion, the map $\widetilde\tau^\ast$ is injective and, from \eqref{eq:k-contact-Darboux-1} we have that
		$$ i(\widetilde Z)i(\widetilde Y)\widetilde\omega^\beta = 0\,. $$
		Notice that this fact does not depend on the choice of the representative vector fields $Y,Z$ used, because any two of them differ in an element of $\ker\widetilde\tau_\ast = \Gamma(\D^\rmR)$. Hence, we have seen that $\restr{\widetilde\omega^\beta}{\widetilde\V\times\widetilde\V} = 0$.

		As a consequence of (ii), we have that
		$$ \ker\widetilde\omega^1\cap\dotsb\cap\ker\widetilde\omega^k = \{0\}\,, $$
		and hence $(\widetilde M,\widetilde\omega^\alpha,\widetilde\V)$ is a $k$-symplectic manifold.
		\item By the Darboux Theorem for $k$-symplectic manifolds (Theorem \ref{thm:k-symplectic-Darboux}), there exists local charts $(\widetilde U;\widetilde q^i,\widetilde p_i^\alpha)$ in $\widetilde M$ such that
		$$ \restr{\widetilde\omega^\alpha}{\widetilde U} = \d\widetilde q^i\wedge\d\widetilde p_i^\alpha\ ,\quad \restr{\widetilde\V}{\widetilde U} = \left\langle \parder{}{\widetilde p_i^\alpha} \right\rangle\,. $$
		With all this in mind, in $U = \widetilde\tau^{-1}(\widetilde U)\subset M$, we can take the coordinates $(y^I,s^\alpha) = (q^i,p_i^\alpha,s^\alpha)$, with $q^i = \widetilde q^i\circ\widetilde\tau$ and $p_i^\alpha = \widetilde p_i^\alpha\circ\widetilde\tau$ satisfying the conditions of the theorem.
	\end{enumerate}
\end{proof}

This theorem allows us to consider the manifold introduced in Example \ref{ex:canonical-k-contact-structure} as the canonical model for this kind of $k$-contact manifold. Moreover, every $k$-contact manifold which is the contactification of a $k$-symplectic manifold (see Example \ref{ex:contactification-k-symplectic-manifold}) has Darboux coordinates.

\begin{rmrk}\label{rmrk:k-precontact-manifolds}\rm
	When some of the conditions stated in Definition \ref{dfn:k-contact-manifold} do not hold, we say that $\eta^1,\dotsc,\eta^k\in\Omega^1(M)$ is a \textbf{$k$-precontact structure} and that $(M,\eta^1,\dotsc,\eta^k)$ is a \textbf{$k$-precontact manifold}. For this kind of manifolds, Reeb vector fields are not uniquely determined. The particular case $k=1$ has been analyzed in \cite{DeLeo2019}, where the properties of the so-called {\it precontact structures} and {\it precontact manifolds} are studied in detail.
\end{rmrk}

\section{Hamiltonian formalism for \texorpdfstring{$k$}--contact systems}
\label{sec:k-contact-Hamiltonian-systems}

Now that we have introduced the geometric framework of $k$-contact geometry, we are ready to deal with the Hamiltonian formulation of field theories with dissipation.

\begin{dfn}\label{dfn:k-contact-Hamiltonian-system}
	A \textbf{$k$-contact Hamiltonian system} is a family $(M, \eta^\alpha,H)$, where $(M,\eta^\alpha)$ is a $k$-contact manifold and $H\in\Cinfty(M)$ is called a \textbf{Hamiltonian function}. Consider a map $\psi\colon D\subset\R^k\to M$. The \textbf{$k$-contact Hamilton--De Donder--Weyl equations} for the map $\psi$ are
	\begin{equation}\label{eq:k-contact-HDW}
		\begin{dcases}
			i(\psi_\alpha')\d\eta^\alpha = \left( \d H - (\Lie_{\Reeb_\alpha}H)\eta^\alpha \right)\circ\psi\,,\\
			i(\psi_\alpha')\eta^\alpha = -H\circ\psi\,.
		\end{dcases}
	\end{equation}
\end{dfn}
Now we are going to look at the expression in coordinates of the Hamilton--De Donder--Weyl equations \eqref{eq:k-contact-HDW}. Consider first the \textbf{adapted coordinates} $(x^I,s^\alpha)$. In these coordinates,
$$ \Reeb_\alpha = \parder{}{s^\alpha}\ ,\quad \eta^\alpha = \d s^\alpha - f_I^\alpha(x)\d x^I\ ,\quad \d\eta^\alpha = \frac{1}{2}\omega^\alpha_{IJ}\d x^I\wedge\d x^J\ ,\mbox{ with } \omega^\alpha_{IJ} = \parder{f_I^\alpha}{x^J} - \parder{f^\alpha_J}{x^I}\,. $$
The map $\psi\colon D\subset\R^k\to M$ has coordinate expression $\psi(t) = (x^I(t),s^\alpha(t))$. Then,
$$ \psi_\alpha' = \left(x^I,s^\beta; \parder{x^I}{t^\alpha},\parder{s^\beta}{t^\alpha}\right)\,. $$
The Hamilton--De Donde--Weyl equations read
\begin{equation*}
	\begin{dcases}
		\parder{x^J}{t^\alpha}\omega_{JI}^\alpha = \left(\parder{H}{x^I} + \parder{H}{s^\alpha}f_I^\alpha\right)\circ\psi\,,\\
		\parder{s^\alpha}{t^\alpha} - f_I^\alpha\parder{x^I}{t^\alpha} = -H\circ\psi\,.
	\end{dcases}
\end{equation*}
On the other hand, in \textbf{Darboux coordinates}, the map $\psi$ has local expression $\psi(t) = (q^i(t),p_i^\alpha(t),s^\alpha(t))$. Hence, equations \eqref{eq:k-contact-HDW} read
\begin{equation}\label{eq:k-contact-HDW-Darboux-coordinates}
	\begin{dcases}
		\parder{q^i}{t^\alpha} = \parder{H}{p_i^\alpha}\circ\psi\,,\\
		\parder{p^\alpha_i}{t^\alpha} = -\left( \parder{H}{q^i} + p_i^\alpha\parder{H}{s^\alpha} \right)\circ\psi\,,\\
		\parder{s^\alpha}{t^\alpha} = \left( p_i^\alpha\parder{H}{p_i^\alpha} - H \right)\circ\psi\,.
	\end{dcases}
\end{equation}

\begin{rmrk}\rm
	If $(M,\eta^\alpha)$ is a $k$-precontact manifold, then $(M,\eta^\alpha,H)$ is said to be a \textbf{$k$-precontact Hamiltonian system}.
\end{rmrk}

\begin{dfn}
	Consider a $k$-contact Hamiltonian system $(M,\eta^\alpha,H)$. The \textbf{$k$-contact Hamilton--De Donder--Weyl equations} for a $k$-vector field $\bfX = (X_\alpha)\in\X^k(M)$ are
	\begin{equation}\label{eq:k-contact-HDW-fields}
		\begin{dcases}
			i(X_\alpha)\d\eta^\alpha = \d H - (\Lie_{\Reeb_\alpha}H)\eta^\alpha\,,\\
			i(X_\alpha)\eta^\alpha = -H\,.
		\end{dcases}
	\end{equation}
	A $k$-vector field solution to these equations is a \textbf{$k$-contact Hamiltonian $k$-vector field}.
\end{dfn}

\begin{prop}\label{prop:k-contact-HDW-have-solutions}
	The $k$-contact Hamilton--De Donder--Weyl equations \eqref{eq:k-contact-HDW-fields} admit solutions. They are not unique if $k > 1$.
\end{prop}
\begin{proof}
	A $k$-vector field $\bfX\in\X^k(M)$ can be decomposed as $\bfX = \bfX^\rmC + \bfX^\rmR$, using the direct sum decomposition $\T M = \D^\rmC\oplus\D^\rmR$. If $\bfX$ is a solution to \eqref{eq:k-contact-HDW-fields}, then $\bfX^\rmC$ is a solution to the first equation and $\bfX^\rmR$ is a solution to the second one.

	At this point we need to introduce two vector bundle maps:
	\begin{align*}
		& \rho\colon\T M\to\oplus^k\cT M\ ,\qquad \rho(v) = \left( \widehat{\d\eta^1}(v),\dotsc,\widehat{\d\eta^k}(v) \right)\,,\\
		& \tau\colon\oplus^k\T M\to\cT M\ ,\qquad \tau(v_1,\dotsc,v_k) = \widehat{\d\eta^\alpha}(v_\alpha)\,.
	\end{align*}
	Now we have to remark some facts:
	\begin{itemize}
		\item $\ker\rho = \D^\rmR$ is the Reeb distribution.
		\item Using the canonical identification $(E\oplus F)^\ast = E^\ast\oplus F^\ast$, the transposed morphism of $\tau$ is $\transp{\tau} = -\rho$. The proof of this fact uses that $\transp{ \widehat{\d\eta^\alpha} } = -\widehat{\d\eta^\alpha}$.
		\item The first Hamilton--De Donder--Weyl equation for a $k$-vector field $\bfX$ can be written as
		$$ \tau\circ\bfX = \d H - (\Lie_{\Reeb_\alpha}H)\eta^\alpha\,. $$
	\end{itemize}
	A sufficient condition for this linear equation to have solutions $\bfX$ is that the right-hand side must be in the image of the morphism $\tau$, that is, to be anihilated by any section of $\ker\transp{\tau} = \D^\rmR$. Using that $i(\Reeb_\beta)(\d H - (\Lie_{\Reeb_\alpha}H)\eta^\alpha) = 0$ for any $\beta$, we can conclude that the first Hamilton--De Donder--Weyl equation has solutions. In particular, it has solutions $\bfX^\rmC$ belonging to the contact distribution.

	On the other hand, the second Hamilton--De Donder--Weyl equation for a $k$-vector field $\bfX$ has solutions belonging to the Reeb distribution, for instance $\bfX^\rmR = -H\Reeb_1$. Non-uniqueness for $k>1$ is obvious.
\end{proof}

Consider a $k$-vector field $\bfX = (X_1,\dotsc,X_k)\in\X^k(M)$ with local expression in \textbf{adapted coordinates}
$$ X_\alpha = (X_\alpha)^I\parder{}{x^I} + (X_\alpha)^\beta\parder{}{s^\beta}\,. $$
Hence, equation \eqref{eq:k-contact-HDW-fields} yields the conditions
\begin{equation*}
	\begin{dcases}
		(X_\alpha)^J\omega_{JI}^\alpha = \parder{H}{x^I} + \parder{H}{s^\alpha}f_I^\alpha\,,\\
		(X_\alpha)^\alpha - f_I^\alpha(X_\alpha)^I = -H\,.
	\end{dcases}
\end{equation*}
On the other hand, consider a $k$-vector field $\bfX = (X_1,\dotsc,X_k)\in\X^k(M)$ with local expression in \textbf{Darboux coordinates}
$$ X_\alpha = (X_\alpha)^i\parder{}{q^i} + (X_\alpha)^\beta_i\parder{}{p_i^\beta} + (X_\alpha)^\beta\parder{}{s^\beta}\,. $$
Now, equation \eqref{eq:k-contact-HDW-fields} gives the conditions
\begin{equation}\label{eq:k-contact-HDW-fields-Darboux-coordinates}
	\begin{dcases}
		(X_\alpha)^i = \parder{H}{p_i^\alpha}\,,\\
		(X_\alpha)^\alpha_i = -\left( \parder{H}{q^i} + p_i^\alpha\parder{H}{s^\alpha} \right)\,,\\
		(X_\alpha)^\alpha = p_i^\alpha\parder{H}{p_i^\alpha} - H\,.
	\end{dcases}
\end{equation}

\begin{prop}\label{prop:k-contact-equiv-fields-sections}
	Consider an integrable $k$-vector field $\bfX\in\X^k(M)$. Then, every integral section $\psi\colon D\subset\R^k\to M$ of $\bfX$ satisfies the $k$-contact Hamilton--De Donder--Weyl equation \eqref{eq:k-contact-HDW} if, and only if, $\bfX$ is a solution to \eqref{eq:k-contact-HDW-fields}.
\end{prop}
\begin{proof}
	This proposition is a direct consequence of equations \eqref{eq:k-contact-HDW} and \eqref{eq:k-contact-HDW-fields}, and of the fact that if $\bfX$ is integrable, then every point of $M$ is in the image of an integral section of $\bfX$. 
\end{proof}

\begin{rmrk}\label{rmrk:k-contact-difference-fields-sections}\rm
	It is important to point out that, as in the $k$-symplectic case, equations \eqref{eq:k-contact-HDW} and \eqref{eq:k-contact-HDW-fields} are not fully equivalent because a solution to \eqref{eq:k-contact-HDW} may not be an integral section of an integrable $k$-vector field solution to \eqref{eq:k-contact-HDW-fields}. This fact will be of interest when studying symmetries and dissipated quantities.
\end{rmrk}

\begin{prop}
	The $k$-contact Hamilton--De Donder--Weyl equations \eqref{eq:k-contact-HDW-fields} are equivalent to
	\begin{equation}\label{eq:k-contact-HDW-alternative}
		\begin{dcases}
			\Lie_{X_\alpha}\eta^\alpha = -(\Lie_{\Reeb_\alpha}H)\eta^\alpha\,,\\
			i(X_\alpha)\eta^\alpha = -H\,.
		\end{dcases}
	\end{equation}
\end{prop}
To end this section, we are going to offer a sufficient for a $k$-vector field to be a solution to the Hamilton--De Donder--Weyl equations \eqref{eq:k-contact-HDW-fields} without making use of the Reeb vector fields $\Reeb_\alpha$. This may be useful when dealing with singular systems, where the Reeb vector fields are not uniquely defined.

\begin{thm}\label{thm:equivalent-k-contact-Hamilton-equations-no-Reeb}
	Consider a $k$-contact Hamiltonian system $(M,\eta^\alpha,H)$ and the 2-forms
	$$ \Omega^\alpha = -H\d\eta^\alpha + \d H\wedge\eta^\alpha\,. $$
	Let $\O$ be the open set $\O = \{p\in M\ \vert\ H(p)\neq 0\}\subset M$. Then, if a $k$-vector field $\bfX = (X_\alpha)\in\X^k(M)$ satisfies equations
	\begin{equation}\label{eq:k-contact-no-Reeb-fields}
		\begin{dcases}
			i(X_\alpha)\Omega^\alpha = 0\,,\\
			i(X_\alpha)\eta^\alpha = -H\,,
		\end{dcases}
	\end{equation}
	on $\O$, it is also a solution of the Hamilton--De Donder--Weyl equations \eqref{eq:k-contact-HDW-fields} on the open set $\O$.
\end{thm}
\begin{proof}
	Let $\bfX$ be a $k$-vector field satisfying equations \eqref{eq:k-contact-no-Reeb-fields}. Then,
	$$ 0 = i(X_\alpha)\Omega^\alpha = -Hi(X_\alpha)\d\eta^\alpha + (i(X_\alpha)\d H)\eta^\alpha + H\d H\,, $$
	and thus,
	\begin{equation}\label{eq:k-contact-no-Reeb-eq1}
		Hi(X_\alpha)\d\eta^\alpha = (i(X_\alpha)\d H)\eta^\alpha + H\d H\,.
	\end{equation}
	Now, if we contract this last equation with every Reeb vector field $\Reeb_\beta$, we get
	\begin{align*}
		0 &= Hi(\Reeb_\beta)i(X_\alpha)\d\eta^\alpha\\
		&= (i(X_\alpha)\d H)i(\Reeb_\beta)\eta^\alpha + Hi(\Reeb_\beta)\d H\\
		&= (i(X_\alpha)\d H)\delta^\alpha_\beta + Hi(\Reeb_\beta)\d H\,,
	\end{align*}
	and hence,
	$$ i(X_\beta)\d H = -Hi(\Reeb_\beta)\d H $$
	for every $\beta$. Using this fact in equation \eqref{eq:k-contact-no-Reeb-eq1}, we obtain
	$$ Hi(X_\alpha)\d\eta^\alpha = H(\d H - (i(\Reeb_\alpha)\d H)\eta^\alpha) = H(\d H - (\Reeb_\alpha(H))\eta^\alpha)\,, $$
	and hence $i(X_\alpha)\d\eta^\alpha = \d H - (\Reeb_\alpha(H))\eta^\alpha$ wherever $H\neq 0$.
\end{proof}

Taking into account Definition \ref{dfn:k-contact-Hamiltonian-system} and Proposition \ref{prop:k-contact-equiv-fields-sections}, we have the following result.

\begin{prop}
	On the open subset $\O = \{p\in M\ \vert\ H(p)\neq 0\}\subset M$, if a map $\psi\colon D\subset\R^k\to M$ is an integral section of a $k$-vector field solution to equations \eqref{eq:k-contact-no-Reeb-fields}, then it is a solution to
	\begin{equation}\label{eq:k-contact-no-Reeb-sections}
		\begin{dcases}
			i(\psi'_\alpha)\Omega^\alpha = 0\,,\\
			i(\psi'_\alpha)\eta^\alpha = -H\circ\psi\,.
		\end{dcases}
	\end{equation}
\end{prop}

\section{Symmetries of \texorpdfstring{$k$}--contact Hamiltonian systems}
\label{sec:k-contact-Hamiltonian-symmetries}

There are many different notions of symmetry of a given problem, depending on the structure preserved. In some cases, one puts the emphasis on the transformations preserving the underlying geometric structures of the problem, or on the transformations that preserve its solutions \cite{Gra2002}. In particular, this has been done in the case of $k$-symplectic Hamiltonian systems \cite{Rom2007}. We will apply these ideas to the case of $k$-contact Hamiltonian systems. We will begin by defining those symmetries preserving the solutions of the systems.

\begin{dfn}
	Consider a $k$-contact Hamiltonian system $(M,\eta^\alpha,H)$.
	\begin{itemize}
		\item A \textbf{dynamical symmetry} is a diffeomorphism $\Phi\colon M\to M$ such that if $\psi$ is a solution to the $k$-contact Hamilton--De Donder--Weyl equations \eqref{eq:k-contact-HDW}, then so is $\Phi\circ\psi$.
		\item An \textbf{infinitesimal dynamical symmetry} is a vector field $Y\in\X(M)$ such that its local flow is made of local dynamical symmetries.
	\end{itemize}
\end{dfn}

Before giving a characterization of symmetries in terms of $k$-vector fields, we need to recall a fact about $k$-vector fields and integral sections.

\begin{lem}
	Consider a $k$-vector field $\bfX = (X_\alpha)\in\X^k(M)$ and a diffeomorphism $\Phi\colon M\to M$. If a map $\psi$ is an integral section of $\bfX$, then $\Phi\circ\psi$ is an integral section of the $k$-vector field $\Phi_\ast\bfX = (\Phi_\ast X_\alpha)$. In particular, if $\bfX$ is integrable, so is $\Phi_\ast\bfX$.
\end{lem}

With this in mind, we have the following result.

\begin{prop}
	If $\Phi\in\Diff(M)$ is a dynamical symmetry, then, if $\bfX$ is an integrable $k$-vector field solution to the $k$-contact Hamilton--De Donder--Weyl equations for fields \eqref{eq:k-contact-HDW-fields}, $\Phi_\ast\bfX$ is another solution.

	Conversely, if $\Phi$ transforms every $k$-vector field $\bfX$ solution to \eqref{eq:k-contact-HDW-fields} into another solution, then for every integral section $\psi$ of $\bfX$, we have that $\Phi\circ\psi$ is a solution to the $k$-contact Hamilton--De Donder--Weyl equations for sections \eqref{eq:k-contact-HDW}.
\end{prop}
\begin{proof}
	($\Rightarrow$) Let $x\in M$ and let $\psi$ be an integral section of the $k$-vector field $\bfX$ passing through the point $\Phi^{-1}(x)$, i.e., $\psi(t_0) = \Phi^{-1}(x)$. The map $\psi$ is a solution to equations \eqref{eq:k-contact-HDW} and, as $\Phi$ is a dynamical symmetry, so is $\Phi\circ\psi$. By the preceding lemma, it is an integral section of $\Phi_\ast\bfX$ through the point $\Phi(\psi(t_0)) = \Phi(\Phi^{-1}(x)) = x$ and thus we have that $\Phi_\ast\bfX$ has to be a solution to \eqref{eq:k-contact-HDW-fields} at the points $(\Phi\circ\psi)(t)$ and, in particular, at the point $(\Phi\circ\psi)(t_0) = x$.

	($\Leftarrow$) Let $\bfX\in\X^k(M)$ be a solution to \eqref{eq:k-contact-HDW-fields} and let $\psi\colon D\subset\R^k\to M$ be an integral section of the $k$-vector field $\bfX$. By hypothesis, $\Phi_\ast\bfX$ is also a solution to \eqref{eq:k-contact-HDW-fields}. Then, by the previous lemma, we have that $\Phi\circ\psi$ is a solution to equations \eqref{eq:k-contact-HDW}.
\end{proof}

Another kind of symmetry are those preserving the geometric structures of the problem.

\begin{dfn}
	Consider a $k$-contact Hamiltonian system $(M,\eta^\alpha,H)$.
	\begin{itemize}
		\item A \textbf{Hamiltonian $k$-contact symmetry} is a diffeomorphism $\Phi\in\Diff(M)$ such that
		$$ \Phi^\ast\eta^\alpha = \eta^\alpha\ ,\quad\Phi^\ast H = H\,. $$
		\item An \textbf{infinitesimal Hamiltonian $k$-contact symmetry} is a vector field $Y\in\X(M)$ whose local flow is a local Hamiltonian $k$-contact symmetry:
		$$ \Lie_Y\eta^\alpha = 0\ ,\quad\Lie_YH = 0\,. $$
	\end{itemize}
\end{dfn}

\begin{prop}\label{prop:k-contact-symmetry-preserves-Reeb}
	Every (infinitesimal) Hamiltonian $k$-contact symmetry preserves the Reeb vector fields:
	$$ \Phi_\ast\Reeb_\alpha = \Reeb_\alpha \quad \mbox{(or $[Y,\Reeb_\alpha] = 0$)}\,. $$
\end{prop}
\begin{proof}
	We have that
	\begin{gather*}
		i(\Phi_\ast^{-1}\Reeb_\alpha)(\Phi^\ast\d\eta^\alpha) = \Phi^\ast i(\Reeb_\alpha)\d\eta^\alpha = 0\,,\\
		i(\Phi_\ast^{-1}\Reeb_\alpha)(\Phi^\ast\eta^\alpha) = \Phi^\ast i(\Reeb_\alpha)\eta^\alpha = 1\,,
	\end{gather*}
	and, since $\Phi^\ast\eta^\alpha = \eta^\alpha$ and the Reeb vector fields are unique, we conclude that $\Phi_\ast\Reeb_\alpha = \Reeb_\alpha$.

	The proof of the infinitesimal case is straightforward from the definition.
\end{proof}

The following proposition stablishes the relation between Hamiltonian $k$-contact symmetries and dynamical symmetries.

\begin{prop}\label{prop:k-contact-symmetries-are-dynamical}
	(Infinitesimal) Hamiltonian $k$-contact symmetries are (infinitesimal) dynamical symmetries.
\end{prop}
\begin{proof}
	Consider a solution $\psi$ of the $k$-contact Hamilton--De Donder--Weyl equations \eqref{eq:k-contact-HDW} and a Hamiltonian $k$-contact symmetry $\Phi$. Then,
	\begin{align*}
		i((\Phi\circ\psi)'_\alpha)\eta^\alpha &= i(\Phi_\ast(\psi'_\alpha))((\Phi^{-1})^\ast\eta^\alpha)\\
		&= (\Phi^{-1})^\ast i(\psi'_\alpha)\eta^\alpha\\
		&= (\Phi^{-1})^\ast(-H\circ\psi)\\
		&= -H\circ(\Phi\circ\psi)\,,\\
		i((\Phi\circ\psi)'_\alpha)\d\eta^\alpha &= i(\Phi_\ast(\psi'_\alpha))((\Phi^{-1})^\ast\d\eta^\alpha)\\
		&= (\Phi^{-1})^\ast i(\psi'_\alpha)\d\eta^\alpha\\
		&= (\Phi^{-1})^\ast\big( (\d H - (\Lie_{\Reeb_\alpha}H)\eta^\alpha)\circ\psi \big)\\
		&= \Big( \d(\Phi^{-1})^\ast H - \big(\Lie_{(\Phi^{-1})^\ast\Reeb_\alpha}(\Phi^{-1})^\ast H\big)(\Phi^{-1})^\ast\eta^\alpha \Big)\circ(\Phi\circ\psi)\\
		&= \big( \d H - (\Lie_{\Reeb_\alpha}H)\eta^\alpha \big)\circ(\Phi\circ\psi)\,.
	\end{align*}
	The proof of the infinitesimal case is straightforward from the definition.
\end{proof}

\section{Dissipation laws}
\label{sec:k-contact-Hamiltonian-dissipation-laws}

When working with conservative mechanical systems, it is often of great interest to find quantities which are preserved along a solution. These quantities are called conserved quantities. Some examples of usual conserved quantities in mechanics are the energy or the different momenta. In the case of systems with dissipation, these quantities are not preserved, but dissipated. This behaviour was described in the case of contact systems in the so-called energy dissipation theorem \ref{thm:energy-dissipation-contact}, which says that, if $X_H$ is a contact Hamiltonian vector field of a contact Hamiltonian vector field $(M,\eta,H)$, then
$$ \Lie_{X_H}H = -(\Lie_\Reeb H)H\,. $$
This last equations tells us that, in a contact system, the dissipations are exponentials with rate $-\Lie_\Reeb H$.

In the case of dissipative field theories, a similar structure can be observed in the first equation of \eqref{eq:k-contact-HDW-alternative}, which can be understood as the dissipation of the contact 1-forms $\eta^\alpha$. Now, taking into account the definition of conservation law for field theories stated in \cite{Olv1986} and Remark \ref{rmrk:k-contact-difference-fields-sections}, we can define:

\begin{dfn}
	Consider a $k$-contact Hamiltonian system $(M,\eta^\alpha,H)$ and a map $F\colon M\to\R^k$ given by $F = (F^1,\dotsc,F^k)$. Then, $F$ is said to satisfy
	\begin{itemize}
		\item the \textbf{dissipation law for sections} if, for every solution $\psi$ to the $k$-contact Hamilton--De Donder--Weyl equations \eqref{eq:k-contact-HDW}, the divergence of $(F\circ\psi) = (F^\alpha\circ\psi)\colon\R^k\to\R^k$, defined as
		$$ \Div(F\circ\psi) = \parder{(F^\alpha\circ\psi)}{t^\alpha}\,, $$
		satisfies that
		\begin{equation}\label{eq:k-contact-dissipation-law-sections}
			\Div(F\circ\psi) = -\big( (\Lie_{\Reeb_\alpha}H)F^\alpha \big)\circ\psi\,.
		\end{equation}
		\item the \textbf{dissipation law for $k$-vector fields} if, for every $\bfX = (X_\alpha)\in\X^k(M)$ solution to the $k$-contact Hamilton--De Donder--Weyl equations \eqref{eq:k-contact-HDW-fields}, we have that
		\begin{equation}\label{eq:k-contact-dissipation-law-fields}
			\Lie_{X_\alpha}F^\alpha = -(\Lie_{\Reeb_\alpha}H)F^\alpha\,.
		\end{equation}
	\end{itemize}
\end{dfn}

The relationship between these two different notions of dissipation law is given by the following proposition.

\begin{prop}\label{prop:relation-dissipation-laws-Hamiltonian}
	Let $F = (F^\alpha)\colon M\to\R^k$ be a map satisfying the dissipation law for sections \eqref{eq:k-contact-dissipation-law-sections}. Then, for every integrable $k$-vector field $\bfX = (X_\alpha)$ solution to the $k$-contact Hamilton--De Donder--Weyl equations for fields \eqref{eq:k-contact-HDW-fields}, we have that \eqref{eq:k-contact-dissipation-law-fields} holds for $\bfX$.

	Conversely, if $F$ satisfies the dissipation law for fields \eqref{eq:k-contact-dissipation-law-fields} for a $k$-vector field $\bfX = (X_\alpha)\in\X^k(M)$, then the dissipation law for sections \eqref{eq:k-contact-dissipation-law-sections} holds for every integral section $\psi$ of $\bfX$.
\end{prop}
\begin{proof}
	Let $\bfX = (X_\alpha)\in\X^k(M)$ be an integrable $k$-vector field solution to \eqref{eq:k-contact-HDW-fields}, let $\psi\colon\R^k\to M$ be an integral section of $\bfX$ and consider a map $F = (F^\alpha)\colon M\to\R^k$ satisfying the dissipation law for section \eqref{eq:k-contact-dissipation-law-sections}. Then, by Proposition \ref{prop:k-contact-equiv-fields-sections}, $\psi$ is a solution to the $k$-contact Hamilton--De Donder--Weyl equations \eqref{eq:k-contact-HDW}. Hence,
	$$ (\Lie_{X_\alpha}F^\alpha)\circ\psi = \frac{\d}{\d t^\alpha}(F^\alpha\circ\psi) = \Div(F\circ\psi) = -\big( (\Lie_{\Reeb_\alpha}H)F^\alpha \big)\circ\psi\,, $$
	and as $\bfX$ is integrable, there exists an integral section through every point.

	Conversely, if \eqref{eq:k-contact-dissipation-law-fields} holds, then the statement is a direct consequence of the above expression.
\end{proof}

\begin{lem}\label{lem:previous-k-contact-dissipation-theorem}
	Let $Y\in\X(M)$ be an infinitesimal dynamical symmetry. Then, for every $k$-vector field $\bfX = (X_\alpha)$ solution to the $k$-contact Hamilton--De Donder--Weyl equations \eqref{eq:k-contact-HDW-fields}, we have
	$$ i([Y,X_\alpha])\eta^\alpha = 0\ ,\quad i([Y,X_\alpha])\d\eta^\alpha = 0\,. $$
\end{lem}
\begin{proof}
	Let $F_\varepsilon$ be the local 1-parameter group of diffeomorphisms generated by $Y$. As $Y$ is an infinitesimal dynamical symmetry, we have that
	$$ i(F_\varepsilon^\ast X_\alpha)\eta^\alpha = i(X_\alpha)\eta^\alpha\,, $$
	because both are solutions to the Hamilton--De Donder--Weyl equations \eqref{eq:k-contact-HDW-fields}. Then, as the contraction is continuous, we have
	$$ i([Y,X_\alpha])\eta^\alpha = i\left( \lim_{\varepsilon\to 0}\frac{F_\varepsilon^\ast X_\alpha - X_\alpha}{\varepsilon} \right)\eta^\alpha = \lim_{\varepsilon\to 0}\frac{ i(F^\ast_\varepsilon X_\alpha)\eta^\alpha - i(X_\alpha)\eta^\alpha }{\varepsilon} = 0\,. $$
	The proof of the second equality is completely analogous.
\end{proof}

The last theorem of this section relates dissipation laws for $k$-vector fields with infinitesimal dynamical symmetries.

\begin{thm}[Dissipation theorem for $k$-contact Hamiltonian systems]\label{thm:k-contact-Hamiltonian-dissipation-theorem}
	Let $Y$ be an infinitesimal dynamical symmetry. Then, $F^\alpha = -i(Y)\eta^\alpha$ satisfies the dissipation law for $k$-vector fields.
\end{thm}
\begin{proof}
	Let $\bfX = (X_\alpha)\in\X^k(M)$ be a solution to the $k$-contact Hamilton--De Donder--Weyl equations \eqref{eq:k-contact-HDW-fields}. By Lemma \ref{lem:previous-k-contact-dissipation-theorem}, we have that $i([Y,X_\alpha])\eta^\alpha = 0$. Hence,
	$$ \Lie_{X_\alpha}(i(Y)\eta^\alpha) = i([X_\alpha,Y])\eta^\alpha + i(Y)\Lie_{X_\alpha}\eta^\alpha = -(\Lie_{\Reeb_\alpha}H)i(Y)\eta^\alpha\,. $$
\end{proof}

	\chapter[\texorpdfstring{$k$}--contact Lagrangian systems]{\texorpdfstring{$k$}--contact Lagrangian systems}
	\label{ch:k-contact-Lagrangian}


In this chapter we are going to develop a geometric formalism to deal with dissipative Lagrangian field theories. We will use the geometric framework of $k$-contact geometry introduced in Section \ref{sec:k-contact-geometry}. In Section \ref{sec:k-contact-Lagrangian-systems} we begin by extending the canonical structures of the bundle $\oplus^k\T Q$ to $\oplus^k\T Q\times\R^k$. These structures permit us to introduce the notions of second-order partial differential equation and holonomic section. Given a Lagrangian function, we define its associated Lagrangian energy, Cartan forms and contact forms. With all these geometric tools we can finally introduce $k$-contact Lagrangian systems. Finally, we define the Legendre map associated to a Lagrangian function, which allow us to classify Lagrangians as regular (or hyperregular) and singular. In Section \ref{sec:k-contact-Euler-Lagrange-equations} we use the notions introduced in the previous section and in Chapter \ref{ch:k-contact-Hamiltonian} to write the $k$-contact Lagrangian equations for $k$-vector fields and the $k$-contact Euler--Lagrange equations. Section \ref{sec:singular-case-k-precontact} deals with the $k$-contact formalism for singular Lagrangian functions. This will be of special interest when developing the Skinner--Rusk formalism for $k$-contact systems in Chapter \ref{ch:Skinner-Rusk-k-contact}. In Sections \ref{sec:k-contact-Lagrangian-symmetries} and \ref{sec:k-contact-Lagrangian-dissipation-laws} we adapt the different notions of symmetry and dissipation laws for $k$-contact Hamiltonian systems introduced in Sections \ref{sec:k-contact-Hamiltonian-symmetries} and \ref{sec:k-contact-Hamiltonian-dissipation-laws} to $k$-contact Lagrangian systems. Finally, Section \ref{sec:k-contact-symmetries-Lagrangian-function} is devoted to study the symmetries of the Lagrangian function of $k$-contact Lagrangian systems. The main reference on this topic is \cite{Gas2021}.


\section{Lagrangian formalism for \texorpdfstring{$k$}--contact systems}
\label{sec:k-contact-Lagrangian-systems}

Throughout this chapter, our phase space we will be the bundle $\oplus^k\T Q\times\R^k$ with natural projections
$$ \bar\tau_1\colon\oplus^k\T Q\times\R^k\to\oplus^k\T Q\ ,\quad \bar\tau^\alpha\colon\oplus^k\T Q\times\R^k\to\T Q\ ,\quad s^\alpha\colon\oplus^k\T Q\times\R^k\to\R\,, $$
and equipped with coordinates $(q^i,v^i_\alpha,s^\alpha)$. As $\oplus^k\T Q\times\R^k\to\oplus^k\T Q$ is a trivial bundle, the canonical structures in $\oplus^k\T Q$ defined in the beginning of Section \ref{sec:k-symplectic-Lagrangian-systems} (the canonical $k$-tangent structure and the Liouville vector field) can be extended to $\oplus^k\T Q\times\R^k$ in a natural way, and are denoted with the same notation $(J^\alpha)$ and $\Delta$. Their coordinate expressions remain the same:
$$ J^\alpha = \parder{}{v^i_\alpha}\otimes\d q^i\ ,\quad\Delta = v^i_\alpha\parder{}{v^i_\alpha}\,. $$
Using these canonical structures, one can also extend the notion of {\sc sopde} to $\oplus^k\T Q\times\R^k$:
\begin{dfn}
	A $k$-vector field $\mathbf{\Gamma} = (\Gamma_\alpha)\in\X^k(\oplus^k\T Q\times\R^k)$ is a \textbf{second-order partial differential equation} ({\sc sopde} for short) if
	$$ J^\alpha(\Gamma_\alpha) = \Delta\,. $$
\end{dfn}
In local coordinates, a {\sc sopde} reads
$$ \Gamma_\alpha = v^i_\alpha\parder{}{q^i} + (\Gamma_\alpha)_\beta^i\parder{}{v^i_\beta}(\Gamma_\alpha)^\beta\parder{}{s^\beta}\,. $$

\begin{dfn}\label{dfn:k-contact-holonomic-section}
	Consider a section $\psi\colon\R^k\to Q\times\R^k$ os the projection $Q\times\R^k\to\R^k$ with $\psi = (\phi,s^\alpha)$, where $\phi\colon\R^k\to Q$. The \textbf{first prolongation} of $\psi$ to $\oplus^k\T Q\times\R^k$ is the map $\psi'\colon\R^k\to\oplus^k\T Q\times\R^k$ given by $\psi' = (\phi',s^\alpha)$, where $\phi'$ is the first prolongation of $\phi$ to $\oplus^k\T Q$ defined in \ref{dfn:first-prolongation-k-tangent-bundle}. The map $\psi'$ is said to be \textbf{holonomic}.
\end{dfn}

\begin{prop}
	A $k$-vector field $\mathbf{\Gamma}\in\X^k(\oplus^k\T Q\times\R^k)$ is a {\sc sopde} if, and only if, its integral sections are holonomic.
\end{prop}

With these geometric tools in mind, we can now state the Lagrangian formalism for field theories with dissipation.

\begin{dfn}
	A \textbf{Lagrangian function} is a function $\L\in\Cinfty(\oplus^k\T Q\times\R^k)$. We can define:
	\begin{itemize}
		\item The \textbf{Lagrangian energy} associated to $\L$ is the function
		$$ E_\L = \Delta(\L) - \L\in\Cinfty(\oplus^k\T Q\times\R^k)\,. $$
		\item The \textbf{Cartan forms} associated to $\L$ are
		$$ \theta^\alpha_\L = \transp{J^\alpha}\circ\d\L\in\Omega^1(\oplus^k\T Q\times\R^k)\ ,\quad\omega^\alpha_\L = -\d\theta^\alpha_\L\in\Omega^2(\oplus^k\T Q\times\R^k)\,. $$
		\item The \textbf{contact forms} associated to $\L$ are
		$$ \eta^\alpha_\L = \d s^\alpha - \theta^\alpha_\L\in\Omega^1(\oplus^k\T Q\times\R^k)\,. $$
		\item The couple $(\oplus^k\T Q\times\R^k,\L)$ is a \textbf{$k$-contact Lagrangian system}.
	\end{itemize}
\end{dfn}
Notice that $\d\eta^\alpha_\L = \omega^\alpha_\L$. Taking natural coordinates $(q^i,v^i_\alpha,s^\alpha)$ in $\oplus^k\T Q\times\R^k$, the local expressions of the elements introduced in the previous definition are
\begin{align*}
	E_\L &= v^i_\alpha\parder{\L}{v^i_\alpha} - \L\,,\\
	\theta^\alpha_\L &= \parder{\L}{v^i_\alpha}\d q^i\,,\\
	\omega^\alpha_\L &=  -\parderr{\L}{q^j}{v^i_\alpha}\d q^j\wedge\d q^i - \parderr{\L}{v^j_\beta}{v^i_\alpha}\d v^j_\beta\wedge\d q^i - \parderr{\L}{s^\beta}{v^i_\alpha}\d s^\beta\wedge\d q^i \,,\\
	\eta^\alpha_\L &= \d s^\alpha - \parder{\L}{v^i_\alpha}\d q^i\,.
\end{align*}
Now, taking into account Definitions \ref{dfn:fibre-derivative} and \ref{dfn:contact-legendre-map}, we can define the Legendre map for $k$-contact Lagrangian systems.
\begin{dfn}
	Consider a Lagrangian function $\L\in\Cinfty(\oplus^k\T Q\times\R^k)$. The \textbf{Legendre map} associated to $\L$ is the fibre derivative of $\L$, considered as a function on the vector bundle $\oplus^k\T Q\times\R^k\to Q\times\R^k$; that is, the map
	$$ \F\L\colon \oplus^k\T Q\times\R^k\to\oplus^k\cT Q\times\R^k\,, $$
	given by
	$$ \F\L({v_1}_q,\dotsc,{v_k}_q; s^\alpha) = \left( \F\L(\cdot,s^\alpha)({v_1}_q,\dotsc,{v_k}_q) ; s^\alpha\right)\,, $$
	where $({v_1}_q,\dotsc,{v_k}_q)\in\oplus^k\T Q$ and $\L(\cdot,s^\alpha)$ denotes the Lagrangian function with $s^\alpha$ freezed.
\end{dfn}
The local expression of the Legendre map defined above is
$$ \F\L(q^i,v^i_\alpha,s^\alpha) = \left( q^i,\parder{\L}{v^i_\alpha},s^\alpha \right)\,. $$
The Legendre map allows us to give an alternative definition of the Cartan forms. We have that
$$ \theta^\alpha_\L = \F\L^\ast\theta^\alpha\ ,\quad\omega^\alpha_\L = \F\L^\ast\omega^\alpha\,, $$
where $\theta^\alpha,\omega^\alpha$ are the extensions to $\oplus^k\cT Q\times\R^k$ of the canonical forms of $\oplus^k\cT Q$ defined in Example \ref{ex:canonical-model-k-symplectic}.
\begin{prop}\label{prop:k-contact-regular-Lagrangian}
	Let $\L\in\Cinfty(\oplus^k\T Q\times\R^k)$ be a Lagrangian function. Then, the following are equivalent:
	\begin{enumerate}[{\rm(1)}]
		\item The Legendre map $\F\L$ is a local diffeomorphism.
		\item The fibre Hessian of $\L$
		$$ \F^2\L\colon \oplus^k\T Q\times\R^k\to(\oplus^k\cT Q\times\R^k)\otimes(\oplus^k\cT Q\times\R^k)\,, $$
		is everywhere nondegenerate (the tensor product is of vector bundles over $Q\times\R^k$).
		\item The couple $(\oplus^k\T Q\times\R^k,\eta^\alpha_\L)$ is a $k$-contact manifold.
	\end{enumerate}
\end{prop}
\begin{proof}
	Taking natural coordinates $(q^i,v^i_\alpha,s^\alpha)$ in $\oplus^k\T Q\times\R^k$, it is clear that
	\begin{align}
		\F\L(q^i,v^i_\alpha,s^\alpha) &= \left( q^i,\parder{\L}{v^i_\alpha},s^\alpha \right)\,,\label{eq:k-contact-Legendre-coordinates}\\
		\F^2\L(q^i,v^i_\alpha,s^\alpha) &= \left(q^i,W_{ij}^{\alpha\beta},s^\alpha\right)\,,\mbox{ where }W_{ij}^{\alpha\beta} = \left(\parderr{\L}{v^i_\alpha}{v^j_\beta}\right)\,.
	\end{align}
	Then, the conditions in the proposition mean that the matrix $W = \big(W_{ij}^{\alpha\beta}\big)$ is everywhere nonsingular.
\end{proof}

\begin{dfn}
	Let $\L\in\Cinfty(\oplus^k\T Q\times\R^k)$ be a Lagrangian function. The Lagrangian $\L$ is said to be \textbf{regular} if the equivalent statements in Proposition \ref{prop:k-contact-regular-Lagrangian} hold. Otherwise, $\L$ is a \textbf{singular} Lagrangian. In particular, if $\F\L$ is a global diffeomorphism, $\L$ is said to be a \textbf{hyperregular} Lagrangian.
\end{dfn}

Consider a regular $k$-contact Lagrangian system $(\oplus^k\T Q\times\R^k,\L)$. By Theorem \ref{thm:k-contact-Reeb}, we have that the Reeb vector fields $(\Reeb_\L)_\alpha\in\X(\oplus^k\T Q\times\R^k)$ for this Lagrangian system are the unique solution to
\begin{equation*}
	\begin{dcases}
		i\big((\Reeb_\L)_\alpha\big)\d\eta^\beta_\L = 0\,,\\
		i\big((\Reeb_\L)_\alpha\big)\eta^\beta_\L = \delta_\alpha^\beta\,.\\
	\end{dcases}
\end{equation*}
The local expression of the Reeb vector fields $(\Reeb_\L)_\alpha$ is
$$ (\Reeb_\L)_\alpha = \parder{}{s^\alpha} - W_{\gamma\beta}^{ji}\parderr{\L}{s^\alpha}{v^j_\gamma}\parder{}{v^i_\beta}\,, $$
where $W_{\alpha\beta}^{ij}$ is inverse of the Hessian matrix, namely
$$ W_{\alpha\beta}^{ij}\parderr{\L}{v^j_\beta}{v^k_\gamma} = \delta^i_k\delta^\gamma_\alpha\,. $$

\section{\texorpdfstring{$k$}--contact Euler--Lagrange equations}
\label{sec:k-contact-Euler-Lagrange-equations}

As a consequence of the results and definitions of the previous section, we have that every regular (resp. singular) $k$-contact Lagrangian system $(\oplus^k\T Q\times\R^k,\L)$ has associated the $k$-contact (resp. $k$-precontact) Hamiltonian system $(\oplus^k\T Q\times\R^k,\eta^\alpha_\L,E_\L)$. With this fact in mind, we can define

\begin{dfn}
	Consider a $k$-contact Lagrangian system $(\oplus^k\T Q\times\R^k,\L)$. The \textbf{$k$-contact Euler--Lagrange equations for a holonomic map} $\psi\colon\R^k\to\oplus^k\T Q\times\R^k$ are
	\begin{equation}\label{eq:k-contact-Euler-Lagrange-section}
		\begin{dcases}
			i(\psi'_\alpha)\d\eta^\alpha_\L = \left( \d E_\L - (\Lie_{(\Reeb_\L)_\alpha}E_\L)\eta^\alpha_\L \right)\circ\psi\,,\\
			i(\psi'_\alpha)\eta^\alpha_\L = -E_\L\circ\psi\,.
		\end{dcases}
	\end{equation}
	The \textbf{$k$-contact Lagrangian equations for a $k$-vector field} $\bfX = (X_\alpha)\in\X^k(\oplus^k\T Q\times\R^k)$ are
	\begin{equation}\label{eq:k-contact-Lagrangian-equations}
		\begin{dcases}
			i(X_\alpha)\d\eta^\alpha_\L = \d E_\L - (\Lie_{(\Reeb_\L)_\alpha}E_\L)\eta^\alpha_\L \,,\\
			i(X_\alpha)\eta^\alpha_\L = -E_\L\,.
		\end{dcases}
	\end{equation}
	A $k$-vector field $\bfX$ solution to the $k$-contact Lagrangian equations \eqref{eq:k-contact-Lagrangian-equations} is called a \textbf{$k$-contact Lagrangian $k$-vector field}.
\end{dfn}
The following proposition ensures the existence of solutions to equations \ref{eq:k-contact-Lagrangian-equations}.
\begin{prop}
	Consider a regular $k$-contact Lagrangian system $(\oplus^k\T Q\times\R^k,\L)$. Then, the $k$-contact Lagrangian equations \eqref{eq:k-contact-Lagrangian-equations} admit solutions. They are not unique if $k>1$.
\end{prop}
\begin{proof}
	The proof of this result is the same as that of Proposition \ref{prop:k-contact-HDW-have-solutions}.
\end{proof}
Taking natural coordinates $(q^i,v^i_\alpha,s^\alpha)$ in $\oplus^k\T Q\times\R^k$, equations \eqref{eq:k-contact-Euler-Lagrange-section} read
\begin{equation}\label{eq:k-contact-Euler-Lagrange-section-coordinates}
	\begin{dcases}
		\parder{}{t^\alpha}\left( \parder{\L}{v^i_\alpha} \circ\psi\right) = \left( \parder{\L}{q^i} + \parder{\L}{s^\alpha}\parder{\L}{v^i_\alpha} \right)\circ\psi\,,\\
		\parder{(s^\alpha\circ\psi)}{t^\alpha} = \L\circ\psi\,.
	\end{dcases}
\end{equation}
On the other hand, for a $k$-vector field $\bfX = (X_\alpha)\in\X^k(\oplus^k\T Q\times\R^k)$ with local expression
$$ X_\alpha = (X_\alpha)^i\parder{}{q^i} + (X_\alpha)^i_\beta\parder{}{v^i_\beta} + (X_\alpha)^\beta\parder{}{s^\beta}\,, $$
the $k$-contact Lagrangian equations \eqref{eq:k-contact-Lagrangian-equations} read
\begin{align}
	0 &= \left( (X_\alpha)^j - v_\alpha^j \right)\parderr{\L}{v^j_\alpha}{s^\beta}\,,\label{eq:k-contact-Lagrangian-1}\\
	0 &= \left( (X_\alpha)^j - v_\alpha^j \right)\parderr{\L}{v^i_\beta}{v^j_\alpha}\,,\label{eq:k-contact-Lagrangian-2}\\
	0 &= \left( (X_\alpha)^j - v_\alpha^j \right)\parderr{\L}{q^i}{v^j_\alpha} + \parder{\L}{q^i} - \parderr{\L}{s^\beta}{v^i_\alpha}(X_\alpha)^\beta \nonumber\\
	& \qquad\qquad\qquad\qquad\qquad - \parderr{\L}{q^j}{v^i_\alpha}(X_\alpha)^j - \parderr{\L}{v^j_\beta}{v^i_\alpha}(X_\alpha)^j_\beta + \parder{\L}{s^\alpha}\parder{\L}{v^i_\alpha} \,,\label{eq:k-contact-Lagrangian-3}\\
	0 &= \L + \parder{\L}{v^i_\alpha}\left( (X_\alpha)^i - v^i_\alpha \right) - (X_\alpha)^\alpha\,.\label{eq:k-contact-Lagrangian-4}
\end{align}
If the Lagrangian $\L$ is regular, equations \eqref{eq:k-contact-Lagrangian-2} lead to $(X_\alpha)^i = v^i_\alpha$, which are the {\sc sopde} conditions for the $k$-vector field $\bfX$. In this case, \eqref{eq:k-contact-Lagrangian-1} holds identically and \eqref{eq:k-contact-Lagrangian-3} and \eqref{eq:k-contact-Lagrangian-4} give
\begin{align}
	-\parder{\L}{q^i} + \parderr{\L}{s^\beta}{v^i_\alpha}(X_\alpha)^\beta + \parderr{\L}{q^j}{v^i_\alpha}v^j_\alpha + \parderr{\L}{v^j_\beta}{v^i_\alpha}(X_\alpha)^j_\beta &= \parder{\L}{s^\alpha}\parder{\L}{v^i_\alpha}\,,\label{eq:k-contact-Lagrangian-regular-1}\\
	(X_\alpha)^\alpha &= \L\,.\label{eq:k-contact-Lagrangian-regular-2}
\end{align}
Notice that, if the {\sc sopde} $\bfX$ is integrable, equations \eqref{eq:k-contact-Lagrangian-regular-1} and \eqref{eq:k-contact-Lagrangian-regular-2} are the Euler--Lagrange equations \eqref{eq:k-contact-Euler-Lagrange-section-coordinates} for its integral maps. Hence, we have proved the following proposition.
\begin{prop}
	If $\L\in\Cinfty(\oplus^k\T Q\times\R^k)$ is a regular Lagrangian, the corresponding Lagrangian $k$-vector fields $\bfX$ (solutions to the $k$-contact Lagrangian equations \eqref{eq:k-contact-Lagrangian-equations}) are {\sc sopde}s and if, in addition, $\bfX$ is integrable, its integral maps are solutions to the $k$-contact Euler--Lagrange field equations \eqref{eq:k-contact-Euler-Lagrange-section}.

	This {\sc sopde} $\bfX$ is called the \textbf{Euler--Lagrange $k$-vector field} associated to the Lagrangian $\L$.
\end{prop}

\begin{rmrk}\rm
	Notice that, in the Lagrangian formalism of dissipative field theories, the second equation in \eqref{eq:k-contact-Euler-Lagrange-section-coordinates} relates the variation of the coordinates $s^\alpha$ to the Lagrangian function $\L$.
\end{rmrk}

\begin{rmrk}\rm
	If the Lagrangian $\L$ is regular or hyperregular, the Legendre map $\F\L$ is a (local) diffeomorphism between $(\oplus^k\T Q\times\R^k,\eta^\alpha_\L)$ and $(\oplus^k\cT Q\times\R^k,\eta^\alpha)$, where $\F\L^\ast\eta^\alpha = \eta^\alpha_\L$. Moreover, there exists, at least locally, a function $H\in\Cinfty(\oplus^k\cT Q\times\R^k)$ such that $H = E_\L\circ\F\L^{-1}$. Then, we have the $k$-contact Hamiltonian system $(\oplus^k\cT Q\times\R^k,\eta^\alpha,H)$, for which $\F\L_\ast(\Reeb_\L)_\alpha = \Reeb_\alpha$. Hence, if $\mathbf{\Gamma}$ is an Euler--Lagrange $k$-vector field associated to the Lagrangian $\L$ in $\oplus^k\T Q\times\R^k$, we have that $\F\L_\ast\mathbf{\Gamma}=\bfX$ is a $k$-contact Hamiltonian $k$-vector field associated to $H$ in $\oplus^k\cT Q\times\R^k$, and conversely.
\end{rmrk}

\begin{rmrk}\rm
	If the Lagrangian $\L$ is not regular, equations \eqref{eq:k-contact-Euler-Lagrange-section} and \eqref{eq:k-contact-Lagrangian-equations} do not have solutions everywhere in $\oplus^k\T Q\times\R^k$ but, in the most favourable cases, they do have solutions in a submanifold $S\hookrightarrow\oplus^k\T Q\times\R^k$, which can be obtained by applying an appropiate constraint algorithm. However, solutions to \eqref{eq:k-contact-Lagrangian-equations} need not to be {\sc sopde}s. In these cases, the {\sc sopde} condition has to be imposed as an additional condition. In the next section we will study this case in more detail.
\end{rmrk}

\begin{rmrk}\rm
	In the particular case $k=1$, we obtain the contact Lagrangian formalism for mechanical systems with dissipation \cite{DeLeo2019,Gas2019}.
\end{rmrk}

\section{The singular case: \texorpdfstring{$k$}--precontact Lagrangian and Hamiltonian systems}
\label{sec:singular-case-k-precontact}

In the case of singular Lagrangians, most of the results and properties stated in the above sections do not hold.

In this case, for the Lagrangian formalism, the couple $(\oplus^k\T Q\times\R^k,\eta^\alpha_\L)$ is not a $k$-contact manifold, but a $k$-precontact one (see Remark \ref{rmrk:k-precontact-manifolds}, and hence the Reeb vector fields are not uniquely defined. Nevertheless, the $k$-contact Euler--Lagrange and Lagrangian equations \eqref{eq:k-contact-Euler-Lagrange-section} and \eqref{eq:k-contact-Lagrangian-equations} for the system $(\oplus^k\T Q\times\R^k,\eta^\alpha_\L,E_\L)$ are independent on the family of Reeb vector fields $\Reeb_\alpha$ used (as it is proved in \cite{DeLeo2019} for the case $k=1$). In any case, in the singular case, solutions to the $k$-contact Lagrangian are not necessarily {\sc sopde}s and this condition must be added to the $k$-contact Lagrangian equation \eqref{eq:k-contact-Lagrangian-equations}. Moreover, the field equations are not necessarily consistent everywhere on $\oplus^k\T Q\times\R^k$ and we have to implement a suitable constraint algorithm in order to find a submanifold $S_f\hookrightarrow\oplus^k\T Q\times\R^k$ (if it exists) where there are {\sc sopde} $k$-vector fields in $\oplus^k\T Q\times\R^k$, tangent to the submanifold $S_f$, which are solutions to equations \eqref{eq:k-contact-Lagrangian-equations} on $S_f$.

In order to state the Hamiltonian formalism for the singular case, we need to assume some minimal regularity conditions. So, following \cite{Got1979}, we define:

\begin{dfn}
	A singular Lagrangian $\L$ is said to be \textbf{almost-regular} if the following conditions hold:
	\begin{enumerate}[{\rm(1)}]
		\item The submanifold $\P = \F\L(\oplus^k\T Q\times\R^k)\subset\oplus^k\cT Q\times\R^k$ is closed.
		\item The Legendre map $\F\L$ is a submersion onto its image $\P$.
		\item For every $p\in\P$, the fibre $\F\L^{-1}(p)\subset\oplus^k\T Q\times\R^k$ is a connected submanifold.
	\end{enumerate}
\end{dfn}

Then, if $j_\P\colon\P\hookrightarrow\oplus^k\cT Q\times\R^k$ is the natural embedding and $\eta^\alpha_\P = j_\P^\ast\eta^\alpha\Omega^1(\P)$, we have that $(\P,\eta^\alpha_\P)$ is a $k$-precontact manifold (see Remark \ref{rmrk:k-precontact-manifolds}. Moreover, the Lagrangian energy $E_\L$ is $\F\L$-projectable and there is a unique $H_\P\in\Cinfty(\P)$ such that $E_\L = \F\L_0^\ast H_\P$, where $\F\L_0\colon\oplus^k\T Q\times\R^k\to\P$ is defined by $\F\L = j_\P\circ\F\L_0$. Hence, on the submanifold $\P$ there is a Hamiltonian formalism associated to the Lagrangian system, and the $k$-contact Hamilton--De Donde--Weyl equations for a $k$-vector field $\bfY = (Y_\alpha)\in\X^k(\P)$ are
\begin{equation}\label{eq:k-contact-HDW-fields-singular}
	\begin{dcases}
		i(Y_\alpha)\d\eta^\alpha_\P = \d H_\P - (\Lie_{\Reeb_\alpha}H_\P)\eta^\alpha_\P\,,\\
		i(Y_\alpha)\eta^\alpha_\P = -H_\P\,.
	\end{dcases}
\end{equation}
As in the Lagrangian formalism, equations \eqref{eq:k-contact-HDW-fields-singular} are not necessarily compatible everywhere in $\P$ and the constraint algorithm should also be implemented to find a submanifold $P_f\hookrightarrow\P$ (if possible) where there are $k$-vector fields tangent to $P_f$ which are solutions to equations \eqref{eq:k-contact-HDW-fields-singular} on $P_f$.

As a final remark, we will recall the guidelines of the constraint algorithm (see Chapter \ref{ch:constraint-algorithms} for more details). Let $(M,\eta^\alpha,H)$ be a $k$-precontact Hamiltonian system and consider the $k$-contact Hamiltonian field equations \eqref{eq:k-contact-HDW-fields}.

\begin{itemize}
	\item The first step consists in finding the \textbf{compatibility conditions}: Let $M_1$ be the subset of $M$ where a solution to \eqref{eq:k-contact-HDW-fields} exists, namely,
	$$ M_1 = \{ p\in M\ \vert\ \exists(Y_1,\dotsc,Y_k)\in\oplus^k\T_pM \mbox{ solution to \eqref{eq:k-contact-HDW-fields}} \}\,. $$
	If $M_1\hookrightarrow M$ is a submanifold, there exists a section of the natural projection $\tau^1_M\colon\oplus^k\T M\to M$ defined on $M_1$ solution to \eqref{eq:k-contact-HDW-fields}, but which may not be a $k$-vector field on $M_1$.
	\item Then we apply the \textbf{tangency condition}: define a new subset $M_2\subset M_1$ as
	$$ M_2 = \{ p\in M_1\ \vert\ \exists(Y_1,\dotsc,Y_k)\in\oplus^k\T_pM_1 \mbox{ solution to \eqref{eq:k-contact-HDW-fields}} \}\,. $$
	Assuming that $M_2\hookrightarrow M_1$ is a submanifold, then there exists a section of the projection $\tau^1_{M_1}\colon\oplus^k\T M_1\to M_1$ defined on $M_2$ solution to equations \eqref{eq:k-contact-HDW-fields} which may not be a $k$-vector field on $M_2$.

	Taking a basis of independent constraint functions $\{\zeta^I\}$ locally defining $M_1$, the constraints defining $M_2$ are given by
	$$ \restr{(\Lie_{Y_\alpha}\zeta^I)}{M_1} = 0\,. $$
	\item Iterating this procedure, we can obtain a sequence of \textbf{constraint submanifolds}
	$$
		\dotsb\hookrightarrow M_i\hookrightarrow\dotsb\hookrightarrow M_2\hookrightarrow M_1\hookrightarrow M\,.
	$$
	If this procedure stabilizes, that is, there exists a natural number $f\in\mathbb{N}$ such that $M_{f+1} = M_f$ and $\dim M_f>0$, we say that $M_f$ is the \textbf{final constraint submanifold},  where we can find solutions to equations \eqref{eq:k-contact-HDW-fields}. Notice that the $k$-vector field solution may not be unique and, in general, they are not integrable.
\end{itemize}

\section{Symmetries of \texorpdfstring{$k$}--contact Lagrangian systems}
\label{sec:k-contact-Lagrangian-symmetries}

In this section we are going to define and study different notions of symmetry of a $k$-contact Lagrangian system. There are many kinds of symmetries, depending on the structure that they preserve. The most important types of symmetries are those transformations that preserve the solutions of the system and those preserving its geometric structure (see \cite{Bua2015,Gra2002,Rom2007}). We are going to follow the same guidelines as in Section \ref{sec:k-contact-Hamiltonian-symmetries}. Hence, the following definitions and properties are adapted from the ones stated for $k$-contact Hamiltonian systems to the case of a regular $k$-contact Lagrangian system $(\oplus^k\T Q\times\R^k,\L)$. The ommited proofs are completely analogous to the ones in Section \ref{sec:k-contact-Hamiltonian-symmetries}.

We will begin by introducing the notion of dynamical symmetries for $k$-contact Lagrangian systems, namely, the transformations preserving the solutions of the system.
\begin{dfn}
	Consider a regular $k$-contact Lagrangian system $(\oplus^k\T Q\times\R^k,\L)$.
	\begin{itemize}
		\item A \textbf{Lagrangian dynamical symmetry} is a diffeomorphism $\Phi\colon\oplus^k\T Q\times\R^k\to\oplus^k\T Q\times\R^k$ such that, for every solution $\psi$ to the $k$-contact Euler--Lagrange equations \eqref{eq:k-contact-Euler-Lagrange-section}, $\Phi\circ\psi$ is also a solution.
		\item An \textbf{infinitesimal Lagrangian dynamical symmetry} is a vector field $Y\in\X(\oplus^k\T Q\times\R^k)$ such that its local flow is made of local Lagrangian dynamical symmetries.
	\end{itemize}
\end{dfn}

\begin{lem}
	Consider a diffeomorphism $\Phi\in\Diff(\oplus^k\T Q\times\R^k)$ and $k$-vector field $\bfX=(X_1,\dotsc,X_k)\in\X^k(\oplus^k\T Q\times\R^k)$. If a map $\psi$ is an integral section of $\bfX$, then $\Phi\circ\psi$ is an integral map of $\Phi_\ast\bfX = (\Phi_\ast X_\alpha)$. In particular, if $\bfX$ is integrable, so is $\Phi_\ast\bfX$.
\end{lem}

\begin{prop}
	Let $\Phi\in\Diff(\oplus^k\T Q\times\R^k)$ be a Lagrangian dynamical symmetry. Then, for every integrable $k$-vector field $\bfX\in\X^k(\oplus^k\T Q\times\R^k)$ solution to the $k$-contact Lagrangian equations \eqref{eq:k-contact-Lagrangian-equations}, $\Phi_\ast\bfX$ is also a solution to \eqref{eq:k-contact-Lagrangian-equations}.

	Conversely, if $\Phi\in\Diff(\oplus^k\T Q\times\R^k)$ transforms every solution $\bfX$ to the $k$-contact Lagrangian equations \eqref{eq:k-contact-Lagrangian-equations} into another solution, for every integral map $\psi$ of $\bfX$, we have that $\Phi\circ\psi$ is a solution to the $k$-contact Euler--Lagrange equations \eqref{eq:k-contact-Euler-Lagrange-section}.
\end{prop}

The second kind of symmetries we will be dealing with are $k$-contact symmetries for Lagrangian systems, that is, the transformations preserving the $k$-contact underlying structure.

\begin{dfn}
	Let $(\oplus^k\T Q\times\R^k,\L)$ be a regular $k$-contact Lagrangian system.
	\begin{itemize}
		\item A \textbf{Lagrangian $k$-contact symmetry} is a diffeomorphism $\Phi\in\Diff(\oplus^k\T Q\times\R^k)$ such that
		$$ \Phi^\ast\eta^\alpha_\L = \eta^\alpha_\L\ ,\quad \Phi^\ast E_\L = E_\L\,. $$
		\item An \textbf{infinitesimal Lagrangian $k$-contact symmetry} is a vector field $Y\in\X(\oplus^k\T Q\times\R^k)$ whose local flow is a Lagrangian $k$-contact symmetry; namely,
		$$ \Lie_Y\eta^\alpha_\L = 0\ ,\quad\Lie_YE_\L = 0\,. $$
	\end{itemize}
\end{dfn}

\begin{prop}\label{prop:Lagrangian-k-contact-symmetry-preserves-Reeb}
	Every (infinitesimal) Lagrangian $k$-contact symmetry preserves the Reeb vector fields, that is,
	$$ \Phi_\ast(\Reeb_\L)_\alpha = (\Reeb_\L)_\alpha \quad\mbox{(or $[Y,(\Reeb_\L)_\alpha = 0$).} $$
\end{prop}

As a consequence of this last result, we have the relation between these two types of symmetries.
\begin{prop}\label{prop:Lagrangian-k-contact-sym-implies-Lagrangian-dynamical-sym}
	(Infinitesimal) Lagrangian $k$-contact symmetries are (infinitesimal) Lagrangian dynamical symmetries.
\end{prop}

\section{Dissipation laws}
\label{sec:k-contact-Lagrangian-dissipation-laws}

\begin{dfn}
	Consider a map $\oplus^k\T Q\times\R^k\to\R^k$, $F = (F^1,\dotsc,F^k)$. Then, $F$ is said to satisfy
	\begin{itemize}
		\item the \textbf{dissipation law for sections} if, for every map $\psi$ solution to the $k$-contact Euler--Lagrange equations \eqref{eq:k-contact-Euler-Lagrange-section}, the divergence of $F\circ\psi = (F^\alpha\circ\psi)\colon\R^k\to\R^k$, defined as
		$$ \Div(F\circ\psi) = \parder{(F^\alpha\circ\psi)}{t^\alpha}\,, $$
		satisfies that
		\begin{equation}\label{eq:k-contact-Lagrangian-dissipation-law-sections}
			\Div(F\circ\psi) = -\left( \left(\Lie_{(\Reeb_\L)_\alpha}E_\L \right)F^\alpha \right)\circ\psi\,.
		\end{equation}
		\item the \textbf{dissipation law for $k$-vector fields} if, for every $k$-vector field $\bfX=(X_\alpha)$ solution to the $k$-contact Lagrangian equations \eqref{eq:k-contact-Lagrangian-equations}, we have that
		\begin{equation}\label{eq:k-contact-Lagrangian-dissipation-law-fields}
			\Lie_{X_\alpha}F^\alpha = -(\Lie_{(\Reeb_\L)_\alpha}E_\L)F^\alpha\,.
		\end{equation}
	\end{itemize}
\end{dfn}
These two concepts are partially related by the following proposition.
\begin{prop}\label{prop:relation-dissipation-laws-k-contact-Lagrangian}
	If $F = (F^\alpha)$ fulfills the dissipation law for sections \ref{eq:k-contact-Lagrangian-dissipation-law-sections}, for every integrable $k$-vector field $\bfX$ solution to the $k$-contact Lagrangian equations \eqref{eq:k-contact-Lagrangian-equations}, we have that equation \eqref{eq:k-contact-Lagrangian-dissipation-law-fields} holds for $\bfX$.

	Conversely, if \eqref{eq:k-contact-Lagrangian-dissipation-law-fields} holds for a $k$-vector fields $\bfX$, then \eqref{eq:k-contact-Lagrangian-dissipation-law-sections} holds for every integral map $\psi$ of $\bfX$.
\end{prop}

\begin{prop}
	Let $Y\in\X(\oplus^k\T Q\times\R^k)$ be an infinitesimal Lagrangian dynamical symmetry. Then, for every solution $\bfX = (X_\alpha)\in\X^k(\oplus^k\T Q\times\R^k)$ to the $k$-contact Lagrangian equations \eqref{eq:k-contact-Lagrangian-equations}, we have
	$$ i([Y, X_\alpha])\eta^\alpha_\L = 0\ ,\quad i([Y, X_\alpha])\d\eta^\alpha_\L = 0\,. $$
\end{prop}

To end this section, we are going to state the Dissipation Theorem for $k$-contact Lagrangian systems, which is the analogous of Theorem \ref{thm:k-contact-Hamiltonian-dissipation-theorem} to Lagrangian systems.

\begin{thm}[Dissipation theorem for $k$-contact Lagrangian systems]\label{thm:k-contact-Lagrangian-dissipation-theorem}
	Let $Y\in\X(\oplus^k\T Q\times\R^k)$ be an infinitesimal Lagrangian dynamical symmetry. Then, $F^\alpha = -i(Y)\eta^\alpha_\L$ satisfies the dissipation law for $k$-vector fields \eqref{eq:k-contact-Lagrangian-dissipation-law-fields}.
\end{thm}

\section{Symmetries of the Lagrangian function}
\label{sec:k-contact-symmetries-Lagrangian-function}

\begin{dfn}
	Given a diffeomorphism $\varphi\colon Q\to Q$, the diffeomorphism
	$$ \Phi = (\T^k\varphi,\Id_{\R^k})\colon\oplus^k\T Q\times\R^k\to\oplus^k\T Q\times\R^k\,, $$
	where $\T^k\varphi\colon\oplus^\T Q\to\oplus^k\T Q$ denotes the canonical lift of $\varphi$ to $\oplus^k\T Q$, is called the \textbf{canonical lift} of $\varphi$ to $\oplus^k\T Q\times\R^k$. Any transformation $\Phi$ of this type is called a \textbf{natural transformation} $\oplus^k\T Q\times\R^k$.
\end{dfn}

\begin{dfn}
	Given a vector field $Z\in\X(Q)$, we define its \textbf{complete lift} to $\oplus^k\T Q\times\R^k$ as the vector field $Z^\rmC\in\X(\oplus^k\T Q\times\R^k)$ whose local flow is the canonical lift of the local flow of $Z$ to $\oplus^k\T Q\times\R^k$. Any infinitesimal transformation of this kind is said to be a \textbf{natural infinitesimal transformation} of $\oplus^k\T Q\times\R^k$.
\end{dfn}

Let $(\oplus^k\T Q\times\R^k,\L)$ be a regular $k$-contact Lagrangian system. It is well known that the canonical $k$-tangent structure $(J^\alpha)$ and the Liouville vector field $\Delta$ in $\oplus^k\T Q$ are invariant under the action of canonical lifts of diffomorphisms and vector fields from $Q$ to $\oplus^k\T Q$. Then, it is easy to prove that the canonical structures $(J^\alpha)$ and $\Delta$ and the Reeb vector fields $(\Reeb_\L)_\alpha$ in $\oplus^k\T Q\times\R^k$ are also preserved by canonical lifts of diffeomorphisms and vector fields from $Q$ to $\oplus^k\T Q\times\R^k$.

Hence, we have the following relation between Lagrangian-preserving natural transformations and Lagrangian $k$-contact symmetries.

\begin{prop}
	If $\Phi\in\Diff(\oplus^k\T Q\times\R^k)$ (resp. $Y\in\X(\oplus^k\T Q\times\R^k)$) is a canonical lift to $\oplus^k\T Q\times\R^k$ of a diffeomorphism $\varphi\in\Diff(Q)$ (resp. of a vector field $Z\in\X(Q)$) that leaves the Lagrangian $\L$ invariant, then it is an (infinitesimal) Lagrangian $k$-contact symmetry, that is,
	$$ \Phi^\ast\eta^\alpha_\L = \eta^\alpha_\L\ ,\quad \Phi^\ast E_\L = E_\L\qquad \mbox{(resp. $\Lie_Y\eta^\alpha_\L = 0$, $\Lie_Y E_\L = 0$).} $$
	Hence, by Proposition \ref{prop:Lagrangian-k-contact-sym-implies-Lagrangian-dynamical-sym}, it is also an (infinitesimal) Lagrangian dynamical symmetry.
\end{prop}

As a direct consequence of the above result, we have the following {\it momentum dissipation theorem}.

\begin{prop}\label{prop:k-contact-symmetries-of-the-Lagrangian}
	If $\dparder{\L}{q^i} = 0$, then $\dparder{}{q^i}$ is an infinitesimal contact symmetry and its associated dissipation law is given by the ``momenta'' $\dparder{\L}{v^i_\alpha}$. That is, for every $k$-vector field $\bfX = (X_\alpha)$ solution to the $k$-contact Lagrangian equations \eqref{eq:k-contact-Lagrangian-equations}, we have
	$$ \Lie_{X_\alpha}\left(\parder{\L}{v^i_\alpha}\right) = -(\Lie_{(\Reeb_\L)_\alpha}E_\L)\parder{\L}{v^i_\alpha} = \parder{\L}{s^\alpha}\parder{\L}{v^i_\alpha}\,. $$
\end{prop}

	\chapter[Skinner--Rusk formalism for \texorpdfstring{$k$}--contact field theories]{Skinner--Rusk formalism for \phantom{n} \texorpdfstring{$k$}--contact field theories}
	\label{ch:Skinner-Rusk-k-contact}


In this chapter we develop the Skinner--Rusk formalism \cite{Kam1982,Ski1983} for $k$-contact systems. This formalism is particularly interesting when dealing with singular Lagrangians. One of the reasons for this is that it includes the second-order condition even if the Lagrangian function is singular, in contrast with the Lagrangian formalism, where the holonomy condition must be imposed as an additional requirement. Over the years, this formalism developed by R. Skinner and R. Rusk has been generalized to many different types of systems (nonautonomous, vakonomic and holonomic, control, first-order field theories and higher order mechanical systems and field theories) \cite{Bar2007,Bar2008,Cam2009,Can2002,Col2010,Cor2002,DeLeo2003,Ech2004,Gas2018,Gra2005,Gra1991,Pri2011,Pri2012,Pri2015,Rey2005,Rey2012,Vit2010}.

In Chapter \ref{ch:Skinner-Rusk-contact}, we already developed this unified formalism in the case of contact systems (see \cite{DeLeo2020} for more details). In order to generalize the Skinner--Rusk for contact systems to $k$-contact systems, we will follow the work done in \cite{Rey2005} and \cite{Rey2012}, where the authors generalized the Skinner--Rusk formalism to $k$-symplectic and $k$-cosymplectic field theories.

In Section \ref{sec:extended-Pontryagin-bundle-k-contact} we present the extended Pontryagin bundle and its canonical geometric structures: the coupling function, the canonical 1-forms, the canonical 2-forms and the contact 1-forms. We also define the notion of {\sc sopde} in the extended Pontryagin bundle and state the existence of Reeb vector fields. Finally, we define the Hamiltonian function associated to a Lagrangian function. With this geometric framework, in Section \ref{sec:k-contact-dynamical-equations} we are able to state the Lagrangian--Hamiltonian problem and apply the constraint algorithm described in Section \ref{sec:singular-case-k-precontact} to solve it. To end this chapter, in Section \ref{sec:SR-k-contact-recovering} we show that we can recover both the Lagrangian and Hamiltonian formalisms from the Skinner--Rusk formalism developed in this chapter. This chapter is based in \cite{Gra2021}.

\section{The extended Pontryagin bundle: \texorpdfstring{$k$}--precontact canonical structure}
\label{sec:extended-Pontryagin-bundle-k-contact}

Consider a $k$-contact field theory with configuration manifold $Q\times\R^k$ with $\dim Q = n$ and coordinates $(q^i,s^\alpha)$. Consider the bundles $\oplus^k\T Q\times\R^k$ and $\oplus^k\cT Q\times\R^k$ endowed with canonical coordinates $(q^i,v^i_\alpha,s^\alpha)$ and $(q^i,p_i^\alpha,s^\alpha)$ respectively. In these bundles, we have the following natural projections
\begin{align*}
	\tau_1 &\colon \oplus^k\T Q\times\R^k \to \oplus^k\T Q\ , & \tau_0&\colon \oplus^k\T Q\times\R^k \to Q\times\R^k\,,\\
	\pi_1 &\colon \oplus^k\cT Q\times\R^k \to \oplus^k\cT Q\ , & \pi_0&\colon \oplus^k\cT Q\times\R^k \to Q\times\R^k \ .
\end{align*}
Denoting by $\d s^\alpha$ the volume forms of the different copies of $\R$ and its pull-backs to all the manifolds by the corresponding projections, we can consider the canonical forms $\theta\in\Omega^1(\cT Q)$ and $\omega\in\Omega^2(\cT Q)$ with local expressions $\theta = p_i\d q^i$ and $\omega = \d q^i\wedge\d p_i$ in $\cT Q$. We will denote by $\theta^\alpha$ and $\omega^\alpha$ their pull-backs to $\oplus^k\cT Q$ and $\oplus^k\cT Q\times\R^k$, which have coordinate expressions
$$ \theta^\alpha = p_i^\alpha\d q^i\quad,\quad\omega^\alpha = \d q^i\wedge\d p_i^\alpha\,. $$

\begin{dfn}
	The \textbf{extended Pontryagin bundle} or \textbf{extended unified bundle} of a manifold $Q$ is the bundle $\W = \oplus^k\T Q\times_Q\oplus^k\cT Q\times\R^k$, and it is equipped with the natural projections
	\begin{align*}
		\rho_1&\colon \W\to\oplus^k\T Q\times\R^k\ , & 
		\rho_2&\colon \W\to\oplus^k\cT Q\times\R^k\ , \\
		\rho_0&\colon \W\to Q\times\R^k\ , & 
		s^\alpha&\colon\W\to\R \ .
	\end{align*}
\end{dfn}

The following diagram summarizes the projections described above:
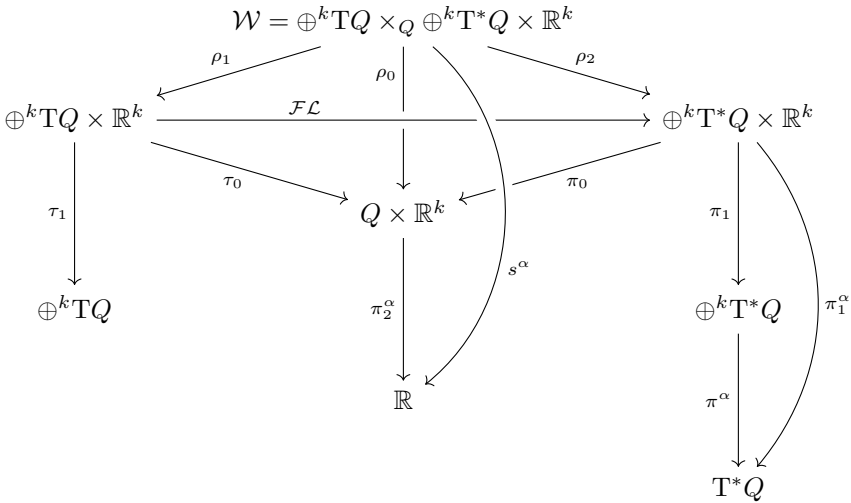
\begin{center}
\begin{tikzcd}
	& \W = \oplus^k\T Q\times_Q\oplus^k\cT Q\times\R^k \arrow[dl, swap, "\rho_1"] \arrow[dr, "\rho_2"] \arrow[dd, swap, pos=0.2, "\rho_0"]\\
\oplus^k\T Q\times\R^k \arrow[rr, crossing over, pos=0.3, "\F\L"] \arrow[dr, swap, "\tau_0"] \arrow[dd, swap, "\tau_1"] & & \oplus^k\cT Q\times\R^k \arrow[dl, "\pi_0"] \arrow[dd, swap, "\pi_1"] \arrow[dddd, bend left=40, "\pi_1^\alpha"] \\
	& Q\times\R^k \arrow[dd, swap, "\pi_2^\alpha"] \\
	\oplus^k\T Q & & \oplus^k\cT Q \arrow[dd, swap, "\pi^\alpha"] \\
	& \R  \arrow[uuuu, crossing over, bend right=50, leftarrow, swap, pos=0.4, "s^\alpha"]\\
	& & \cT Q
\end{tikzcd}
\end{center}

The extended Pontryagin bundle of a manifold $Q$ endowed with coordinates $(q^i)$ has natural coordinates $(q^i,v^i_\alpha,p_i^\alpha,s^\alpha)$.


\begin{dfn}
	A map $\psi\colon\R^k\to\W$ is said to be \textbf{holonomic} if $\rho_1\circ\psi\colon\R^k\to\oplus^k\T Q\times\R^k$ is holonomic (see Definition \ref{dfn:k-contact-holonomic-section}). A $k$-vector field $\bfZ\in\X^k(\W)$ is a \textbf{second-order partial differential equation} in $\W$ ({\sc sopde} for short) if its integral sections are holonomic in $\W$.
\end{dfn}

A holonomic map $\psi\colon\R^k\to\W$ has local expression
$$ \psi(t) = \left( \psi^i(t),\parder{\psi^i}{t^\alpha}(t),\psi^\alpha_i(t),\psi^\alpha(t) \right)\,, $$
while a {\sc sopde} $k$-vector field $\bfZ = (Z_1,\dotsc,Z_k)\in\X^k(\W)$ has local expression
$$ Z_\alpha = v^i_\alpha\parder{}{q^i} + (Z_\alpha)^i_\beta\parder{}{v^i_\beta} + (Z_\alpha)^\beta_i\parder{}{p_i^\beta} + (Z_\alpha)^\beta\parder{}{s^\beta}\,. $$

The extended Pontryagin bundle has the following canonical structures:

\begin{dfn}
	Let $\W = \oplus^k\T Q\times_Q\oplus^k\cT Q\times\R^k$ be the extended Pontryagin bundle of a manifold $Q$. 
	\begin{itemize}
		\item The \textbf{coupling function} in $\W$ is the map $\mathcal{C}\colon\W\to\R$ defined as
		$$ \mathcal{C}(v_{1q},\dotsc,v_{kq},\vartheta_q^1,\dotsc,\vartheta_q^1, s^\alpha) = \vartheta_q^\alpha(v_{\alpha q})\,. $$
		\item The \textbf{canonical 1-forms} are the forms $\Theta^\alpha = \rho_2^\ast\:\theta^\alpha\in\Omega^1(\W)$.
		\item The \textbf{canonical 2-forms} are $\Omega^\alpha = \rho_2^\ast\:\omega^\alpha = -\d\Theta^\alpha\in\Omega^2(\W)$.
		\item The \textbf{contact 1-forms} are the forms $\eta^\alpha = \d s^\alpha - \Theta^\alpha\in\Omega^1(\W)$. Notice that $\d\eta^\alpha = \Omega^\alpha$.
	\end{itemize}
\end{dfn}

Taking canonical coordinates $(q^i,v^i_\alpha,p_i^\alpha, s^\alpha)$ in $\W$, these natural structures have local expressions
\begin{equation*}
	\Theta^\alpha = p_i^\alpha\d q^i\ ,\quad\Omega^\alpha = \d q^i\wedge\d p_i^\alpha\ ,\quad\eta^\alpha = \d s^\alpha - p_i^\alpha\d q^i\,.
\end{equation*}

The family $(\W,\eta^\alpha)$ is a $k$-precontact manifold.

\begin{prop} There exists a family of Reeb vector fields $\Reeb_1,\dotsc,\Reeb_k\in\X(\W)$ satisfying the conditions
$$
	\begin{cases}
		i(\Reeb_\alpha)\d\eta^\beta = 0\,,\\
		i(\Reeb_\alpha)\eta^\beta = \delta_\alpha^\beta\,.
	\end{cases}
$$
\end{prop}

\begin{rmrk}\rm
	Notice that, as the manifold $\W$ is $k$-precontact, the family $(\Reeb_\alpha)$ of Reeb vector fields is not unique. In fact, when written in natural coordinates, coordinates, $\Reeb_\alpha$ are
	$$
		\Reeb_\alpha = \parder{}{s^\alpha} + (\Reeb_\alpha)^i_\beta\parder{}{v^i_\beta}\,,
	$$
	where $(\Reeb_\alpha)^i_\beta\in\Cinfty(\W)$ are arbitrary functions in $\W$.
\end{rmrk}

\begin{dfn}
	Consider a Lagrangian function $L\colon\oplus^k\T Q\times\R^k\to\R$ and let $\L=\rho_1^\ast L\colon\W\to\R$. We define the \textbf{Hamiltonian function} associated to $L$ by
	\begin{equation}\label{eq:k-contact-SR-Hamiltonian-function}
		\H = \mathcal{C} - \L = p_i^\alpha v^i_\alpha - L(q^j,v^j_\alpha,s^\alpha)\in\Cinfty(\W)\,.
	\end{equation}
\end{dfn}

\begin{rmrk}\rm
	As the manifold $\W$ along with the contact 1-forms $\eta^\alpha$ is a $k$-precontact manifold, we have that $(\W,\eta^\alpha,\H)$ is a $k$-precontact Hamiltonian system.
\end{rmrk}

\section{\texorpdfstring{$k$}--contact dynamical equations}
\label{sec:k-contact-dynamical-equations}

\begin{dfn}
	Consider the $k$-precontact Hamiltonian system $(\W,\eta,\H)$. Its associated \textbf{Lagrangian--Hamiltonian problem} consists in finding the integral sections $\psi\colon\R^k\to\W$ of a $k$-vector field $\bfZ = (Z_1,\dotsc,Z_k)\in\X^k(\W)$ such that
	\begin{equation}\label{eq:k-contact-fields-SR}
		\begin{cases}
			i(Z_\alpha)\d\eta^\alpha = \d\mathcal{H} - (\Lie_{\Reeb_\alpha}\mathcal{H})\eta^\alpha\,,\\
			i(Z_\alpha)\eta^\alpha = -\mathcal{H}\,,
		\end{cases}
	\end{equation}
	or, what is equivalent,
	$$ \begin{cases}
		\Lie_{Z_\alpha}\eta^\alpha = -(\Lie_{\Reeb_\alpha}\mathcal{H})\eta^\alpha\,,\\
		i(Z_\alpha)\eta^\alpha = -\mathcal{H}\,.
	\end{cases} $$
\end{dfn}

Given that $(\W,\eta^\alpha,\H)$ is a $k$-precontact Hamiltonian system, equations \eqref{eq:k-contact-fields-SR} are not consistent everywhere in $\W$. In this case, we need to use the constraint algorithm described in Section \ref{sec:singular-case-k-precontact} 
in order to find (when possible) a final constraint submanifold of $\W$ 
where the existence of consistent solutions to equations \eqref{eq:k-contact-fields-SR} is ensured.

Taking canonical coordinates $(q^i,v^i_\alpha,p_i^\alpha,s^\alpha)$ in $\W$, the local expression of a $k$-vector field $\bfZ = (Z_1,\dotsc,Z_k)$ in $\W$ is
$$ Z_\alpha = (Z_\alpha)^i\parder{}{q^i} + (Z_\alpha)^i_\beta\parder{}{v^i_\beta} + (Z_\alpha)^\beta_i\parder{}{p_i^\beta} + (Z_\alpha)^\beta\parder{}{s^\beta}\,. $$
Hence, we have
\begin{align*}
	i(Z_\alpha)\d\eta^\alpha &= (Z_\alpha)^i\d p_i^\alpha - (Z_\alpha)_i^\alpha\d q^i\,,\\
	i(Z_\alpha)\eta^\alpha &= (Z_\alpha)^\alpha - p_i^\alpha (Z_\alpha)^i\,.
\end{align*}
and,
\begin{align*}
	\d\H &= v_\alpha^i\d p_i^\alpha + \left( p_i^\alpha - \parder{\L}{v_\alpha^i} \right)\d v_\alpha^i - \parder{\L}{q^i}\d q^i - \parder{\L}{s^\alpha}\d s^\alpha\,,\\
	(\Reeb_\alpha(\H))\eta^\alpha &= -\parder{\L}{s^\alpha}(\d s^\alpha - p_i^\alpha\d q^i)\,.
\end{align*}
Taking all this into account, the second equation \eqref{eq:k-contact-fields-SR} gives
\begin{equation}\label{eq:k-contact-fields-SR-second}
	(Z_\alpha)^\alpha = \left((Z_\alpha)^i - v_\alpha^i\right)p_i^\alpha + L\circ\rho_1\,,
\end{equation}
while the first equation \eqref{eq:k-contact-fields-SR} gives
\begin{align}
	(Z_\alpha)^i &= v_\alpha^i & &(\mbox{coefficients in }\d p_i^\alpha)\,,\label{eq:k-contact-fields-SR-first-one}\\
	p_i^\alpha &= \parder{\L}{v_\alpha^i} = \parder{L}{v_\alpha^i}\circ\rho_1 & &(\mbox{coefficients in }\d v_\alpha^i)\,,\label{eq:k-contact-fields-SR-first-two}\\
	(Z_\alpha)_i^\alpha &= \parder{L}{q^i}\circ\rho_1 + p_i^\alpha\left(\parder{L}{s^\alpha}\circ\rho_1\right) & &(\mbox{coefficients in }\d q^i)\,.\label{eq:k-contact-fields-SR-first-three}
\end{align}
From these equations we have that
\begin{itemize}
	\item Conditions \eqref{eq:k-contact-fields-SR-first-two} and \eqref{eq:k-contact-fields-SR-first-one} imply that $(Z_\alpha)^\alpha = L\circ\rho_1\,.$
	\item Equations \eqref{eq:k-contact-fields-SR-first-one} are the {\sc sopde} conditions for the $k$-vector field $\bfZ$. Hence, as usual, we obtain straightforwardly the \textsc{sopde} condition from the Skinner--Rusk formalism \cite{Ski1983}. This is an important difference with the Lagrangian formalism, where we need to impose the second order condition in the case of singular Lagrangians.
	\item The algebraic equations \eqref{eq:k-contact-fields-SR-first-two} are consistency conditions which define a {\sl first constraint submanifold} $\W_1\hookrightarrow\W$. In fact, $\W_1$ is essentially the graph of $\F\L$:
	$$ \W_1 = \left\{ (v_q,\F\L(v_q))\in\W\ \vert\  v_q\in\oplus^k\T Q\times\R^k \right\}\,.$$
	Notice that this means that the Skinner--Rusk formalism includes the definition of the Legendre map as a consequence of the constraint algorithm.
\end{itemize}

With all this in mind, if a $k$-vector field $\bfZ = (Z_1,\dotsc,Z_k)$ is a solution to equations \eqref{eq:k-contact-fields-SR} then, $Z_\alpha$ has coordinate expression
$$ Z_\alpha = v^i_\alpha\parder{}{q^i} + (Z_\alpha)^i_\beta\parder{}{v^i_\beta} + (Z_\alpha)^\beta_i\parder{}{p_i^\beta} + (Z_\alpha)^\beta\parder{}{s^\beta}\qquad\mbox{(on $\W_1$)}\,, $$
where we have the restrictions
$$
	\begin{dcases}
		(Z_\alpha)^\alpha = \L\,,\\
		(Z_\alpha)_i^\alpha = \parder{\L}{q^i} + p_i^\alpha\parder{\L}{s^\alpha}\,.
	\end{dcases}
$$

\begin{rmrk}\rm
	It is important to point out that the $k$-vector field $\bfZ$ does not depend on the arbitrary functions $(\Reeb_\alpha)^i_\beta$, that is, on the family of Reeb vector fields $\Reeb_\alpha$ of $\W$ chosen.
\end{rmrk}

Now the constraint algorithm continues by imposing the tangency of $\bfZ$ to the first constraint submanifold $\W_1$. We denote by $\xi_j^\beta$ the constraint functions defining $\W_1$,
$$ \xi_j^\beta = p_j^\beta - \parder{\L}{v_\beta^j}\,. $$
Imposing the conditions $X_\alpha(\xi_j^\beta)=0$ we get
\begin{multline}\label{eq:k-contact-SR-tangency-condition}
	0 = X_\alpha(\xi_j^\beta) = X_\alpha\bigg(p_j^\beta - \parder{\L}{v_\beta^j}\bigg) \\
	= (Z_\alpha)_j^\beta - \parder{^2\L}{q^i\partial v_\beta^j}v^i_\alpha - \parder{^2\L}{v_\gamma^i\partial v_\beta^j}(Z_\alpha)_\gamma^i - \parder{^2\L}{s^\gamma\partial v_\beta^j}(Z_\alpha)^\gamma \qquad\mbox{(on $\W_1$)}\,.
\end{multline}

\begin{rmrk}\rm
	As we are imposing the tangency of the solution to the submanifold $\W_1$, sometimes it may be interesting also to demand the tangency of the Reeb vector fields to the first constraint submanifold $\W_1$. The Reeb vector fields $\Reeb_\alpha$ are tangent to $\W_1$ if, and only if,
	$$
		\parder{^2\L}{s^\alpha\partial v_\gamma^j} + (\Reeb_\alpha)_\beta^i\parder{^2\L}{v_\beta^i\partial v_\gamma^j} = 0\,,
	$$
	which are conditions for the functions $(\Reeb_\alpha)_\beta^i$.
	(It is important to remark that this system of equations might be incompatible).
\end{rmrk}

Notice that, in general, equations \eqref{eq:k-contact-fields-SR} do not have a unique solution. In fact, the solutions to \eqref{eq:k-contact-fields-SR} have the form
$$ 
(Z_1,\dotsc,Z_k) + (\ker\Omega^\sharp\cap\ker\eta^\sharp)\ , 
$$
where $\bfZ=(Z_1,\dotsc,Z_k)$ is a particular solution, $\Omega^\sharp$ is the morphism given by
\begin{align*}
	\Omega^\sharp \colon \oplus^k\T\W &\longrightarrow \cT\W\\
	(Z_1,\dotsc,Z_k) &\longmapsto \Omega^\sharp(Z_1,\dotsc,Z_k) = i(Z_\alpha)\d\eta^\alpha.
\end{align*}
and $\eta^\sharp$ is defined as
$$ \eta^\sharp(Z_1,\dotsc,Z_k) = \eta^\alpha(Z_\alpha)\,. $$

At this point we have to distinguish two cases:
\begin{itemize}
	\item If $\L$ is a regular Lagrangian, equations \eqref{eq:k-contact-SR-tangency-condition} allow us to compute the functions $(Z_\alpha)_\gamma^i$. Notice that, although we can ensure the existence of solutions, we do not have uniqueness of solutions to equations \eqref{eq:k-contact-fields-SR}.
	\item If the Lagrangian $\L$ is singular, equations \eqref{eq:k-contact-SR-tangency-condition} establish some relations among the functions $F_{\alpha\gamma}^i$. In addition, some new constraints may appear defining a new constraint submanifold $\W_2\hookrightarrow\W_1\hookrightarrow\W$. We must now implement the constraint algorithm described in Section \ref{sec:singular-case-k-precontact} in order to obtain a constraint submanifold (if it exists) where we can assure the existence of solutions tangent to this submanifold.
\end{itemize}

\section{Recovering the Lagrangian and the Hamiltonian formalisms}
\label{sec:SR-k-contact-recovering}

Consider the restriction of the projections $\rho_1\colon\W\to\oplus^k\T Q\times\R^k$, $\rho_2\colon\W\to\oplus^k\cT Q\times\R^k$
restricted to $\W_1\subset\W$,
\begin{equation*}
	\rho_1^0\colon\W_1\to\oplus^k\T Q\times\R^k\ ,\qquad \rho_2^0\colon\W_1\to\oplus^k\cT Q\times\R^k\,.
\end{equation*}
Since $\W_1$ is the graph of the Legendre transformation $\F\L$, it is clear that the projection $\rho_1^0$ is really a diffeomorphism.

Consider an integrable $k$-vector field $\bfZ=(Z_1,\dotsc,Z_k)$ solution to equations \eqref{eq:k-contact-fields-SR}. Every integral section $\psi\colon\R^k\to\W$, given by
$\psi(t) = (\psi^i(t),\psi^i_\alpha(t), \psi_i^\alpha(t),\psi^\alpha(t))$,
is of the form
$$ \psi = (\psi_L,\psi_H)\,, $$
with $\psi_L = \rho_1\circ\psi\colon\R^k\to\oplus^k\T Q\times\R^k$, and if $\psi$ takes values in $\W_1$, we also have that $\psi_H = \F\L\circ\psi_L$:
\begin{align*}
	\psi_H(t) &= (\rho_2\circ\psi)(t)\\
	&= (\psi^i(t),\psi_i^\alpha(t),\psi^\alpha(t)) \\
	&= \left( \psi^i(t), \parder{\L}{v_\alpha^i}(\psi_L(t)), \psi^\alpha(t) \right) \\
	&= (\F\L\circ\psi_L)(t)\,,
\end{align*}
where we have used \eqref{eq:k-contact-fields-SR-first-two}. Notice that in this way, we can always project from the Skinner--Rusk formalism onto the Lagrangian or the Hamiltonian formalisms by restricting to the first or second factor of the extended Pontryagin bundle $\W$. 
In particular, relations \eqref{eq:k-contact-fields-SR-first-two} define the image of the Legendre transformation $\F\L(\oplus^k\T Q\times\R^k)\subset\oplus^k\cT Q\times\R^k$. These relations are called \textbf{primary Hamiltonian constraints}.

The following theorem establishes how we can recover the Euler--Lagrange equations \eqref{eq:k-contact-Euler-Lagrange-section} from the Skinner--Rusk formalism.

\begin{thm}\label{thm:k-contact-SR-T0}
	Consider an integrable $k$-vector field $\bfZ=(Z_1,\dotsc,Z_k)$ in $\W$, solution to equations \eqref{eq:k-contact-fields-SR}. Let $\psi\colon\R^k\to\W_1\subset\W$ be an integral section of $\bfZ$ given by $\psi = (\psi_L,\psi_H)$, with $\psi_H = \F\L\circ\psi_L$. Then, $\psi_L$ is the first prolongation of the projected section $\phi = \tau_0\circ\rho_1^0\circ\psi\colon\R^k\to Q\times\R^k$, and $\phi$ is a solution to the $k$-contact Euler--Lagrange equations \eqref{eq:k-contact-Euler-Lagrange-section}.
\end{thm}
\begin{proof}
	Consider an integral section $\psi(t) = \left( \psi^i(t), \psi_\alpha^i(t), \psi_i^\alpha(t), \psi^\alpha(t) \right)$ of the $k$-vector field $\bfZ$. Then, we have that
	\begin{equation}\label{eq:k-contact-SR-sections-field}
		Z_\alpha(\psi(t)) = \parder{\psi^i}{t^\alpha}(t)\left.\parder{}{q^i}\right\vert_{\psi(t)} + \parder{\psi^i_\beta}{t^\alpha}(t)\left.\parder{}{v_\beta^i}\right\vert_{\psi(t)} + \parder{\psi_i^\beta}{t^\alpha}(t)\left.\parder{}{p_i^\beta}\right\vert_{\psi(t)} + \parder{\psi^\beta}{t^\alpha}(t)\left.\parder{}{s^\beta}\right\vert_{\psi(t)}\,.
	\end{equation}
	Now, from \eqref{eq:k-contact-fields-SR-second}, \eqref{eq:k-contact-fields-SR-first-one}, \eqref{eq:k-contact-fields-SR-first-two} and \eqref{eq:k-contact-SR-sections-field} we get
	\begin{align}
		\parder{\psi^\alpha}{t^\alpha}(t) &= (L\circ\rho_1)(\psi(t)) = L(\psi_L(t))\,,\label{eq:l-contact-SR-one-sections-L}\\
		\psi_i^\alpha(t) &= p_i^\alpha(\psi(t)) = \left( \parder{L}{v_\alpha^i}\circ\rho_1 \right)(\psi(t)) = \parder{L}{v_\alpha^i}(\psi_L(t))\,,\\
		\psi_\alpha^i(t) &= v_\alpha^i(\psi(t)) = (Z_\alpha^i)(\psi(t)) = \parder{\psi^i}{t^\alpha}(t)\,,\label{eq:k-contact-SR-three-sections-L}\\
		\parder{\psi_i^\beta}{t^\alpha}(t) &= (Z_\alpha)^\beta_i(\psi(t))\,.\label{eq:k-contact-SR-four-sections-L}
	\end{align}
	Using the conditions above and equation \eqref{eq:k-contact-fields-SR-first-three}, we obtain
	$$ \parder{\psi^\alpha_i}{t^\alpha}(t) = \left( \parder{L}{q^i}\circ\rho_1 \right)(\psi(t)) + p_i^\alpha(\psi(t))\left( \parder{L}{s^\alpha}\circ\rho_1 \right)(\psi(t))\,, $$
	and hence,
	$$ \parder{}{t^\alpha}\parder{L}{v^i_\alpha}(\psi_L(t)) = \parder{L}{q^i}(\psi_L(t)) + \parder{L}{v_i^\alpha}(\psi_L(t))\parder{L}{s^\alpha}(\psi_L(t))\,, $$
	$$ \psi_L = \left(\psi^i,\parder{\psi^i}{t^\alpha},\psi^\alpha\right)\,. $$
	It is clear that $\psi_L$ is the first prolongation of the map $\phi = \tau_0\circ\rho_1\circ\psi\colon\R^k\to Q\times\R^k$ given by $\phi = (\psi^i,\psi^\alpha)$, which is a solution to the $k$-contact Euler--Lagrange field equations \eqref{eq:k-contact-Euler-Lagrange-section-coordinates}, which is the expression in coordinates of equation \eqref{eq:k-contact-Euler-Lagrange-section}.
\end{proof}

The following theorem shows how to recover the $k$-contact Hamilton field equations \eqref{eq:k-contact-HDW-Darboux-coordinates} from the Skinner--Rusk formalism.

\begin{thm}\label{thm:k-contact-SR-T1}
	Let $\bfZ=(Z_1,\dotsc,Z_k)$ be an integrable $k$-vector field in $\W$ solution to equations \eqref{eq:k-contact-fields-SR} and $\psi\colon\R^k\to\W_1\subset\W$ be an integral section of $\bfZ$ given by $\psi = (\psi_L,\psi_H)$, with $\psi_H = \F\L\circ\psi_L$. If the Lagrangian $L$ is regular, $\psi_H$ is a solution to the $k$-contact Hamilton field equations \eqref{eq:k-contact-HDW-Darboux-coordinates}, where the Hamiltonian function $H$ is given by $E_L = H\circ\F L$.
\end{thm}
\begin{proof}
	We have that $L$ is a regular Lagrangian and hence, $\F\L$ is a local diffeomorphism. Then, for every point $p\in\oplus^k\T Q\times\R^k$, there exists an open subset $U\subset\oplus^k\T Q\times\R^k$ containing the point $p$ such that the restriction $\restr{\F\L}{U}\colon U\to\F\L(U)$ is a diffeomorphism. Using this, we can define a function $\widetilde H = \restr{E_\L}{U}\circ(\restr{\F\L}{U})^{-1}$. From now on, we will consider that the maps $E_\L$ and $\F\L$ are restricted to the open set $U$.
	Now, using that $E_\L = \widetilde H\circ\F\L$, it is clear that
	\begin{equation}\label{eq:k-contact-SR-Hamilton-sections-U}
			\parder{\widetilde H}{p^\alpha_i}\circ\F\L = v_\alpha^i\ ,\qquad
			\parder{\widetilde H}{q^i} \circ\F\L = -\parder{L}{q^i}\,.
	\end{equation}
	We consider now the subset $V = \psi_L^{-1}(U)\subset\R^k$ and restrict $\psi$ to $V$, so we have
	\begin{equation*}
		\begin{array}{rcl}
			\restr{\psi}{V}\colon V\subset\R^k & \to & U \oplus_{\R^k}\F\L(U)\\
			t & \mapsto & (\psi_L(t),\psi_H(t)) = (\psi_L(t),(\F\L\circ\psi_L)(t))
		\end{array}
	\end{equation*}
	Taking into account \eqref{eq:k-contact-fields-SR-first-three}, \eqref{eq:k-contact-SR-three-sections-L}, \eqref{eq:k-contact-SR-four-sections-L} and \eqref{eq:k-contact-SR-Hamilton-sections-U},
	\begin{align*}
		\parder{\widetilde H}{p_i^\alpha}(\psi_H(t)) &= \left( \parder{\widetilde H}{p_i^\alpha}\circ\F\L \right)(\psi_L(t)) = v_\alpha^i(\psi_L(t)) = \parder{\psi^i}{t^\alpha}(t)\,,\\
		\parder{\widetilde H}{q^i}(\psi_H(t)) &= \left( \parder{\widetilde H}{q^i}\circ\F\L \right)(\psi_L(t)) = -\parder{L}{q^i}(\psi_L(t)) = -\left( \parder{L}{q^i}\circ\rho_1 \right)(\psi(t))\\
		&= \left( p_i^\alpha\left( \parder{L}{s^\alpha}\circ\rho_1 \right) - (Z_\alpha)^\alpha_i \right)(\psi(t)) \\
		&= p_i^\alpha\left( \parder{L}{s^\alpha}\circ\rho_1 \right)(\psi(t)) - (Z_\alpha)^\alpha_i(\psi(t))\\
		&= p_i^\alpha \parder{L}{s^\alpha}(\psi_L(t)) - \parder{\psi^\alpha_i}{t^\alpha}(t) = -p_i^\alpha\parder{E_\L}{s^\alpha}(\psi_L(t)) - \parder{\psi^\alpha_i}{t^\alpha}(t)\\
		&= -p_i^\alpha\parder{(\widetilde H\circ\F\L)}{s^\alpha}(\psi_L(t)) - \parder{\psi^\alpha_i}{t^\alpha}(t) = -p_i^\alpha\parder{\widetilde H}{s^\alpha}(\psi_H(t)) - \parder{\psi^\alpha_i}{t^\alpha}(t)\,,
	\end{align*}
	and then
	$$
		\begin{dcases}
			\parder{\psi^i}{t^\alpha}(t) = \parder{\widetilde H}{p_i^\alpha}(\psi_H(t))\,,\\
			\parder{\psi^\alpha_i}{t^\alpha}(t) = -\left( \parder{\widetilde H}{q^i} + p_i^\alpha\parder{\widetilde H}{s^\alpha} \right)(\psi_H(t))\,.
		\end{dcases}
	$$
	Finally, considering equation \eqref{eq:l-contact-SR-one-sections-L}, we deduce that
	$$
		\parder{\psi^\alpha}{t^\alpha}(t) = L\circ\psi_	L(t) = p_i^\alpha\parder{\psi^i}{t^\alpha}(t) - \widetilde H\circ\psi_H = \left( p_i^\alpha\parder{\widetilde H}{p_i^\alpha} - \widetilde H \right)(\psi_H(t))\,.
	$$
	In conclusion, we have that $\psi_H$ is a solution of the $k$-contact Hamilton field equations \eqref{eq:k-contact-HDW-Darboux-coordinates} on $V$.
\end{proof}

We have seen that we can recover the Euler--Lagrange field equations and Hamilton field equations from the Skinner--Rusk formalism. Conversely, we have the following result:

\begin{thm}\label{thm:k-contact-SR-T2}
	Let $L\in\Cinfty(\oplus^k\T Q\times\R^k)$ be a regular Lagrangian function and consider a $k$-vector field $\mathbf{X}=(X_1,\dotsc,X_k)$
	in $\oplus^k\T Q\times\R^k$, solution to the $k$-contact Lagrangian equations \eqref{eq:k-contact-Lagrangian-equations}. Then, the $k$-vector field $\bfZ=(Z_\alpha)$ in $\W$ defined as $Z_\alpha = (\mathrm{Id}_{\oplus^k\T Q\times\R^k}\times\F\L)_\ast(X_\alpha)$ is a solution to equations \eqref{eq:k-contact-fields-SR}. Moreover, if $\psi_L\colon\R^k\to\oplus^k\T Q\times\R^k$ is an integral section of $\mathbf{X}$, $\psi=(\psi_L,\F\L\circ\psi_L)\colon\R^k\to\W$ is an integral section of $\bfZ$.
\end{thm}
\begin{proof}
	Consider a regular Lagrangian function $L\in\Cinfty(\oplus^k\T Q\times\R^k)$ and let $\mathbf{X}=(X_1,\dotsc,X_k)$ be a $k$-vector field in $\oplus^k\T Q\times\R^k$ solution to equations \eqref{eq:k-contact-Lagrangian-equations}. Hence, $X_\alpha$ is written in coordinates as
	$$ X_\alpha = v_\alpha^i\parder{}{q^i} + (X_\alpha)^i_\beta\parder{}{v_\beta^i} + (X_\alpha)^\beta\parder{}{s^\beta}\,, $$
	where the functions $(X_\alpha)^\beta$ and $(X_\alpha)^i_\beta$ satisfy the conditions
	\begin{align}
		(X_\alpha)^\alpha &= L\,,\label{eq:k-contact-SR-cond-1}\\
		-\parder{L}{q^i} + \parder{^2L}{s^\beta\partial v_\alpha^i}(X_\alpha)^\beta + \parder{^2L}{q^j\partial v_\alpha^i}v_\alpha^j + \parder{^2L}{v^j_\beta\partial v^i_\alpha}(X_\alpha)^j_\beta &= \parder{L}{s^\alpha}\parder{L}{v_\alpha^i}\label{eq:k-contact-SR-cond-2}\,.
	\end{align}
	Now, using the coordinate expression \eqref{eq:k-contact-Legendre-coordinates} of the Legendre map $\F L$ and taking into account that $Z_\alpha = (\mathrm{Id}_{\oplus^k\T Q\times\R^k}\times\F\L)_\ast(X_\alpha)$, we have
	\begin{multline}\label{eq:k-contact-SR-local-expr}
		Z_\alpha = v_\alpha^i\parder{}{q^i} + (X_\alpha)^i_\beta\parder{}{v_\beta^i} \\ + \left(v^j_\alpha\parder{^2L}{q^j\partial v_\gamma^i} + (X_\alpha)^j_\beta\parder{^2L}{v^j_\beta\partial v_\gamma^i} + (X_\alpha)^\beta\parder{^2L}{s^\beta\partial v^i_\gamma}\right)\parder{}{p_i^\gamma} + (X_\alpha)^\beta\parder{}{s^\beta}\,.
	\end{multline}
	From \eqref{eq:k-contact-SR-cond-1}, \eqref{eq:k-contact-SR-cond-2} and \eqref{eq:k-contact-SR-local-expr}, it is clear that $\bfZ = (Z_\alpha)$ fulfills conditions \eqref{eq:k-contact-fields-SR-second}, \eqref{eq:k-contact-fields-SR-first-one}, \eqref{eq:k-contact-fields-SR-first-three} and \eqref{eq:k-contact-SR-tangency-condition} and hence, the $k$-vector field $\bfZ$ is a solution of \eqref{eq:k-contact-fields-SR} tangent to $\W_1$.

	It is also clear from the definition of integral section that $\psi = (\psi_L,\F\L\circ\psi_L)$ is an integral section of $\bfZ$.
\end{proof}

\begin{rmrk}\rm
	In the case of singular Lagrangians, the results in Theorems \ref{thm:k-contact-SR-T0}, \ref{thm:k-contact-SR-T1}, and \ref{thm:k-contact-SR-T2} hold on the corresponding final constraint submanifolds of the Lagrangian, Hamiltonian and Skinner--Rusk formalisms.
	
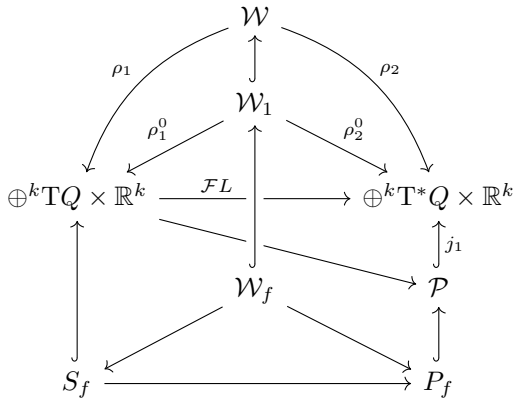
\begin{center}
		\begin{tikzcd}
			& \W \arrow[ddr, bend left=25, "\rho_2"] \arrow[ddl, bend right=25,  swap, "\rho_1"] \\
			& \W_1 \arrow[dl, swap, "\rho_1^0"] \arrow[dr, "\rho_2^0"] \arrow[u, hookrightarrow] \\
			\oplus^k\T Q\times\R^k \arrow[rr, pos=0.3, "\F L"] \arrow[rrd] & & \oplus^k\cT Q\times\R^k \\
			& \W_f \arrow[dl] \arrow[dr] \arrow[uu, hookrightarrow, crossing over, swap, pos=0.8]  & \P \arrow[u, hookrightarrow, swap, "j_1"] \\
			S_f \arrow[rr] \arrow[uu, hookrightarrow] & & P_f \arrow[u, hookrightarrow]
		\end{tikzcd}
	\end{center}
\end{rmrk}

	\chapter{Examples in field theory}
	\label{ch:examples-k-contact}


In this last chapter we study several examples of $k$-contact field theories. In each of them we analyze different aspects of the theory depending on their interest and relevance in each example.

The first example \ref{sec:damped-vibrating-string} to be studied is the \textbf{damped vibrating string}. For this example we give a complete description of the Lagrangian, Hamiltonian and Skinner--Rusk formalisms, as well as a brief analysis of some of its symmetries.

The second example \ref{sec:two-coupled-vibrating-strings} consists of \textbf{two coupled vibrating strings with damping}. We develop its Hamiltonian formulation and find an infinitesimal $k$-contact symmetry of the system and its associated dissipation law.

In Example \ref{sec:Burgers-equation} we deal with the well-known \textbf{Burgers' equation} \cite{Bat1915,Sal2015}. This equation is closely related to the heat equation. We show that, although the heat equation is not variational, we can find a variational formulation of it by adding an additional dependent variable. Then, we see that Burgers' equation can be seen as a contactification of the heat equation.

The problem of finding a Lagrangian function that yields a certain partial differential equation is known as the \textbf{inverse problem}. Example \ref{sec:inverse-problem-elliptic-hyperbolic-equations} provides a method of obtaining $k$-contact Lagrangians whose Euler--Lagrange equations coincide with a certain type of partial differential equations. We also apply this method to find a Lagrangian for a vibrating membrane with damping. The Lagrangian obtained is very similar to the one used in the first example for the damped vibrating string. In addition, a $k$-contact Lagrangian symmetry for the damped vibrating membrane is found and, from it, we deduce its associated dissipation law.

We have already seen that when dealing with $k$-contact Lagrangian functions, we usually get terms linear in the velocities in the Euler--Lagrange equations that produce a dissipation of the energy. However, it is possible to obtain similar terms when dealing with $k$-symplectic Lagrangians. In this case, the terms have a specific form, arising from the coefficients of a closed 2-form, and do not dissipate the energy. In this fifth example \ref{sec:vibrating-string-Lorentz-vs-dissipation}, we study an academic example consisting on a \textbf{nonconducting vibrating string with charge}, where we have both terms. This brings up the differences between the additional terms. We also study a symmetry of this system.

The sixth example \ref{sec:Klein-Gordon} consists in contactifying the well-known \textbf{Klein--Gordon equation} \cite{Itz1980}. We see that, with the appropiate dissipation term in the Klein--Gordon Lagrangian, we can obtain the \textbf{telegrapher's equation} \cite{Hay2018,Sal2015}. We also develop the Skinner--Rusk formalism for this system.

The seventh and last example \ref{sec:Maxwell-dissipation} of $k$-contact field theories consists in contactifying the well-known \textbf{Maxwell's equations} of electromagnetism \cite{Jac1999,Pan1962}. We do this by adding an additional term to the Maxwell's Lagrangian. We give a complete study of the Skinner--Rusk formalism of this system and obtain a set of equations quite similar to Maxwell's equations but with additional terms. It is well known that one can derive the equations of electromagnetic waves from Maxwell's equations. In our case, we see that with the appropiate dissipative term in the Maxwell's Lagrangian, we can derive an equation that represents \textbf{damped electromagnetic waves}.

\section{The damped vibrating string}
\label{sec:damped-vibrating-string}

It is well known that a vibrating string can be described using the Lagrangian formalism. Consider the coordinates $(t,x)$ for the time and the space. Denote by $u$ the separation of a point in the string from its equilibrium point, and hence $u_t$ and $u_x$ will denote the derivative of $u$ with respect to the two independent variables. The Lagrangian function for this system is
\begin{equation}\label{eq:Lagrangian-vibrating-string}
    L(u,u_t,u_x) = \frac{1}{2}\rho u_t^2 - \frac{1}{2}\tau u_x^2\,,
\end{equation}
where $\rho$ is the linear mass density of the string and $\tau$ is the tension of the string. We will assume that these quantities are constant. the Euler--Lagrange equation for this Lagrangian density is
$$ u_{tt} = c^2 u_{xx}\,, $$
where $c^2 = \dfrac{\tau}{\rho}$, which is the 1-dimensional wave equation.

\subsection*{Lagrangian formalism}

In order to model a vibrating string with linear damping, we can modify the Lagrangian function \eqref{eq:Lagrangian-vibrating-string} so that it becomes a $k$-contact Lagrangian \cite{Gas2020}. The new Lagrangian function $\L$ will be defined in the phase bundle $\oplus^2\T Q\times \R^2$, where $Q = \R$, equipped with coordinates $(u; u_t, u_x; s^t, s^x)$:
\begin{equation}\label{eq:Lagrangian-vibrating-string-dis}
	\L(u, u_t, u_x, s^t, s^x) = L - \gamma s^t = \frac{1}{2}\rho u_t^2 - \frac{1}{2}\tau u_x^2 - \gamma s^t\,.
\end{equation}
The canonical structures of the bundle $\oplus^2\T \R\times\R^2$ have coordinates expressions
\begin{equation*}
	J^t = \parder{}{u_t}\otimes\d u\ ,\quad
	J^x = \parder{}{u_x}\otimes\d u\ ,\quad
	\Delta = u_t\parder{}{u_t} + u_x\parder{}{u_x}\,.
\end{equation*}
The Lagrangian energy associated to the Lagrangian \eqref{eq:Lagrangian-vibrating-string-dis} is
$$ E_\L = \Delta(\L) - \L = \frac{1}{2}\rho u_t^2 - \frac{1}{2}\tau u_x^2 + \gamma s^t\,, $$
the differentials of $\L$ and $E_\L$ are
\begin{align*}
	\d\L &= \rho u_t\d u_t - \tau u_x\d u_x - \gamma\d s^t\,,\\
	\d E_\L &= \rho u_t\d u_t - \tau u_x\d u_x + \gamma\d s^t\,,
\end{align*}
the Cartan 1-forms are
\begin{align*}
	\theta^t_\L &= \transp{(J^t)}\circ\d\L = \rho u_t\d u\,,\\
	\theta^x_\L &= \transp{(J^x)}\circ\d\L = -\tau u_x\d u\,,
\end{align*}
and the contact 1-forms are
\begin{align*}
	\eta^t_\L &= \d s^t - \rho u_t\d u\,,\\
	\eta^x_\L &= \d s^x + \tau u_x\d u\,.
\end{align*}
The differentials of the contact 1-forms are
\begin{align*}
	\d\eta^t_\L &= \rho\d u\wedge\d u_t\,,\\
	\d\eta^x_\L &= -\tau\d u\wedge\d u_x\,.
\end{align*}
With all these, we can compute the Reeb vector fields:
$$ (\Reeb_\L)_t = \parder{}{s^t}\ ,\quad(\Reeb_\L)_x = \parder{}{s^x}\,. $$
Consider now a 2-vector field $\bfX = (X_1,X_2)\in\X(\oplus^2\T \R\times\R^2)$ with local expression
\begin{align*}
	X_1 &= f_1\parder{}{u} + F_{1t}\parder{}{u_t} + F_{1x}\parder{}{u_x} + g_1^t\parder{}{s^t} + g_1^x\parder{}{s^x}\,,\\
	X_2 &= f_2\parder{}{u} + F_{2t}\parder{}{u_t} + F_{2x}\parder{}{u_x} + g_2^t\parder{}{s^t} + g_2^x\parder{}{s^x}\,.
\end{align*}
We have that
\begin{align*}
	i(X_1)\d\eta^t_\L &= \rho f_1\d u_t - \rho F_{1t}\d u\,,\\
	i(X_2)\d\eta^x_\L &= -\tau f_2\d u_x + \tau F_{2x}\d u\,,\\
	(\Lie_{(\Reeb_\L)_t}E_\L)\eta^t_\L &= \gamma\d s^t - \gamma\rho u_t\d u\,,\\
	(\Lie_{(\Reeb_\L)_x}E_\L)\eta^x_\L &= 0\,.
\end{align*}
Hence, the first equation in \eqref{eq:k-contact-Lagrangian-equations} reads
$$ \rho f_1\d u_t - \rho F_{1t}\d u - \tau f_2\d u_x + \tau F_{2x}\d u = \rho u_t\d u_t - \tau u_x\d u_x + \gamma\rho u_t\d u\,, $$
which yields the conditions
\begin{align}
	-\rho\, F_{1t} + \tau F_{2x} &= \gamma\rho u_t & & \mbox{(coefficients in $\d u$)}\,,\label{eq:vibrating-string-lagrangian-condition-1}\\
	f_1 &= u_t & & \mbox{(coefficients in $\d u_t$)}\,,\label{eq:vibrating-string-lagrangian-condition-2}\\
	f_2 &= u_x & & \mbox{(coefficients in $\d u_x$)}\,.\label{eq:vibrating-string-lagrangian-condition-3}\\
\end{align}
Notice that the equations \eqref{eq:vibrating-string-lagrangian-condition-2} and \eqref{eq:vibrating-string-lagrangian-condition-3} above are the {\sc sopde} conditions for the 2-vector field $\bfX$.
On the other hand, the second equation in \eqref{eq:k-contact-Lagrangian-equations} gives the condition
$$ g_1^t + g_2^x = \frac{1}{2}\rho u_t^2 - \frac{1}{2}\tau u_x^2 - \gamma s^t = \L\,. $$
With all this, the 2-vector field $\bfX$ has local expression
\begin{align*}
	X_1 &= u_t\parder{}{u} + \left( \frac{\tau}{\rho}F_{2x} - \gamma u_t \right)\parder{}{u_t} + F_{1x}\parder{}{u_x} + \left( \L - g_2^x \right)\parder{}{s^t} + g_1^x\parder{}{s^x}\,,\\
	X_2 &= u_x\parder{}{u} + F_{2t}\parder{}{u_t} + F_{2x}\parder{}{u_x} + g_2^t\parder{}{s^t} + g_2^x\parder{}{s^x}\,,
\end{align*}
and the functions $F_{1x},F_{2t},F_{2x},g_1^x,g_2^t,g_2^x$ remain undetermined. Notice that equation \eqref{eq:vibrating-string-lagrangian-condition-1} leads to
\begin{equation}\label{eq:damped-vibrating-string-edp}
	u_{tt} - c^2u_{xx} + \gamma u_t = 0\,,
\end{equation}
which is the equation for a vibrating string with damping.

\subsubsection*{Symmetries}

We can see that the vector field $Y = \tparder{}{u}$ is an infinitesimal Lagrangian $k$-contact symmetry because $\tparder{\L}{u} = 0$ (see Proposition \ref{prop:k-contact-symmetries-of-the-Lagrangian}). Then, by Theorem \ref{thm:k-contact-Lagrangian-dissipation-theorem} the map $F = (F^t, F^x)\colon M\to\R^2$ given by
$$ F^t = -i\left(\parder{}{u}\right)\eta^t_\L = \rho u_t\ ,\quad F^x = -i\left(\parder{}{u}\right)\eta^x_\L = -\tau u_x $$
satisfies the dissipation law for $k$-vector fields \eqref{eq:k-contact-Lagrangian-dissipation-law-fields}.

\subsection*{Hamiltonian formalism}

The Legendre map associated to the Lagrangian \eqref{eq:Lagrangian-vibrating-string-dis} $\F\L\colon\oplus^2\T \R\times\R^2\to\oplus^2\cT \R\times\R^2$ is given by
$$ \F\L(u,u_t,u_x,s^t,s^x) = (u, p^t, p^x, s^t, s^x)\,, $$
where $p^t = \rho u_t$ and $p^x = -\tau u_x$. Hence, the contact forms $\eta^t,\eta^x$ of $\oplus^2\cT \R\times\R^2$ are
\begin{align*}
	\eta^t = \d s^t - p^t\d u\,,\\
	\eta^x = \d s^x - p^x\d u\,,
\end{align*}
and the Hamiltonian function is
$$ H = \frac{1}{2\rho}(p^t)^2 - \frac{1}{2\tau}(p^x)^2 + \gamma s^t\,. $$
We have that
$$ \d\eta^t = \d u\wedge\d p^t\ ,\quad \d\eta^x = \d u\wedge\d p^x\,, $$
and
$$ \d H = \frac{p^t}{\rho}\d p^t - \frac{p^x}{\tau}\d p^x + \gamma\d s^t\,. $$
The Reeb vector fields are
$$ \Reeb_t = \parder{}{s^t}\ ,\quad\Reeb_x = \parder{}{s^x}\,. $$
Consider now a 2-vector field $\bfY = (Y_1,Y_2)\in\X^2(\oplus^2\T\R^\ast\times\R^2)$ with local expression
\begin{align*}
	Y_1 &= f_1\parder{}{u} + G_1^t\parder{}{p^t} + G_1^x\parder{}{p^x} + g_1^t\parder{}{s^t} + g_1^x\parder{}{s^x}\,,\\
	Y_2 &= f_2\parder{}{u} + G_2^t\parder{}{p^t} + G_2^x\parder{}{p^x} + g_2^t\parder{}{s^t} + g_2^x\parder{}{s^x}\,.
\end{align*}
We have that
\begin{align*}
	i(Y_1)\d\eta^t + i(Y_2)\d\eta^x &= f_1\d p^t - G_1^t\d u + f_2\d p^x - G_2^x\d u\,,\\
	\d H - (\Lie_{\Reeb_t}H)\eta^t - (\Lie_{\Reeb_x}H)\eta^x &= \frac{p^t}{\rho}\d p^t - \frac{p^x}{\tau}\d p^x + \gamma p^t\d u\,.
\end{align*}
Hence, the first equation in \eqref{eq:k-contact-HDW-fields} gives the conditions
\begin{align}
	G_1^t + G_2^x &= -\gamma p^t & & \mbox{(coefficients in $\d u$)}\,,\label{eq:vibrating-string-hamiltonian-condition-1}\\
	f_1 &= \frac{p^t}{\rho} & & \mbox{(coefficients in $\d p^t$)}\,,\label{eq:vibrating-string-hamiltonian-condition-2}\\
	f_2 &= -\frac{p^x}{\tau} & & \mbox{(coefficients in $\d p^x$)}\,.\label{eq:vibrating-string-hamiltonian-condition-3}
\end{align}
On the other hand, as $i(Y_1)\eta^t + i(Y_2)\eta^x = g_1^t - p^tf_1 + g_2^x - p^xf_2$, the second equation in \eqref{eq:k-contact-HDW-fields} gives
$$ g_1^t + g_2^x = \frac{1}{2\rho}(p^t)^2 - \frac{1}{2\tau}(p^x)^2 - \gamma s^t\,. $$
Hence, the 2-vector field $\bfY$ reads
\begin{align*}
	Y_1 &= \frac{p^t}{\rho}\parder{}{u} - \left(G_2^x + \gamma p^t\right)\parder{}{p^t} + G_1^x\parder{}{p^x}\\
	&\quad + \left(\frac{1}{2\rho}(p^t)^2 - \frac{1}{2\tau}(p^x)^2 - \gamma s^t - g_2^x\right)\parder{}{s^t} + g_1^x\parder{}{s^x}\,,\\
	Y_2 &= -\frac{p^x}{\tau}\parder{}{u} + G_2^t\parder{}{p^t} + G_2^x\parder{}{p^x} + g_2^t\parder{}{s^t} + g_2^x\parder{}{s^x}\,,
\end{align*}
and the functions $G_1^x,G_2^t,G_2^x,g_1^x,g_2^t,g_2^x$ remain undetermined. Given a map $\psi(t) = (u(t),p^t(t),p^x(t),s^t(t),s^x(t))$, the Hamilton--De Donder--Weyl equations \eqref{eq:k-contact-HDW} for it read
\begin{equation*}
	\begin{dcases}
		\parder{u}{t} = \frac{1}{\rho}p^t\,,\\
		\parder{u}{x} = -\frac{1}{\tau}p^x\,,\\
		\parder{p^t}{t} + \parder{p^x}{x} = -\gamma p^t\,,\\
		\parder{s^t}{t} + \parder{s^x}{x} = \frac{1}{2\rho}(p^t)^2 - \frac{1}{2\tau}(p^x)^2 - \gamma s^t\,.
	\end{dcases}
\end{equation*}
Using the first and second equations above into the third equation, we obtain
$$ \rho u_{tt} - \tau u_{xx} + \gamma\rho u_t = 0\,, $$
which is the equation of the damped vibrating string \eqref{eq:damped-vibrating-string-edp}.

\subsubsection*{Symmetries}

It is easy to see that, as in the Lagrangian formalism, the vector field $\tparder{}{u}$ is a $k$-contact symmetry. It induces the map
$$ F = \left( -i\left(\parder{}{u}\right)\eta^t,-i\left(\parder{}{u}\right)\eta^x \right) = (p^t, p^x)\,. $$
This map $F$ satisfies the dissipation law for $k$-vector fields \eqref{eq:k-contact-dissipation-law-fields}:
$$ \Lie_{Y_1}p^t + \Lie_{Y_2}p^x = -(\Lie_{\Reeb_t}H)p^t - (\Lie_{\Reeb_x}H)p^x = -2\gamma p^t\,. $$
Along a solution $(Y_1,Y_2)$, this law is $p_t^t + p_x^x = -2\gamma p^t$.

\subsection*{Skinner--Rusk formalism}

Consider the extended Pontryagin bundle
$$ \W = \oplus^2\T\R\times_\R\oplus^2\cT\R\times\R^2 $$
equipped with natural coordinates $(u, u_t, u_x, p^t, p^x, s^t, s^x)$. In this bundle, the coupling function is
$$ \C = p^xu_x + p^tu_t\,, $$
and we have the canonical forms
\begin{align*}
	\Theta^t = p^t\d u\quad &,\quad \Omega^t = -\d\Theta^2 = \d u\wedge\d p^t\,,\\
	\Theta^x = p^x\d u\quad &,\quad \Omega^x = -\d\Theta^1 = \d u\wedge\d p^x\,,
\end{align*}
and the canonical contact 1-forms
$$ \eta^t = \d s^t - p^t\d u\ ,\quad\eta^x = \d s^x - p^x\d u\,. $$
We can take the vector fields
$$ \Reeb_t = \parder{}{s^t}\ ,\quad\Reeb_x = \parder{}{s^x} $$
as Reeb vector fields. Given the Lagrangian function $\L\colon\oplus^2\T\R\times\R^2\to\R$ defined in \eqref{eq:Lagrangian-vibrating-string-dis}, we can build the Hamiltonian function $\H = \C - \L$, which has coordinate expression
$$ \H = p^tu_t + p^xu_x - \frac{1}{2}\rho u_t^2 + \frac{1}{2}\tau u_x^2 + \gamma s^t\,. $$
To solve the Lagrangian--Hamiltonian problem for the 2-precontact Hamiltonian system $(\W,\eta^\alpha,\H)$ means to find a 2-vector field $\mathbf{Z} = (Z_1,Z_2)$ in $\W$ satisfying equations \eqref{eq:k-contact-fields-SR}.
For our Hamiltonian function $\H$, we have
$$ \d\H - \Reeb_\alpha(\H)\eta^\alpha = u_t\d p^t + u_x\d p^x + (p^t - \rho u_t)\d u_t + (p^x + \tau u_x)\d u_x + \gamma p^t\d u\,. $$
Let $\mathbf{Z} = (Z_\alpha)$ be a 2-vector field with local expression
$$ Z_\alpha = f_\alpha\parder{}{u} + F_{\alpha t}\parder{}{u_t} + F_{\alpha x}\parder{}{u_x} + G_\alpha^t\parder{}{p^t} + G_\alpha^x\parder{}{p^x} + g_\alpha^t\parder{}{s^t} + g_\alpha^x\parder{}{s^x}\,. $$
Now,
$$ i(Z_\alpha)\d\eta^\alpha = f_1\d p^t + f_2\d p^x - (G_1^t + G^2_x)\d u\,, $$
and hence, the first equation in \eqref{eq:k-contact-fields-SR} gives the conditions
\begin{align*}
	G_1^t + G_2^x &= -\gamma p^t & & \mbox{(coefficients in $\d u$)}\,,\\
	p^t &= \rho u_t & & \mbox{(coefficients in $\d u_t$)}\,,\\
	p^x &= -\tau u_x & & \mbox{(coefficients in $\d u_x$)}\,,\\
	f_1 &= u_t & & \mbox{(coefficients in $\d p^t$)}\,,\\
	f_2 &= u_x & & \mbox{(coefficients in $\d p^x$)}\,.
\end{align*}
Notice that combining the first three conditions we recover the damped wave equation \eqref{eq:damped-vibrating-string-edp}. Furthermore, the last two equations are the holonomy conditions.
The second equation in \eqref{eq:k-contact-fields-SR} gives the condition
$$ g_1^t + g_2^x = \frac{1}{2}\rho u_t^2 - \frac{1}{2}\tau u_x^2 - \gamma s^t = \L\,. $$
In addition, we have obtained the constraints
$$ \xi_1 = p^t - \rho u_t = 0\quad,\quad\xi_2 = p^x + \tau u_x = 0 \ , $$
which define the submanifold $\W_1\hookrightarrow\W$. Imposing the tangency of the 2-vector field $\mathbf{Z}$ to the submanifold $\W_1$ we get the conditions
\begin{align*}
	0 = Z_1(\xi_1) = G_1^t - \rho F_{1t}\quad &,\quad
	0 = Z_2(\xi_1) = G_2^t - \rho F_{2t}\,,\\
	0 = Z_1(\xi_2) = G_1^x + \tau F_{1x}\quad &,\quad
	0 = Z_2(\xi_2) = G_2^x + \tau F_{2x}\,,
\end{align*}
which determine partially some of the arbitrary functions and no new constraints appear, so the constraint algorithm finishes with the submanifold $\W_f = \W_1$, giving the solutions $\mathbf{Z}=(Z_1,Z_2)$ with
\begin{align*}
	Z_1 &= u_t\parder{}{u} - \frac{\gamma p^t + G_2^x}{\rho}\parder{}{u_t} - \frac{G_1^x}{\tau}\parder{}{u_x}\\
	&\quad - \left( \gamma p^t + G_2^x \right)\parder{}{p^t} + G_1^x\parder{}{p^x} + (\L - g_2^x)\parder{}{s^t} + g_1^x\parder{}{s^x}\,,\\
	Z_2 &= u_x\parder{}{u} + \frac{G_2^t}{\rho}\parder{}{u_t} - \frac{G_2^x}{\tau}\parder{}{u_x} + G_2^t\parder{}{p^t} + G_1^x\parder{}{p^x} + g_2^t\parder{}{s^t} + g_2^x\parder{}{s^x}\,,
\end{align*}
where $G_1^x,G_2^t,G_2^x,g_1^x,g_2^t,g_2^x$ are arbitrary functions.

It is important to point out that we can project on each factor of the product manifold $\W = \oplus^2\T\R\times_\R\oplus^2\cT\R\times\R^2$ with the projections $\rho_1$ and $\rho_2$ to recover the Lagrangian and Hamiltonian formalisms. In the Lagrangian formalism we have the holonomic 2-vector field $\bfX = (X_1,X_2)$ given by
\begin{align*}
	X_1 &= u_t\parder{}{u} + \left( \frac{\tau}{\rho}F_{2x} - \gamma u_t \right)\parder{}{u_t} + F_{1x}\parder{}{u_x} + (\L - g_2^x)\parder{}{s^t} + g_1^x\parder{}{s^x}\,,\\
	X_2 &= u_x\parder{}{u} + F_{2t}\parder{}{u_t} + F_{2x}\parder{}{u_x} + g_2^t\parder{}{s^t} + g_2^x\parder{}{s^x}\,,
\end{align*}
where $F_{1x},F_{2t},F_{2x},g_1^x,g_2^t,g_2^x$ are arbitrary functions. On the other side, in the Hamiltonian formalism we have the Hamiltonian 2-vector field $\bfY = (Y_1,Y_2)$ given by
\begin{align*}
	Y_1 &= \frac{p^t}{\rho}\parder{}{u} - \left( \gamma p^t + G_2^x \right)\parder{}{p^t} + G_1^x\parder{}{p^x} + \left(\frac{(p^t)^2}{2\rho} - \frac{(p^x)^2}{2\tau} - \gamma s^t - g_2^x\right)\parder{}{s^t} + g_1^x\parder{}{s^x}\,,\\
	Y_2 &= -\frac{p^x}{\tau}\parder{}{u} + G_2^t\parder{}{p^t} + G_1^x\parder{}{p^x} + g_2^t\parder{}{s^t} + g_2^x\parder{}{s^x}\,,
\end{align*}
where $G_1^x,G_2^t,G_2^x,g_1^x,g_2^t,g_2^x$ are arbitrary functions.

\section{Two coupled vibrating strings with damping}
\label{sec:two-coupled-vibrating-strings}

Consider a system of two coupled strings with damping. The configuration manifold of the system is $Q = \R^2$ equipped with coordinates $(q^1,q^2)$, where each coordinate represent the displacement of each string. The Hamiltonian phase bundle of this system is $M = \oplus^2\cT\R^2\times\R^2$ endowed with natural coordinates $(q^1,q^2,p_1^t,p_2^t,p_1^x,p_2^x,s^t,s^x)$. The 1-forms
\begin{align*}
	\eta^t &= \d s^t - p_1^t\d q^1 - p_2^t\d q^2\,,\\
	\eta^x &= \d s^x - p_1^x\d q^1 - p_2^x\d q^2
\end{align*}
define a 2-contact structure on $M$. The Reeb vector fields are $\Reeb_t = \tparder{}{s^t}$, $\Reeb_x = \tparder{}{s^x}$. Consider now the Hamiltonian function
$$ H = \frac{1}{2}\left( (p_1^t)^2 + (p_2^t)^2 + (p_1^x)^2 + (p_2^x)^2 \right) + C(z) + \gamma s^t\,, $$
where $C$ is a function that represents a coupling of the two strings, and we assume that depends only on $z = \sqrt{(q^1)^2 + (q^2)^2}$.
Consider a 2-vector field $\bfX = (X_1,X_2)\in\X^2(M)$ with local expression
\begin{align*}
	X_1 &= f_1^1\parder{}{q^1} + f_1^2\parder{}{q^2} + G_{11}^t\parder{}{p_1^t} + G_{12}^t\parder{}{p_2^t} + G_{11}^x\parder{}{p_1^x} + G_{12}^x\parder{}{p_2^x} + g_1^t\parder{}{s^t} + g_1^x\parder{}{s^x}\,,\\
	X_2 &= f_2^1\parder{}{q^1} + f_2^2\parder{}{q^2} + G_{21}^t\parder{}{p_1^t} + G_{22}^t\parder{}{p_2^t} + G_{21}^x\parder{}{p_1^x} + G_{22}^x\parder{}{p_2^x} + g_2^t\parder{}{s^t} + g_2^x\parder{}{s^x}\,.
\end{align*}
We have that
\begin{gather*}
	\d H = p_1^t\d p_1^t + p_2^t\d p_2^t + p_1^x\d p_1^x + p_2^x\d p_2^x + C'(z)\frac{q^1}{z}\d q^1 + C'(z)\frac{q^2}{z}\d q^2 + \gamma\d s^t\,,\\
	(\Lie_{\Reeb_t}H)\eta^t + (\Lie_{\Reeb_x}H)\eta^x = \gamma\d s^t - \gamma p_1^t\d q^1 - \gamma p_2^t\d q^2\,.
\end{gather*}
Hence, the first equation in \eqref{eq:k-contact-HDW-fields} yield the conditions
\begin{align*}
	-G_{11}^t - G_{21}^x &= C'(z)\frac{q^1}{z} + \gamma p_1^t & & \mbox{(coefficients in $\d q^1$)}\,,\\
	-G_{12}^t - G_{22}^x &= C'(z)\frac{q^2}{z} + \gamma p_2^t & & \mbox{(coefficients in $\d q^2$)}\,,\\
	f_1^1 &= p_1^t & & \mbox{(coefficients in $\d p_1^t$)}\,,\\
	f_1^2 &= p_2^t & & \mbox{(coefficients in $\d p_2^t$)}\,,\\
	f_2^1 &= p_1^x & & \mbox{(coefficients in $\d p_1^x$)}\,,\\
	f_2^2 &= p_2^x & & \mbox{(coefficients in $\d p_2^x$)}\,.
\end{align*}
On the other hand, the second equation in \eqref{eq:k-contact-HDW-fields} gives
$$ g_1^t + g_2^x = \frac{1}{2}\left( (p_1^t)^2 + (p_2^t)^2 + (p_1^x)^2 + (p_2^x)^2 \right) - C(z) - \gamma s^t\,. $$
Hence, the 2-vector field $\bfX$ solution to \eqref{eq:k-contact-HDW-fields} has local expression
\begin{align*}
	X_1 &= \  p_1^t\parder{}{q^1} + p_2^t\parder{}{q^2} + \left( -G_{21}^x - C'(z)\frac{q^1}{z} - \gamma p_1^t \right)\parder{}{p_1^t}\\
	&\quad + \left( -G_{22}^x - C'(z)\frac{q^2}{z} - \gamma p_2^t \right)\parder{}{p_2^t} + G_{11}^x\parder{}{p_1^x} + G_{12}^x\parder{}{p_2^x} \\
	&\quad + \left(\frac{1}{2}\left( (p_1^t)^2 + (p_2^t)^2 + (p_1^x)^2 + (p_2^x)^2 \right) - C(z) - \gamma s^t - g_2^x\right)\parder{}{s^t} + g_1^x\parder{}{s^x}\,,\\
	X_2 &= \  p_1^x\parder{}{q^1} + p_2^x\parder{}{q^2} + G_{21}^t\parder{}{p_1^t} + G_{22}^t\parder{}{p_2^t} + G_{21}^x\parder{}{p_1^x} + G_{22}^x\parder{}{p_2^x} + g_2^t\parder{}{s^t} + g_2^x\parder{}{s^x}\,,
\end{align*}
where $G_{11}^x,G_{12}^x,G_{21}^t,G_{22}^t,G_{21}^x,G_{22}^x,g_1^x,g_2^t,g_2^x$ are arbitrary functions. The Hamilton--De Donder--Weyl equations \eqref{eq:k-contact-HDW-Darboux-coordinates} for a map $\psi(t,x) = (q^i(t,x),p_i^\alpha(t,x),s^\alpha(t,x))$ are
\begin{equation*}
	\begin{dcases}
		\parder{q^1}{t} = p_1^t\,,\\
		\parder{q^1}{x} = p_1^x\,,\\
		\parder{q^2}{t} = p_2^t\,,\\
		\parder{q^2}{x} = p_2^x\,,\\
		\parder{p_1^t}{t} + \parder{p_1^x}{x} = - C'(z)\frac{q^1}{z} - p_1^t\gamma\,,\\
		\parder{p_2^t}{t} + \parder{p_2^x}{x} = - C'(z)\frac{q^2}{z} - p_2^t\gamma\,,\\
		\parder{s^t}{t} + \parder{s^x}{x} = \frac{1}{2}\left( (p_1^t)^2 + (p_2^t)^2 + (p_1^x)^2 + (p_2^x)^2 \right) - C(z) - \gamma s^t\,.
	\end{dcases}
\end{equation*}
Combining the first six equations above, we obtain the system
\begin{equation*}
	\begin{dcases}
		\parder{^2 q^1}{t^2} + \parder{^2 q^1}{x^2} + \gamma p_1^t + C'(z)\frac{q^1}{z} = 0\,,\\
		\parder{^2 q^2}{t^2} + \parder{^2 q^2}{x^2} + \gamma p_2^t + C'(z)\frac{q^2}{z} = 0\,,
	\end{dcases}
\end{equation*}
which corresponds to two coupled strings with damping with coupling function $C$.

\subsubsection*{Symmetries}

It is easy to see that the vector field
$$ Y = q^1\parder{}{q^2} - q^2\parder{}{q^1} + p_1^t\parder{}{p_2^t} - p_2^t\parder{}{p_1^t} + p_1^x\parder{}{p_2^x} - p_2^x\parder{}{p_1^x} $$
is an infinitesimal $k$-contact symmetry of the system. It induces the map $F = (F^t,F^x)$ given by
$$ F^t = -i(Y)\eta^t = q^1p_2^t - q^2p_1^t\ ,\quad F^x = -i(Y)\eta^x = q^1p_2^x - q^2p_1^x\,, $$
which satisfies the dissipation law for $k$-vector fields \eqref{eq:k-contact-dissipation-law-fields} along a the solution $(X_1,X_2)$:
\begin{align*}
	\Lie_{X_1}F^t + \Lie_{X_2}F^x &= \Lie_{X_1}(q^1p_2^t - q^2p_1^t) + \Lie_{X_2}(q^1p_2^x - q^2p_1^x)\\
	&= q^1\left(\parder{p_2^t}{t} + \parder{p_2^x}{x}\right) - q^2\left(\parder{p_1^t}{t} + \parder{p_1^x}{x}\right)\\
	&= -\gamma(q^1p_2^t - q^2p_1^t)\,.
\end{align*}

\section{Burgers' equation}
\label{sec:Burgers-equation}

The Burgers' equation \cite{Bat1915,Sal2015} is a remarkable nonlinear partial differential equation. It appears in many areas of applied mathematics. It reads
\begin{equation}\label{eq:burgers-equation}
	u_t + u u_x = k u_{xx}\,,
\end{equation}
where $t,x$ are the independent variables, $u = u(t,x)$ is the dependent variable and $k\geq 0$ a diffusion coefficient. Burgers' equation is closely related to the heat equation
\begin{equation}\label{eq:heat-equation}
	u_t = ku_{xx}\,.
\end{equation}
In fact, we will show that Burgers' equation \eqref{eq:burgers-equation} can be formulated as a contactification of the heat equation \eqref{eq:heat-equation}. This will be done in several steps.

\subsection*{Lagrangian formulation of the heat equation}

In order to contactify the heat equation, we will need a Hamiltonian formulation of it. This Hamiltonian formulation can be obtained via the Legendre map if we have a Lagrangian formulation. Although the heat equation is not variational, it can be made variational by considering an additional dependent variable $v$, and taking as Lagrangian the function \cite{Ibr2004}
\begin{equation}\label{eq:Lagrangian-heat}
	L = -ku_xv_x - \frac{1}{2}(vu_t - uv_t)\,,
\end{equation}
whose Euler--Lagrange equations are
\begin{equation}\label{eq:Euler-Lagrange-heat}
	[L]_u = kv_{xx} + v_t = 0\ ,\quad [L]_v = ku_{xx} - u_t = 0\,.
\end{equation}
The first equation is linear homogeneous and therefore it always has solutions, for instance $v = 0$. Hence, there is a bijection between the solutions to the heat equation \eqref{eq:heat-equation} and the solutions to the Euler--Lagrange equations \eqref{eq:Euler-Lagrange-heat} with $v = 0$.

\subsection*{Hamiltonian formulation of the heat equation}

Now we are going to apply the Hamilton--De Donder--Weyl formalism to the Lagrangian $L$ defined in \eqref{eq:Lagrangian-heat}. The Legendre map associated to the Lagrangian $L$ is a map
$$ \F L\colon\oplus^2\T\R^2\to P = \oplus^2\cT\R^2\,. $$
The phase bundle is $P = \R^6$, equipped with coordinates $(u,v,p^t,p^x,q^t,q^x)$, where $p^t,p^x$ are the momenta of the variable $u$ and $q^t,q^x$ are the momenta of the variable $v$ with respect to the independent variables. The Legendre map relates these momenta with the configuration fields and their velocities:
\begin{align*}
	\F L^\ast p^t &= -\frac{1}{2}v\ ,\quad \F L^\ast p^x = -kv_x\ ,\\
	\F L^\ast q^t &= \frac{1}{2}u\ ,\quad \F L^\ast q^x = -k u_x\,.
\end{align*}
Hence the image of the Legendre map $P_0 = \F L(\oplus^2\T\R^2)\subset P$ is given by the two constraints
$$ p^t + \frac{1}{2}v = 0\ ,\quad q^t - \frac{1}{2}u = 0\,. $$
We will use coordinates $(u,v,p^x,q^x)$ on $P_0$. Hence, the Hamiltonian function on $P_0$ is
$$ H_0 = -\frac{1}{k}p^xq^x\,. $$
The manifold $P$ is equipped with an exact 2-symplectic structure defined by the 1-forms
$$ p^t\d u + q^t \d v\ ,\quad p^x\d u + q^x\d v\,. $$
The pull-backs to of these forms to $P_0$ no longer define a 2-symplectic structure, but nevertheless we have two 1-forms
$$ \theta^t = \frac{1}{2}(-v\d u + u\d v)\ ,\quad \theta^x = p^x\d u + q^x\d v $$
such that
$$ \omega^t = -\d\theta^t = -\d u\wedge\d v\ ,\quad \omega^x = -\d\theta^x = \d u\wedge\d p^x + \d v\wedge\d q^x\,. $$
Consider a map $\psi\colon\R^2\to P_0$ with local expression $\psi = (u,v,p^x,q^x)$. The Hamilton--De Donder--Weyl equation $i(\psi_t')\omega^t + i(\psi_x')\omega^x = \d H_0\circ\psi$ for the map $\psi$ reads
$$ \partial_t v - \partial_x p^x = 0\ ,\quad -\partial_t u - \partial_x q^x = 0\ ,\quad \partial_x u = -\frac{1}{k}q^x\ ,\quad \partial_x v = -\frac{1}{k}p^x\,. $$
Combining these equations, we get the heat equation for $u$ and its complementary equation for the additional variable $v$:
$$ \partial_t = k\partial_x^2 u\ ,\quad \partial_t v = -k\partial_x^2 v\,. $$
As in the Lagrangian formulation, the equation for $v$ is linear homogeneous, and hence there is a correspondence the solutions of this system with $v = 0$ and the solutions to the heat equation.

\subsection*{Contact Hamiltonian formulation of the Burgers' equation}

Consider now the manifold $P_0$ defined above and its two differential 1-forms $\theta^t,\theta^x$. In order to construct a 2-contact manifold, we define the product manifold $M = P_0\times\R^2$, with coordinates $(u,v,p^x,q^x;s^t,s^x)$. In $M$, we can construct the contact forms
$$ \eta^t = \d s^t - \theta^t\ ,\quad \eta^x = \d s^x - \theta^x\,. $$
Their differentials are $\d\eta^t = \omega^t$ and $\d\eta^x = \omega^x$.

With the notations introduced in Section \ref{sec:k-contact-geometry}, since $\eta^t,\eta^x$ are linearly independent at every point, we have:
\begin{itemize}
	\item $\C^\rmC = \langle\eta^t,\eta^x\rangle$ is a regular codistribution of rank 2,
	\item $\D^\rmR = \langle\Reeb_t,\Reeb_x\rangle$, is a regular distribution of rank 2, where
	$$ \Reeb_t = \parder{}{s^t}\ ,\quad\Reeb_x = \parder{}{s^x}\,. $$
	\item $\D^\rmC\cap\D^\rmR = \{0\}$, since no nonzero linear combination of the Reeb vector fields is anihilated by the contact forms $\eta^t,\eta^x$.
\end{itemize}
Hence, $(M,\eta^t,\eta^x)$ is a 2-contact manifold. Notice that it coincides with the 2-contact manifold in Example \ref{ex:2-contact-R6}. Consider in this manifold the 2-contact Hamiltonian function
$$ H = H_0 + \gamma u s^x = -\frac{1}{k}p^xq^x + \gamma u s^x\,. $$
Now we have a 2-contact Hamiltonian system $(M,\eta^t,\eta^x,H)$.
The Hamilton--De Donder--Weyl equations \eqref{eq:k-contact-HDW} for this system are
\begin{equation}\label{eq:HDW-Burgers}
	\begin{dcases}
		i(\psi_t')\d\eta^t + i(\psi_x')\d\eta^x = \d H - (\Lie_{\Reeb_t}H)\eta^t - (\Lie_{\Reeb_x}H)\eta^x\,,\\
		i(\psi_t')\eta^t + i(\psi_x')\eta^x = -H\,.
	\end{dcases}
\end{equation}
Computing the first one, we obtain the set of equations
\begin{equation*}
	\begin{dcases}
		\partial_t v - \partial_x p^x = \gamma(s^x + up^x)\,,\\
		-\partial_t u - \partial_x q^x = \gamma uq^x\,,\\
		\partial_x u = -\frac{1}{k}q^x\,,\\
		\partial_x v = -\frac{1}{k}p^x\,.
	\end{dcases}
\end{equation*}
Using the latter two equations in the former ones, we get
\begin{equation*}
	\begin{dcases}
		\partial_t u - \gamma k u \partial_x u = k\partial_x^2 u\,,\\
		\partial_t v + \gamma k u \partial_x v = -k\partial_x^2 v + \gamma s^x\,.
	\end{dcases}
\end{equation*}
Setting now the value of the constant $\gamma$ as
$$ \gamma = -\frac{1}{k}\,, $$
the first equation is Burgers' equation for the variable $u$:
$$ \partial_t u + u\partial_x u = k\partial_x^2 u\,. $$
The second Hamilton--De Donder--Weyl equation \eqref{eq:HDW-Burgers} yields
$$ \partial_t s^t - \frac{1}{2}(-v\partial_tu + u\partial_t v) + \partial_xs^x - p^x\partial_x u - q^x\partial_xv = \frac{1}{k}p^xq^x - \gamma us^x\,. $$
Notice that this equation admits solutions $(u,v,p^x,q^x,s^t,s^x)$ with $u$ a solution to Burger's equation \eqref{eq:burgers-equation}, $v = 0$, $p^x = 0$, $q^x = -k\partial_x u$, $s^t = 0$, $s^x = 0$.

Hence, we can conclude that the Burgers' equation \eqref{eq:burgers-equation} can be described by the 2-contact Hamiltonian system $(M,\eta^t,\eta^x,H)$.



\section[Inverse problem for elliptic and hyperbolic equations]{Inverse problem for elliptic and hyperbolic \phantom{n} equations}
\label{sec:inverse-problem-elliptic-hyperbolic-equations}

This example provides a way of obtaining $k$-contact Lagrangians for certain partial differential equations.

A generic second-order linear partial differential equation in $\R^2$ has the form
\begin{equation}\label{eq:generic-second-order-PDE-R2}
	Au_{xx} + 2BU_{xy} + Cu_{yy} + Du_x + Eu_y + Fu + G = 0\,,
\end{equation}
where $A,B,C,D,F,G\in\Cinfty(\R^2)$ are functions such that $A > 0$.
\begin{itemize}
	\item If $B^2 - AC > 0$, equation \eqref{eq:generic-second-order-PDE-R2} is said to be \textbf{hyperbolic}.
	\item If $B^2 - AC < 0$, equation \eqref{eq:generic-second-order-PDE-R2} is said to be \textbf{elliptic}.
	\item If $B^2 - AC = 0$, equation \eqref{eq:generic-second-order-PDE-R2} is said to be \textbf{parabolic}.
\end{itemize}

In $\R^n$, consider the equation
\begin{equation}\label{eq:generic-second-order-PDE}
	A^{\alpha\beta}u_{\alpha\beta} + D^\alpha u_\alpha + G(u) = 0\,,
\end{equation}
where $1\leq\alpha,\beta\leq n$. Consider the particular case where the matrix $A^{\alpha\beta}$ is constant and invertible, (i.e., equation \eqref{eq:generic-second-order-PDE} is not parabolic), $D^\alpha$ is constant and $G$ is an arbitrary function in the variable $u$.

In order to find a $k$-contact Lagrangian formulation of this kind of partial differential equations, consider the manifold $M = \oplus^n\T\R\times\R^n$, equipped with natural coordinates $(u,u_\alpha,s^\alpha)$ and a generic Lagrangian function $L\in\Cinfty(\oplus^n\T\R\times\R^n)$ of the form
\begin{equation}\label{eq:lagrangian-generic-second-order-PDE}
	L = \frac{1}{2}a^{\alpha\beta}(u)u_\alpha u_\beta + b(u) u_\alpha s^\alpha + d(u,s)\,.
\end{equation}
The $k$-contact structure associated to this Lagrangian function is
$$ \eta^\alpha = \d s^\alpha - \parder{L}{u_\alpha}\d u = \d s^\alpha - (a^{\alpha\beta}u_\beta + bs^\alpha + c^\alpha)\d u\,. $$
The $k$-contact Euler--Lagrange equations associated to the Lagrangian $L$ is
\begin{equation}\label{eq:Euler-Lagrange-generic-second-order-PDE}
	a^{\alpha\beta}u_{\alpha\beta} + \frac{1}{2}\left( \parder{a^{\alpha\beta}}{u} - b a^{\alpha\beta} \right)u_\alpha u_\beta - \parder{d}{s^\beta}a^{\beta\alpha}u_\alpha + \left( -\parder{d}{s^\alpha}bs^\alpha + bd - \parder{d}{u} \right) = 0\,.
\end{equation}
Comparing equations \eqref{eq:generic-second-order-PDE} and \eqref{eq:Euler-Lagrange-generic-second-order-PDE}, we obtain the conditions
\begin{equation}\label{eq:inverse-problem-conditions}
	\begin{dcases}
		a^{\alpha\beta} = A^{\alpha\beta}\,,\\
		b = 0\,,\\
		d = -(a^{-1})_{\alpha\beta}D^\beta s^\alpha - \bar g\,,
	\end{dcases}
\end{equation}
where $a = (a^{\alpha\beta})$ and $\dparder{\bar g}{u} = G$.

\subsection*{The damped vibrating membrane}

We are going to apply this method of finding Lagrangian functions to the partial differential equation
\begin{equation}\label{eq:damped-vibrating-membrane}
	u_{tt} - c^2(u_{xx} + u_{yy}) + \gamma u_t = 0\,,
\end{equation}
which models a vibrating membrane with damping. In this case,
$$ A^{\alpha\beta} = \begin{pmatrix}
	1 & 0 & 0 \\
	0 & -c^2 & 0 \\
	0 & 0 & -c^2
\end{pmatrix}\ ,\quad D^\alpha = \begin{pmatrix}
	\gamma \\ 0 \\ 0
\end{pmatrix}\ ,\quad G = 0\,. $$
Hence, according to relations \eqref{eq:inverse-problem-conditions}, we have
$$ a^{\alpha\beta} = \begin{pmatrix}
	1 & 0 & 0 \\
	0 & -c^2 & 0 \\
	0 & 0 & -c^2
\end{pmatrix}\ ,\quad b = 0\ ,\quad d = -\gamma s^t\,. $$
Then a Lagrangian that gives equation \eqref{eq:damped-vibrating-membrane} is
$$ L = \frac{1}{2}u_t^2 - \frac{c^2}{2}(u_x^2 + u_y^2) - \gamma s^t\,, $$
for which
$$ \eta^t = \d s^t - u_t\d u\ ,\quad \eta^x = \d s^x + c^2 u_x\d u\ ,\quad \eta^y = \d s^y + c^2 u_y\d u\,. $$

\subsubsection*{Symmetries}

Notice that the vector field
$$ Y = \parder{}{u} $$
is a 2-contact Lagrangian symmetry. It induces a map $F = (F^t,F^x,F^y)$, given by
$$ F^t = -i(Y)\eta^t = u_t\ ,\quad F^x = -i(Y)\eta^x = -c^2u_x\ ,\quad F^y = -i(Y)\eta^y = -c^2u_y\,, $$
which satisfies the dissipation law for $k$-vector fields \eqref{eq:k-contact-Lagrangian-dissipation-law-fields}.

\section{A vibrating string: Lorentz-like forces versus dissipation forces}
\label{sec:vibrating-string-Lorentz-vs-dissipation}

In the Euler--Lagrange equations \eqref{eq:k-symp-Euler-Lagrange-field-equations} arising from $k$-symplectic systems we may found terms linear in the velocities. Nevertheless, these terms have a specific form, arising from the coefficients of a closed 2-form in the configuration manifold. The most characteristic example of this is the force of a magnetic field acting on a moving charged particle. Such forces do not dissipate energy. On the other hand, other kinds of linear forces in the velocities, such as damping forces, do dissipate energy.

In this example we are going to illustrate the difference betweeen the equations arising from magnetic-like terms in the Lagrangian and the equations given by a $k$-contact formulation of a linear damping. We will analyze the following academic example.

Consider an infinite vertical string aligned with the $z$-axis. Each point of the string can vibrate horizontally. Hence, the independent variables are $(t,z)\in\R^2$, the time and the vertical coordinates, and the phase bundle in the manifold $\oplus^2\T\R^2$, endowed with coordinates $(x, y, x_t, x_z, y_t, y_z)$. We will suppose that the string is nonconducting but it is charged with linear density charge $\lambda$. Now, inspired by the Lagrangian formulation of the Lorentz force, we define the Lagrangian
$$ L = \frac{1}{2}\rho(x_t^2 + y_t^2) - \frac{1}{2}\tau(x_z^2 + y_z^2) - \lambda(\phi - A_1 x_t - A_2 y_t)\,, $$
where $A_1(x,y),A_2(x,y)$ and $\phi(x,y)$ are fixed functions. The Euler--Lagrange equations \eqref{eq:k-symp-Euler-Lagrange-field-equations} for the Lagrangian $L$ are
\begin{equation}\label{eq:euler-lagrange-vibrating-string-lorentz}
	\begin{dcases}
		\rho x_{tt} - \tau x_{zz} = -\lambda\left(\parder{A_2}{x} - \parder{A_1}{y}\right) y_t + \lambda \parder{\phi}{x}\,,\\
		\rho y_{tt} - \tau y_{zz} = \lambda\left(\parder{A_2}{x} - \parder{A_1}{y}\right) x_t + \lambda \parder{\phi}{y}\,.
	\end{dcases}
\end{equation}
Notice that the left-hand side of the equations is the string equation with two vibration modes in the plane XY, while in the right-hand side there is an electromagnetic-like term.

Consider now the 2-contact phase bundle $\oplus^2\T\R^2\times\R^2$, endowed with canonical coordinates $(x, y, x_t, x_z, y_t, y_z, s^t, s^z)$ and modify the Lagrangian $L$ by adding a simple dissipation term,
$$ \L = L + \gamma s^t = \frac{1}{2}\rho(x_t^2 + y_t^2) - \frac{1}{2}\tau(x_z^2 + y_z^2) - \lambda(\phi - A_1 x_t - A_2 y_t) + \gamma s^t\,. $$
This Lagrangian $\L$ induces the 2-contact structure
\begin{align*}
	\eta^t &= \d s^t - (\rho x_t + \lambda A_1)\d x - (\rho y_t + \lambda A_2)\d y\,,\\
	\eta^z &= \d s^z + \tau x_z\d x + \tau y_z \d y\,.
\end{align*}
Hence, the 2-contact Euler--Lagrange equations \eqref{eq:k-contact-Euler-Lagrange-section-coordinates} give
\begin{equation}\label{eq:euler-lagrange-vibrating-string-lorentz-contact}
	\begin{dcases}
		\rho x_{tt} - \tau x_{zz} = -\lambda\left(\parder{A_2}{x} - \parder{A_1}{y}\right) y_t + \lambda \parder{\phi}{x} + \gamma\rho x_t + \gamma\lambda A_1\,,\\
		\rho y_{tt} - \tau y_{zz} = \lambda\left(\parder{A_2}{x} - \parder{A_1}{y}\right) x_t + \lambda \parder{\phi}{y} + \gamma\rho y_t + \gamma\lambda A_2\,.
	\end{dcases}
\end{equation}
When we compare equations \eqref{eq:euler-lagrange-vibrating-string-lorentz} and \eqref{eq:euler-lagrange-vibrating-string-lorentz-contact}, we see that the contact dissipation produces two additional terms: a dissipation force proportional to the velocity and an extra term proportional to $(A_1,A_2)$. This last term appears because the 2-contact Euler--Lagrange equations are not linear with respect to the Lagrangian.

\subsubsection*{Symmetries}

The 2-contact Lagrangian system considered has the infinitesimal Lagrangian 2-contact symmetry
$$ Y = \parder{A_2}{x}\parder{}{x} + \parder{A_1}{y}\parder{}{y}\,. $$
This symmetry induces the map $F = (F^t, F^z)$, given by
\begin{align*}
	F^t &= -i(Y)\eta^t = \rho x_t\parder{A_2}{x} + \lambda\parder{A_2}{x}A_1 + \rho y_t\parder{A_1}{y} + \lambda\parder{A_1}{y}A_2\,,\\
	F^z &= -i(Y)\eta^z = -\tau x_z\parder{A_2}{x} - \tau y_z\parder{A_1}{y}\,.
\end{align*}
which satisfies the dissipation law for 2-vector fields \eqref{eq:k-contact-Lagrangian-dissipation-law-fields}.

\section[Klein--Gordon and the telegrapher's equation]{Klein--Gordon equation with dissipation and the telegrapher's equation}
\label{sec:Klein-Gordon}

The current and voltage on a uniform electrical transmission line is described by the so-called {\it telegrapher's equation} \cite[p.\,306]{Hay2018}\cite[p.\,653]{Sal2015}:
\begin{equation*}
    \begin{dcases}
        \parder{V}{x} = -L\parder{I}{t} - RI\,,\\
        \parder{I}{x} = -C\parder{V}{t} - GV\,.
    \end{dcases}
\end{equation*}
This system can be uncoupled, obtaining the system
\begin{equation*}
    \begin{dcases}
        \parder{^2V}{x^2} = LC\parder{^2V}{t^2} + (LG + RC)\parder{V}{t} + RGV\,,\\
        \parder{^2I}{x^2} = LC\parder{^2I}{t^2} + (LG + RC)\parder{I}{t} + RGI\,.\\
    \end{dcases}
\end{equation*}
Notice that the two equations in the system above are identical, and also known as telegrapher's equations. Both of them can be written as
\begin{equation}\label{eq:telegraph-equation}
	\square u + \gamma\parder{u}{t} + m^2 u = 0\,,
\end{equation}
where $\square$ is the d'Alembertian operator in 1+1 dimensions, and $\gamma$ and $m^2$ are adequate constants. Written this way, we can see the telegrapher's equation as a modified Klein--Gordon equation. We will show that the telegrapher's equation \eqref{eq:telegraph-equation} can be obtained by adding a standard dissipative term to the Klein--Gordon Lagrangian and treating it as a 4-contact Lagrangian.

\subsection*{The Klein--Gordon equation}

The Klein--Gordon equation \cite{Itz1980} is one of the most relevant equations in field theory, either classical or quantum. It can be written as
\begin{equation}\label{eq:klein-gordon}
	(\square + m^2) \phi = 0 \,,
\end{equation}
where $\phi$ is a scalar field in the Minkowski space and $m^2$ is a constant. The Klein--Gordon equation \eqref{eq:klein-gordon} can be derived from the Lagrangian function
\begin{equation}\label{eq:klein-gordon-Lagrangian-0}
	L = \frac{1}{2} (\partial\phi)^2 - \frac{1}{2} m^2 \phi^2 \,.
\end{equation}
Although this Lagrangian can be generalized to include a potential, $L = \frac{1}{2}(\partial\phi)^2 - V(\phi)$, we will stick ourselves to the simple case \eqref{eq:klein-gordon-Lagrangian-0}.

The Lagrangian \eqref{eq:klein-gordon-Lagrangian-0} is autonomous and the space-time is the Minkowski space $\R^4$. Then, it can be described as a 4-symplectic field theory. We will take space-time coordinates $(x^0,x^1,x^2,x^3)$, consider $q$ the field variable and $v_i = \tparder{q}{x^i}$ its corresponding velocities. Hence, the Lagrangian $L\colon\oplus^4\T\R\to\R$ given in \eqref{eq:klein-gordon-Lagrangian-0} is
\begin{equation} \label{eq:klein-gordon-Lagrangian}
	L(q,v_0,v_1,v_2,v_3) = \frac{1}{2}\left(v_0^2 - v_1^2 - v_2^2 - v_3^2\right) - \frac{1}{2} m^2 q^2 
\end{equation}
and the Klein--Gordon equation is
$$
	\parder{^2\phi}{(x^0)^2} - \parder{^2\phi}{(x^1)^2} - \parder{^2\phi}{(x^2)^2} - \parder{^2\phi}{(x^4)^2} + m^2\phi = 0 \,. 
$$

\subsection*{From the Klein--Gordon to the telegrapher's equation}

In order to contactify the Klein--Gordon equation, we consider the Lagrangian $\L\colon\oplus^4\T\R\times\R^4\to\R$ given by
\begin{equation}\label{eq:klein-gordon-Lagrangian-contact}
	\L(q, v_\alpha, s^\alpha) = L(q, v_\alpha) + \gamma_\mu s^\mu = \frac{1}{2} \left(v_0^2 - v_1^2 - v_2^2 - v_3^2\right) - \frac{1}{2} m^2 q^2 + \gamma_\mu s^\mu \,,
\end{equation}
which is defined in the 4-contact manifold $\oplus^4\T\R\times\R^4$, $L$ is the Klein--Gordon Lagrangian \eqref{eq:klein-gordon-Lagrangian} and $\gamma = (\gamma_\mu)\in\R^4$ is constant.

We are going to describe the Skinner--Rusk formalism for this 4-contact Lagrangian. Consider the extended Pontryagin bundle $\W = \oplus^4\T\R\times_{\R}\oplus^4\cT\R\times\R^4$ endowed with natural coordinates $(q, v_0, v_1, v_2,v_3,p^0,p^1,p^2,p^3,s^0,s^1,s^2,s^3)$. The coupling function of $\W$ is
$$ \C = p^0v_0 + p^1v_1 + p^2v_2 + p^3v_3\,. $$
The bundle $\W$ has the canonical forms
\begin{align*}
	\Theta^0 &= p^0\d q\ , \qquad \eta^0 = \d s^0 - p^0\d q\ ,\qquad \Omega^0 = -\d\Theta^0 = \d q\wedge\d p^0 = \d\eta^0\,,\\
	\Theta^1 &= p^1\d q\ , \qquad \eta^1 = \d s^1 - p^1\d q\ ,\qquad \Omega^1 = -\d\Theta^1 = \d q\wedge\d p^1 = \d\eta^1\,,\\
	\Theta^2 &= p^2\d q\ , \qquad \eta^2 = \d s^2 - p^2\d q\ ,\qquad \Omega^2 = -\d\Theta^2 = \d q\wedge\d p^2 = \d\eta^2\,,\\
	\Theta^3 &= p^3\d q\ , \qquad \eta^3 = \d s^3 - p^3\d q\ ,\qquad \Omega^3 = -\d\Theta^3 = \d q\wedge\d p^3 = \d\eta^3\,. 
\end{align*}
The vector fields $\Reeb_\alpha = \tparder{}{s^\alpha}$ are Reeb vector fields in $\W$. The 4-contact Lagrangian function \eqref{eq:klein-gordon-Lagrangian-contact} allows us to construct the Hamiltonian function
$$
	\H = \C - \L = p^\alpha v_\alpha - \frac{1}{2} \left(v_0^2 - v_1^2 - v_2^2 - v_3^2\right) + \frac{1}{2} m^2 q^2 - \gamma_\mu s^\mu\,,
$$
with
$$ \d\H = m^2 q\d q + (p^0-v_0)\d v_0 + (p^1 + v_1)\d v_1 + (p^2 + v_2)\d v_2 + (p^3 + v_3)\d v_3 + v_\alpha\d p^\alpha - \gamma_\mu \d s^\mu\,. $$

To solve the Lagrangian--Hamiltonian problem for the 4-precontact Hamiltonian system $(\W,\eta^\alpha,\H)$ consists in finding a 4-vector field $\bfZ = (Z_0,Z_1,Z_2,Z_3)\in\X^4(\W)$ solution to equations \eqref{eq:k-contact-fields-SR}. Consider a 4-vector field $\bfZ = (Z_0,Z_1,Z_2,Z_3)$ in $\W$ with local expression
$$ 
	Z_\alpha = f_\alpha\parder{}{q} + F_{\alpha\beta}\parder{}{v_\beta} + G_\alpha^\beta\parder{}{p^\beta} + g_\alpha^\beta\parder{}{s^\beta}\,. 
$$
We have that
\begin{equation}\label{eq:KG-left-hand-side}
	i(Z_\alpha)\d\eta^\alpha = f_\alpha\d p^\alpha - G_\alpha^\alpha\d q\,.
\end{equation}
and that
\begin{multline}\label{eq:KG-right-hand-side}
	\d\H - \Reeb_\alpha(\H)\eta^\alpha = (p^0 - v_0)\d v_0 + (p^1 + v_1)\d v_1 \\+ (p^2 + v_2)\d v_2 + (p^3 + v_3)\d v_3 + v_\alpha\d p^\alpha + (m^2q - \gamma_\mu p^\mu)\d q\,,
\end{multline}
Equating \eqref{eq:KG-left-hand-side} and \eqref{eq:KG-right-hand-side}, we obtain the conditions
\begin{align}
	G_\alpha^\alpha &= -m^2q + \gamma_\mu p^\mu & & \mbox{(coefficients in $\d q$)}\,,\label{eq:G-kg}\\
	p^0 &= v_0 & & \mbox{(coefficients in $\d v_0$)}\,,\label{eq:0-kg}\\
	p^1 &= -v_1 & & \mbox{(coefficients in $\d v_1$)}\,,\label{eq:1-kg}\\
	p^2 &= -v_2 & & \mbox{(coefficients in $\d v_2$)}\,,\label{eq:2-kg}\\
	p^3 &= -v_3 & & \mbox{(coefficients in $\d v_3$)}\,,\label{eq:3-kg}\\
	f_\alpha &= v_\alpha & & \mbox{(coefficients in $\d p^\alpha$)}\,,\label{eq:f-kg}
\end{align}
and the second condition in \eqref{eq:k-contact-fields-SR} yields
$$ g_\alpha^\alpha = \L\,. $$

Notice that condition \eqref{eq:f-kg} is the {\sc sopde} condition for the 4-vector field $\bfZ$, which is recovered from the Skinner--Rusk formalism as usual. In addition, we have obtained the constraints
\begin{align*}
	\xi_0 &= p^0 - v_0 = 0\quad ,\quad \xi_1 = p^1 + v_1 = 0\,,\\
	\xi_2 &= p^2 + v_2 = 0\quad ,\quad \xi_3 = p^3 + v_3 = 0\,,
\end{align*}
defining the submanifold $\W_1\hookrightarrow\W$. Imposing the tangency condition of $\bfZ$ to this submanifold $\W_1$, we get the relations
\begin{align*}
	0 = Z_\alpha(\xi_0) = G_\alpha^0 - F_{\alpha 0}\quad &,\quad
	0 = Z_\alpha(\xi_1) = G_\alpha^1 + F_{\alpha 1}\,,\\
	0 = Z_\alpha(\xi_2) = G_\alpha^2 + F_{\alpha 2}\quad &,\quad
	0 = Z_\alpha(\xi_3) = G_\alpha^3 + F_{\alpha 3}\,.
\end{align*}
These conditions partially determine some of the arbitrary functions and no new constraints appear. 
Hence, the constraint algorithm finishes with the submanifold $\W_f = \W_1$ 
and gives the solutions 
$\mathbf{Z} = (Z_0,Z_1,Z_2,Z_3)$, where
\begin{align*}
	Z_0 =& \ v_0\parder{}{q} + \left(-m^2 q + \gamma_\mu p^\mu - G_1^1 - G_2^2 - G_3^3\right)\parder{}{v_0} - G_0^1\parder{}{v_1} - G_0^2\parder{}{v_2} + G_0^3\parder{}{v_3}\\
	& + \left(-m^2 q + \gamma_\mu p^\mu - G_1^1 - G_2^2 - G_3^3\right)\parder{}{p^0} + G_0^1\parder{}{p^1} + G_0^2\parder{}{p^2} + G_0^3\parder{}{p^3} \\
	& + \left(\L - g_1^1 - g_2^2 - g_3^3\right)\parder{}{s^0} + g_0^1\parder{}{s^1} + g_0^2\parder{}{s^2} + g_0^3\parder{}{s^3}\,,\\
	Z_1 =& \ v_1\parder{}{q} + G_1^0\parder{}{v_0} - G_1^1\parder{}{v_1} - G_1^2\parder{}{v_2} - G_1^3\parder{}{v_3} + G_1^0\parder{}{p^0} + G_1^1\parder{}{p^1}\\
	&\quad + G_1^2\parder{}{p^2} + G_1^3\parder{}{p^3} + g_1^0\parder{}{s^0} + g_1^1\parder{}{s^1} + g_1^2\parder{}{s^2} + g_1^3\parder{}{s^3}\,,\\
	Z_2 =& \ v_2\parder{}{q} + G_2^0\parder{}{v_0} - G_2^1\parder{}{v_1} - G_2^2\parder{}{v_2} - G_2^3\parder{}{v_3} + G_2^0\parder{}{p^0} + G_2^1\parder{}{p^1}\\
	&\quad + G_2^2\parder{}{p^2} + G_2^3\parder{}{p^3} + g_2^0\parder{}{s^0} + g_2^1\parder{}{s^1} + g_2^2\parder{}{s^2} + g_2^3\parder{}{s^3}\,,\\
	Z_3 =& \ v_3\parder{}{q} + G_3^0\parder{}{v_0} - G_3^1\parder{}{v_1} - G_3^2\parder{}{v_2} - G_3^3\parder{}{v_3} + G_3^0\parder{}{p^0} + G_3^1\parder{}{p^1}\\
	&\quad + G_3^2\parder{}{p^2} + G_3^3\parder{}{p^3} + g_3^0\parder{}{s^0} + g_3^1\parder{}{s^1} + g_3^2\parder{}{s^2} + g_3^3\parder{}{s^3}\,,
\end{align*}
where $G_\alpha^\beta,g_\alpha^\beta$, for $(\alpha,\beta)\in\left(\{0,1,2,3\}\times\{0,1,2,3\}\right)\setminus\{(0,0)\}$, are arbitrary functions.

Now we can project onto each factor of the manifold $\W$ using the projections $\rho_1,\rho_2$ to recover the Lagrangian and Hamiltonian formalisms. In the Lagrangian formalism we obtain the holonomic 4-vector field $\mathbf{X} = (X_0,X_1,X_2,X_3)$ given by
\begin{align*}
	X_0 =& \ v_0\parder{}{q} + \left( -m^2 q + \gamma_0v_0 - \gamma_1v_1 - \gamma_2v_2 - \gamma_3v_3 + F_1^1 + F_2^2 + F_3^3 \right)\parder{}{v_0}  \\
	&+ F_{01}\parder{}{v_1} + F_{02}\parder{}{v_2} + F_{03}\parder{}{v_3}
	 + \left( \L - g_1^1 - g_2^2 - g_3^3 \right)\parder{}{s^0} + g_0^1\parder{}{s^1} + g_0^2\parder{}{s^2} + g_0^3\parder{}{s^3}\,,\\
	X_1 =& \ v_1\parder{}{q} + F_{10}\parder{}{v_0} + F_{11}\parder{}{v_1} + F_{12}\parder{}{v_2} + F_{13}\parder{}{v_3} + g_1^0\parder{}{s^0} + g_1^1\parder{}{s^1} + g_1^2\parder{}{s^2} + g_1^3\parder{}{s^3}\,,\\
	X_2 =& \ v_2\parder{}{q} + F_{20}\parder{}{v_0} + F_{21}\parder{}{v_1} + F_{22}\parder{}{v_2} + F_{23}\parder{}{v_3} + g_2^0\parder{}{s^0} + g_2^1\parder{}{s^1} + g_2^2\parder{}{s^2} + g_2^3\parder{}{s^3}\,,\\
	X_3 =& \ v_3\parder{}{q} + F_{30}\parder{}{v_0} + F_{31}\parder{}{v_1} + F_{32}\parder{}{v_2} + F_{33}\parder{}{v_3} + g_3^0\parder{}{s^0} + g_3^1\parder{}{s^1} + g_3^2\parder{}{s^2} + g_3^3\parder{}{s^3}\,,
\end{align*}
where $F_\alpha^\beta,g_\alpha^\beta$ for $(\alpha,\beta)\in\left(\{0,1,2,3\}\times\{0,1,2,3\}\right)\setminus\{(0,0)\}$, are arbitrary functions. In the Hamiltonian counterpart, we get the Hamiltonian 4-vector field $\mathbf{Y} = (Y_0,Y_1,Y_2,Y_3)$ given by
\begin{align*}
	Y_0 =& \ v_0\parder{}{q} + \left(-m^2q + \gamma_\mu p^\mu - G_1^1 - G_2^2 - G_3^3\right)\parder{}{p^0} + G_0^1\parder{}{p^1} + G_0^2\parder{}{p^2} + G_0^3\parder{}{p^3} \\
	& + \left(\L - g_2^2 - g_3^3 - g_4^4\right)\parder{}{s^0} + g_0^1\parder{}{s^1} + g_0^2\parder{}{s^2} + g_0^3\parder{}{s^3}\,,\\
	Y_1 =& \ v_1\parder{}{q} + G_1^0\parder{}{p^0} + G_1^1\parder{}{p^1} + G_1^2\parder{}{p^2} + G_1^3\parder{}{p^3} + g_1^0\parder{}{s^0} + g_1^1\parder{}{s^1} + g_1^2\parder{}{s^2} + g_1^3\parder{}{s^3}\,,\\
	Y_2 =& \ v_2\parder{}{q} + G_2^0\parder{}{p^0} + G_2^1\parder{}{p^1} + G_2^2\parder{}{p^2} + G_2^3\parder{}{p^3} + g_2^0\parder{}{s^0} + g_2^1\parder{}{s^1} + g_2^2\parder{}{s^2} + g_2^3\parder{}{s^3}\,,\\
	Y_3 =& \ v_3\parder{}{q} + G_3^0\parder{}{p^0} + G_3^1\parder{}{p^1} + G_3^2\parder{}{p^2} + G_3^3\parder{}{p^3} + g_3^0\parder{}{s^0} + g_3^1\parder{}{s^1} + g_3^2\parder{}{s^2} + g_3^3\parder{}{s^3}\,,
\end{align*}
where the functions $G_\alpha^\beta,g_\alpha^\beta$ with $(\alpha,\beta)\in\left(\{0,1,2,3\}\times\{0,1,2,3\}\right)\setminus\{(0,0)\}$ are arbitrary.

Notice that conditions \eqref{eq:G-kg}, \eqref{eq:0-kg}, \eqref{eq:1-kg}, \eqref{eq:2-kg}, \eqref{eq:3-kg} and \eqref{eq:f-kg} lead to
$$ \left( \square + m^2 - \gamma_0\parder{}{x^0} + \gamma_1\parder{}{x^1} + \gamma_2\parder{}{x^2} + \gamma_3\parder{}{x^3}\right)\phi = 0\,, $$
which represents a ``damped'' Klein--Gordon equation. Clearly, taking $\gamma_\mu = 0$, we recover the Klein--Gordon equation \eqref{eq:klein-gordon}. An important particular case arises when taking $\gamma_\mu = (-\gamma, 0, 0, 0)$. In this case, the telegrapher's equation
$$ \square\phi + \gamma \parder{\phi}{x^0} + m^2\phi = 0 $$
happens to be a particular case of the ``damped'' Klein--Gordon equation.

\section{Maxwell's equations with dissipation}
\label{sec:Maxwell-dissipation}

The behaviour of the electromagnetic field in vacuum is described by Maxwell's equations \cite[p.\,2]{Jac1999}:
\begin{align}
	& \nabla\cdot E = \frac{\rho}{\epsilon_0}\label{maxwell-1}\,,\\
	& \nabla\cdot B = 0\label{maxwell-2}\,,\\
	& \nabla\times E = -\parder{B}{t}\label{maxwell-3}\,,\\
	& \nabla\times B = \mu_0 J + \mu_0\epsilon_0\parder{E}{t}\label{maxwell-4}\,,
\end{align}
where $E$ is the electric field, $B$ is the magnetic field, $\rho$ is the charge density, $J$ is the current density, $\epsilon_0$ is the permitivity of free space, $\mu_0$ is the permeability of free space and the speed of light $c = \frac{1}{\sqrt{\epsilon_0\mu_0}}$.

It is well known that we can rewrite Maxwell's equations in the Minkowski space $\mathbb{M}$ equipped with the Minkowski metric $g_{\mu\nu}$,
$$ g_{\mu\nu} = \begin{pmatrix}
1 & 0 & 0 & 0\\
0 & -1 & 0 & 0\\
0 & 0 & -1 & 0\\
0 & 0 & 0 & -1 \end{pmatrix}\,, $$
by defining the electromagnetic tensor $F_{\mu\nu}$ given by
$$ F_{\mu\nu} = \parder{A_\nu}{x^\mu} - \parder{A_\mu}{x^\nu} = \partial_\mu A_\nu - \partial_\nu A_\mu = A_{\nu,\,\mu} - A_{\mu,\,\nu}\,, $$
where $A^\mu = \left(\dfrac{\phi}{c},A_1,A_2,A_3\right)$ is the electromagnetic 4-potential.
The electromagnetic tensor $F_{\mu\nu}$ can be written in matrix form as
$$ F_{\mu\nu} = \begin{pmatrix}
0 & E_x/c & E_y/c & E_z/c\\
-E_x/c & 0 & -B_z & B_y\\
-E_y/c & B_z & 0 & -B_x\\
-E_z/c & -B_y & B_x & 0 \end{pmatrix}\,. $$
We can also define de current 4-vector as $\mathcal{J}^\mu = (c\rho,J)$. With these objects, the first pair of Maxwell's equations \eqref{maxwell-1} and \eqref{maxwell-4} are written as
\begin{equation}\label{maxwell-1st-pair}
	\partial_\mu F^{\mu\nu} = \mu_0\mathcal{J}^\mu\,,
\end{equation}
while the second pair of Maxwell's equations \eqref{maxwell-2} and \eqref{maxwell-3} become
\begin{equation}\label{maxwell-2nd-pair}
	\partial_\alpha F_{\mu\nu} + \partial_\mu F_{\nu\alpha} + \partial_\nu F_{\alpha\mu} = 0\,,
\end{equation}
also known as Bianchi identity. Equations \eqref{maxwell-2nd-pair} are a direct consequence of the definition of $F_{\mu\nu}$, while the first pair of Maxwell's equations \eqref{maxwell-1st-pair} can be obtained as the Euler--Lagrange equations for the Lagrangian
\begin{equation*}
	L = -\frac{1}{4\mu_0}F_{\mu\nu}F^{\mu\nu} - A_\mu\mathcal{J}^\mu\,.
\end{equation*}
From now on, we are going to consider Maxwell's equations without charges and currents ($\mathcal{J}^\mu = 0$),
\begin{gather*}
	\partial_\mu F^{\mu\nu} = 0\,,\\
	\partial_\alpha F_{\mu\nu} + \partial_\mu F_{\nu\alpha} + \partial_\nu F_{\alpha\mu} = 0\,.
\end{gather*}

\subsection*{Skinner--Rusk formalism}

Now we are going to develop the Skinner--Rusk formalism for the Lagrangian with dissipation \cite{Laz2018,Gas2021b}
\begin{equation}\label{eq:lag-maxwell-dis}
	\L = -\frac{1}{4\mu_0}F_{\mu\nu}F^{\mu\nu} - \gamma_\alpha s^\alpha\,,
\end{equation}
defined on the manifold $\oplus^4\T \R^4\times \R^4$ equipped with coordinates $(A_\mu,A_{\mu,\,\nu}\,; s^\alpha)$, where $\mu,\nu,\alpha = 0,1,2,3$ and $\gamma_\alpha = (\gamma_0, \bm{\gamma})$ is a constant 4-vector.

We begin by considering the extended Pontryagin bundle
$$ \W = \oplus^4\T\R^4\times_{\R^4}\oplus^4\cT\R^4\times\R^4\,, $$
equipped with natural coordinates $(A_\mu, A_{\mu,\,\nu}, P^{\mu,\,\nu}, s^\alpha)$. We have the coupling function
$$ \C = P^{\mu,\,\nu} A_{\mu,\,\nu}\,, $$
the canonical forms
$$
	\Theta^\alpha = P^{\mu,\,\alpha}\d A_\mu\quad ,\quad
	\Omega^\alpha = -\d\Theta^\alpha = \d A_\mu\wedge\d P^{\mu,\,\alpha}\,,
$$
and the contact forms
\begin{align*}
	\eta^\alpha = \d s^\alpha - P^{\mu,\,\alpha}\d A_\mu\,.
\end{align*}
Using the Lagrangian \eqref{eq:lag-maxwell-dis}, we define the Hamiltonian function
$$ \H = \C - \L = P^{\mu,\,\nu} A_{\mu,\,\nu} + \frac{1}{4\mu_0}F_{\mu\nu}F^{\mu\nu} + \gamma_\alpha s^\alpha\,. $$
It is easy to check that the vector fields $\Reeb_\alpha = \dparder{}{s^\alpha}$ are Reeb vector fields of $\W$. 
To solve the Lagrangian--Hamiltonian problem for the 4-precontact system $(\W,\eta^\alpha, \H)$ means to find a 4-vector field $\mathbf{Z} = (Z_0,Z_1,Z_2,Z_3)\in\X^4(\W)$ satisfying equations \eqref{eq:k-contact-fields-SR}. We have that
$$ \d\H - \Reeb_\alpha(\H)\eta^\alpha = \left( P^{\mu,\,\nu} - \frac{1}{\mu_0}F^{\mu\nu} \right)\d A_{\mu,\,\nu} + A_{\mu,\,\nu}\d P^{\mu,\,\nu} - \gamma_\alpha P^{\mu,\,\alpha}\d A_\mu\,. $$
Then, consider a 4-vector field $\mathbf{Z} = (Z_0,Z_1,Z_2,Z_3)$ in $\W$ with local expression
$$ Z_\alpha = (Z_\alpha)_\mu\parder{}{A_\mu} + (Z_\alpha)_{\mu\beta}\parder{}{A_{\mu,\,\beta}} + (Z_\alpha)^{\mu\beta}\parder{}{P^{\mu,\,\beta}} + (Z_\alpha)^\beta\parder{}{s^\beta}\,. $$
For this vector field, we have
\begin{align*}
	i(Z_\alpha)\d\eta^\alpha &= (Z_\alpha)_\mu\d P^{\mu,\,\alpha} - (Z_\alpha)^{\mu,\,\alpha}\d A_\mu\,,\\
	i(Z_\alpha)\eta^\alpha &= (Z_\alpha)^\alpha - P^{\mu,\,\alpha}(Z_\alpha)_\mu\,,
\end{align*}
and thus the first equation in \eqref{eq:k-contact-fields-SR} gives the conditions
\begin{align}
	(Z_\alpha)^{\mu\alpha} &= -\gamma_\alpha P^{\mu,\,\alpha} & & \mbox{(coefficients in $\d A_\mu$)}\,,\label{max-cond-1}\\
	P^{\mu,\,\nu} &= \frac{1}{\mu_0}F^{\mu\nu} & & \mbox{(coefficients in $\d A_{\mu,\,\nu}$)}\,,\label{max-cond-2}\\
	A_{\mu,\,\alpha} &= (Z_\alpha)_\mu & & \mbox{(coefficients in $\d P_{\mu,\,\alpha}$)}\,.\label{max-cond-3}
\end{align}
Furthermore, the second equation in \eqref{eq:k-contact-fields-SR} gives
$$ (Z_\alpha)^\alpha = P^{\mu,\,\alpha}\left( (Z_\alpha)_\mu - A_{\mu,\,\alpha} \right) + \L\,, $$
and hence, using \eqref{max-cond-3},
$$ (Z_\alpha)^\alpha = \L\,. $$
We have obtained the constraint functions
$$ \xi^{\mu\nu} = P^{\mu,\,\nu} - \frac{1}{\mu_0}F^{\mu\nu}\,, $$
defining a submanifold $\W_1\hookrightarrow\W$. Now we have to impose the tangecy of the 4-vector field $\mathbf{Z}$ to this submanifold $\W_1$:
\begin{align*}
	0 &= Z_\alpha(\xi^{\mu\nu})
	= Z_\alpha\left(P^{\mu,\,\nu} - \frac{1}{\mu_0}F^{\mu\nu}\right)
	= (Z_\alpha)^{\mu\nu} - \frac{1}{\mu_0}\parder{F^{\mu\nu}}{A_{\tau\beta}}(Z_\alpha)_{\tau\beta}\\
	&= (Z_\alpha)^{\mu\nu} - \frac{1}{\mu_0}\left( g^{\mu\tau}g^{\nu\beta} - g^{\mu\beta}g^{\nu\tau} \right)(Z_\alpha)_{\tau\beta}
\end{align*}
which partially determine some of the coefficients of the 4-vector field $\mathbf{Z}$. Notice that no new constraints appear and hence the constraint algorithm ends with the submanifold $\W_f = \W_1$ and gives the solutions $\mathbf{Z} = (Z_0,Z_1,Z_2,Z_3)$, where
\begin{align*}
	Z_\alpha &= A_{\mu,\,\alpha}\parder{}{A_\mu} + (Z_\alpha)_{\mu\nu}\parder{}{A_{\mu,\,\nu}} + (Z_\alpha)^{\mu\nu}\parder{}{P^{\mu,\,\nu}} + (Z_\alpha)^\beta\parder{}{s^\beta}\,,
\end{align*}
satisfying the conditions
\begin{equation*}
	\begin{dcases}
		(Z_\alpha)^\alpha = \L\,,\\
		(Z_\alpha)^{\mu\alpha} = -\gamma_\alpha P^{\mu,\,\alpha}\,,\\
		(Z_\alpha)^{\mu\nu} = \frac{1}{\mu_0}\left( g^{\mu\tau}g^{\nu\beta} - g^{\mu\beta}g^{\nu\tau} \right)(Z_\alpha)_{\tau\beta}\,.
	\end{dcases}
\end{equation*}

\subsection*{4-contact Maxwell equations and damped electromagnetic waves}

Notice that, combining equations \eqref{max-cond-1} and \eqref{max-cond-2}, we obtain
$$ \partial_\alpha F^{\alpha\mu} = -\gamma_\alpha F^{\alpha\mu}\,, $$
which is the dissipative version of the first pair of Maxwell's equations. Together with the Bianchi identity \eqref{maxwell-2nd-pair}, we can write the 4-contact Maxwell's equations without charges and currents:
\begin{align}
	& \nabla\cdot E = -\bm{\gamma}\cdot E\,\label{maxwell-damped-1}\\
	& \nabla\cdot B = 0\,\label{maxwell-damped-2}\\
	& \nabla\times E = -\parder{B}{t}\,\label{maxwell-damped-3}\\
	& \nabla\times B = \mu_0\epsilon_0\parder{E}{t} - \bm{\gamma}\times B + \frac{\gamma_0}{c}E\,.\label{maxwell-damped-4}
\end{align}

Applying the curl operator to the third and fourth equations \eqref{maxwell-damped-3}, \eqref{maxwell-damped-4}, we get
\begin{align*}
	\mu_0\epsilon_0\parder{^2E}{t^2} - \nabla^2E + \frac{\gamma_0}{c}\parder{E}{t} = -\bm{\gamma}\times(\nabla\times E)\,,\\
	\mu_0\epsilon_0\parder{^2B}{t^2} - \nabla^2B + \frac{\gamma_0}{c}\parder{B}{t} = -\nabla\times(\bm{\gamma}\times B)\,.
\end{align*}
Taking $\gamma_\mu = (\gamma_0,\bm{0})$, we obtain
\begin{align*}
	\parder{^2E}{t^2} - c^2\nabla^2E + c\gamma_0\parder{E}{t} = 0\,,\\
	\parder{^2B}{t^2} - c^2\nabla^2B + c\gamma_0\parder{B}{t} = 0\,,
\end{align*}
which are the 3-dimensional analogues of the damped wave equation \eqref{eq:damped-vibrating-string-edp} studied in Example \ref{sec:damped-vibrating-string}.

\backmatter

\chapter{Conclusions}


In this final chapter we summarize the results obtained in the development of this thesis. We also give a list of the publications derived from this work. Finally, we point out several interesting lines of future research.

\subsection*{Summary of contributions}

The starting point of this work has been the contact formulation of Hamiltonian and Lagrangian mechanical systems \cite{Bra2017a,Bra2017b,DeLeo2019b}. This thesis has been devoted to enlarge the knowledge on these systems and generalize this theory to the case of first-order field theories.

The new results presented in this thesis are the following ones:

\begin{itemize}
	\item We show an almost equivalent alternative form of writing the contact Hamiltonian equations \eqref{eq:contact-hamilton-equations-fields} without using the Reeb vector field (Proposition \ref{prop:equivalent-contact-Hamiltonian-equations}). We give the corresponding version of the contact Lagrangian equations without Reeb vector field (Section \ref{sec:contact-Lagrangian-systems}). This way of writing the dynamical equations of the system might be useful when dealing with singular systems, because in such cases we do not have a uniquely determined Reeb vector field.

	\item We have defined several notions of symmetry of a contact Hamiltonian system and stated some relations between them (Section \ref{sec:dissipated-conserved-quantities-contact-Hamiltonian}). We have introduced the concept of dissipated quantity of a contact Hamiltonian system and proved that every infinitesimal dynamical symmetry has associated a dissipated quantity (Theorem \ref{thm:dissipation-contact}). In particular, we state the energy dissipation theorem for contact Hamiltonian systems (Theorem \ref{thm:energy-dissipation-contact}). We also define the concept of conserved quantity and we have proved that the quotient of two dissipated quantities is a conserved quantity and that the product of a conserved and a dissipated quantity is a new dissipated quantity (Proposition \ref{prop:conserved-dissipated-contact}). We have studied the symmetries of canonical contact Hamiltonian systems (\ref{sec:symmetries-canonical-contact-Hamiltonian-systems}). In particular, we have proved the momentum dissipation theorem \ref{thm:momentum-dissipation-contact}. We have studied the symmetries of contact Lagrangian systems (Section \ref{sec:symmetries-contact-Lagrangian-systems}). In particular, we have proved that if the Lagrangian function does not depend on the position $q^i$, then the vector field $\tparder{}{q^i}$ is an infinitesimal contact symmetry and its associated dissipated quantity is the momentum $\tparder{\L}{v^i}$ (Theorem \ref{thm:contact-Lagrangian-not-depending-position}). We have also compared the symmetries of a Hamiltonian system on a symplectic manifold and its corresponding contactified system (Section \ref{sec:symmetries-contactified-system}).

	\item Chapter \ref{ch:Skinner-Rusk-contact} generalizes the Skinner--Rusk formalism \cite{Ski1983} to contact systems. We define the extended Pontryagin bundle and describe its canonical precontact structure (Section \ref{sec:extended-Pontryagin-bundle-contact}) and developed the Skinner--Rusk formalism for contact systems is developed in detail (Section \ref{sec:contact-dynamical-equations}). In particular, we show that the holonomy condition is recovered from the formalism even in the singular case. We also see that the Legendre map arises as a set of constraints. We have proved that we can recover both the Lagrangian and Hamiltonian formalisms from the Skinner--Rusk formalism as usual (Section \ref{sec:SR-contact-recovering}).

	\item In order to deal with singular dissipative field theories, we first studied the case of singular nonautonomous field theories in the $k$-cosymplectic setting (Section \ref{sec:k-precosymplectic-geometry}). In particular, we have defined the notion of $k$-precosymplectic manifold (Definition \ref{dfn:k-precosymplectic-manifold}) and proved the existence of global Reeb vector fields in every $k$-precosymplectic manifold (Proposition \ref{prop:Reeb-k-precosymplectic}). We have described in full detail the constraint algorithm for singular nonautonomous field theories. In particular, we characterize the constraints arising and give an operational way to compute the constraint submanifolds (Section \ref{sec:constraint-algorithm-k-precosymplectic}).

	\item We have introduced the notion of $k$-contact manifold (Definition \ref{dfn:k-contact-manifold}). This concept is a generalization of the notion of contact manifold and $k$-symplectic structure. We have shown that the Reeb distribution of a $k$-contact manifold is involutive, and therefore integrable (Lemma \ref{lem:Reeb-distribution-involutive}), and proved the existence and uniqueness of a family of Reeb vector fields spanning the Reeb distribution (Theorem \ref{thm:k-contact-Reeb}). We end this section stating the Darboux theorem for $k$-contact manifolds (Theorem~\ref{thm:k-contact-Reeb}). We have presented the Hamiltonian formalism for $k$-contact systems, stated the $k$-contact Hamilton--De Donder--Weyl equations and prove that they have solutions, although they are not unique if $k>1$ (Section \ref{sec:k-contact-Hamiltonian-systems}). We also offer an alternative way of writing Hamilton--De Donder--Weyl equations without making use of the Reeb vector fields (Theorem \ref{thm:equivalent-k-contact-Hamilton-equations-no-Reeb}). We have introduced several notions of symmetries of $k$-contact Hamiltonian systems and proved some of their properties (Section \ref{sec:k-contact-Hamiltonian-symmetries}). We have generalized the concept of dissipated quantity to the notion of dissipation law and present two types dissipation laws (Section \ref{sec:k-contact-Hamiltonian-dissipation-laws}). We have stablished the relation between the two types of dissipation laws (Proposition \ref{prop:relation-dissipation-laws-Hamiltonian}) and proved that every infinitesimal dynamical symmetry has associated a map $F\colon M\to\R^k$ satisfying the dissipation law for $k$-vector fields (Theorem \ref{thm:k-contact-Hamiltonian-dissipation-theorem}).

	\item In order to develop a $k$-contact formalism for Lagrangian field theories, we have described the canonical structures of the bundle $\oplus^k\T Q\times\R^k$ (Section \ref{sec:k-contact-Lagrangian-systems}). We have also shown that, given a regular Lagrangian function $\L\in\Cinfty(\oplus^k\T Q\times\R^k)$, we can define a $k$-contact structure $(\eta^\alpha_\L)$ in the manifold $\oplus^k\T Q\times\R^k$ (Proposition \ref{prop:k-contact-regular-Lagrangian}). Thus, $(\oplus^k\T Q\times\R^k, \eta^\alpha_\L,E_\L)$ is a $k$-contact Hamiltonian system. We have pointed out the differences between regular and singular Lagrangians and described the difficulties that arise when dealing with singular systems. We also summarize the constraint algorithm that allows to find (if it exists) a submanifold where the $k$-contact Lagrangian equations are consistent and have solutions tangent to this submanifold (Section \ref{sec:singular-case-k-precontact}). We have introduced several notions of symmetries of $k$-contact Lagrangian systems and stated some of their properties (Section \ref{sec:k-contact-Lagrangian-symmetries}). We have extended the notion of dissipated quantity of a mechanical system and thus defining two notions of dissipation law and stablished the relation between them (Section \ref{sec:k-contact-Lagrangian-dissipation-laws}). We have also studied the symmetries of the Lagrangian function of a $k$-contact system and, in particular, we have stated the momentum dissipation theorem \ref{prop:k-contact-symmetries-of-the-Lagrangian}.

	\item We have generalized the Skinner--Rusk formalism presented in Chapter \ref{ch:Skinner-Rusk-contact} to the case of field theories. We have defined the extended Pontryagin bundle and describe its canonical $k$-precontact structure (Section \ref{sec:extended-Pontryagin-bundle-k-contact}). We have presented the Skinner--Rusk formalism for $k$-contact systems and applied the constraint algorithm to it. In particular, we showed that the holonomy condition is re\-covered from the formalism even if the Lagrangian function is singular and that the Legendre map arises as a set of constraint functions (Section \ref{sec:k-contact-dynamical-equations}). Finally, we see that we can recover both the Lagrangian and Hamiltonian formalisms from the Skinner--Rusk formalism as usual (Section \ref{sec:SR-k-contact-recovering}).
\end{itemize}

Along the present thesis, several examples have been worked out, including both regular and singular systems in mechanics and field theory. In Chapter \ref{ch:contact-examples} we analyze the following mechanical systems: the damped harmonic oscillator, the motion of a particle in a constant gravitational field with friction, the parachute equation, Lagrangians with holonomic dissipation term, a central force with dissipation, the damped simple pendulum using the Lagrange multipliers method and Cawley's Lagrangian with dissipation. In Section \ref{sec:constraint-algorithm-examples} we apply the $k$-precosymplectic constraint algorithm to systems described by Lagrangian functions which are affine in the velocities and to a singular quadratic Lagrangian. Chapter \ref{ch:examples-k-contact} is devoted to analyze several dissipative field theories: the damped vibrating string, two coupled vibrating strings with damping, Burgers' equation as a contactification of the heat equation, the inverse problem for a type of elliptic and hyperbolic partial differential equations, a comparison between Lorentz-like forces and dissipative forces on a vibrating string, Klein--Gordon and the telegrapher's equation, Maxwell's equations with dissipation and damped electromagnetic waves.

\subsection*{Further research}

There are some lines of future research and open problems derived from this thesis:

\begin{itemize}
	\item To develop a geometric formalism to deal with nonautonomous dissipative mechanical systems. We will need to define some notion of \emph{cocontact} manifold, in the same way as cosymplectic geometry is the natural framework for nonautonomous mechanical systems.
	\item The $k$-contact formalism described in this thesis is useful when dealing with dissipative field theories described by autonomous Lagrangians. It is necessary to develop a \emph{$k$-cocontact} formalism (in the same way as the $k$-cosymplectic formalism generalizes the $k$-symplectic formalism) to work with nonautonomous dissipative field theories.
	\item It would be interesting to generalize the multisymplectic formalism in order to model dissipative field theories, thus developing a \emph{multicontact formalism}. This formalism will have to be extended in order to deal with singular dissipative field theories. In the case of nondissipative field theories, the multisymplectic formalism is of great interest as it has both the $k$-symplectic and the $k$-cosymplectic formalism as particular cases.
	\item The Herglotz variational principle \cite{Her1930,DeLeo2021b} for dissipative mechanical systems could be generalized to a new variational principle yielding the $k$-contact Euler--Lagrange equations.
	\item It would be interesting to find new examples of mechanical systems and field theories described by contact or $k$-contact Lagrangians. In particular, one could study the meaning of adding a dissipative term to the Lagrangian of the relativistic free-particle or to gravitational Lagrangians, such as the Einstein--Palatini or the Hilbert--Einstein Lagrangians (see \cite{Gas2018b} and references therein).
	\item It would be worthwhile to use the contact formalism to deal with dynamical systems not necessarily mechanical, such as population dynamics. In particular, it might be of interest the study of reversible systems from the contact point of view.
\end{itemize}

\subsection*{List of publications}

The publications derived from this work are \cite{DeLeo2020,Gas2020,Gas2019,Gas2021,Gra2020,Gra2021}. In addition, there have been 10 contributions to national and international congresses and workshops, 4 of them being talks and 6 of them being posters.

The list of publications, in chronological order, is the following:
\begin{itemize}
	\item \cite{Gra2020} X. Gràcia, X. Rivas and N. Román-Roy. ``Constraint algorithm for singular field theories in the $k$-cosymplectic framework''. {\it J. Geom. Mech.}, {\bf 12}:1--23, 2020. \url{https://doi.org/10.3934/jgm.2020002}.

	-- Sections \ref{sec:k-precosymplectic-geometry}, \ref{sec:constraint-algorithm-k-precosymplectic} and \ref{sec:constraint-algorithm-examples}.

	\item \cite{Gas2019} J. Gaset, X. Gràcia, M. C. Muñoz-Lecanda, X. Rivas and N. Román-Roy. ``New contributions to the Hamiltonian and Lagrangian contact formalisms for dissipative mechanical systems and their symmetries''. {\it Int. J. Geom. Methods Mod. Phys.}, {\bf 16}(6):2050090, 2020. \url{https://doi.org/10.1142/S0219887820500905}.

	-- Section \ref{sec:contact-Lagrangian-systems} and Chapters \ref{ch:symmetries-contact-systems} and \ref{ch:contact-examples}.

	\item \cite{Gas2020} J. Gaset, X. Gràcia, M. C. Muñoz-Lecanda, X. Rivas and N. Román-Roy. ``A contact geometry framework for field theories with dissipation''. {\it Ann. Phys.}, {\bf 414}:168092, 2020. \url{https://doi.org/10.1016/j.aop.2020.168092}.

	-- Chapters \ref{ch:k-contact-Hamiltonian} and \ref{ch:examples-k-contact}.

	\item \cite{Gas2021} J. Gaset, X. Gràcia, M. C. Muñoz-Lecanda, X. Rivas and N. Román-Roy. ``A $k$-contact Lagrangian formalism for nonconservative field theories''. {\it Rep. Math. Phys.}, {\bf 87}(3):347--368, 2021. \url{https://doi.org/10.1016/S0034-4877(21)00041-0}.

	-- Chapters \ref{ch:k-contact-Lagrangian} and \ref{ch:examples-k-contact}.

	\item \cite{DeLeo2020} M. de León, J. Gaset, M. Lainz-Valcázar, X. Rivas and N. Román-Roy. ``Unified Lagrangian-Hamiltonian formalism for contact systems''. {\it Fortschritte der Phys.}, {\bf 68}(8):2000045, 2020. \url{https://doi.org/10.1002/prop.202000045}.

	-- Chapters \ref{ch:Skinner-Rusk-contact} and \ref{ch:contact-examples}.

	\item \cite{Gra2021} X. Gràcia, X. Rivas and N. Román-Roy. ``Skinner--Rusk formalism for $k$-contact systems'', preprint, 2021. \url{https://arxiv.org/abs/2109.07257}

	-- Chapters \ref{ch:Skinner-Rusk-k-contact} and \ref{ch:examples-k-contact}.

\end{itemize}





\bibliographystyle{abbrv}

\cleardoublepage
\fancyhead[LO]{Bibliography}
\bibliography{bibliografia.bib}



\end{document}